\documentclass{pasa}

\usepackage[dvips]{color}
\usepackage[normalem]{ulem}
\usepackage{hyperref}
\hypersetup{
    colorlinks=false,
    pdfborder={0 0 0},
}

\title[AGB nucleosynthesis and yields]{The Dawes Review 2: Nucleosynthesis and stellar yields of low and
intermediate-mass single stars}
\author[Karakas \& Lattanzio]{Amanda I. Karakas$^{1}$, \and John C. Lattanzio$^{2}$\\
\affil{$^1$Research School of Astronomy \& Astrophysics, Australian National University, Canberra ACT 2611, Australia}%
\affil{$^2$Monash Centre for Astrophysics, School of Mathematical Sciences, Monash University, Clayton VIC 3800, 
Australia}}%
\jid{PASA}
\doi{10.1017/pas.\the\year.xxx}
\jyear{\the\year}

\usepackage[authoryear]{natbib}
\bibpunct{(}{)}{;}{a}{}{,}
\newcommand{\Msun}{{\rm M}_{\odot}}
\newcommand{\Lsun}{{\rm L}_{\odot}}
\newcommand{\iso}[2]{\hbox{${}^{#1}{\rm #2}$}}

\begin{document}%
\begin{abstract}
The chemical evolution of the Universe is governed by the chemical yields 
from stars, which in turn is determined primarily by the initial stellar mass. 
Even stars as low as 0.9$\Msun$ can,  at low metallicity, contribute to the 
chemical evolution of elements.    Stars less massive than about 10$\Msun$
experience recurrent mixing events that can significantly change the 
surface composition of the envelope, with observed enrichments in carbon, 
nitrogen, fluorine, and heavy elements synthesized by the slow neutron 
capture process (the $s$-process). Low and intermediate mass
stars release their nucleosynthesis
products through stellar outflows or winds, in contrast to massive stars that 
explode as core-collapse supernovae. Here we review the stellar evolution
and nucleosynthesis for single stars up to $\sim10\Msun$ from the main
sequence through to the tip of the asymptotic giant branch (AGB). 
We include a discussion of the main uncertainties that affect theoretical
calculations and review the latest observational data, which are 
used to constrain uncertain details of the stellar models. We finish with a 
review of the stellar yields available for stars less massive than about 10$\Msun$
and discuss efforts by various groups to 
address these issues and provide homogeneous yields for low and 
intermediate-mass stars covering a broad range of metallicities.  
\end{abstract}
\maketitle%

\begin{quote}
{\it The Dawes Reviews are substantial reviews of topical areas in astronomy, published by authors of international 
standing at the invitation of the PASA Editorial Board. The reviews recognise William Dawes (1762-1836), second 
lieutenant in the Royal Marines and the astronomer on the First Fleet. Dawes was not only an accomplished astronomer, 
but spoke five languages, had a keen interest in botany, mineralogy, engineering, cartography and music, compiled 
the first Aboriginal-English dictionary, and was an outspoken opponent of slavery.}
\end{quote}

\section{Introduction} \label{sec:intro}

%1.1 Diversity, light, dust etc

Stars with initial masses between about $0.8\Msun$ and 10$\Msun$
dominate the stellar population in our Milky Way Galaxy.  
This mass interval spans a huge range in stellar lifetimes, from the longest lived
low-mass stars, that have existed for as long as our Galaxy ($\approx 1.2
\times 10^{10}$ years) to the most massive of this range, whose lives are
over in the blink of a cosmic eye ($\lesssim 20$ million years). 
These stars are numerous because of the
shape of the initial mass function which peaks at $\approx
1\Msun$. Their importance is underlined by that fact
that they experience a diversity of rich nucleosynthesis, 
making them crucial contributors to the chemical evolution 
of elements in our Universe \citep[e.g.,][]{travaglio01a,romano10,kobayashi11a}. 
When these stars evolve they lose mass through strong stellar outflows or winds and it
has been estimated that they have produced nearly 90\% of the dust injected 
into the interstellar medium (ISM) of our Galaxy, with massive stars 
accounting for the rest \citep{sloan08}.  Furthermore, galaxies dominated by intermediate-age
stellar populations have a significant fraction of their starlight
emitted by low and intermediate mass stars, especially when they evolve off the main sequence 
to the giant branches \citep{mouchine02,maraston05,maraston06,tonini09,melbourne12}.

%1.2 Nucleosynthesis
For low and intermediate-mass stars the most important nucleosynthesis
occurs when the stars reach the giant branches. 
It is during the ascent of the red giant branch (RGB) that the first dredge-up occurs.
This changes the surface composition by mixing to the surface material 
from the interior that has been  exposed
to partial hydrogen (H) burning. It is also on the upper part of the
RGB where {\em extra mixing} processes occur in the envelopes of low-mass giants.
These are processes not included in traditional
calculations which assume convection is the only mixing mechanism present. Such processes
may include meridional circulation, shear mixing, and various hydrodynamic and magnetic 
mixing processes. Empirically we know that something occurs that results in 
further products of H-burning nucleosynthesis 
becoming visible at the surface. 

The more massive stars in our selected mass range will also
experience a second dredge-up, which occurs following core helium (He) exhaustion
as the star begins its ascent of the giant branch for the second time, now 
called the asymptotic giant branch, or AGB.  It is on the AGB where
we expect the largest changes to the surface composition. These are driven by a
complex interplay of nucleosynthesis and mixing.  The
nucleosynthesis is driven by thermal instabilities in the He-burning shell, known as
shell flashes or thermal pulses.
The products of this burning, mostly carbon, may be mixed to the 
stellar surface by recurrent convective mixing episodes. These
mixing episodes can occur after each thermal pulse and are known
as third dredge-up events. 

Thermal pulses are responsible for a large variety of stellar spectral types.
Stars begin their lives with an atmosphere that is oxygen rich, in the sense that the
ratio of the number of carbon to oxygen atoms $n($C$)/n($O$)$ is less than unity. 
Recurring third dredge-up on the AGB can add enough carbon to the envelope
that the star becomes carbon rich with $n($C$)/n($O$)\ge 1$, hence 
becoming a ``carbon star'' (or C star). There are many different types of C
stars including both {\em intrinsic}, meaning that they result from 
internal evolution (as described above, e.g., C(N) stars)
or {\em extrinsic}, where it is mass transfer from a close binary C star that produces  
$n($C$)/n($O$)\ge 1$ in a star that is not sufficiently evolved to have thermal pulses itself 
\citep[e.g., CH stars and dwarf C stars, ][]{wallerstein98}.
It is also the third dredge-up that mixes to the surface the heavy
elements such as barium and lead that are produced by the slow neutron 
capture process (the $s$-process). This can result in S-stars, barium stars and 
technetium-rich stars \citep{wallerstein98}.  Strong stellar winds then expel this enriched
material into the ISM, where it can contribute to the
next generation of star formation.

Intermediate-mass AGB stars may also
experience hot bottom burning (HBB), where the bottom of the 
convective envelope penetrates into the top of the H-burning shell. 
Proton-capture nucleosynthesis occurs at the base of the mixed envelope
\citep{bloecker91,lattanzio92,boothroyd95}. Third dredge up 
operates alongside HBB and this can lead to some interesting
results, such as substantial production of primary nitrogen, together with other 
hydrogen burning products including sodium and aluminium. 

The short lifetime of those AGB stars that experience 
HBB ($\tau \lesssim 100$Myr) has implicated them as potential polluters of
Galactic globular clusters (GCs), which show abundance trends consistent with
hot H burning \citep{gratton04,gratton12,prantzos07}. The ability of detailed models to
match the observed abundance trend depends on highly uncertain assumptions
about the treatment of convection and mass loss in stellar models
\citep[e.g.][]{fenner04,karakas06b,ventura09a}. 

Not so long ago there was a belief that if you were interested
in the chemical evolution of the Galaxy, or indeed the Universe, then all you needed
was yields from core-collapse supernovae (SNe), and perhaps Type~I supernovae \citep[e.g.,][]{timmes95}.
But with an increased understanding of the breadth and depth of nucleosynthesis in AGB stars has 
come clear evidence that the picture is simply incomplete without them.
It has been shown by \citet{kobayashi11a} that AGB models are essential to reproduce the solar
system abundances of carbon, nitrogen, and the neutron-rich 
isotopes of oxygen and neon. Similarly, \citet{renda04} and \citet{kobayashi11b}
showed the importance of AGB stars for fluorine. \citet{fenner04} performed
a similar study for magnesium, highlighting the contribution from intermediate-mass
AGB stars of low metallicity to the chemical evolution of \iso{25}Mg and \iso{26}Mg.
The importance of AGB stars to understanding the composition
anomalies seen in globular clusters is just another reason why they are of 
such interest to contemporary astrophysics.

%1.3 Definitions

\subsection{Definitions and Overview of Evolution}\label{sec:definitions}

We will here be concerned with stars with masses between about $0.8\Msun$ and $10\Msun$.
Stars more massive than this proceed through all nuclear burning phases
and end their lives as core-collapse supernovae.
While these massive stars are relatively rare they inject considerable energy
and nucleosynthesis products into the galaxy per event. For this reason
they are extremely important when considering the evolution of galaxies.
Their remnants are either neutron stars or black holes 
\citep[for the evolution and nucleosynthesis of massive stars we refer the reader to][]{langer12,nomoto13}.

\begin{figure*}
\begin{center}
\includegraphics[width=0.99\textwidth,angle=0]{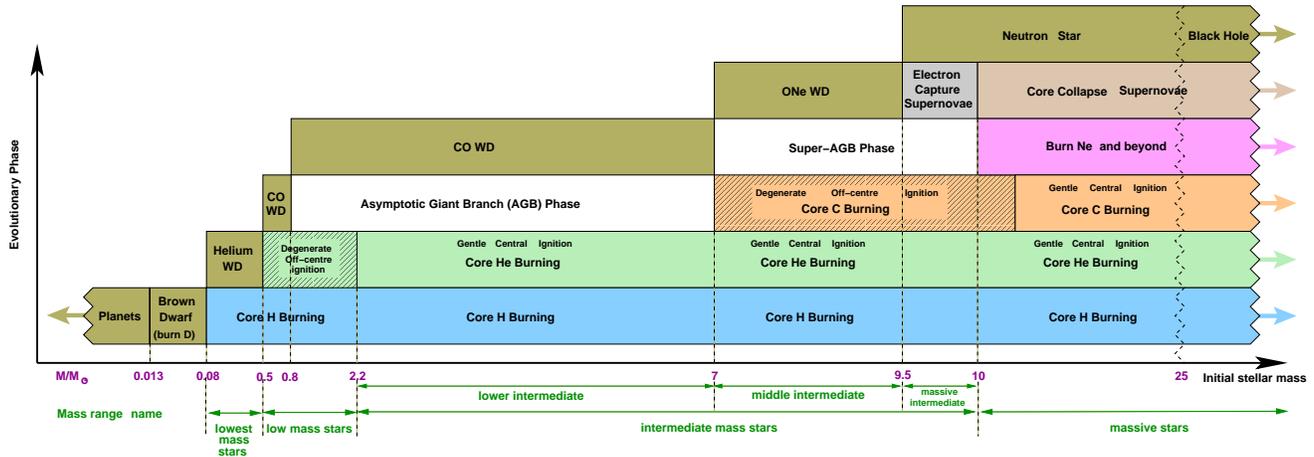}
\caption{Schematic showing how stellar mass determines the main nuclear burning
phases at solar metallicity, as well as the fate of the final remnant.  This defines the 
different mass intervals we will deal with  in this paper. Note that the borders are often 
not well determined theoretically, depending on details such as mass loss and the implementation 
of mixing, for example. This is particularly true for the borders around the region of the 
electron-capture supernovae. Likewise, all numbers are rough estimates, and depend on 
composition in addition to details of the modelling process.}
\label{masses}
\end{center}
\end{figure*}

Stars that will become AGB stars begin their journey with core H and He burning 
(and possibly C burning on the ``super-AGB''; see below),
before they lose their outer envelopes to a stellar wind during the AGB
phase of stellar evolution. It is convenient to define
mass ranges according to the evolutionary behaviour the stars will experience.
The exact numerical values will, of course,  depend on the star's composition
and possibly other effects (such as rotation, which we ignore for now).

The definitions we will use are given below and shown in Figure~\ref{masses} for 
solar metallicity. A reduction in the global stellar metallicity will shift the
borders introduced here to a lower mass (e.g., core C burning will ignite at about 
7$\Msun$ at $Z = 10^{-4}$ instead of about 8$\Msun$ at $Z=Z_{\odot} \approx 0.014$).
We introduce some new nomenclature in this diagram, while maintaining the definitions
of ``low'' and ``intermediate'' mass as given in the existing literature.

\subsubsection{The Lowest Mass Stars}
We define the ``lowest mass stars'' as those that burn H in their core but take part in
no further (significant) nuclear burning processes.

\subsubsection{The Low Mass Stars}

We have defined ``low-mass stars'' to be those with initial masses between about
0.8 and 2$\Msun$ which experience He ignition under degenerate conditions, known
as the core He flash \citep{demarque71,despain81,deupree84,dearborn06,mocak09}.
Stars more massive than this succeed in igniting He gently. These low-mass stars will 
experience core He burning and then all but the least massive of these will go on to the
AGB (without an appreciable second dredge-up),
ending their lives as C-O white dwarfs (WDs, see Figure~\ref{masses}).

\subsubsection{The Intermediate Mass Stars}

We then enter the domain of ``intermediate mass'' stars, a name well
known in the literature. Here we have broken this mass range into three
distinct sub-classes, based on C ignition and their final fate.
We will only use these new names when the sub-divisions are important, 
otherwise we simply call them ``intermediate-mass stars''.

\subsubsection{The Lower Intermediate Mass Stars}

These stars are not sufficiently massive to ignite the C in their core, which is now composed 
primarily of C and O following He burning. We say the star is of ``lower intermediate mass'', 
being about $2$--$7\Msun$. These stars will proceed to the AGB following core He exhaustion, and the
more massive of them will experience the second dredge-up as they begin their ascent of the AGB.
They will end their lives as C-O white dwarfs.

\subsubsection{The Middle Intermediate Mass Stars}

At slightly higher masses we find C ignites (off centre)
under degenerate conditions. We have defined these stars as ``middle intermediate mass stars''.
These stars go on to experience thermal pulses on what is called the ``super-AGB'';  %(or SAGB); 
they are distinguished from genuine massive stars by the fact that they do not experience further 
nuclear burning in their cores. Super-AGB stars
were first studied by Icko Iben and collaborators \citep[e.g.,][]{ritossa96}, and
their final fate depends on the competition between mass loss and core growth.
If they lose their envelope before the core reaches the Chandrasekhar mass, as is the usual case,
then the result is an O-Ne white dwarf.

\subsubsection{The Massive Intermediate Mass Stars}

If, on the other hand, the core grows to exceed the Chandrasekhar mass then these stars 
may end their lives as electron-capture supernovae. Stars in this
very narrow mass range (perhaps less than $0.5\Msun$) we shall
call  ``massive intermediate mass stars''.

It is still unclear what fraction of super-AGB stars explode as electron-capture supernovae, the 
details being dependent on uncertain input physics and implementation choices 
\citep{poelarends08}. The existence of massive white dwarfs \citep{gansicke10}, with masses 
above the C-O core limit of $\approx 1.1\Msun$ lends some support to the scenario that at 
least some fraction evade exploding as supernovae. The super-AGB stars that do explode as 
electron-capture supernovae have been proposed as a potential site for the formation of 
heavy elements via the rapid neutron capture process \citep[the $r$-process;][]{wanajo09,wanajo11}.
A review of this field is therefore particularly timely as we are only now becoming aware of the
nucleosynthesis outcomes of super-AGB stars and their
progeny \citep{siess10,doherty14a,doherty14b}.

\subsubsection{The Massive Stars}

Stars with masses greater than about $10\Msun$ we call
``massive stars'' and these will proceed through Ne/O
burning and beyond, and end their lives as iron core
collapse supernovae. Note that there is a rich variety of
outcomes possible, depending on the way one models
mixing and other processes, and we do not show all of
the different sub-cases here.
We have tried to maintain the existing definitions in
the literature, while adding some divisions that we think
are useful. We also reserve the use of the word ``massive''
for those stars that end their lives as supernovae,
being either ``massive intermediate stars'' in the case of
electron-capture supernovae, or the traditional ``massive
stars'' for the case of iron core-collapse supernovae.

%1.4 Yields
\subsection{Stellar Yield Calculations}\label{sec:yieldcals}

Stellar yields are an essential ingredient of chemical evolution models.  Prior to 2001,
the only stellar yields available for low and intermediate-mass stars were for 
synthetic AGB evolution models or from a combination of detailed and synthetic models 
\citep{vandenhoek97,forestini97,marigo01,izzard04b}. 

Synthetic AGB models are generally calculated by constructing fitting 
formulae to the results of detailed models, rather than
by solving the equations of stellar evolution. This approach was originally motivated
by the linear core-mass versus luminosity relation noted by \cite{pacz70}. It was soon realised
that many other important descriptors and properties of AGB evolution could be
similarly parameterised, saving the huge effort that goes into a fully consistent 
solution of the equations of stellar evolution, 
with all of the important input physics that 
is required (opacities, equations of state, thermonuclear reaction rates, 
convective mixing, etc).  These models can be used to examine rapidly the 
effect of variations in some stellar physics or model inputs. One must remember
of course that there is no feedback on the structure.  Any
change that would alter the stellar structure such that the parameterised relations 
also change is not going to be included in the results. Nevertheless, even within
this limitation there are many uses for synthetic models.
Further, we are now starting to see the next generation
of synthetic codes. These sophisticated codes are more like hybrids, 
combining parameterised evolution with detailed envelope integrations. 
An example is the {\sc Colibri} code of \citet{colibri}.

With the growth of cluster computing it is now common to have access to
thousands of CPU nodes. It is possible for stellar 
models of many different masses and compositions to
be calculated in detail on modern computer clusters. In this way we can obtain results from detailed 
models in reasonable times.  The first stellar yields from detailed AGB models were published by
\citet{ventura01} and \citet{herwig04b} but for a limited ranges of masses and/or metallicities.
The first stellar yields for a large range of masses and metallicities from detailed AGB models were 
published by \citet{karakas07b}, with an update by \citet{karakas10a}. 
In \S\ref{sec:yields} we provide an updated list of the latest 
AGB stellar yields that are available in the literature.

%1.5 Calculations

In Figure~\ref{HRD1} and in what follows we show stellar 
evolutionary sequences that were computed using 
the Mount Stromlo/Monash Stellar Structure code.
This code has undergone various revisions and updates over the past
decades \citep[e.g.,][]{lattanzio86,lattanzio89,lattanzio91a,frost96,karakas07b,campbell08,karakas10b,karakas12}.
We will highlight particular improvements that affect the nucleosynthesis in \S\ref{sec:agb}.
We note that the stellar evolutionary sequences described here are 
calculated using a reduced nuclear network that
includes only  H, He, C, N, and O. The wealth of data on abundances from stars
necessitates the inclusion of more nuclear species. Most of these are 
involved in reactions that produce negligible energy \citep[e.g., the higher 
order H burning Ne-Na and Mg-Al reactions,][]{arnould99}.
For this reason, a post-processing nucleosynthesis code is usually
sufficient, provided there is no feedback on the structure from the reactions not included 
in the evolutionary calculations. This is indeed usually the case.

We take the results from our evolutionary calculations and use them as input
for our post-processing nucleosynthesis code {\sc Monsoon}
\citep{cannon93,frost98a} in order to calculate the abundances 
of many elements and isotopes \citep[for a selection of
recent papers we refer the interested reader to][]{campbell08,lugaro12,kamath12,karakas12,shingles13,doherty14a}. 
In {\sc Monsoon} we require initial abundances (usually scaled solar) along with 
nuclear reaction rates and $\beta$-decay lifetimes,
and include time-dependent convection using an advective algorithm. 
We couple the nuclear burning with convective mixing in relevant regions of the star.
It is important to remember that the results presented here 
depend on the input physics and numerical
procedure, with different codes sometimes finding different results. For example, 
the inclusion of core overshoot during the main sequence and core He-burning 
will lower the upper mass limit for a C-O core AGB star from 
$\approx 8\Msun$ to $\approx 6\Msun$ \citep[e.g.,][]{bertelli86,bertelli86b,lattanzio91b,bressan93,fagotto94}.

%1.6 Previous Reviews
The most recent reviews of AGB evolution and nucleosynthesis include \citet{busso99}, 
with a focus on nucleosynthesis and the operation of the $s$-process, and
\citet{herwig05}, who  reviewed the evolution and nucleosynthesis of AGB stars in general,
including a discussion of multi-dimensional hydrodynamical
simulations relevant to AGB star evolution. Since 2005 there have been many advances,
including insights into AGB mass loss provided by the Spitzer and Herschel
Space Observatories, new theoretical models of AGB stars covering a 
wide range of masses and compositions, and the publication of stellar yields from detailed 
AGB star models.  In this review we focus on theoretical models of low and intermediate-mass stars
and in particular on recent progress in calculating AGB yields.  Not only are yields needed
for chemical evolution modelling but they are also needed to interpret the wealth of observational
data coming from current surveys such as SkyMapper \citep{keller07} and SEGUE \citep{yanny09}, which are
geared toward discovering metal-poor stars in the Galactic halo. Future surveys and instruments 
(e.g. the GAIA-ESO survey, HERMES on the AAT, APOGEE, LAMOST) will also require accurate stellar 
yields from stars in all mass ranges in order to disentangle the Galactic substructure revealed 
through chemical tagging \citep{freeman02}.

Finally we note that, as we will discuss in \S\ref{sec:needextramix}, there is compelling evidence
for some form of mixing on the RGB that is needed explain the abundances seen at the tip of the RGB. 
The standard models simply fail in this regard. 
While the number of isotopes affected is reasonably small (e.g., \iso{3}He, \iso{7}Li, \iso{13}C)
it is essential to include the effect of this mixing
in order to properly model the chemical evolution of those few isotopes.
Usually, a set of stellar yields is calculated based on standard models, which we know are
wrong because they fail to match the observed abundances along the RGB.

\section{Evolution and Nucleosynthesis prior to the Asymptotic Giant Branch} \label{sec:preagb}

\subsection{Illustrative Examples}

In the following we describe the evolution and nucleosynthesis for representative low 
and intermediate mass stars. All have a metallicity $Z=0.02$\footnote{where $Z$ is the global mass fraction 
of all elements heavier than H and He, with mass fractions $X$ and $Y$ respectively.}. According to the most recent 
solar abundances from \citet{asplund09} the global solar metallicity is $Z=0.0142$, which makes 
these models slightly super solar, with a  [Fe/H] = $+0.14$\footnote{using the standard spectroscopic notation 
  [X/Y] = $\log_{10}$(X/Y)$_{*} - \log_{10}$(X/Y)$_{\odot}$.}.

\begin{figure*}
\begin{center}
\includegraphics[width=10cm, angle=0]{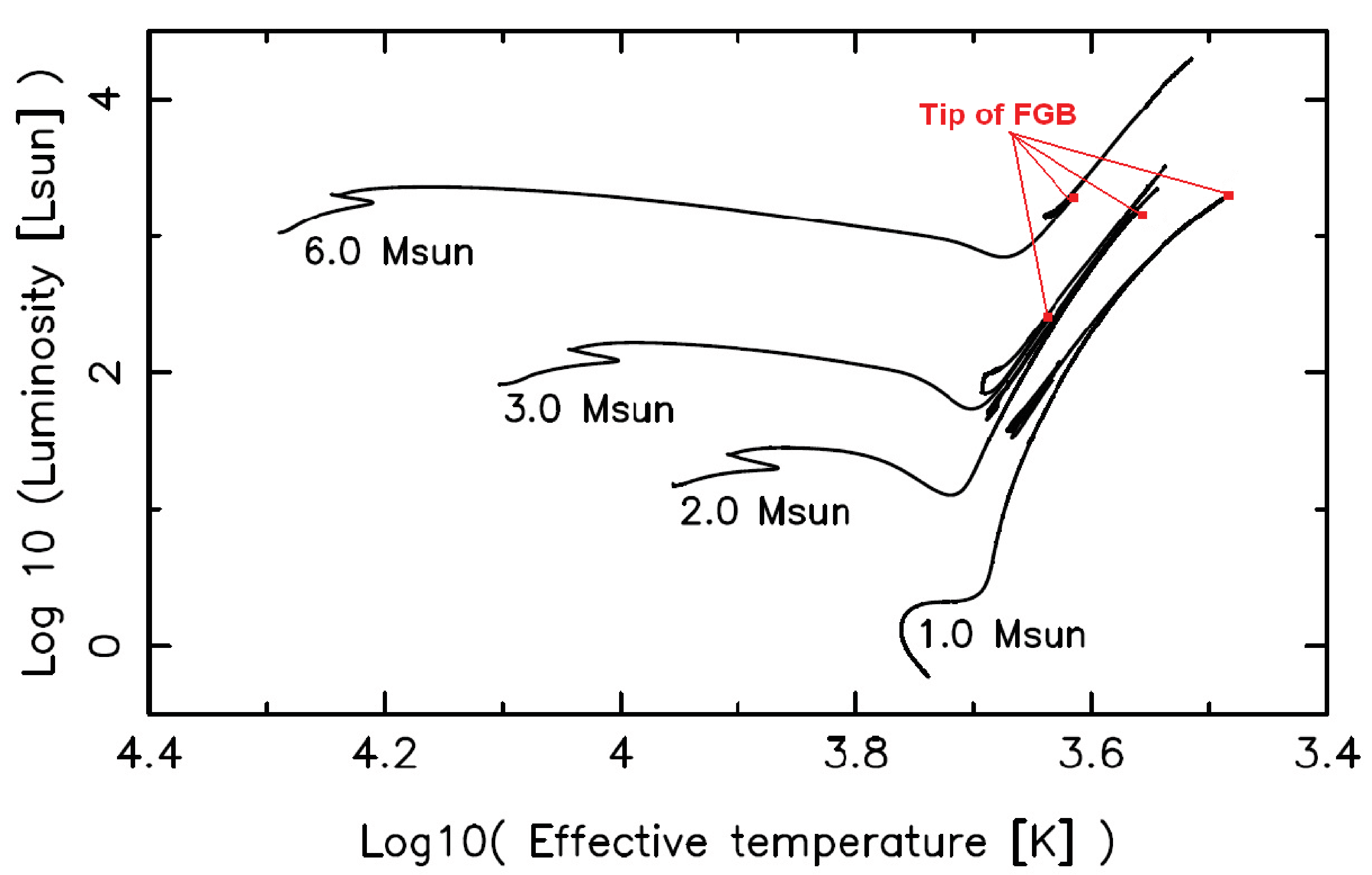}
\caption{A Hertzsprung-Russell (HR) diagram showing the evolutionary tracks for masses of 1, 2, 3, and 6$\Msun$
with a global metallicity of $Z$ = 0.02.
The evolutionary tracks show the evolution from the ZAMS through to the start of thermally-pulsing AGB.
The thermally-pulsing phase has been removed for clarity. The location of the tip of the 
RGB is indicated by the asterisk.}\label{HRD1}
\end{center}
\end{figure*}

To illustrate the evolution of low and intermediate-mass stars we 
use new stellar evolutionary sequences with masses between 1$\Msun$ and 8$\Msun$ and $Z=0.02$ calculated with
the same version of the Mt Stromlo/Monash Stellar Structure code described in \citet{kamath12}. 
Within this grid of models, the divide between low-mass stellar evolution and intermediate is at
about 2$\Msun$ as we discuss below.
These models will also introduce the basic principles of the evolution of all stars that evolve up to the
AGB. The theoretical evolutionary tracks for a sample of these models
are shown in Figure~\ref{HRD1} and include all evolutionary phases from the zero age main 
sequence through to the AGB. 

All stars begin their nuclear-burning life on the main sequence, where fusion converts H to He
in the stellar core. The majority of a star's nuclear-burning life is spent on 
the main sequence, which is why most stars in the night sky and most stars in a typical colour-magnitude
diagram are in this phase of stellar evolution.

The 1$\Msun$ model in Figure~\ref{HRD1} burns H on the main sequence via the pp chains.
In contrast, the models with $M \ge 2\Msun$ shown in Figure~\ref{HRD1} mostly burn H in the 
core via CNO cycling.  The higher temperature dependence of the CNO cycles, with a 
rate roughly $\propto T^{16-20}$ (compared to a rate approximately $\propto T^4$ 
for the pp chains at $Z=0.02$), produces a steep energy generation rate and results in 
the formation of a convective core. It is convenient to divide the zero-age main sequence (ZAMS) 
into an ``upper'' and ``lower'' 
main sequence, which is reflected in slightly different mass-radius and mass-luminosity 
relationships for these two regions. The division between the two is instructive: stars on the lower 
main sequence have convective envelopes, radiative cores and are primarily powered by the pp chains. 
Conversely, stars on the upper main sequence show radiative envelopes, convective cores and 
are powered primarily through the CN (and ON) cycles.
The dividing mass between the upper and lower main sequence is at about 1.2$\Msun$ for $Z=0.02$. 

The larger a star's  mass the  larger is its gravity. Hence it requires a substantial pressure gradient 
to maintain hydrostatic equilibrium. The central pressure is therefore higher and in turn so is the temperature.
This means that more massive stars burn at much higher luminosities and given that fusing  four protons into 
one $^4$He nucleus produces a constant amount of energy, then the duration of the H 
burning phase must be correspondingly lower as the stellar mass increases. 
From a quick  inspection of Figure~\ref{HRD1} it is clear that the 6$\Msun$ model is much brighter on the main 
sequence than the 2$\Msun$ model by almost two orders of magnitude.  Core 
H exhaustion takes  place after approximately $1 \times 10^{9}$ years (or 1Gyr) 
for the 2$\Msun$ model but only 53 million years for the 6$\Msun$ model. 

\begin{figure*}
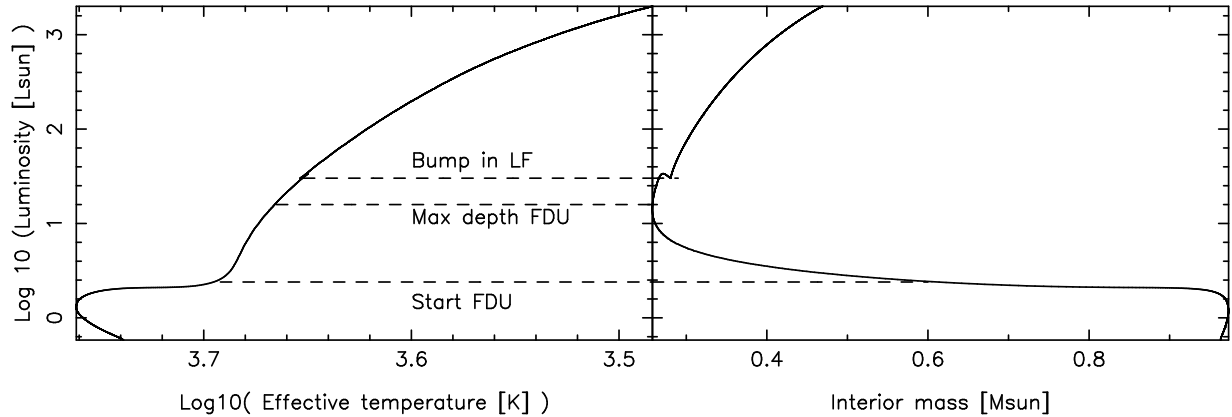

\begin{center}
\begin{tabular}{cc}
\includegraphics[width=0.65\columnwidth, angle=270]{Fig3a.ps} & \hspace{-0.6cm}
\includegraphics[width=0.65\columnwidth, angle=270]{Fig3b.ps} \\
\end{tabular}
\caption{First dredge-up in the 1$\Msun$, $Z=0.02$ model. The left panel 
shows the HR diagram and the right panel shows the luminosity as a function of the
mass position of the inner edge of the convective envelope. We can clearly see that the 
envelope begins to deepen just as the star leaves the main sequence, and 
reaches its deepest extent on the RGB, marking the end of FDU. Further evolution sees the star
reverse its evolution and descend the RGB briefly before resuming the climb. This corresponds to 
the observed bump in the luminosity function of stellar clusters (see text for details).
}\label{FDUM3}
\end{center}
\end{figure*}

Following core H exhaustion the core contracts and the star crosses the Hertzsprung Gap. 
Nuclear burning is now established in a shell 
surrounding the contracting \iso{4}He core. Simultaneously, the outer layers expand 
and cool and as a consequence become convective, due to an increase in the opacity at lower temperatures. 
The star runs up against the Hayashi limit, where the coolest envelope solution corresponds to 
complete convection. The star cannot cool further and it begins its rise 
up the giant branch while the convective envelope grows deeper and deeper (in mass).
This is shown in Figure~\ref{FDUM3} for the $1\Msun$ model.
This deepening of the convective envelope leads 
to mixing of the outer envelope with regions that have
experienced some nuclear processing, with the result that the products of H
burning are mixed to the surface. This is called the ``first dredge-up'', hereafter FDU.

The star is now very big (up $\sim 100$ times its radius 
on the main sequence) but most of the mass in the core is within a small fraction of the total radius. 
A consequence of this is that the outer layers are only tenuously held onto the star and 
can be lost through an outflow of gas called a stellar wind.  At present we do not know how much mass is 
lost during the ascent of the RGB. This may be perhaps as 
much as 30\% of the star's total mass for the lowest mass
stars that spend the longest time on the RGB. Kepler data for metal-rich old open clusters have provided some
constraints, with the amount of mass lost on the RGB estimated to be less than results from applying 
the commonly used Reimer's mass-loss prescription \citep{reimers75,kudritzki78,miglio12}. While there are refinements 
to the Reimer's mass-loss law \citep{catelan00,schroder05,schroder07} we are still lacking a 
detailed understanding of the physical mechanism responsible for mass loss on the RGB.

During the ascent of the RGB our low and intermediate-mass stars experience the FDU, which we will 
address in detail in \S\ref{sec:FDU}. Simultaneously, the He core continues to 
contract and heat and in the case of low-mass 
stars becomes electron degenerate. Neutrino energy losses become important, and since they
are very dependent on density, they dominate in the centre. This produces a cooling 
and can cause the mass location of the temperature maximum to move outward. The RGB lifetime is terminated 
when the necessary temperatures for 
central He ignition are reached, at about 100~million~K. For our low-mass stars the 
triple alpha reactions are ignited at the point of maximum temperature and under 
degenerate conditions \citep{despain81,deupree84}.

The electron degenerate equation of state results in the temperature and density 
being essentially decoupled. When He does begin to fuse into C, the energy released does not go 
into expansion but stays as thermal energy, raising the temperature of the plasma locally. 
This leads to a much higher burning rate and a runaway results, leading to a violent and explosive  
He ignition that is known as the ``core He flash.''

The maximum initial mass for the core He-flash to occur is about $2.1\Msun$ at $Z=0.02$ using
the new grid of models presented here, which include no convective overshoot 
\citep[similar to the models of][where the maximum mass is at about 2.25$\Msun$]{karakas02}.  This is the dividing
line between low and intermediate mass stars. In contrast, models which include overshooting from 
the convective H-burning core find that this division  occurs at a lower mass of
$M \approx 1.6\Msun$ \citep{bertelli86}. 

For intermediate-mass stars, the cores are not degenerate and He is ignited under quiescent conditions.
These stars often do not proceed as far up the RGB as do low-mass stars, prior to He ignition. 
As a consequence their RGB phase is shorter and the FDU phase can be terminated relatively early 
for these more  massive stars.
This has consequences for colour-magnitude diagrams and is demonstrated in Figure~\ref{HRD1}.
For example, the 2$\Msun$ model has a long RGB lifetime of $\approx 200 \times 10^{6}$ years or 200 Myr. 
This means that while the minimum effective temperature attained by the 2$\Msun$ red-giant 
model is less ($T_{\rm eff} \approx 3600$~K) than the 6$\Msun$ model ($T_{\rm eff} \approx 4100$~K), 
the peak RGB luminosity is similar, at $\log_{10}(L/{\rm L}_{\odot}) \approx 3.2$.  
This means that old low-mass RGB populations are observable out to great distances 
(e.g., Galactic GCs and dwarf spheroidal galaxies, which are dominated by
old low-mass stars). Note the contrast to the 3$\Msun$ model, which has 
a peak RGB luminosity that is more than 10 times lower, at only $140L_{\odot}$ (due to ignition of 
He under non-degenerate conditions). 

Following core He ignition the star 
settles down to a period of central He burning, where He burns in a convective core and H 
in a shell, which provides most of the luminosity. The Coulomb repulsion is larger for He than 
for H and more particle (kinetic) energy is required to sustain 
the triple-alpha process. This then requires that  the temperature is 
higher for He burning.  Note also that about a factor of 10 less energy is
produced by the triple alpha process per gram of fuel than during H burning. The overall
result is the core He burning phase is shorter than the main sequence. For example,
for the 2$\Msun$ model core He burning lasts 124~Myrs (a figure of about 100 Myr is typical for low-mass stars), 
compared to $\approx 13$Myrs for the 
6$\Msun$ model. Helium burning increases the content of \iso{12}C, which in turn 
increases the abundance of \iso{16}O from the reaction \iso{12}C($\alpha,\gamma$)\iso{16}O. 

The details of He burning are subject to uncertainties that are all too often ignored or
dismissed. We have known for decades that the fusion of He into C and O produces a discontinuity
in the opacity  at the edge of the convective core \citep{castellani71a}. 
This means that there is an acceleration at the edge
of the core. In other words there is no neutrally stable point which would be the edge of the 
core if one were using the Schwarzschild or Ledoux criterion for determining the borders of convective regions.
The result is that the convective core grows with time. 
The next complication is that the variation of temperature and density is such that there
is a local minimum in the radiative gradient in the convective region and this causes
the region to split into a convective core and a ``semi-convective''region \citep{castellani71b}.

This semi-convection extends the duration of the core He burning phase by mixing more fuel into the core. 
Star counts in clusters clearly show that observations require this extension to the core He
burning phase, and models constructed without semi-convection are a poor match \citep{buzzoni83,buonanno85}. 

There is yet another complication that arises as the star approaches the exhaustion of its core He supply.
Theoretical models show that as the He content decreases, the convective core is
unstable to rapid growth into the semi-convective region. 
This results in ``breathing pulses''  of the convective core \citep{gingold76,castellani85}. 
These mix more He into the core and further extend
the lifetime in this phase. While this behaviour shows many of the signs of a numerical instability,
an analytic study by \citet{sweidem73} showed that there is a genuine physical basis for the instability, and 
indeed verified that it should only occur when the central He mass fraction reduces below about 0.12.
Nevertheless, appealing again to star counts as a proxy for timescales, the data seem
to argue against the reality of these pulses \citep[][but see also \citealt{campbell13}]{renzini88,caputo89}.

This leaves us in a most unsatisfactory position. We have an instability shown by models, which 
theory can explain and indeed argues to be real, but that the data do not support. Further, we have
no obvious way to calculate through this phase in a way that removes the breathing pulses 
\citep[although see][]{dorman93}. What is worse is that the details of the evolution through this phase
determine the size of the He exhausted core and the position of the H-burning shell as the 
star arrives on the early AGB.  The star must now rapidly adopt the structure of a 
thermally-pulsing AGB star, by which we mean that burning shells will burn through the fuel profile
resulting from the earlier evolution until they reach the thermally-pulsing AGB structure.  
This results in removing some of the uncertainty in the structure that exists at the end of core He burning.
But it is still true that the subsequent evolution on 
the AGB is critically dependent on the core size which is poorly understood because of 
uncertainties during the prior core He burning phase.

Following exhaustion of the core He supply, low and intermediate-mass stars proceed toward the red giant branch,
now called the ``asymptotic giant branch'' because the colour-magnitude diagrams of old clusters show 
this population seemingly joining the first giant branch almost asymptotically. A better name may have been
the less commonly used ``second giant branch'', but AGB is now well established.

For stars more massive than about $4\Msun$ or with H-exhausted core
masses $\gtrsim 0.8\Msun$ (depending on the composition) the convective envelope extends quite 
some distance into the H-exhausted region. It usually reaches deeper than during the 
FDU \citep[e.g.][]{boothroyd99}. This event is called the ``second dredge-up'', hereafter SDU. 
In both cases (FDU and SDU) we are mixing to the surface the products of H burning, 
so qualitatively the changes are similar. However, there are substantial quantitative differences, 
as we discuss below in \S\ref{sec:FDU}.

\begin{figure*}
\begin{center}
\includegraphics[width=1.7\columnwidth]{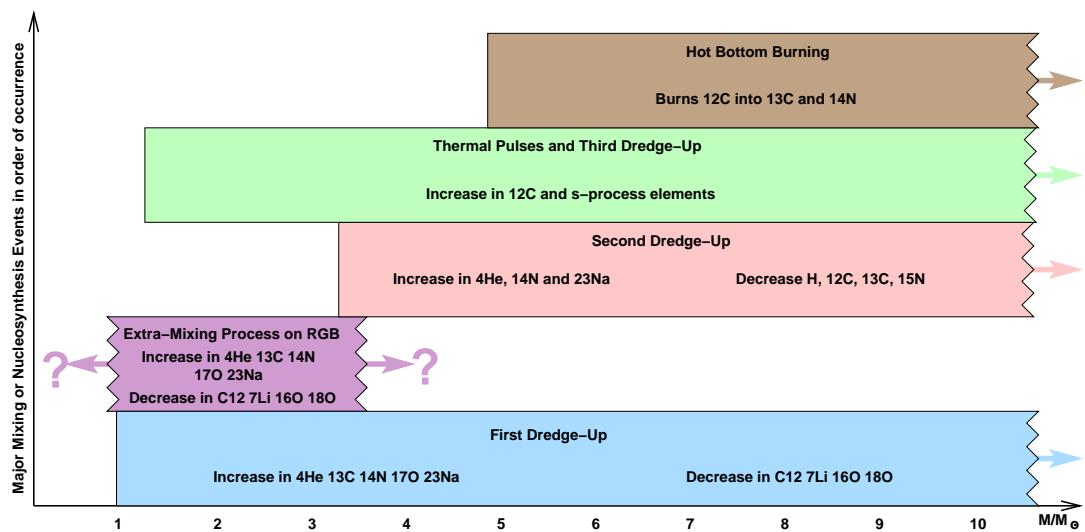} 
\caption{A schematic diagram showing the mass dependence of the different dredge-up, mixing and 
nucleosynthesis events. The species most affected are also indicated. The lower mass limits for
the onset of the SDU, TDU, and HBB depend on metallicity and we show approximate values for $Z=0.02$.
Note that the ``extra-mixing'' band has a very uncertain upper mass-limit, because the 
mechanism of the mixing is at present unknown.}\label{DU}
\end{center}
\end{figure*}

\subsection{First Dredge-Up}\label{sec:FDU}

We have outlined the evolution of low and intermediate-mass stars above. Now we look in 
more detail at the first and second dredge-up prior to the thermally-pulsing AGB. Figure~\ref{DU} shows the
different dredge-up processes that stars experience as a function of their mass.  It also shows 
the rough qualitative changes in surface abundances that result. 

\subsubsection{Abundance Changes due to FDU}\label{sec:FDUabunds}

\begin{figure}
\begin{center}
\includegraphics[width=0.8\columnwidth, angle=90]{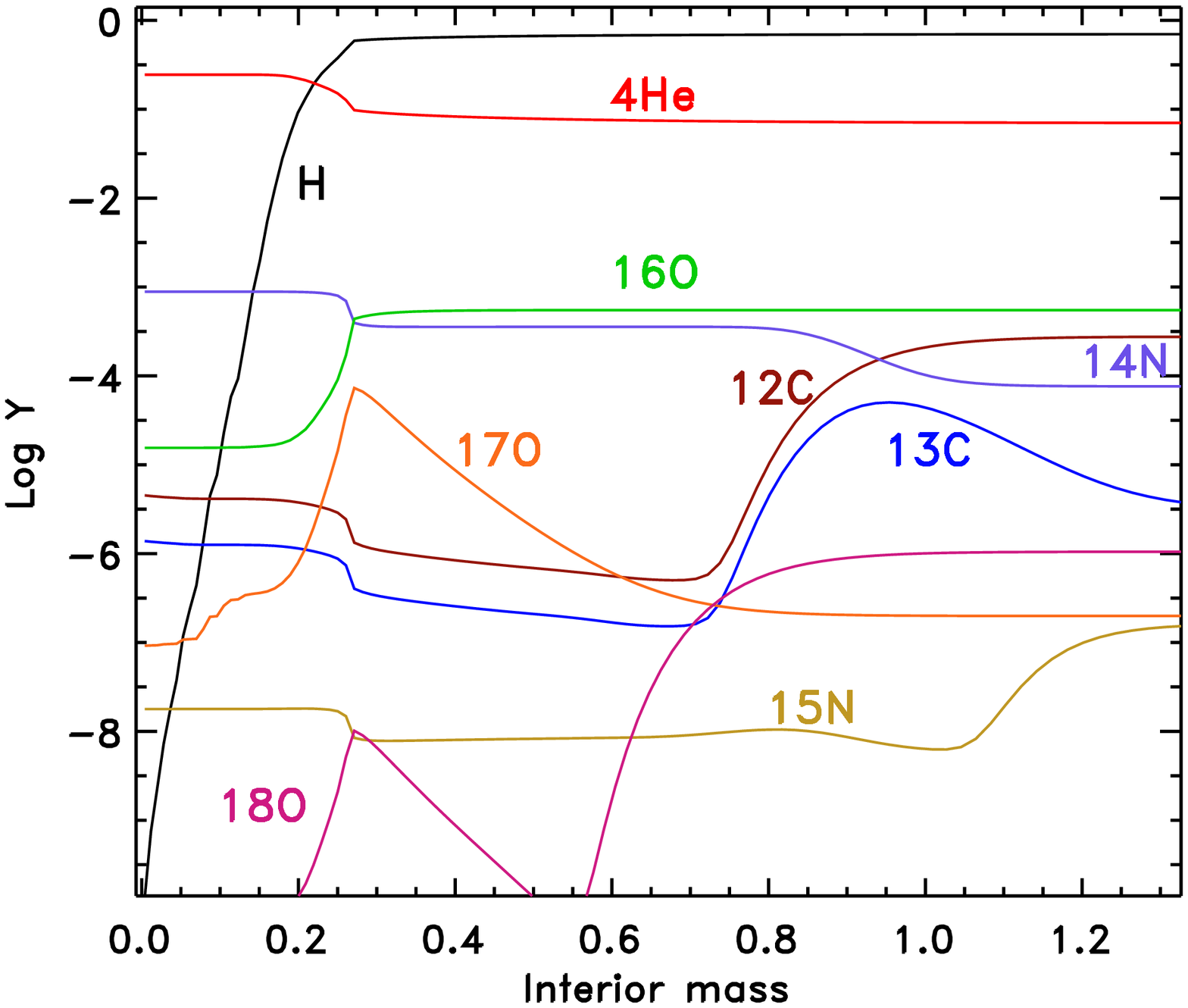} 
\includegraphics[width=0.8\columnwidth, angle=90]{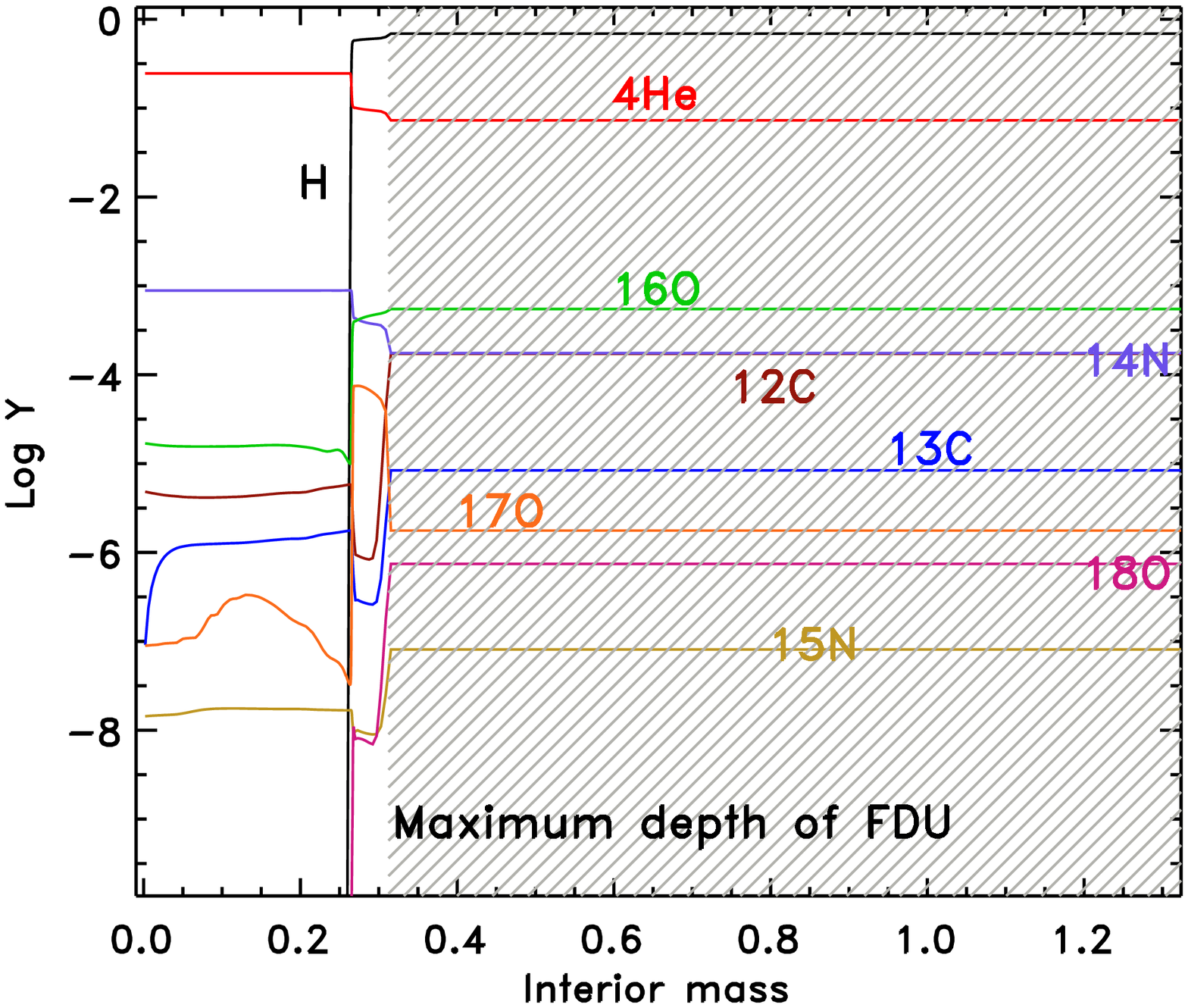} 
\caption{Composition profiles from the 2$\Msun$, $Z=0.02$ model. The top panel illustrates
the interior composition after the main sequence and before the FDU takes place, 
showing mostly CNO isotopes. The lower panel shows the composition at the deepest extent of the 
FDU (0.31$\Msun$), where the shaded region is the convective envelope. 
Surface abundance changes after the FDU include: a reduction in the C/O ratio from 0.50 to 0.33,
in the \iso{12}C/\iso{13}C ratio from 86.5 to 20.5, an increase in the isotopic ratio of 
\iso{14}N/\iso{15}N  from 472 to  2188,  a decrease in \iso{16}O/\iso{17}O from 2765 to 266,
and an increase in \iso{16}O/\iso{18}O from 524 to 740.
Elemental abundances also change:~[C/Fe] decreases by about 0.20, [N/Fe] increases by about $0.4$, and
[Na/Fe] increases by about 0.1. The helium abundance increases by $\Delta Y \approx 0.012$.}\label{m2z02fdu}
\end{center}
\end{figure}

The material mixed to the envelope by the FDU has been subjected to partial
H burning, which means it is still mostly H but with some added 
\iso{4}He and the products of CN cycling. Figure~\ref{m2z02fdu} shows the 
situation for $2\Msun$ model.  The upper panel shows the abundance profile 
after the star has departed the main sequence and prior to the FDU. We have
plotted the major species and some selected species involved in the CNO cycles. 
The lower panel is taken at the time of the maximum depth of the convective envelope. The
timescale for convective mixing is much shorter than the evolution timescale so 
mixing essentially homogenises the region instantly, as far as we are concerned.

Typical surface abundance changes from FDU are an increase in the \iso{4}He abundance by 
$\Delta Y \approx 0.03$ (in mass fraction), a decrease in the 
\iso{12}C abundance by about 30\%, and an increase in the \iso{14}N and \iso{13}C abundances. 
In Table~\ref{fdusduvalues} we provide the predicted post-FDU and SDU values for model 
stars with masses between 1 and 8$\Msun$ at $Z = 0.02$. We include the He mass fraction, 
$Y$, the isotopic ratios of C, N, and O, and the mass fraction of Na.

\begin{figure}
\begin{center}
\includegraphics[width=0.7\columnwidth, angle=270]{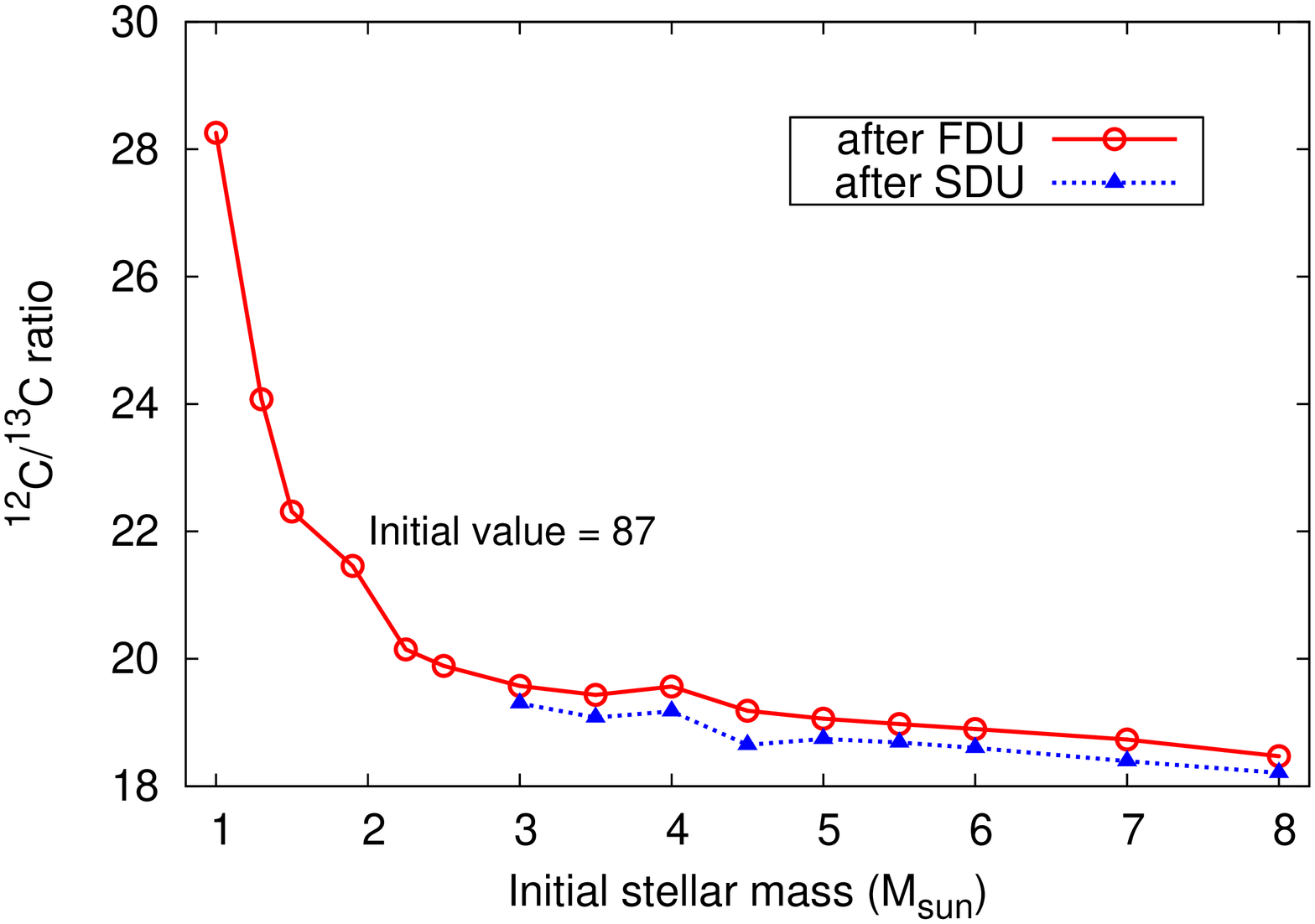}
\includegraphics[width=0.7\columnwidth, angle=270]{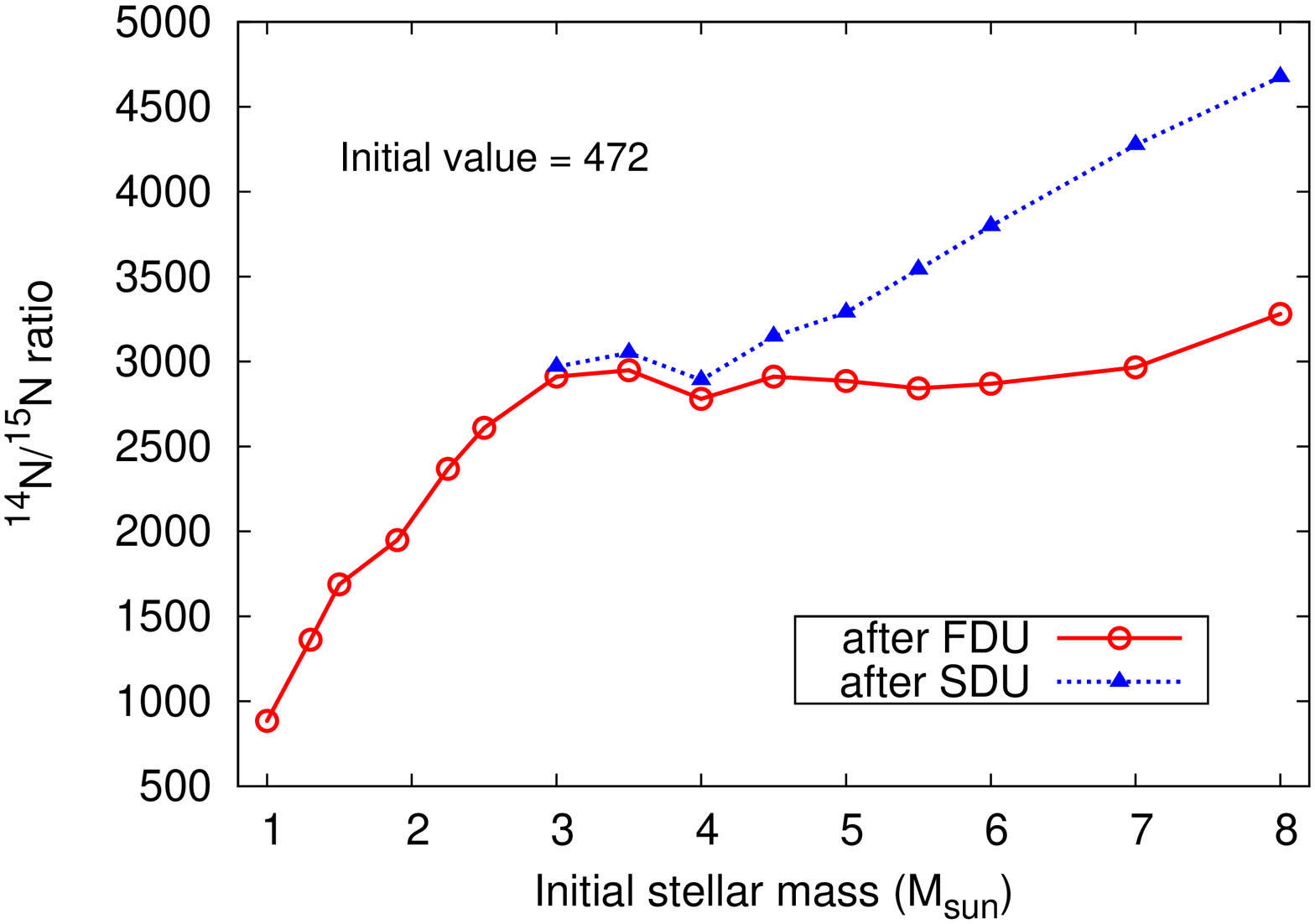} 
\includegraphics[width=0.7\columnwidth, angle=270]{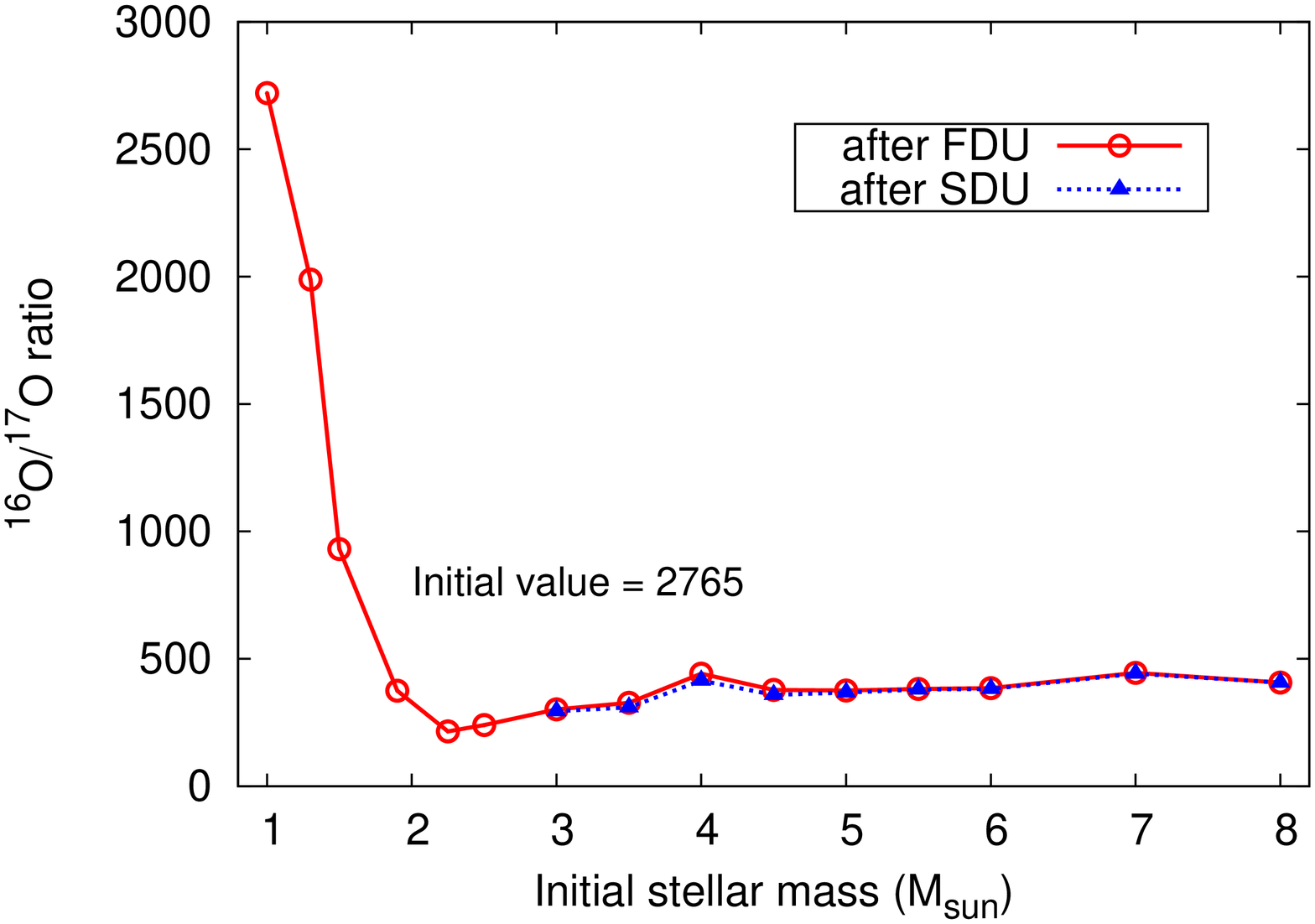} 
\caption{Surface abundance predictions from $Z=0.02$ models showing the ratio (by number) 
of \iso{12}C/\iso{13}C, \iso{14}N/\iso{15}N, and \iso{16}O/\iso{17}O
after the first dredge-up (red solid line) and second dredge-up (blue dashed line).
}\label{fdumasses}
\end{center}
\end{figure}

The C isotopic ratio is a very useful tracer of stellar evolution
and nucleosynthesis in low and intermediate-mass stars. Firstly, this is because 
the \iso{12}C/\iso{13}C ratio is one of the few isotopic ratios that can be readily derived 
from stellar spectra which means that there are large samples of stars for comparison
to theoretical calculations \citep[e.g.,][]{gilroy91,gratton00,smiljanic09,miko10,taut13}. 
Second, the C isotope ratio is predicted  to vary significantly at the surface as a 
result of the FDU (and SDU) as shown in Figure~\ref{fdumasses} and in Table~\ref{fdusduvalues}. 
This figure  shows that the number ratio of \iso{12}C/\iso{13}C drops 
from its initial value (typically about 89 for the Sun) to lie between 18 and 
26 \citep[see also][]{charbonnel94,boothroyd99}. 
Comparisons for intermediate-mass stars are in relatively good agreement with the observations, 
to within $\sim \! 25$\% \citep{eleid94,charbonnel94,boothroyd99,santrich13}. 
But the agreement found at low luminosities is not seen further up the 
RGB \citep[e.g.,][]{charbonnel95}, as discussed in \S\ref{sec:needextramix}.

\begin{figure}
\begin{center}
\includegraphics[width=0.6\columnwidth, angle=270]{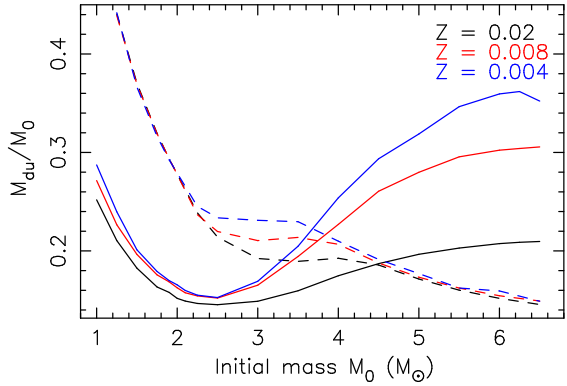}
\caption{Innermost mass layer reached by the convective 
envelope during the first dredge-up (solid lines) and second dredge-up 
(dashed lines) as a function of the initial stellar mass and metallicity. The mass 
co-ordinate on the $y$-axis is given as a fraction of the total stellar mass ($M_{\rm du}/M_{0}$).
From \citet{karakasThesis}.
}
\label{depthfdu}
\end{center}
\end{figure}

In Figure~\ref{depthfdu} we show the 
innermost mass layer reached by the convective envelope during
FDU (solid lines) and second dredge-up (dashed lines) 
as a function of the initial stellar mass and metallicity. The deepest FDU
occurs in the $Z=0.02$ models at approximately 2.5$\Msun$ with
a strong metallicity dependence for models with masses over about 3$\Msun$
\citep{boothroyd99}. In contrast there is little difference in the depth of
the second dredge-up for models with $\gtrsim 3.5\Msun$ regardless of metallicity.
In even lower metallicity intermediate-mass stars, the RGB phase is skipped altogether 
because core He burning is ignited before the model star reaches the RGB
so that the first change to the surface composition is actually due to 
the second dredge up.

\begin{table*}
\begin{center}
\caption{Predicted post-FDU and SDU values for model stars with masses between 1 and 8$\Msun$
at $Z = 0.02$. We include the helium mass fraction, $Y$, the isotopic ratios of carbon, nitrogen,
and oxygen, and the mass fraction of sodium. Initial values are given in the first row.}
\begin{tabular}{cccccccccc}
\hline Mass & Event & $Y$ & C/O & \iso{12}C/\iso{13}C & \iso{14}N/\iso{15}N & \iso{16}O/\iso{17}O & 
\iso{16}O/\iso{18}O &  X(\iso{23}Na) \\
\hline 
\hline
Initial &       & 0.280 & 0.506 & 86.50 & 472  & 2765  & 524 & 3.904($-$5) \\ \hline
1.00    & FDU   & 0.304 & 0.449 & 28.26 & 884  & 2720  & 556 & 3.904($-$5) \\ 
        & SDU   & 0.304 & 0.445 & 26.75 & 931  & 2617  & 560 & 3.911($-$5) \\
1.30    & FDU   & 0.303 & 0.392 & 24.07 & 1362 & 1989  & 627 & 3.941($-$5) \\
        & SDU   & 0.303 & 0.390 & 23.43 & 1395 & 1977  & 629 & 3.942($-$5) \\
1.50    & FDU   & 0.300 & 0.362 & 22.31 & 1688 & 930.6 & 674 & 4.157($-$5) \\
        & SDU   & 0.300 & 0.360 & 21.74 & 1736 & 913.8 & 678 & 4.165($-$5) \\
1.90    & FDU   & 0.292 & 0.343 & 21.46 & 1948 & 374.6 & 710 & 4.630($-$5) \\
        & SDU   & 0.292 & 0.341 & 21.03 & 1994 & 372.2 & 714 & 4.638($-$5) \\
2.00    & FDU   & 0.292 & 0.326 & 20.49 & 2188 & 265.8 & 741 & 4.840($-$5) \\
        & SDU   & 0.292 & 0.325 & 20.16 & 2224 & 265.2 & 743 & 4.844($-$5) \\
2.25    & FDU   & 0.291 & 0.320 & 20.15 & 2368 & 214.6 & 754 & 5.112($-$5) \\
        & SDU   & 0.291 & 0.320 & 20.00 & 2385 & 214.4 & 754 & 5.114($-$5) \\
2.50    & FDU   & 0.294 & 0.320 & 19.89 & 2610 & 240.0 & 754 & 5.358($-$5) \\
        & SDU   & 0.294 & 0.319 & 19.73 & 2633 & 239.4 & 756 & 5.363($-$5) \\
3.00    & FDU   & 0.300 & 0.322 & 19.57 & 2912 & 301.3 & 751 & 5.643($-$5) \\
        & SDU   & 0.300 & 0.319 & 19.30 & 2970 & 294.7 & 755 & 5.666($-$5) \\
3.50    & FDU   & 0.298 & 0.323 & 19.43 & 2950 & 327.1 & 748 & 5.683($-$5) \\
        & SDU   & 0.298 & 0.319 & 19.08 & 3051 & 309.1 & 756 & 5.752($-$5) \\
4.00    & FDU   & 0.293 & 0.332 & 19.56 & 2780 & 441.6 & 728 & 5.469($-$5) \\
        & SDU   & 0.293 & 0.328 & 19.17 & 2892 & 415.6 & 737 & 5.553($-$5) \\
4.50    & FDU   & 0.293 & 0.325 & 19.18 & 2912 & 377.8 & 743 & 5.603($-$5) \\
        & SDU   & 0.297 & 0.320 & 18.65 & 3147 & 357.9 & 755 & 5.791($-$5) \\
5.00    & FDU   & 0.291 & 0.324 & 19.06 & 2886 & 375.5 & 745 & 5.567($-$5) \\
        & SDU   & 0.309 & 0.322 & 18.74 & 3289 & 367.5 & 751 & 5.962($-$5) \\
5.50    & FDU   & 0.289 & 0.324 & 18.98 & 2843 & 381.8 & 746 & 5.510($-$5) \\ 
        & SDU   & 0.322 & 0.325 & 18.68 & 3542 & 379.3 & 747 & 6.223($-$5) \\
6.00    & FDU   & 0.289 & 0.324 & 18.90 & 2870 & 384.5 & 746 & 5.529($-$5) \\
        & SDU   & 0.333 & 0.325 & 18.60 & 3798 & 381.5 & 747 & 6.472($-$5) \\
7.00    & FDU   & 0.291 & 0.324 & 18.68 & 3035 & 397.4 & 748 & 5.665($-$5) \\
        & SDU   & 0.350 & 0.325 & 18.40 & 4275 & 395.4 & 750 & 6.885($-$5) \\
8.00    & FDU   & 0.296 & 0.324 & 18.47 & 3281 & 406.9 & 750 & 5.881($-$5) \\
        & SDU   & 0.362 & 0.325 & 18.21 & 4675 & 406.4 & 637 & 7.205($-$5) \\
\hline
\hline
\end{tabular}
\label{fdusduvalues}
\end{center}
\end{table*}

%O isotopes?
One check on the models concerns the predictions for O isotopes. Broadly, the CNO cycles
produce $^{17}$O but significantly destroy $^{18}$O, with the result that the FDU should
increase the observed ratio of $^{17}$O/$^{16}$O and
decrease the observed ratio of $^{18}$O/$^{16}$O \citep[e.g., Table~\ref{fdusduvalues} and][]{dearborn92}. 
Spectroscopic data, where available, seem to agree reasonably 
well with the FDU predictions \citep{dearborn92,boothroyd94},
but see also \S\ref{sec:needextramix}.

\begin{figure}
\begin{center}
\includegraphics[width=0.7\columnwidth, angle=270]{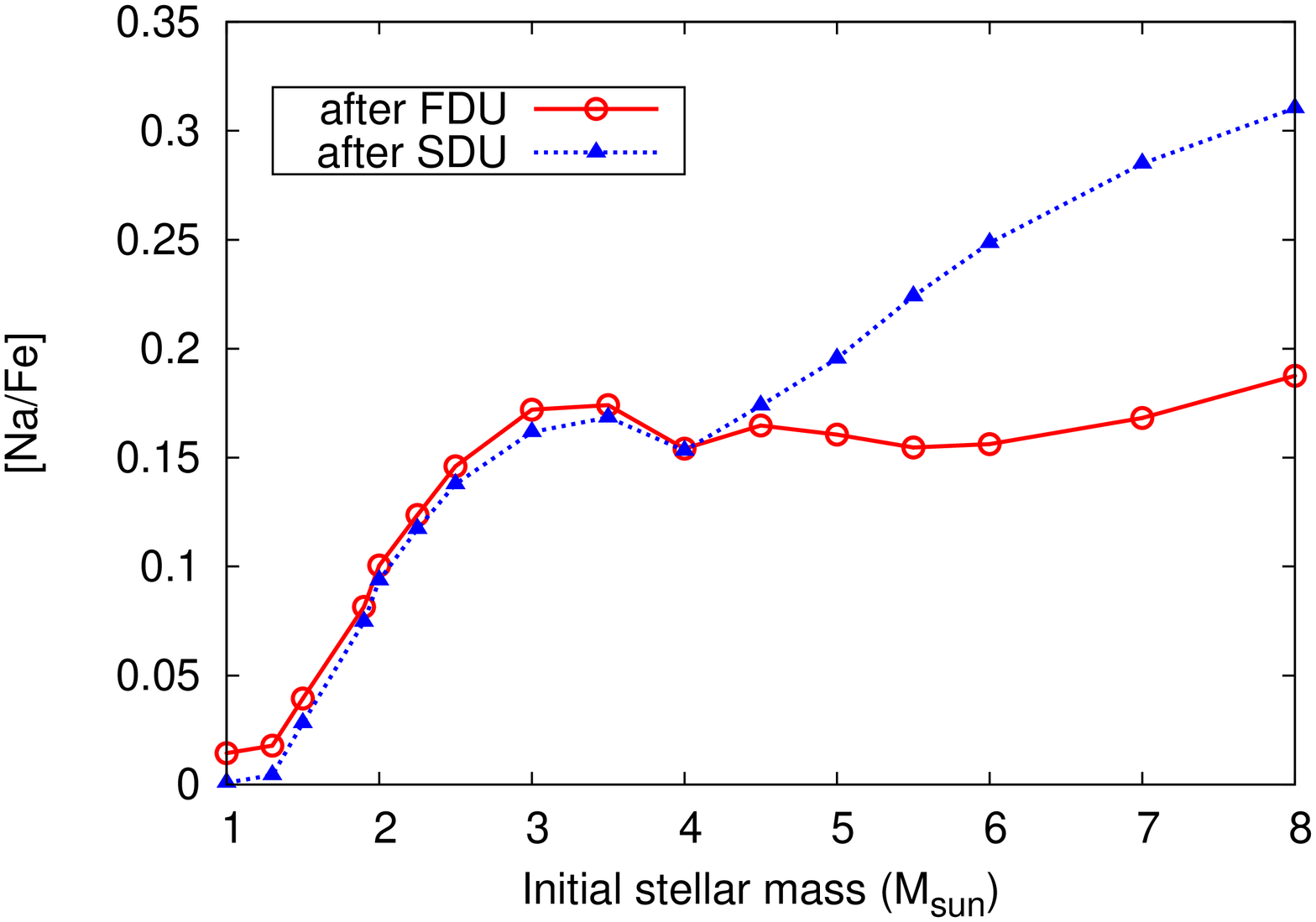}
\caption{Predicted [Na/Fe] after the FDU and SDU for the $Z=0.02$ models.}
\label{sodium}
\end{center}
\end{figure}

The effect of the first dredge up on other elements
(besides lithium, which we discuss in \S\ref{sec:lithium} below) is relatively
minor. It is worth noting that there is some dispute about the status of sodium.  
Figure~\ref{sodium} shows that sodium is not predicted to be significantly enhanced 
in low-mass stars by the FDU or by extra-mixing 
processes \citep{charbonnel10} at disk metallicities. This agrees with \citet{eleid95} 
who find modest
enhancements in low mass (0.1 dex) and intermediate
mass ($0.2$--$0.3$ dex) stars. 
Observations are in general agreement with models
with masses $\gtrsim 2\Msun$, where enhancements of up to [Na/Fe] $\lesssim 0.3$ are 
found in stars up to 8$\Msun$ at $Z = 0.02$ (Figure~\ref{sodium}).  
This result is confirmed by observations that show only mild enhancements of
[Na/Fe] $\lesssim  +0.2$ \citep{hamdani00,smiljanic09}.
But this is in contradiction with other
studies showing typical overabundances of $+0.5$
\citep{bragaglia01,jacobson07,schuler09}. 
The reasons for the conflicting
results are not well understood; we refer to \citet{smiljanic12} 
for a detailed discussion of observational uncertainties.

\subsubsection{The Onset of FDU}

Theoretical models must confront observations for us to
verify that they are reliable or identify where they need
improvements. In the current context there are two related
comparisons to be made: the expected nucleosynthesis, which we addressed above,
and the structural aspects. In this section we look
at the predictions for the location of the start of FDU
and in the next section we compare the observed
location of the bump in the luminosity function to 
theoretical models.

During the evolution up the giant branch there are three significant
points, as illustrated in Figure ~\ref{FDUM3}:
\begin{itemize}
\item{}the luminosity at which the surface abundances
start to change;
\item{}the maximum depth of the FDU;
\item{}the luminosity of the bump.
\end{itemize}
Note that these are not independent -- the maximum depth of the FDU determines 
where the abundance discontinuity occurs, and that determines the position 
of the bump. Similarly, the resulting compositions are dependent on the
depth of the FDU and we are not free to adjust that without consequences for
both the observed abundances and the location of the bump.

The obvious question is how closely do these points
match the observations? Perhaps the best place to look
is in stars clusters, as usual. \citet{gilroy91} found that
the ratios of C/N and \iso{13}C/\iso{13}C 
at the onset of the FDU matched the models quite well, as
reported in \citet{charbonnel94}. \citet{mishenina06}
also looked at C/N and various other species, and the
data again seem to match models for the onset of the FDU.
Of course the onset is very rapid and the data are
sparse. This is also seen in the study by \citet{chaname05}
who found that the C isotopic ratio in M67
fitted the models rather well (see their Figures 13 and
14). For field giants (with $-2 <$[Fe/H]$< -1$) the data
are not so good, with mostly lower limits for $^{12}$C/$^{13}$C
making it hard to identify the exact onset of the FDU 
\citep[see][Figure~16]{chaname05}. 
Better data over a larger range in luminosity in many clusters are needed to
check that the models are not diverging from reality at this
early stage in the evolution.

\subsubsection{The Bump in the Luminosity Function}\label{sec:bump}

We now move to an analysis of the luminosity function (LF) bump. There is much more literature here,
dating back to \citet{sweigart78} where it was shown that
the bump reduced in size and appeared at higher luminosities
as either the He content increased or
the metallicity decreased. Indeed, at low
metallicity (say [Fe/H] $\lesssim -1.6$) it can be hard to identify
the LF bump, and \citet{fusipecci90} combined
data for the GCs M92, M15 and NGC~5466 so that they could
reliably identify the bump in these clusters (all of similar
metallicity). Their conclusion, based on a study of 11
clusters, was that the theoretical position of the bump
was 0.4 mag too bright.

The next part of the long history of this topic was
the study by \citet{cassisi97} with newer models
who concluded that there was no discrepancy within
the theoretical uncertainty. Idealized models show a discontinuity
in composition at the mass where the convective
envelope reached its maximum inward extent. But
in reality this discontinuity is likely to be a steep profile,
with a gradient determined by many things, such as the
details of mixing at the bottom of the convective envelope.
It is likely that gravity waves and the possibility
of partial overshoot would smooth this profile through
entrainment, and \citet{cassisi02} showed that such
uncertainties do cause a small shift in the position of
the bump but they are unlikely to be significant.

\citet{riello03} looked at 54 Galactic GCs
and found good agreement between theory and
observation, both for the position of the bump itself as
well as the number of stars (i.e., evolutionary timescales)
in the bump region. The only caveat was that for low
metallicities there seemed to be a discrepancy but it was
hard to quantify due to the low number of stars available.
A Monte Carlo study by \citet{bjork06} concluded that the difference between theory and observation
was no larger than the uncertainties in both of
those quantities. In what seems to be emerging as a consensus,
\citet{dicecco10} found that the metal-poor
clusters showed a discrepancy of about 0.4 magnitudes,
and that variations in CNO and $\alpha$-elements (e.g., O, Mg, Si, Ti)
did not improve the situation. These authors did point out that the position
of the bump is sensitive to the He content
and since we now believe that there are multiple populations
in most GCs, this is going to cause a spread in the 
position of the LF bump.

It would appear that a reasonable conclusion is that
the theoretical models are in good agreement, while perhaps
being about 0.2 magnitudes too bright \citep{cassisi11},
except for the metal-poor regime where
the discrepancy may be doubled to 0.4 mag, although
this is plagued by the bump being small and harder
to observe. \citet{nataf13} find evidence for a second
parameter, other than metallicity, being involved.
In their study of 72 globular clusters there were some
that did not fit the models well, and this was almost
certainly due to the presence of multiple populations.
This reminds us again that quantitative studies must
include these different populations and that this could
be the source of some of the discrepancies found in the
literature.

The obvious way to decrease the luminosity of the
bump is to include some overshooting inward from 
the bottom of the
convective envelope. This will push the envelope deeper
and will shift the LF bump to lower luminosities. 
Of course this deeper mixing alters the predictions of 
the FDU; however it appears that there is a saturation of 
composition changes such that the small increase 
needed in the depth of the FDU does not produce an observable
difference in the envelope abundances 
\citep[][Angelou, 2013, private communication and]{kamath12}.

\subsubsection{The Need for Extra-Mixing}\label{sec:needextramix}

\begin{figure*}
\begin{center}
\includegraphics[width=1.8\columnwidth]{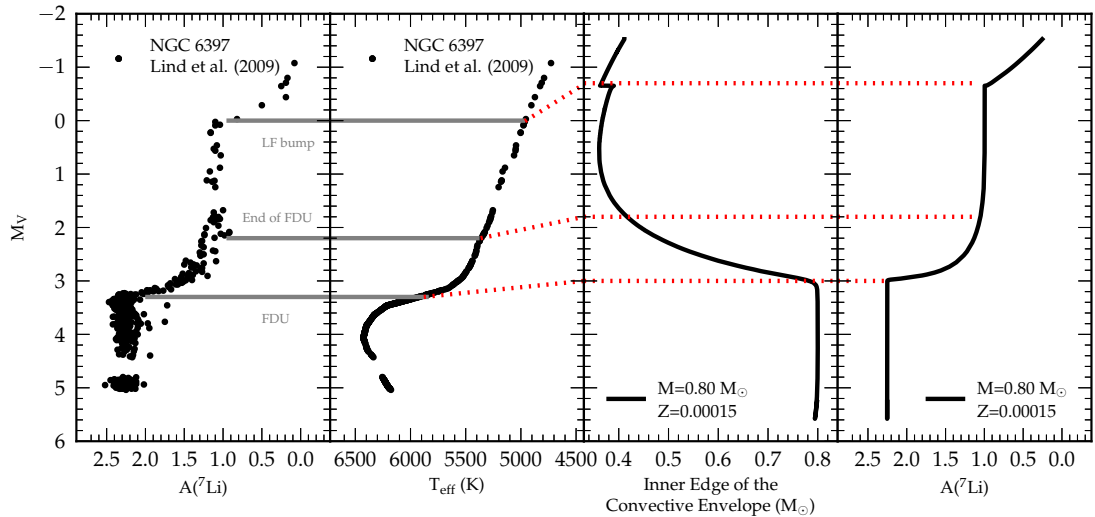}
\caption{Observed behaviour of Li in stars in NGC~6397, from \citet{lind09}, and as modelled by
Angelou et al. (2014, in preparation). The leftmost panel shows 
A(Li) $=\log_{10}$(Li/H) $+12$ plotted against luminosity for NGC~6397. 
Moving to the right the next panel shows the
HR diagram for the same stars. The next panel to the right shows the inner edge
of the convective envelope in a model of a typical red-giant star in NGC~6397. The
rightmost panel shows the resulting predictions for A(Li) using thermohaline
mixing and $C = 120$ (see \S\ref{sec:thm}). Grey lines identify the positions on the RGB
where the major mixing events take place. Dotted red lines identify these
with the theoretical predictions in the rightmost panels.}\label{LindLi}
\end{center}
\end{figure*}

We have seen that the predictions for FDU are largely in agreement with the observations. However
when we look at higher luminosities we find that something has changed the abundances beyond 
the values predicted from FDU. Standard models do not predict any further changes on the RGB once 
FDU is complete. Something must be occurring in real stars that is not predicted by the models.

For low-mass stars the predicted trend of the post-FDU \iso{12}C/\iso{13}C  
ratio is a rapid decrease with increasing initial mass as illustrated in Figure~\ref{fdumasses}.
Yet this  does not agree with the observed trend.   
For example, observations of the \iso{12}C/\iso{13}C ratio in open metal-rich clusters 
reveal values of $\lesssim 20$, sometimes $\lesssim 10$ \citep[e.g.,][]{gilroy89,smiljanic09,miko10} well below 
the predicted values of $\approx 25-30$. 
The deviation between theory and observation is even more striking in metal-poor field stars and in giants 
in GCs \citep[e.g.,][]{pilachowski96,gratton00,gratton04,cohen05b,origlia08,valenti11}. 

The observed trend is in the same direction as the FDU: i.e., as if we are mixing in more material that has 
been processed by CN cycling, so that $^{12}$C decreases just as $^{13}$C and $^{14}$N increase.
Indeed, in some GCs we see a clear decrease in [C/Fe] with increasing luminosity on the RGB
\citep[see][and references therein]{angelou11,angelou12}.
If some form of mixing can connect the hot region at the top of the H-burning shell with the 
convective envelope then the results of the burning can be seen at the surface.
These observations have been interpreted as evidence for extra mixing 
taking place between the base of the convective envelope and the H shell.

Further evidence comes from observations of the fragile element Li \citep{pilachowski93,lind09} which 
essentially drops at the FDU to the predicted value of A(Li)$\approx 1$\footnote{using the notation 
A(X)\ $= \log \epsilon$(X) =  $\log10(N_X/{\rm N_H})+12$ and $N_X$ is the abundance 
(by number) of element X.}, 
but for higher luminosities decreases to much lower abundances of A(Li)$\approx 0$ to $-1$. 
Again, this can be explained by exposing the envelope material
to higher temperatures, where Li is destroyed. So just as for the C isotopes 
the observations argue for some form of extra mixing to join the envelope to the region of the 
H-burning shell (see also \S\ref{sec:lithium}).

%O isotopes and deep-mixing
We discussed O isotopes earlier as a diagnostic of the FDU. Although these 
ratios may not be so easy to determine spectroscopically, the %new 
science of meteorite grain analysis \citep[for reviews see][]{zinner98,lodders05} 
offers beautiful data on O isotopic ratios from Al$_2$O$_3$  grains.
We expect that these grains would be expelled from the star
during periods of mass loss and would primarily sample the tip of the RGB or the AGB.
Some of these data show good agreement with predictions for the FDU, while
a second group clearly require further $^{18}$O destruction. This can be
provided by the deep-mixing models of \citet{wasserburg95} and \citet{nollett03}. 
Hence pre-solar grains contain further evidence for the existence of some kind of extra 
mixing on the RGB.

We note here that a case has been made for some similar form of mixing in AGB envelopes
as we discuss later in \S\ref{sec:uncerts}.

\subsubsection{The $^3$He problem}\label{sec:he3problem}

Another piece of evidence for deep-mixing concerns the stellar yield of $^3$He,
as discussed recently by \citet{lagarde11,lagarde12a}. We now have good constraints on the
primordial abundance of $^3$He, from updated Big Bang Nucleosynthesis calculations together with 
WMAP data. The currently accepted value is $^3$He/H = $1.00 \pm 0.07 \times 10^{-5}$ according to \citet{cyburt08} and
$^3$He/H = $1.04 \pm 0.04 \times 10^{-5}$ according to \citet{coc04}. This is within a factor of 2 or 3 
of the best estimates of the local value in the present interstellar medium of
$^3$He/H = $2.4 \pm 0.7 \times 10^{-5}$ according to \citet{gloeckler96}, a value
also in agreement with the measurements in Galactic HII regions by \citet{bania02}. 
This indicates a very slow growth of the $^3$He content over the Galaxy's lifetime.

However, the evolution of low-mass stars predicts that they produce copious amounts of $^3$He. 
This isotope is produced by the pp chains and when the stars reach the giant branches %made plural - both AGB & RGB
their stellar winds carry the $^3$He into the interstellar medium. Current models for the chemical evolution
of the Galaxy, using standard yields for $^3$He, predict that the local interstellar medium should 
show $^3$He/H $ \approx 5 \times 10^{-5}$ \citep{lagarde12a} which is about twice the observed value.

It has long been recognised that one way to solve this problem is to change the yield of $^3$He in
low-mass stars to almost zero \citep{charbonnel95}. In this case the build up of $^3$He over the lifetime
of the Galaxy will be much slower. One way to decrease the yield of $^3$He is to destroy it in
the star on the RGB while the extra mixing is taking place. 
We discuss this further in \S\ref{sec:extramix}.

\subsection{Non-convective mixing processes on the First Giant Branch}\label{sec:extramix}

\subsubsection{The onset of extra mixing}\label{sec:onset}
So where does the extra-mixing begin and what can it be?
Observations generally indicate that the conflict between theory and observation 
does not arise until the star has reached the luminosity of the bump in the LF.
This is beautifully demonstrated in the Li data from \citet{lind09} which we reproduce in Figure~\ref{LindLi}
together with a theoretical calculation for a model of the 
appropriate mass and composition for NGC~6397 (Angelou, private communication).
The left panel shows the measured A(Li) values for the stars as a function of magnitude. The 
near constant values until $M_V \approx 3.3$ are perfectly consistent with the models. It
is at this luminosity that the convective envelope starts to penetrate into regions 
that have burnt $^7$Li, diluting the surface $^7$Li content. This stops when the
envelope reaches its maximum extent, near $M_V \approx 1.5$. Then there is a sharp
decrease in the Li abundance once the luminosity reaches $M_V \approx 0$, which roughly corresponds to the
point where the H shell has reached the discontinuity left behind by FDU (see right panel). 
This is the same position as the bump in the LF. Note that there is a small
discrepancy between the 
models and the data, as shown in the figure and as discussed in \S\ref{sec:bump}.

This is the common understanding -- that the deep mixing begins once the advancing H shell removes the 
abundance discontinuity left behind by the FDU. The reason for this is that one expects that gradients in 
the composition can inhibit mixing \citep{kippwei90} and once they are removed by burning 
then the mixing is free to develop. It was \citet{mestel53} who first proposed that for a 
large enough molecular weight gradient one could effectively have a barrier to mixing \citep[see also][]{chaname05}.
This simple theory, combined with the very close alignment of the beginning of the extra mixing and the LF bump, 
has led to the two being thought synonymous. However we do note that there are discrepancies with this idea. 
For example it has been pointed out by many authors that there is a serious problem
with the metal-poor GC M92 \citep{chaname05,angelou12}. Here the data
show a clear decrease in [C/Fe] with increasing luminosity on the RGB \citep{bellman01,gsmith03},
starting at $M_V \approx 1$--$2$. 
The problem is that the decrease begins well before the bump in the LF, which \citet{martell08} 
place at $M_V \approx -0.5$. This is a substantial disagreement. A similar 
disagreement was noted by \citet{angelou12} for M15 although possibly not for NGC~5466, 
despite all three clusters having a very similar [Fe/H] of $\approx -2$.

To be sure that we cannot dismiss this disagreement lightly, there is also the work by
\citet{drake11} on $\lambda$~Andromeda, a mildly metal-poor ([Fe/H] $\approx -0.5$) first ascent giant
star that is believed to have recently completed its FDU. It is not yet bright enough for the H shell to
have reached the abundance discontinuity left by the FDU, but it shows $^{12}$C/$^{13}$C $\lesssim 20$,
which is below the prediction for the FDU and more in line with the value expected after
extra mixing has been operating for some time. We know $\lambda$~Andromeda is a binary so we cannot rule out 
contamination from a companion. But finding a companion that can produce the required
envelope composition is not trivial. \citet{drake11} also give the case of a similar star,
29~Draconis, thus arguing further that these exceptions are not necessarily the result of 
some unusual evolution. The problem demands further study because the discrepancy concerns fundamental 
stellar physics.

\subsubsection{Rotation}\label{sec:rotation}

It is well known that rotating stars cannot simultaneously maintain hydrostatic and thermal
equilibrium, because surfaces of constant pressure (oblate spheroids) 
are no longer surfaces of constant temperature. Dynamical motions develop that are known
as ``meridional circulation'' and which cause mixing of chemical species. It is not our
intention to provide a review of rotation in a stellar context. There are many far
more qualified for such a task and we refer the reader to \citet{heger00}, \citet{tassoul07}, 
\citet{maeder10} and the series of eleven papers by Tassoul and Tassoul, ending with 
\citet{tassoul95}.
Note that most studies that discuss the impact of rotation on stellar evolution 
often ignore magnetic fields. Magnetic fields likely play an important role
in the removal of angular momentum from stars as they evolve 
\citep[e.g., through stellar winds][]{gallet13,mathis13,cohen14}.

Most of the literature on rotating stars concerns massive stars because they 
rotate faster than low and intermediate mass stars. 
Nevertheless, there is a substantial history of calculations relevant to our subject.
\citet{sweigart79} were the first to attempt to explain the observed extra mixing with
meridional circulation. Later advances in the theory of rotation and chemical transport
\citep{kawaler88,zahn92,maeder98} led to more sophisticated models for the evolution of
rotating low and intermediate mass stars \citep{palacios03,palacios06}.

With specific regard to the extra mixing problem on the RGB, \citet{palacios06} found that
the best rotating models did not produce enough mixing to explain the decrease seen 
in the \iso{12}C/\iso{13}C ratio on the upper RGB, above the bump in the LF. 
This is essentially the same result as found by \citet{chaname05} and
\citet{charbonnel10}. Although one can never dismiss the possibility that a better
understanding of rotation and related instabilities may solve the problem, the
current belief is that rotating models do not reproduce the observations of RGB stars.

\subsubsection{Parameterised models}\label{sec:param}

With the failure of rotation to provide a solution to extra mixing on the RGB, 
the investigation naturally fell to phenomenological models of the mixing.   One main method used is to
set up a conveyor belt of material that mixes to a specified depth and at a specified rate.
An alternative is to solve the diffusion equation for a specified diffusion co-efficient $D$,
which may be specified by a particular formula or a specified value.

It is common in these models to specify the depth of mixing in terms of the difference
in temperature between the bottom of the mixed region and some reference temperature
in the H shell. The rates of mixing are sometimes given as mass fluxes and sometimes
as a speed. These are usually assumed constant on the RGB, although some models
include prescriptions for variation. In any event, there is no reason to believe that the
depth or mixing rate is really constant. (Note also that a constant mass flux
requires a varying mixing speed during evolution along the RGB, and vice versa!). 
%We desperately need to identify the physical mechanism driving the mixing so that we can 
%remove the ad hoc nature of the parameterised approach.

\citet{gsmith92} showed that such simplified models could reproduce the decrease of [C/Fe] 
seen along the RGB of GCs, provided an appropriate choice was made for the
depth and rate of mixing. More sophisticated calculations within a similar paradigm
are provided by \citet{boothroyd94}, \citet{boothroyd95}, \citet{wasserburg95}, \citet{langer95},
\citet{sackboo99}, \citet{nollett03}, \citet{denissenkov00}, \citet{denissenkov06}, and 
\citet{palmerini09}.

In summary, these models showed that for ``reasonable'' values of the
free parameters one could indeed reproduce the observations for the C and O
isotopic ratios, the decrease in A(Li), the variation of [C/Fe] with luminosity,
and also destroy most of the $^3$He traditionally produced by low-mass stars. 

\subsubsection{Thermohaline mixing}\label{sec:thm}

The phenomenon of thermohaline mixing is not new, having appeared in the astrophysics literature many 
decades ago \citep[e.g.,][]{ulrich72}. What is new is the discovery by \citet{eggleton06} that
it may be the cause of the extra mixing that is required on the RGB. We outline here the
pros and cons of the mechanism.

\begin{figure*}
\begin{center}
\includegraphics[width=1.0\columnwidth,angle=-90]{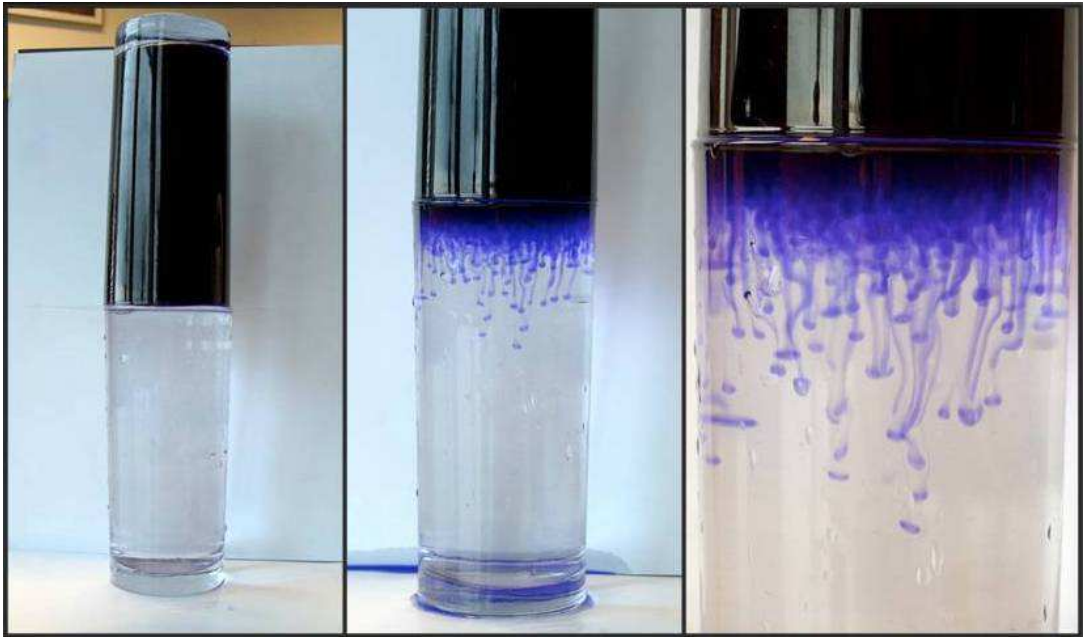}
\caption{A simple experiment in thermohaline mixing that can be performed in the kitchen. Here some 
blue dye has been added to the warm, salty water to make the resulting salt fingers stand out more 
clearly. This experiment was performed by E. Glebbeek and R. Izzard, whom we thank for the picture.}\label{kitchen}
\end{center}
\end{figure*}

The name ``thermohaline mixing'' comes from its widespread occurrence in salt water. Cool water sinks 
while warm water rises. However warm water can hold more salt, making it denser. This means it is
possible to find regions where warm, salty, denser water sits atop cool, fresh, less dense water.
The subsequent development of these layers depends on the relative timescales for the 
two diffusion processes acting in the upper layer -- the diffusion of heat and the diffusion of salt.
For this reason the situation is often called ``doubly diffusive mixing''. In this case the heat diffuses more
quickly than the salt so the denser material starts to form long ``salt fingers'' that penetrate downwards
into the cool, fresh water. Figure~\ref{kitchen} shows a simple example of this form of instability.

We sometimes have an analogous situation in stars. Consider the core He flash. The nuclear burning
begins at the position of the maximum temperature, which is off-centre due to neutrino losses. This fusion
produces $^{12}$C which has a higher molecular weight $\mu$ than the almost pure $^4$He interior to the
ignition point. We have warm, high $\mu$ material sitting atop cool, lower $\mu$ material.
We expect some mixing based on the relative timescales for the heat diffusion and the chemical mixing.
Indeed, this region is Rayleigh-Taylor unstable but stable according 
to the Schwarzschild or Ledoux criteria \citep{grossman93}. This was in fact one of the 
first cases considered in the stellar 
context \citep{ulrich72} although it seems likely that hydrodynamical effects will wipe out 
this $\mu$ inversion before the thermohaline mixing can act \citep{dearborn06,mocak09,mocak10}.
Another common application is mass transfer in a binary system, where nuclearly processed material (of
high $\mu$) is dumped on the envelope of an unevolved companion, which is mostly H 
\citep{stancliffe08}.

\begin{figure*}
\begin{center}
\includegraphics[width=2\columnwidth]{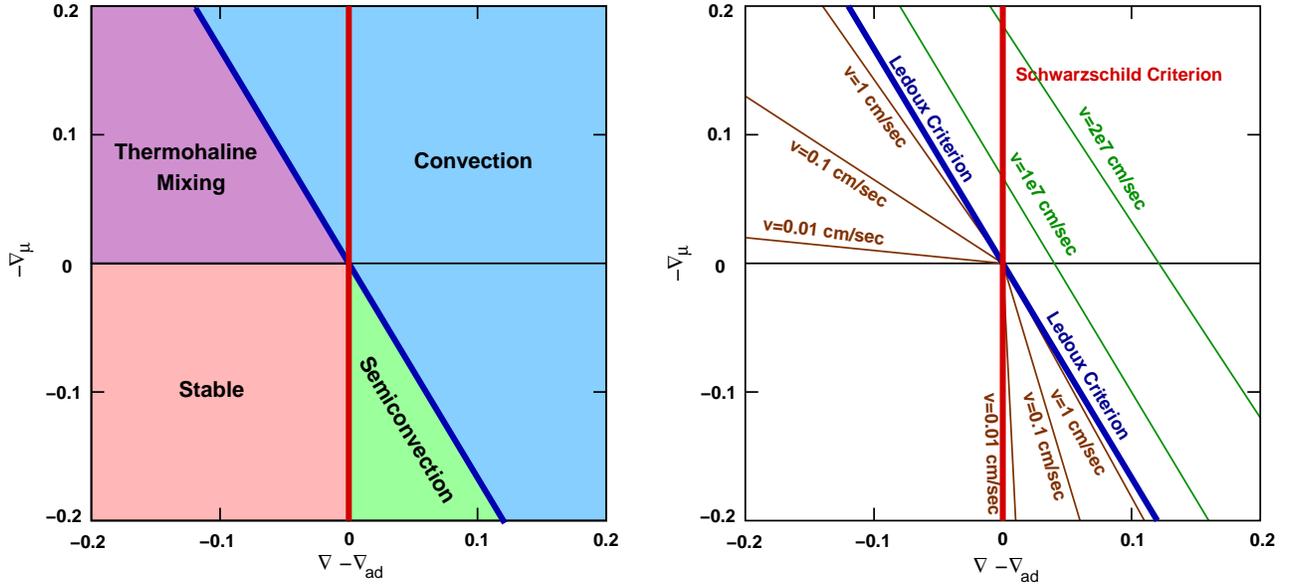}
\caption{These diagrams show the main stability criteria for stellar models, the expected
kind of mixing, and typical velocities. The left panel shows the Schwarzschild criterion
as a red line, with convection expected to the right of the red line. The Ledoux criterion
is the blue line, with convection expected above this line. The green region shows where
the material is stable according to the Ledoux criterion but unstable according to the 
Schwarzschild criterion: this is semiconvection. The magenta region shows that although 
formally stable, mixing can occur if the gradient of the molecular weight is negative. 
The bottom left region is stable with no mixing. The right panel repeats the two stability 
criteria and also gives typical velocities in the convective regime (green lines) as 
well as the thermohaline and semiconvective regions (the brown lines). This figure is based
on Figure~2 in \citet{grossman96}.}\label{grads}
\end{center}
\end{figure*}

One rather nice way of viewing the various mixing mechanisms in stars is given in Figure~\ref{grads},
based on Figure~2 of \citet{grossman96}. In the left panel we give the various stability criteria 
and the types of mixing that result, while the right panel also shows typical mixing speeds.
In both panels the Schwarzschild stability criterion is given by the red line, and it 
predicts convection to the right of this red line. The blue line is the Ledoux criterion, 
with convection expected above the blue line. The green lines show expected velocities 
in the convective regions, while the brown lines show the dramatically reduced 
velocities expected for thermohaline or semi-convective mixing.
For a general hydrodynamic formulation that includes both thermohaline mixing and 
semi-convection we refer the reader to \citet{spiegel72} and \citet{grossman93},
and the series of papers by Canuto \citep{canuto11a,canuto11b,canuto11c,canuto11d,canuto11e}.

\citet{ulrich72} developed a 1D theory for thermohaline mixing that was cast 
in the form of a diffusion co-efficient for use in stellar evolution calculations. This assumed a 
perfect gas equation of state but was later generalised by \citet{kippenhahn80}. These two 
formulations are identical and rely on a single parameter $C$ which is related to the assumed 
aspect ratio $\alpha$ of the resulting fingers via $C = 8/3 \pi^2 \alpha^2$ 
\citep{charbonnel07a}.

\begin{figure*}
\begin{center}
\includegraphics[width=1.8\columnwidth]{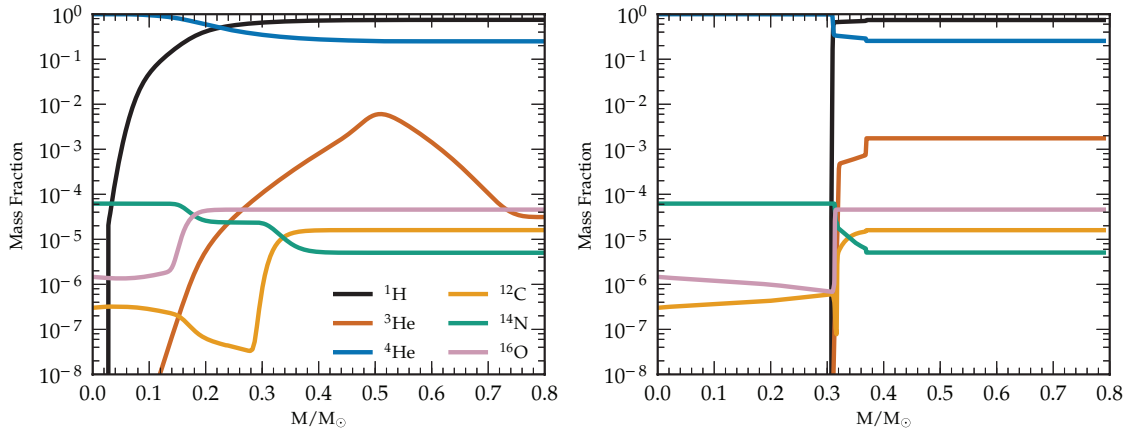}
\caption{Abundance profiles in a $0.8\Msun$ model with $Z=0.00015$ 
(see also Figure~\ref{LindLi}).
The left panel shows a time just after core hydrogen exhaustion and the
right panel shows the situation soon after the maximum inward penetration
of the convective envelope. At this time  the hydrogen burning
shell is at $m(r) \approx 0.31\Msun$ and the convective envelope has
homogenized all abundances beyond $m(r) \approx 0.36\Msun$. The initial
$^3$He profile has been homogenised throughout the mixed region,
resulting in an increase in the surface value, which is then
returned to the interstellar medium through winds, unless some
extra-mixing process can destroy it first.}\label{He3Profile}
\end{center}
\end{figure*}

The relevance of thermohaline mixing for our purposes follows from the work of \citet{eggleton06}.
They found that a $\mu$ inversion developed naturally during the evolution along the RGB. During
main sequence evolution a low-mass star produces $^3$He at relatively low temperatures as a
result of the pp chains. At higher temperatures, closer to the centre, the \iso{3}He is
destroyed efficiently by other reactions in the pp chains. This situation is shown in the
left panel of Figure~\ref{He3Profile}, which shows the profile of $^3$He when the star leaves the
main sequence. The abundance of $^3$He begins very low in the centre, rises to a maximum
about mid-way out (in mass) and then
drops back to the initial abundance at the surface. When the FDU begins it homogenises the composition
profile, as we have seen earlier and as shown in the right panel of Figure~\ref{He3Profile},
and this mixes a significant amount of $^3$He in the
stellar envelope. When the H shell approaches the abundance discontinuity left by the FDU the
first reaction to occur at a significant rate is the destruction of $^3$He, which is at an
abundance that is orders of magnitude higher than the equilibrium value for a region involved
in H burning. The specific reaction is:

\centerline{$^3$He\ $ + ^3$He\ $ \rightarrow$\ $^4$He\ $ + $\ $2$p}

which completes the fusion of He. This reaction is unusual in that it
actually {\em increases\/} the number of particles per volume, starting with two particles 
and producing three. The mass is the same and the reaction reduces $\mu$ locally. This is a 
very small effect, and it is usually swamped by the other fusion reactions that occur in a 
H-burning region. But here, this is the fastest reaction and it rapidly produces a 
$\mu$ inversion. This should initiate mixing.  Further, it occurs just as the H shell 
approaches the discontinuity in the composition left behind by the FDU, i.e., it occurs just 
at the position of the LF bump, in accord with (most) observations. This mechanism has many
attractive features because it is based on well-known physics, occurs in all low-mass stars 
and occurs
at the required position on the RGB \citep{charbonnel07a,eggleton08}.

Having identified a mechanism it remains to determine how to model it.
\citet{eggleton06,eggleton08} preferred to try to determine a mixing
speed from first principles, and try to apply that in a phenomenological way.
\citet{charbonnel07a} preferred to use the existing theory of \citet{ulrich72}
and \citet{kippenhahn80}.
Both groups found that the mechanism had the desired features, in that it
began to alter the surface abundances at the required observed magnitude, it reduced
the C isotope ratio to
the lower values observed, and it destroyed almost all of the $^3$He produced in the star, thus
reconciling the predicted yields of $^3$He with observations \citep{eggleton06,lagarde11,lagarde12a},
provided the
free parameter $C$ was taken to be $C \approx 1000$. Further, mixing reduced the A(Li)
values as required
by observation and it also showed the correct variation in behaviour with metallicity, i.e., the
final $^{12}$C/$^{13}$C ratios were lower for lower metallicity \citep{eggleton08,charbonnel10}.

An extensive study of thermohaline mixing within rotating stars was performed
by \citet{charbonnel10}. They found that thermohaline mixing was far more 
efficient at mixing than meridional circulation, a result also found by
\citet{cantiello10}. Note that the latter authors did not find that thermohaline mixing 
was able to reproduce the observed abundance changes on the RGB, but this is entirely 
due to their choice of a substantially lower value of the free parameter $C$ \citep[see also][]{wachlin11}.
\citet{cantiello10} also showed that thermohaline mixing can continue during 
core He burning as well as on the AGB, and \citet{stancliffe09} found that it
was able to reproduce most of the observed properties of both C-normal and C-rich stars, for
the ``canonical'' value of $C\approx 1000$. 

The interaction between rotation and thermohaline mixing is
a subtle thing, yet both \citet{charbonnel10} and \citet{cantiello10} treated this crudely, 
by simply adding the separate diffusion coefficients. 
This is unlikely to be correct and one can
easily imagine a situation where rotation, or any horizontal turbulence, could decrease the efficiency 
or even remove the thermohaline mixing altogether. Indeed a later study by \citet{maeder13} showed
that simple addition of the coefficients was not correct and these authors provide a formalism for
simultaneously including multiple processes. Calculations using this scheme have 
not yet appeared in the literature.

For a more detailed comparison of predictions with data, \citet{angelou11,angelou12} 
decided to investigate the variation of [C/Fe] with absolute
magnitude in GCs. The C isotopic ratio saturates quickly on the RGB, whereas [C/Fe] and [N/Fe] 
continue to vary along the RGB, providing information on the mixing over a wide range of luminosity.
They found good agreement with the thermohaline mixing mechanism, again provided $C\approx 1000$, although they
also noted that standard models failed to match the FDU found in the more metal-poor GCs, such as M92. 
This is not a failure of the thermohaline mixing paradigm but of the standard 
theory itself (as discussed earlier in \S\ref{sec:onset}).

Clearly the value of the free parameter $C$ is crucial. On the one hand, it is gratifying that
so many observational constraints are matched by a value of $C\approx 1000$. However, the value is
not favoured {\em a priori\/} by some authors. We have seen that within the formalism of the 
idealized 1D theory of \citet{ulrich72} and \citet{kippenhahn80}, the value of $C$ is 
related to the aspect ratio $\alpha$ of the assumed ``fingers'' doing the mixing, 
with $C = 8/3 \pi^2 \alpha^2$. If the mixing is more ``blob-like'' than ``finger-like'' 
then $\alpha \approx 1$ and $C\approx 20$ rather than 1000. This was the case 
preferred by \citet{kippenhahn80}, in fact, whereas \citet{ulrich72} preferred fingers 
with $\alpha\approx 5$ leading to $C\approx 700$, much closer to the value of 1000 
that seems to fit so many constraints. We would caution against a literal interpretation 
of the aspect ratio and finger-like nature of the mixing. The 1D theory is very idealized 
and we feel it is perhaps wise to remember that the diffusion equation is a
convenient, rather than accurate,  description of the mixing.

Even if we assume that the mechanism identified by \citet{eggleton06} is the one driving extra mixing, 
it is unsatisfactory having an idealized, yet approximate, theory that still contains a free parameter.
We need a detailed hydrodynamical understanding of the process. Studies along
these lines have begun but a discussion of that would take us far afield from our main aim in this paper.
We refer the reader to the following papers for 
details: \citet{denissenkov09}, \citet{denissenkov10}, \citet{denissenkov11}, \citet{traxler11},
\citet{rosenblum11}, \citet{mirouh12}, and \citet{brown13}. 
Let us summarise by saying that the models predict more blob-like structures, with
low values of $\alpha$ and $C$ values too small to match the observations. 
The stellar regime is difficult for simulations to model accurately and the final word
is not yet written on the subject.

To summarise, thermohaline mixing occurs naturally at the appropriate magnitude on
the RGB and it provides the right sort of mixing to solve many of the abundance problems
seen on the RGB, as well as the $^3$He problem. One area where the thermohaline mechanism 
is open to criticism is its prediction that low-mass stars should almost completely destroy $^3$He
and yet there are known planetary nebulae (PNe) with large amounts (i.e., consistent 
with the standard models) of $^3$He present, as pointed out by 
\citet[][see also \citealt{guzman13}]{balser07}  immediately
after the paper by \citet{eggleton06}. \citet{charbonnel07b} suggested
that perhaps an explanation could be related to magnetic fields and 
identified the few percent of stars that do not show decreased C 
isotopic ratios with the descendants of Ap stars. They showed that remnant fields of order
$10^4$ -- $10^5$ Gauss, as expected from Ap stars when they become giants, are enough
to inhibit thermohaline mixing. 

\subsubsection{Magnetic Fields and Other Mechanisms}\label{sec:othermech}

Despite its many appealing features, thermohaline mixing still suffers from
at least one major problem:~hydrodynamical models do not support the value 
of $C$ required to match the observations. This leads to the search for other
mechanisms. 

One obvious contender is magnetic fields \citep{busso07b} produced by 
differential rotation just below the convective envelope. This can produce a toroidal
field and mixing by magnetic buoyancy has been investigated by various authors 
\citep[e.g.,][]{nordhaus08,denissenkov09}. 

In contrast, \citet{denissenkov00} suggested that
a combination of meridional circulation and turbulent diffusion could produce the required mixing.
These authors found that the rotation rates required were reasonable but we note that the 
best models of rotating stars at present do not produce enough mixing.

\subsection{Lithium}\label{sec:lithium}

The behaviour of lithium is complex and deserves a special mention. The main isotope of lithium, \iso{7}Li, 
is destroyed by H burning at relatively low temperatures ($T \gtrsim 2.5 \times 10^{6}$K or 2.5~MK) and as such is 
observed to be depleted during the pre-main sequence phase
\citep[see for example,][]{yee10,eggenberger12,jeffries13}. The FDU acts to further reduce the surface 
lithium abundance through dilution with material that has had its lithium previously 
destroyed. It then appears to be further destroyed by extra mixing above the bump on the RGB.
The behaviour of Li in the GC NGC~6397 was discussed earlier in \S\ref{sec:extramix} and 
is shown in Figure~\ref{LindLi}. 

The situation with Li is complicated by the existence of  Li-rich K-giants \citep{charbal00,kumar11,monaco11}.
Approximately 1\% of giants show an enhancement of Li, sometimes a large enhancement to values greater than found 
on the main sequence (say A(Li) $> 2.4$). While some studies argue that these Li-rich giants
are distributed
all along the RGB, others find them clustered predominantly near the bump. \citet{palacios01} suggested that
meridional circulation could lead to a ``Li-flash'' that produces large 
amounts of Li that are only present for a short time, 
making the Li-rich stars themselves relatively rare. Lithium production was found in the
parameterised calculations of extra mixing by \citet{sackboo99}, for the case where the mixing 
(as measured by a mass flux in their case)
was fast enough. \citet{denher04} also found that Li could be produced through sufficiently rapid mixing.
Within the approximation of diffusive mixing, they found that a value
of $D\approx 10^9$cm$^2$s$^{-1}$ is required to explain the usual abundance changes beyond the bump on the RGB, 
but a value about 100 times larger was shown to produce Li.

It seems natural that if the Li-rich stars really are created at
all points along the RGB then the cause is most likely external to the star.
If this requires an increase in the diffusion coefficient then something like a
binary interaction or the engulfing of a planet due to the growth of the stellar envelope
may be involved \citep{siess99a,siess99b,denher04,carlberg09,carlberg10}.

%% Follow up in prep

\citet{kumar11} argue that in fact the Li-rich giants
are predominantly clump stars, involved in core He burning. They argue that the Li may be produced
at the tip of the giant branch at the core flash, and then given the longer evolutionary timescales for 
core He burning, the stars appear to be clustered around the clump (at a similar luminosity to the bump).
In addition to the cases noted above that produce Li, we note here that some implementations of 
thermohaline mixing also produce Li at 
the tip of the giant branch and that there is evidence for Li-rich stars at this phase of 
the evolution \citep{alcala11}. 
%A discussion of Li in this context can be found in Church et al. (2014, in preparation). 

\subsection{Second Dredge-Up}\label{sec:SDU}

We have seen that following core He exhaustion the star begins to ascend the AGB. If the mass is
above about $4\Msun$ then the model will experience the SDU where the convective envelope grows 
into the stellar interior.
In contrast to the FDU, the SDU goes deeper and mixes material exposed to complete H burning. 
The main changes are listed in Table~\ref{fdusduvalues} and include a
substantial increase in the He content by up to $\Delta Y \approx 0.1$ as well as an increase in the 
\iso{14}N/\iso{15}N ratio and the \iso{23}{Na} abundance (Figure~\ref{fdumasses}). 

This huge increase in He is one of the reasons why
intermediate-mass AGB stars have been implicated in the origin of the multiple populations observed 
in Galactic GCs \citep{dantona02,norris04,piotto05}. Increases in He and changes to the 
composition of the light elements C, N, and O are the most likely cause of the multiple main sequence, 
sub-giant, and giant branches observed in all clusters that have Hubble Space Telescope photometry 
\citep[e.g., for 47 Tucanae, NGC~6397, NGC~2808, M22;][]{milone12a,milone12b,milone12c,piotto12} 

Figure~\ref{fdumasses} illustrates the effect of the SDU on the abundance of 
intermediate-mass stars of solar metallicity. Figure~\ref{depthfdu} shows that the depth reached 
by the SDU is approximately the same for all the 5 and 6$\Msun$ models, regardless 
of the initial metallicity \citep{boothroyd99}. The effect of the SDU on other elements 
is small. \citet{boothroyd94} showed that the O isotope 
ratios are essentially unchanged. Small decreases in the surface abundance of fluorine may occur by at 
most 10\% and sodium is predicted to increase by up to a factor of $\approx 2$ at the surface of 
intermediate-mass stars that experience the SDU \citep[e.g., Figure~8][]{eleid95,forestini97}. 

\subsection{Variations at low metallicity} 

\subsubsection{Curtailing First Dredge-Up}

\begin{figure*}
\begin{center}
\includegraphics[width=8cm, angle=270]{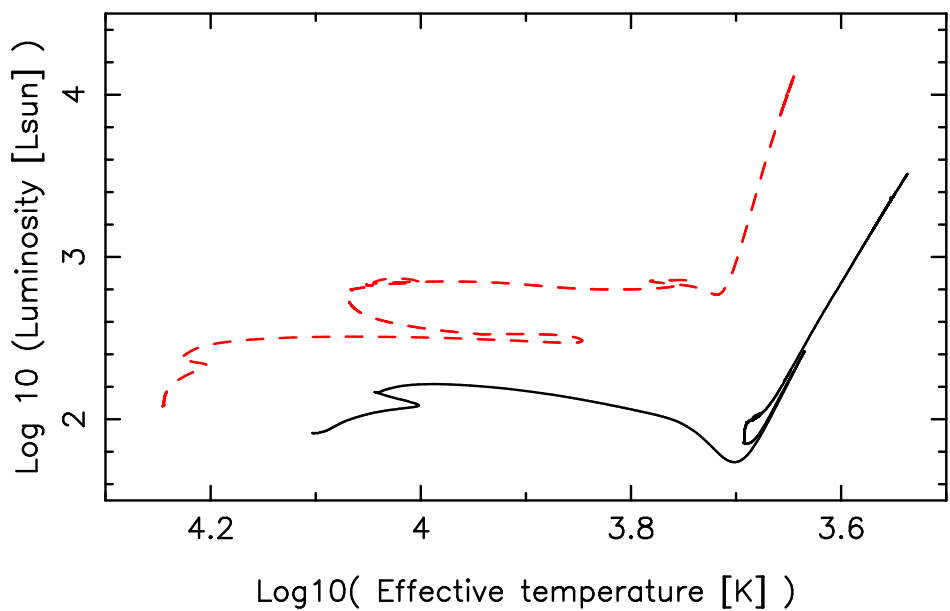}
\caption{Hertzsprung-Russell (HR) diagram showing the evolutionary tracks for 3$\Msun$ models
of $Z=0.02$ (black solid line) and $Z=0.0001$ (red dashed line). The low-metallicity model is 
hotter and brighter at all evolutionary stages and does not experience a RGB phase or the 
first dredge-up.}\label{lowZhr}
\end{center}
\end{figure*}

At metallicities of [Fe/H] $\lesssim -1$, the evolution of post main sequence stars begins to significantly 
differ to that found in disk or near-solar metallicity stars. For intermediate-mass stars the necessary central
temperatures for core He burning are reached while the star crosses the Hertzsprung gap. This means that
the star will ignite He before evolving up the RGB and as a consequence will not experience the FDU
\citep[e.g.,][]{boothroyd99,marigo01b}. In Figure~\ref{lowZhr} we show evolutionary tracks for models of 3$\Msun$ 
at a metallicity of $Z=0.02$ and $Z=0.0001$, respectively. The lowest metallicity model skips the RGB 
altogether.  At a metallicity of $Z=0.001$ or [Fe/H] $\approx -1.2$ the upper mass limit that experiences the 
FDU is 3$\Msun$, by a metallicity of $Z = 0.0001$ or [Fe/H] $\approx -2.3$, this mass has been reduced to 2.25$\Msun$,
and even at that mass the maximum extent of the convective envelope only reaches a depth of $\approx 1\Msun$ from the
centre (compared to a $Z=0.02$ model of 2.25$\Msun$ where the FDU reaches a depth of $\approx 0.35\Msun$ from
the stellar centre). Because the lower metallicity model is hotter and has a larger core during the
main sequence the FDU still causes a 30\% drop in the surface C abundance (compared to a drop 
of 36\% for the $Z=0.02$ model). For intermediate-mass stars that do not experience a FDU, the second
dredge-up event, which takes place during the early ascent of the AGB, is the first mixing episode
that changes the surface composition \citep{boothroyd99,chieffi01,marigo01b,karakas07b,campbell08}.

\subsubsection{The Core Helium Flash}

Metallicity has an important consequence for stars that experience the core He flash. Evolution at
lower metallicities is hotter, owing to a lower opacity. This means that the stars experience a shorter
time on the RGB before reaching temperatures for core He ignition and therefore do not 
become as electron degenerate. This means that the maximum mass for the core He flash decreases with 
decreasing $Z$ \citep{marigo01b}. We mentioned previously that at solar metallicity the maximum
mass for the core He flash is $2.1\Msun$, whereas at [Fe/H] = $-2.3$ the maximum mass is 
1.75$\Msun$, with the 2$\Msun$ model experiencing a fairly quiescent He ignition with only a 
moderate peak in the He luminosity \citep{karakas07b,karakas10a}.

There is now a fairly extensive literature on multi-dimensional studies of the core He flash 
\citep{deupree84,deupree86,deupree87,deupree96,dearborn06,mocak09}.
Early results from two-dimensional hydrodynamic simulations \citep{deupree87} suggest
that the flash could be a relatively quiescent or violent hydrodynamic event, depending on the 
degeneracy of the stellar model. \citet{deupree96} finds for low-mass solar composition
models, using improved but still uncertain input physics, that the core flash is not a violent 
hydrodynamic event and that there is no mixing between the flash-driven H-exhausted core 
and the envelope at this metallicity. More recent multi-dimensional hydrodynamic simulations 
of the core He flash in metal-free stars find that the He-burning convection zone 
moves across the entropy barrier and reaches the H-rich layers  \citep{mocak10,mocak11}.

For the present we assume that stars experiencing the core He flash do not
mix any products into their envelope. This is in accord with standard models and the
observations do not disagree. We do note that this is not necessarily true at lower metallicity, 
as we discuss in the next section.

\subsubsection{Proton Ingestion Episodes}\label{sec:PIE1}

The low entropy barrier between the He- and H-rich layer can lead to the He
flash-driven convective region penetrating the inner edge of the (now extinguished) H-burning shell.
If this happens, protons will be ingested into the hot core during the core He flash. 
If enough protons are ingested, a concurrent secondary flash may occur that is powered by H burning 
and gives rise to further nucleosynthesis in the core. The subsequent dredge-up of matter enriches the
stellar surface with large amounts of He, C, N, and even possibly heavy elements
synthesised by the $s$ process. There has been an extensive number of studies of
the core He flash and resulting nucleosynthesis in low-mass, very metal-poor stars 
\citep[e.g.,][]{dantona82,fujimoto90,hollowell90,schlattl01,picardi04,weiss04,suda07,campbell08,suda10,campbell10}. 
The details of the input physics used in the calculations clearly matter, where the low-mass $Z=0$ 
models of \citet{siess02} find no mixing between the flash-driven convective region and the overlying
H-rich layers. It has been known for sometime that the treatment of the core He flash
in one-dimensional stellar evolutionary codes is approximate at best, owing to the fact that 
the core He flash is a multi-dimensional phenomenon \citep{deupree96,mocak11}. 
We deal more extensively with this in \S\ref{sec:PIE2}.

\section{Evolution and Nucleosynthesis during the Asymptotic Giant Branch} \label{sec:agb}

\begin{figure*}[t]
\centering
\includegraphics[height=8cm,angle=0]{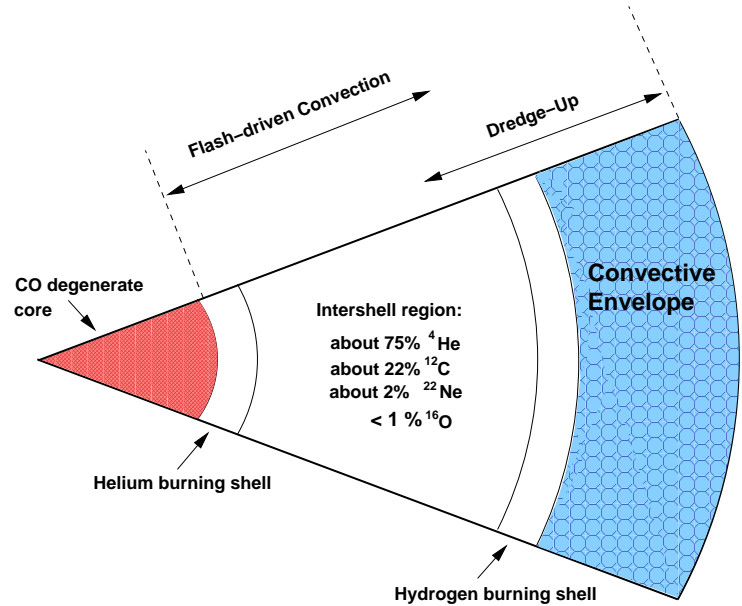}
%\includemovie[
%  poster=agb.eps
%  %text={\small(Loading AGB.avi)}
%]{12cm}{8cm}{AGB.avi}
\caption{Schematic structure of an AGB star showing the electron-degenerate core surrounded by a helium-burning shell 
above the core, and a hydrogen-burning shell below the deep convective envelope. The burning shells are separated by an 
intershell region rich in helium ($\sim 75\%$) and carbon ($\sim 22\%$), with some oxygen and 
\iso{22}Ne. A super-AGB star has an O-Ne degenerate core otherwise the qualitative schematic structure 
remains the same. From \citet*{karakas02}.}
%Click on the image to run an animation of a pulse cycle.}
\label{agb-struct}
\end{figure*}

The mass of the H-exhausted core (hereafter core mass) at the end of core He-burning is 
the prime determinant of many important features of AGB evolution including luminosity and 
nucleosynthesis
\citep[e.g.,][]{dominguez99,imbriani01,straniero03b,ekstrom12,halabi12,valle13}. 
When the star begins to  ascend the AGB the core 
becomes increasingly electron degenerate and the star's energy output is 
mostly provided by He burning, which proceeds through 
the material outside the C-O core as a thin He-burning shell is established.
In intermediate-mass stars the H shell 
is extinguished, which allows the inward movement of the convective envelope and the SDU. 
It is at this time that middle and massive 
intermediate-mass stars ignite C in the C-O core, which results in an O-Ne core prior to the start of the 
thermally-pulsing phase. While the evolution of AGB stars with O-Ne cores (super-AGB stars) 
is qualitatively similar to C-O core AGB stars, we discuss these objects separately in \S\ref{sec:superagb}.

The He shell thins as the star evolves up the AGB and eventually becomes thermally unstable. 
At the first thermal instability of the He shell (also known as a ``thermal pulse'' or ``shell flash'') 
the star is said to have entered the thermally-pulsing-AGB (or TP-AGB) phase. 
The evolution along the AGB prior to the first instability is referred to as the early AGB phase.
The structure of an AGB star, illustrated in Figure~\ref{agb-struct}, is qualitatively the same for all masses.
We now focus on the thermally-pulsing AGB phase of evolution, which alters the surface abundances 
of the models in two distinct and important  ways. The first is through the operation of the 
{\em third dredge-up} (TDU), which can  occur periodically after each thermal pulse and is the mechanism 
for turning (single) stars into C-rich stars.  The second mechanism is {\em hot bottom burning}.

\subsection{The thermally-pulsing Asymptotic Giant Branch}

\begin{figure*}
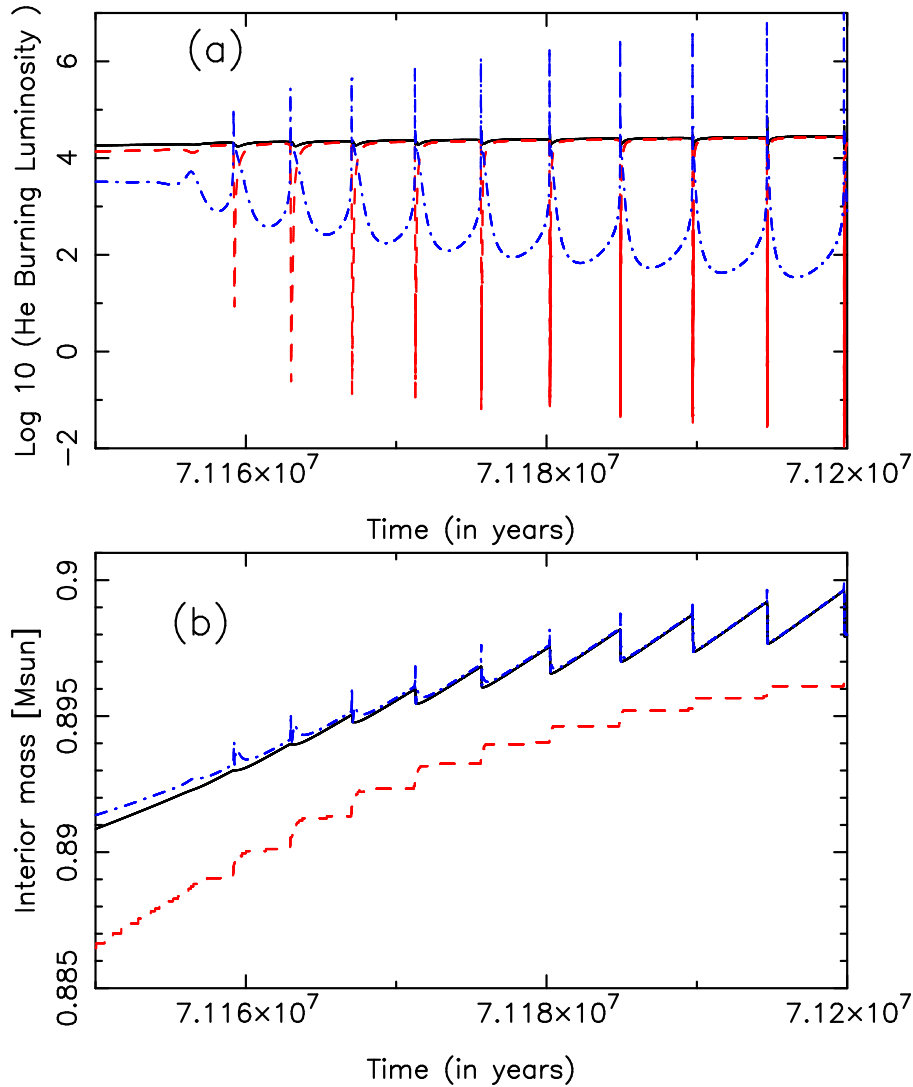

\centering
\includegraphics[height=12cm,angle=270]{m6z02_tp1.ps}
\includegraphics[height=12cm,angle=270]{m6z02-core1.ps}
\caption{Evolution of the luminosities and core masses (in solar units) for a 6$\Msun$, $Z = 0.02$ model 
during the start of the TP-AGB. Each panel shows the evolution during the first 10 thermal pulses.
The panel (a) shows the surface (or radiated) luminosity (black solid line), H-burning shell luminosity 
(blue dot-dashed line), and He-burning shell luminosity (red dashed line). Panel (b) shows the masses of the 
H-exhausted core (black solid line), He-exhausted core (red dashed line), and the
inner edge of the convective envelope (blue dot-dashed line).}
\label{m6z02-tp1}
\end{figure*}

Here we briefly review the main features of AGB evolution.  Previous reviews include \citet{iben91}, \citet{frost96proc}, 
\citet{wood97}, \citet{busso99}, and more recently \citet{herwig05}. 

The thermally-pulsing AGB phase of evolution is 
characterised by relatively long periods of quiescent H-shell burning, known as the interpulse phase, interrupted by 
instabilities of the He-burning shell.  Helium burning is ignited at the base of the He-rich intershell region 
(see Figure~\ref{agb-struct}), which is composed of material exposed to previous He-shell flashes plus the 
ashes of H-shell burning which have accumulated over the previous interpulse phase.  
The He shell burns fiercely and can produce  $\gtrsim 10^{8}\Lsun$ for a short time. 
Figure~\ref{m6z02-tp1} shows the luminosities of the H and He shells, along with the surface luminosity 
for a 6$\Msun$, $Z =0.02$ model star during the first 10 thermal pulses. This figure illustrates the 
beginning of the TP-AGB phase where the strength of thermal pulses grows with time owing to the overall 
contraction of the H-exhausted core, which leads to hotter, more electron
degenerate conditions in the burning shells. For the 6$\Msun$ model shown in Figure~\ref{m6z02-tp1} the luminosity produced 
by the He shell is already $\approx 10^{6}\Lsun$ by the third thermal pulse, 
a figure that grows to over $4 \times 10^{8}\Lsun$ by the final thermal pulses.

\begin{figure*}
\centering
\includegraphics[height=12cm,angle=90]{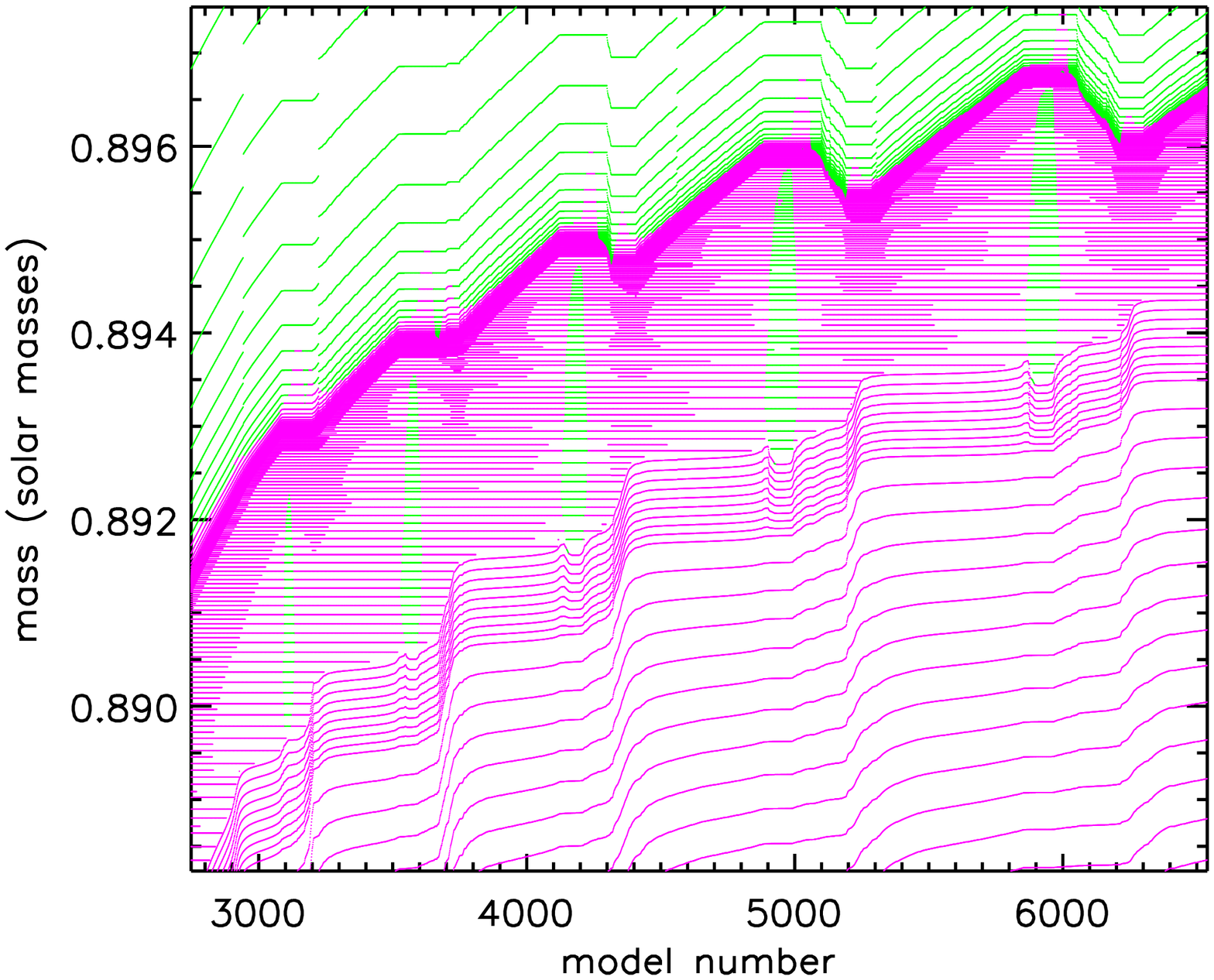}
\caption{Convective regions for the 6$\Msun$, $Z=0.02$ model during the first five thermal pulses. 
The $x$-axis is nucleosynthesis time-step number, which is a proxy for time.
For each model, along the x-axis, a green dot represents a 
convective mass shell and a magenta dot is a radiative shell. The dense magenta regions mark the H and He shells.
The teardrop-shaped pockets correspond to the flash-driven convective region that extends over most of the intershell.
These have the effect of homogenising the abundances within the intershell.  For this model, 
the duration of the convective zones is about 25~years and the interpulse periods about $\approx 4000$ years.}
\label{m6z02-conv}
\end{figure*}

The energy produced by the flash powers a convective region which begins in the He-burning shell and extends 
almost all the way to the H-burning shell. This has the effect of homogenising
abundances in this region.  In Figure~\ref{m6z02-conv} we show the flash-driven convective regions in the intershell during 
the first 5 thermal pulses of the 6$\Msun$, $Z=0.02$ model.   The teardrop-shaped green pockets represent 
flash-driven convection, which lasts for $\lesssim 10^{2}$ years, depending on the core mass. Convection 
in the intershell retreats once the energy from the thermal pulse starts to die down, in what is referred 
to as the power-down phase. During power down, the huge
amount of energy produced by the thermal pulse does not reach the stellar surface but goes into expanding the star, 
which cools the material outwards of the He shell and shuts off the H shell. 
The power-down phase is seen in 
the decrease of the H-shell luminosity in Figure~\ref{m6z02-tp1}. Note that while the H-shell luminosity drops
by many orders of magnitude, the surface luminosity is only seen to dip by about 10\% during the expansion stage.

The cooling of these inner layers leads to an increase in the stellar opacity, which allows the base of the outer 
convective envelope to move inwards in mass, interior to the erstwhile H-burning shell, 
and to regions previously mixed by intershell convection.  
Hence material from the interior, which has been exposed to He burning, is mixed into the envelope where it can
be observed at the surface. This phase is known as third dredge-up and may occur after each thermal pulse. 

In Figure~\ref{m6z02-tp1} we show the evolution of the masses of the
H-exhausted core, of the He-exhausted core, and of the inner edge of the convective envelope for the 6$\Msun$, $Z=0.02$
model during the first 10 thermal pulses. By the third thermal pulse we can see a small temporary 
decrease in the mass of the H-exhausted core, which
is seen more clearly in Figure~\ref{m6z02-conv}. The decrease is caused by the TDU, where 
the inner edge of the convective envelope penetrates into the top layers of the He intershell.
This mixes H into a H-poor region, reducing the mass of the core while mixing the products of 
H and He nucleosynthesis into the envelope. Following the third dredge-up, the star contracts, the H-burning shell 
is re-ignited and the star enters a new interpulse phase. 
The cycle of {\em interpulse--thermal pulse--power-down--dredge-up} may occur many times on the AGB, 
depending on the initial mass, composition, and mass-loss rate. For the 6$\Msun$, $Z = 0.02$ model, this 
cycle occurs 42 times before the model experienced convergence problems of the type discussed by \citet{lau12}
and calculations were terminated. 

In summary, the AGB evolutionary cycle can be broken down into four distinct phases \citep{iben81}:
\begin{enumerate}
\item {\bf Thermal pulse}, which is when the He shell burns brightly producing $\lesssim 10^{8}\Lsun$ for a
short time ($\approx 10^{2}$ years). The energy drives a convective zone in the He intershell.
\item {\bf The power-down phase} when the He shell dies down. The enormous amount of energy from the thermal pulse 
drives an expansion of the whole star and the H shell is extinguished.
\item {\bf Third dredge-up} phase, which is when the outer convective envelope may move 
inwards into regions previously mixed by flash-driven convection. Carbon and other He-burning products 
are mixed to the stellar surface. 
\item {\bf The interpulse phase} is the relatively long ($\approx 10^{4}$ years) phase in between thermal 
pulses where the H shell provides most of the surface luminosity. 
\end{enumerate}

\subsection{Hot bottom burning}

It has been known for some time that intermediate-mass stars over about 5$\Msun$ develop deep convective envelopes 
with very  high temperatures at the base, allowing for nuclear burning and some energy generation 
\citep{scalo75,lattanzio92,lattanzio96}. 
In fact, what happens is that the bottom of the convective envelope is situated near the top of the H-burning shell.
The observational evidence that HBB is occurring in intermediate-mass AGB stars came 
from the lack of optically-bright C-rich stars in the Magellanic Clouds \citep{wood83}. 
Many of these luminous, O-rich stars were later found to be rich in lithium \citep{smith89,smith90b,plez93}.  
The first detailed calculations were made in the early 1990's by \citet{bloecker91} and \citet{lattanzio92}.  
\citet{boothroyd93} found that HBB prevents the formation of a C-rich atmosphere by burning \iso{12}C into 
\iso{14}N thus providing a mechanism for the lack of bright C-stars in the Magellanic Clouds.

HBB can dramatically alter the surface composition. This is because the temperature in a thin region at the very base
of the envelope, hereafter $T_{\rm bce}$,  can exceed $50 \times 10^{6}$K (50 MK), which is hot 
enough for activation of the CNO cycle and also the Ne-Na and Mg-Al chains (if the temperature is high enough). 
In the most massive, lowest metallicity AGB models, and super-AGB models,
the temperature can exceed 100 MK \citep[e.g.,][]{karakas07b,ventura13,doherty14a}. 

\begin{figure}
\centering
\includegraphics[height=8cm,angle=270]{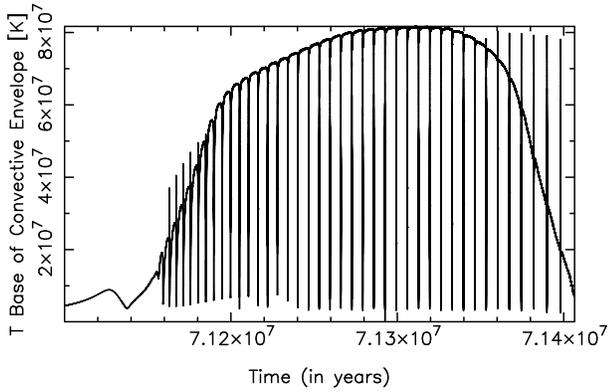}
\caption{The evolution of the temperature at the base of the convective envelope in the 6$\Msun$, $Z = 0.02$ model.}
\label{m6z02-tbce}
\end{figure}

The convective envelopes of AGB stars are well mixed, with a convective turnover time of about 1 year, 
which means that the whole envelope will be exposed to the hot region at least 1000 times per interpulse period.
In Figure~\ref{m6z02-tbce} we show the evolution of the temperature at the base of the convective envelope for the 
6$\Msun$, $Z = 0.02$ model. The temperature reaches a peak of 82 MK at the 28$^{\rm th}$ thermal pulse, 
before decreasing to below 20 MK, at which point HBB has been shut off.  The decrease in temperature is 
caused by mass loss, which slowly erodes the envelope. The minimum envelope mass required to support HBB is 
about 1$\Msun$, depending on the metallicity.

\subsubsection{Dredge-up, HBB and the Brightest C Stars}

Third dredge-up can continue after the cessation of HBB, which allows the C abundance to increase instead of 
being burnt to N. The envelope mass is relatively small at this stage ($\lesssim 1\Msun$), which means that 
dilution is also lower.  It is possible the star will become C-rich at the very tip of the AGB where 
C/O $\ge 1$ \citep{frost98a}, depending on the number of TDU episodes after the end of HBB and the O 
abundance in the envelope (noting that some O can also be destroyed by HBB). \citet{vanloon99a} presented 
observational evidence that supports  this scenario, finding a sample of very luminous, dust-obscured 
AGB stars in the Magellanic Clouds. The existence of very bright, C-rich AGB stars is also evidence 
that stars in this mass range experience TDU, at least down
to the metallicities of the Magellanic Clouds.

\subsubsection{The Core-Mass vs Luminosity Relation}

\begin{figure*}
\centering
\includegraphics[height=12cm,angle=270]{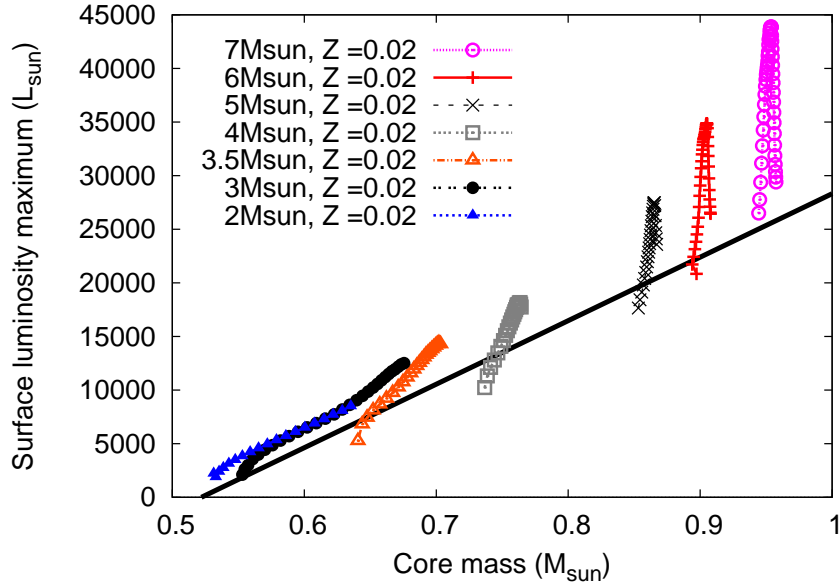}
\caption{The core-mass versus luminosity relationship for a selection of $Z = 0.02$ models between 2$\Msun$ and 7$\Msun$. 
The models with $M \ge 4.5\Msun$ have hot bottom burning and deviate from the 
Paczy\'{n}ski relation, shown by the solid black line.}
\label{coremasslum}
\end{figure*}

The surface luminosity will reach a maximum value during the interpulse and this is reached just before the 
onset of the next thermal pulse. \citet{pacz70} was the first to derive 
a linear relationship between the maximum surface luminosity during the quiescent interpulse phase
and the H-exhausted core mass
\begin{equation}
  L / \Lsun = 59250 \, ( M_{\rm H}/\Msun - 0.522 ). \label{eq:paczynski}
\end{equation}
Paczy\'{n}ski's calculations infer that there is a maximum luminosity an AGB model can have of 52021$\Lsun$, 
determined by the maximum possible core mass of $\approx 1.4\Msun$. Subsequent calculations of intermediate-mass 
AGB stars revealed that hot bottom burning can violate the conditions of the core-mass luminosity relationship 
\citep{bloecker91,lattanzio92,boothroyd92} and that an AGB model can have luminosities larger than predicted by 
Paczy\'{n}ski. The core-mass luminosity relationship on the AGB is also a key ingredient in a synthetic AGB model because 
it determines many fundamental features of AGB evolution including the growth of the  H-exhausted core with time.
The most accurate fits to the core-mass luminosity relationship are those that include a correction for the extra 
luminosity provided by HBB \citep{wagenhuber98,izzard04b,izzard06}.  In Figure~\ref{coremasslum} we show the core-mass 
luminosity relationship for a selection of models at $Z=0.02$. HBB occurs in models with $M \ge 4.5\Msun$ at $Z=0.02$ 
and we can see for these models that the luminosity strongly increases with core mass before peaking and then declining. 
The decline is caused by mass loss, which reduces the temperature at the base of the convective envelope and in turn
the luminosity from the CN cycle. In comparison, the lower mass AGB models do not experience HBB and show 
a reasonably linear core-mass luminosity relationship. 

\subsection{Third dredge up} \label{sec:tduevol}

The chemical enrichment at the surface of AGB stars is governed by the TDU mixing event that follows
a thermal pulse. TDU is responsible for the largest changes to the surface composition of low-mass AGB stars and 
has important consequences for nucleosynthesis in intermediate-mass AGB stars as well owing to the production of primary
C which is converted to primary N by HBB.

\subsubsection{The Dredge-Up Parameter}

\begin{figure}
\centering
\includegraphics[height=5cm,angle=0]{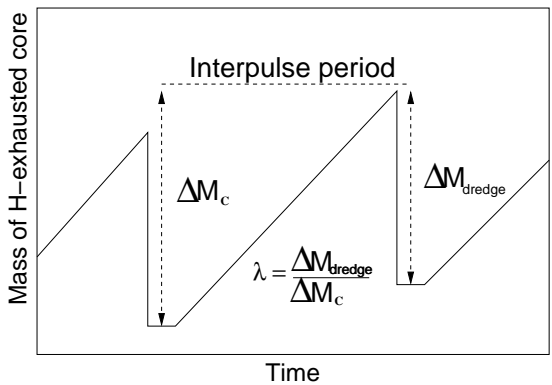}
\caption{The definition of $\lambda$, shown schematically, where the 
$x$--axis represents time and the $y$--axis represents the mass of the 
H-exhausted core.}
\label{lambda}
\end{figure}

If there is dredge-up, then a fraction of the outer-most part of the 
H-exhausted core will be mixed into the envelope according to
\begin{equation}
  \lambda  = \frac{\Delta M_{\rm dredge}} {\Delta M_{\rm core}}, 
\end{equation}
where $\lambda$ is the third dredge-up efficiency parameter, $\Delta M_{\rm dredge}$ is the mass mixed into the envelope,
and $\Delta M_{\rm core}$ is the amount by which the H-exhausted core increases over the previous interpulse phase;
see Figure~\ref{lambda}. From the above definition, when $\lambda = 1$, the core mass does not grow 
from pulse to pulse but remains constant. 
The value of $\lambda$ depends on physical parameters such as the core mass and metallicity of the star. 
Exactly how $\lambda$ depends on these quantities is unknown and reflects our lack of understanding
about how convection operates in stellar interiors. Different stellar codes predict different behaviour, a point
we will come back to in \S\ref{sec:tdu}. Note that there is no {\em a priori\/} reason why $\lambda$ cannot exceed unity.

So in summary, AGB nucleosynthesis depends on 
\begin{enumerate}
\item{} $\lambda$, the efficiency of third dredge-up;
\item{} $M_{\rm c}^{\rm min}$, the minimum core mass at which the TDU begins; this determines how many 
TDU episodes will occur before mass loss removes the envelope;
\item{} the size of the convective envelope, which sets the level of dilution of each TDU episode;
\item{} the mass of the He intershell. 
\end{enumerate}

\citet{karakas02} provided the first parameterisation of 
$\lambda$ and $M_{\rm c}^{\rm min}$ as functions of the total mass, envelope mass, and metallicity 
\citep[see also][]{straniero03}.  The general trend is that $\lambda$ increases with increasing
stellar mass, at a given $Z$. The parameter $\lambda$ also increases with decreasing metallicity, at a given mass 
\citep[e.g.,][]{boothroyd88c}. Naively, this means that it should be easier to make C stars in lower metallicity 
or higher mass models. But there is also a second reason why C-stars are more easily made at lower metallicity.
Carbon production is a primary product of the triple-$\alpha$ process, and hence the C intershell abundance 
does not depend on the global metallicity, $Z$.
In a low metallicity star the amount of C added per pulse is roughly independent of metallicity, 
whereas the amount of O that must be overcome to produce C $>$ O is lower. So fewer pulses are required to make a C-star.
This is true even for $\alpha$-enhanced compositions where [O/Fe] $\approx +0.4$. As we note, stars of lower metallicity 
are predicted to have deeper dredge-up which accelerates the effect further. Both mechanisms 
act to make C-stars easier to form at lower metallicities.

For intermediate-mass stars the situation is more complex. Even though the calculations of \citet{karakas02} predict 
larger values of $\lambda$, the effect of TDU is mitigated by the mass of the He intershell which is approximately a 
factor of 10 smaller in intermediate-mass AGB stars compared to AGB stars of lower mass. 
This means that even if $\lambda \approx 0.9$, the amount of material added to the envelope per TDU event 
($\lambda \times \Delta M_{\rm core}$) is smaller by about an order of magnitude. 
Secondly, the mass of the convective envelope is large in intermediate-mass stars, which means that
the material will be more diluted. Finally, HBB will act to prevent the formation of C-rich luminous 
AGB stars at the highest metallicities \citep[e.g.,][]{karakas12}, which is in agreement with observations of O-rich 
luminous AGB stars in our Galaxy \citep{garcia06,garcia13}. So even though the conventional wisdom is that C-star production
does not happen at intermediate-mass for the above reason, calculations  of low-metallicity ($Z\le 0.001$) 
intermediate-mass stars suggest that C-star formation will occur before HBB ceases 
\citep{herwig04a,herwig04b,karakas10a,lugaro12,fishlock14,straniero14}. 
This is driven by the combination of primary C production plus the effect of HBB destroying some O.

It is important to know if the stellar models are providing an accurate description of the efficiency of mixing in AGB
stars. For example, the models of \citet{karakas02} do not predict any TDU for models less than 2$\Msun$ at $Z=0.02$
and quite efficient TDU for models of intermediate-mass. While it is notoriously difficult to determine the masses of 
stars in our Galaxy, observations suggest that the minimum initial stellar mass for C-star formation is 
$\approx 1.5\Msun$ \citep{wallerstein98}. The Large and Small Magellanic Clouds (LMC, SMC, respectively) are the closest
satellite galaxies of our Milky Way and they both have thousands of known C stars \citep{frogel90,groen04}.  
We know the distances to these two galaxies reasonably well, enabling us to construct C-star luminosity 
functions (CSLFs). 

\subsubsection{The Carbon Star Luminosity Function and other Observational Constraints}

Is it possible to constrain the efficiency of third dredge-up by using the CSLF of the LMC and the SMC?
That the stellar luminosity on the AGB is a nearly linear function of the H-exhausted core mass 
(e.g., Equation~1 and Figure~\ref{coremasslum}) has stimulated the development of {\em synthetic} 
AGB evolution models, as a quick way of simulating populations of AGB stars. The main observational 
constraint which models must face is the CSLF for the Magellanic Clouds. 

\begin{figure}
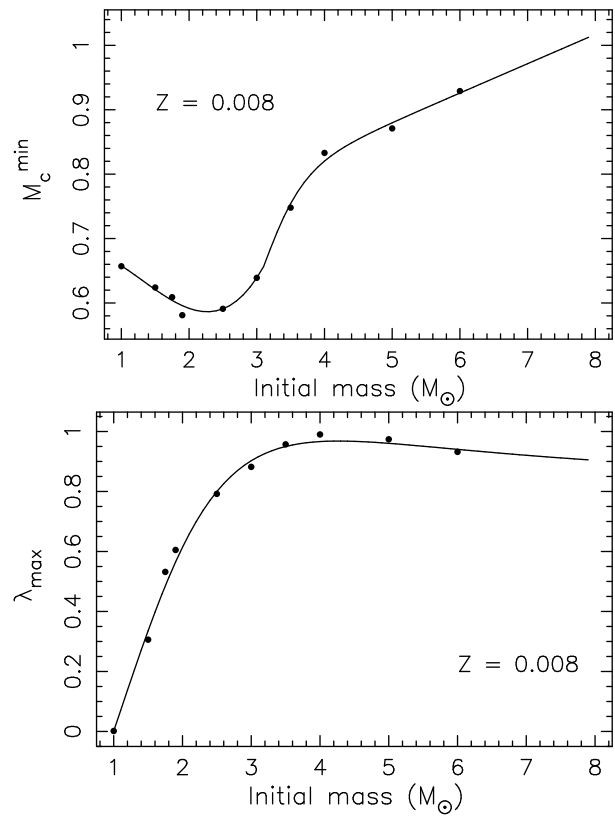

\begin{center}
\includegraphics[height=8cm,angle=270]{mcmin_z008.ps}
\includegraphics[height=8cm,angle=270]{lambdamax_z008.ps}
\caption{The minimum core mass for TDU (upper panel) and the maximum value of $\lambda$ plotted 
against initial mass for the $Z=0.008$ models from \citet{karakas02}. Only models with 
$M \ge 1.9\Msun$ become C-rich. Figure taken from \citet{karakas02}.} 
\label{z008}
\end{center}
\end{figure}

Synthetic AGB evolution calculations performed by \citet{groen93} and \citet{marigo96}, treat $\lambda$ as 
a constant free parameter, calibrated by comparison with the CSLF.  Synthetic AGB calculations designed to 
reproduce the CSLF in the LMC and SMC require an average $\lambda = 0.5$ and $\lambda =0.65$ respectively, 
and $M_{\rm c}^{\rm min}$ $\approx 0.58\Msun$ \citep{groen93,marigo99,izzard04b,marigo07}.
The values for $M_{\rm c}^{\rm min}$ are lower than found in detailed models by e.g., \citet{karakas02} and shown 
in Figure~\ref{z008} and the synthetic best fit values for $\lambda$ are higher than those found for
the low-mass AGB models that become C-rich (e.g., $\lambda \lesssim 0.4$ in Figure~\ref{z008}) unless
considerable overshoot is applied \citep[e.g.,][]{herwig00,cristallo09,weiss09}.

\citet{stancliffe05a} were able to reproduce the CSLF of the LMC by computing AGB models without convective
overshoot using the Cambridge {\em STARS} code, which predicts deeper TDU at smaller core masses than 
\citet{straniero97} or \citet{karakas02}. However the CSLF in the SMC cannot be reproduced from detailed 
{\em STARS} models, indicating that the problem is not yet fully solved.

Star clusters are ideal sites to constrain the TDU in AGB stars as they contain stars of similar age and metallicity.
Open clusters of solar metallicity in the Milky Way Galaxy can be used to study the TDU and the growth of the 
core using observations of white dwarf masses in comparison to theoretical models \citep{kalirai14}.
Star clusters in the Magellanic Clouds in particular prove very valuable as they span a wide range of age, 
which enables us to study the evolution of stars with masses around 1.5 to 5$\Msun$ 
\citep[e.g.][]{girardi95,girardi09,maceroni02,mucciarelli06,mackey08,milone09}. 

The clusters NGC\,1978 and NGC\,1846 in the LMC have been the subject of much study owing to the availability 
of accurate estimates of AGB structural parameters such as pulsation masses, effective temperatures,
and luminosity. \citet{kamath10} obtained (current day) pulsation masses for NGC 1978 
(and the SMC cluster NGC 419),
while \citet{lebzelter07} obtained masses for the LMC cluster NGC 1846. Abundance studies have also been done
\citep{ferraro06,muccia08,lebzelter08,lederer09b}, along with attempts to explain the observed C and O abundances for 
NGC\,1978 and NGC\,1846 using stellar evolution models \citep{lebzelter08,lederer09b}.
 
\citet{kamath12} presented stellar models for AGB stars in NGC\,1978, NGC\,1846, and NGC\,419 with the aim of 
constraining the TDU and mass loss on the AGB.  The stellar evolution 
models were constrained to reflect the observed AGB pulsation mass, cluster metallicity, giant branch effective 
temperature, M-type to C-type transition luminosity and the AGB-tip luminosity. 
A major finding from the study by \citet{kamath12} is that a large amount of convective overshoot 
(up to 3 pressure scale heights) is required at the base of the convective envelope during third dredge-up 
in order to get the correct O-rich to C-rich transition luminosity. Such large overshoot leads 
to $\lambda$ values in the range 0.66 to 0.82 for the best fitting models. The first shell flashes with 
dredge-up occur for core masses of  $M_{\rm c}^{\rm min}$ $\approx 0.56 - 0.58\Msun$. These values are much 
closer to the those values above suggested by synthetic AGB models to fit CSLFs and suggest that considerable 
convective overshoot occurs in low-mass AGB stars 
\citep[see also studies by][]{herwig97,herwig00,cristallo09,weiss09,karakas10b}.

It is important to note that this overshoot is measured from the formal
Schwarzschild boundary. We know that this point is unstable to growth, but it does form
a convenient position from which to measure the required amount of overshoot.
Knowing the Schwarzschild boundary to be unstable has motivated some authors 
to  implement an algorithm to try to search for a neutrally stable point \citep[e.g.][]{lattanzio86}.
We note that this was not able to reproduce the observations in \citet{kamath12}, 
and further mixing was required (although the required depth was not compared to the 
position of the neutral point).

The large spread in the amount of overshoot required to match the observations
could be telling us that is not the best way to quantify the 
required deeper mixing. It may indicate a mass dependence, with lower masses requiring deeper
mixing. It is good to remember that although we call this 
``overshoot'' because it is mixing beyond the Schwarzschild boundary, we are 
not identifying the cause of the deeper mixing as ``convective overshoot'' in 
the usual sense, that is to say, the mixing caused by conservation of momentum 
in the moving gas, which causes it to cross the point of zero acceleration (the Schwarzschild boundary).
Rather, we mean any process that mixes beyond the Schwarzschild border.

\subsection{Nucleosynthesis during Asymptotic Giant Branch Evolution} 

\begin{figure*}
\centering
\includegraphics[height=8.5cm,angle=270]{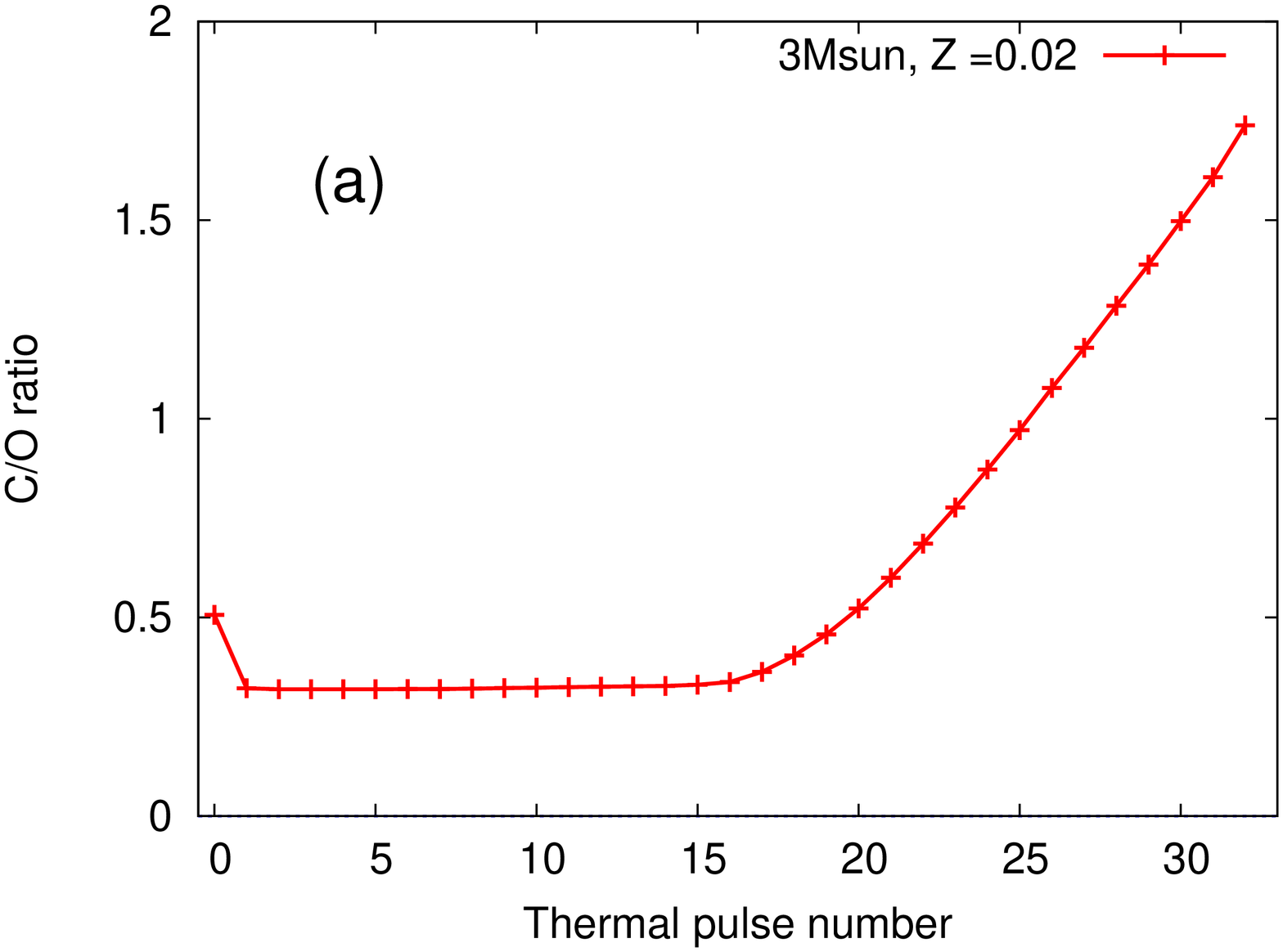}
\includegraphics[height=8.5cm,angle=270]{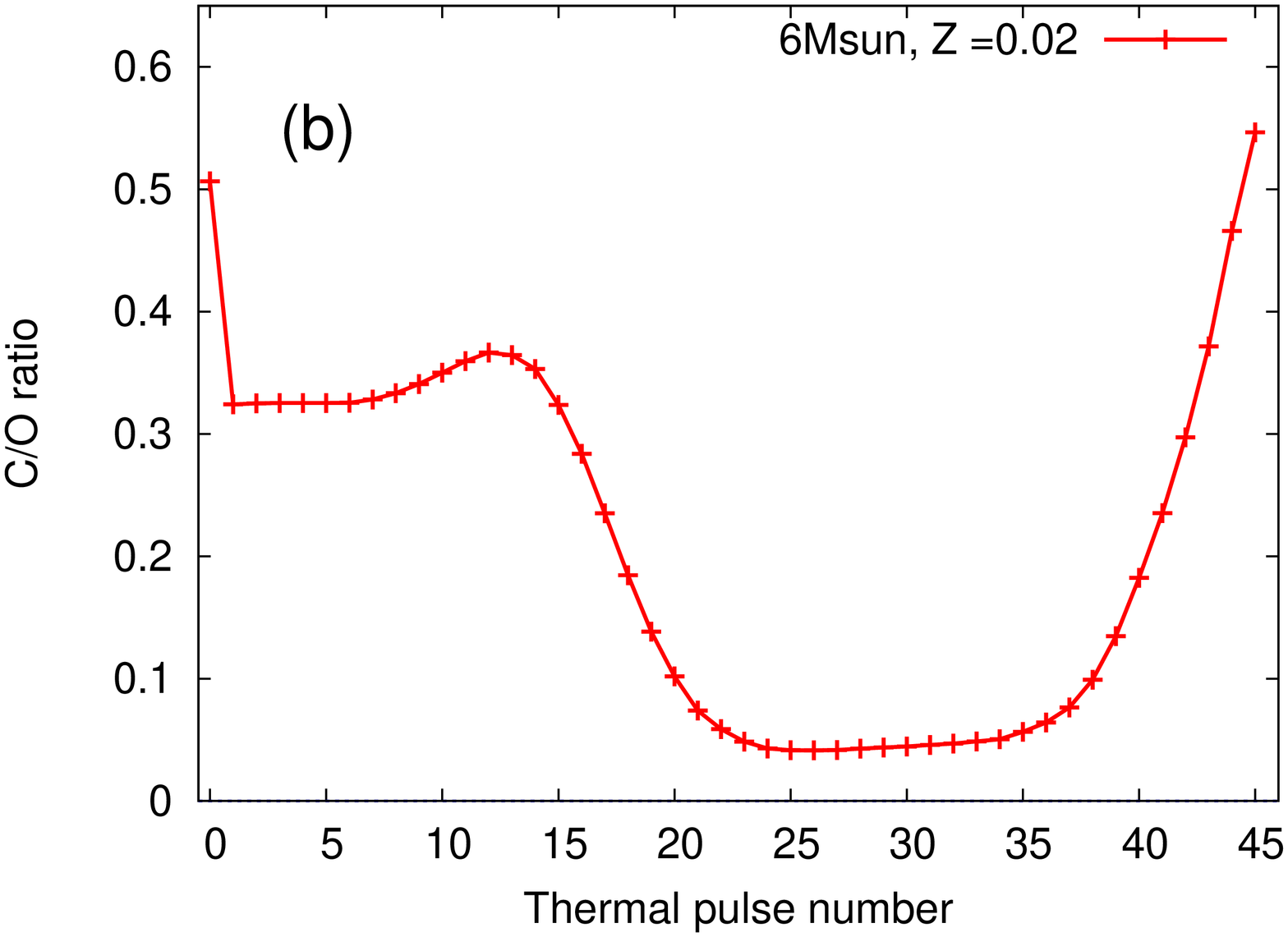}
\caption{The surface C/O ratio as a function of thermal pulse number for (a) a 3$\Msun$, $Z = 0.02$ model AGB star, and 
(b) a 6$\Msun$, $Z = 0.02$ model. The lower mass 3$\Msun$ model does not experience HBB and becomes C-rich. 
In contrast, efficient HBB occurs for the 6$\Msun$ model and the C/O ratio never reaches unity. The C/O ratio 
is given by number, and the initial abundance is the solar ratio at C/O $= 0.506$.}
\label{co-evol}
\end{figure*}

Thermal pulses and dredge-up may occur many times during the TP-AGB phase. Each TDU episode mixes \iso{12}C 
from the He intershell into the envelope and this has the effect of slowly increasing the C/O ratio of the surface,
illustrated in Figure~\ref{co-evol} for models of 3$\Msun$ and 6$\Msun$ at $Z=0.02$. 
Repeated TDU episodes can explain the transition from M-type (C/O $\approx 0.5$, similar to the Sun) to C-type stars:
\begin{equation}
  {\rm M} \rightarrow {\rm MS} \rightarrow {\rm S} \rightarrow {\rm SC}
 \rightarrow {\rm C},
\end{equation}
where SC-type stars have C/O of approximately unity, and C-type stars have C/O $>1$ by definition \citep{wallerstein98}. 

Many C stars also have surface enrichments of heavy elements synthesized by the $s$-process
\citep[e.g., Zr, Y, Sr, Ba, Tc;][]{smith86b,smith87,smith90a,abia02}.  The element technetium has
no stable isotopes. The presence of  this radioactive element in AGB star spectra is a particularly
important indicator of ``recent'' $s$-process nucleosynthesis and mixing
\citep{merrill52,little-marenin79,smith88,vanture91,vaneck99,lebzelter03,vanture07,uttenthaler13}.  
This is because the half-life of \iso{99}Tc (the isotope produced by the $s$-process) is 210,000 years, 
much shorter than the main-sequence lifetime of low-mass stars.

While C and $s$-process elements are the most obvious and easily verifiable examples of He-shell nucleosynthesis
and TDU, there are other elements that are produced during thermal pulses including F, Na, \iso{22}Ne, and the
neutron-rich Mg isotopes. Some of these isotopes are synthesized through a combination of H burning 
and He burning (e.g., Na). Hot bottom burning occurs in the most massive AGB stars and the main observable 
product of H burning is N which is produced by the CNO cycles, although other H-burning products 
may also be made (e.g., Na, Al etc).
The stellar yields of intermediate-mass AGB stars are strongly dependent on the complex interplay between HBB and TDU, 
as TDU is a supplier of primary C and \iso{22}Ne. Overall, the TP-AGB gives rise to a combination of
H and He-processed material that is expelled by the star as its envelope is lost through stellar winds.  

\subsection{Nucleosynthesis via thermal pulses} \label{sec:tps}

A He-shell flash produces heat and mixing throughout the intershell region, 
which is composed mostly of the {\em ashes} of  H-shell burning 
($\approx 98$\% \iso{4}He and 2\% \iso{14}N). The two main He-burning reactions are:
\begin{enumerate}
\item {\bf the triple-alpha process:} effectively 3 \iso{4}He $\rightarrow$ \iso{12}C; the main source of 
energy during thermal pulses;
\item {\bf the \iso{12}C($\alpha,\gamma$)\iso{16}O reaction}, which requires a reservoir of \iso{12}C for efficient
activation; this reaction produces negligible energy during shell flashes.
\end{enumerate}

During a thermal pulse some of the \iso{4}He in the shell is converted into \iso{12}C by partial He burning,
leaving the composition of the well-mixed intershell approximately 70--75\% \iso{4}He (by mass), 
20--25\% \iso{12}C and a few percent \iso{16}O (e.g., Figure~\ref{agb-struct}). These approximate numbers reflect 
the intershell composition of canonical stellar evolution models without overshoot into the C-O core. 
The inclusion of overshoot into the core increases the intershell composition of \iso{12}C and \iso{16}O as 
discussed in \citet{herwig00} and \S\ref{sec:intershellOxy}. There is a few percent (by mass) of \iso{22}Ne 
and trace amounts of  other species including \iso{17}O, \iso{23}Na, \iso{25}Mg, \iso{26}Mg, and \iso{19}F.
Sodium and \iso{27}Al are produced by H-burning during the preceding interpulse phase but are not
destroyed by $\alpha$-capture reactions during the thermal pulse.
The exact composition of the He intershell after a thermal pulse depends on the mass and composition of the 
He shell before the pulse, the duration of the shell flash, as well as the peak temperature and density under which 
the burning takes place. These quantities in turn depend on the stellar mass and metallicity. The core contracts 
with time along the AGB which means that the thermodynamic conditions of the He shell become somewhat more extreme 
toward the end of the TP-AGB compared to the beginning (e.g., higher temperatures and 
densities resulting from a thinner, slightly more electron-degenerate shell).

\begin{figure}
\centering
\includegraphics[height=8.5cm,angle=270]{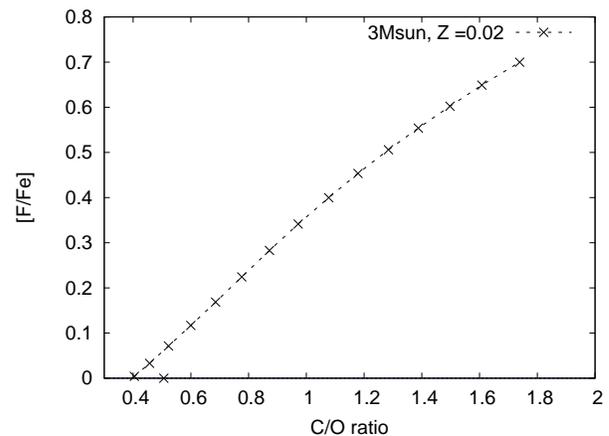}
\caption{The C/O ratio versus the [F/Fe] abundance at the surface of a 3$\Msun$, $Z = 0.02$ AGB model. 
Other products of helium nucleosynthesis include \iso{22}Ne, and 
the final \iso{22}Ne/Ne ratio in this model increases to $\approx 0.4$ 
from 0.068 initially. The total Ne abundance increases from $\log \epsilon$(Ne) = $\log_{10}$(Ne/H) $+12$ = 8.11 
at the main sequence to 8.33 at the tip of the AGB 
where He/H = 0.119, C/O = 1.74 (shown in Figure~\ref{co-evol}), \iso{12}C/\iso{13}C = 119, \iso{14}N/\iso{15}N $\approx 2500$
and N/O = 0.40.}
\label{m3z02-nucleo}
\end{figure}

\subsubsection{The carbon isotopic ratio: \iso{12}C/\iso{13}C} \label{sec:c12c13ratio}

The C isotope ratio \iso{12}C/\iso{13}C is a useful probe of AGB nucleosynthesis. Dredge-up increases
the amount of \iso{12}C, so the ratio will increase from \iso{12}C/\iso{13}C $\approx 10-20$ 
at the tip of the RGB to between 30 and $>$100, depending on the number of TDU episodes and the initial mass. 
For the 3$\Msun$ model star the predicted \iso{12}C/\iso{13}C ratio goes from $\approx 20$ before the AGB
to 119 at the tip of the AGB.

The \iso{12}C/\iso{13}C ratio has been observed in samples of C-rich AGB stars \citep{lambert86,abia97}
as well as PNe \citep{palla00,rubin04}. The C isotopic composition of pre-solar mainstream 
silicon carbide (SiC) grains, which are assumed to form in the extended envelopes of C-rich 
AGB stars, show a well defined distribution where 40 $\lesssim$ \iso{12}C/\iso{13}C $\lesssim$ 100, 
which matches the ratios observed in C(N) stars \citep[e.g.,][]{zinner98}. 

The \iso{12}C/\iso{13}C ratios are difficult to measure in PNe, with
values spanning the range from $\sim 4$, the equilibrium value of the CN cycle, to
upper limits of $\approx 38$ \citep{palla00,palla02,rubin04}. 
These ratios are, in general, lower than measured in C-rich AGB stars and lower than found in pre-solar
SiC grains and suggest efficient mixing of H-burning material with the observed nebula. ALMA observations
of R Sculptoris however show the surprising result that the \iso{12}C/\iso{13}C ratio at the stellar photosphere
is much lower, at 19, compared to the ratio obtained in the present-day mass loss 
\citep[$\gtrsim 60$;][]{vlemmings13}. These authors speculate that the lower C isotopic ratio is due to an 
embedded source of UV-radiation that is primarily photo-dissociating the \iso{13}CO molecule. 
This suggests that we need to be wary of the ratios obtained
from PNe, where the star is a strong UV source illuminating the nebula.

The C and N abundances predicted by models do not match those observed in AGB stars 
\citep[e.g.,][]{lambert86,abia97,milam09}. In particular the observed \iso{12}C/\iso{13}C ratios are lower 
than predicted by standard AGB models \citep[e.g.,][]{forestini97,cristallo09,karakas10a}.
For example, standard AGB models predict that by the time C/O $\ge 1$ then \iso{12}C/\iso{13}C $\ge 80$, 
which is already close to the upper limit observed in AGB stars or measured in SiC grains. 

It is possible to match the observed N and C isotopic ratios by artificially lowering the \iso{12}C/\iso{13}C 
ratio at the tip of the RGB to values observed in RGB stars \citep{kahane00,lebzelter08,karakas10b,kamath12}. 
By adopting a \iso{12}C/\iso{13}C ratio of 12 at the tip of the RGB, \citet{kahane00} 
found that they could match the observed \iso{12}C/\iso{13}C ratio of 45 for CW~Leo.
Similarly \citet{karakas10b} were able to reproduce
the observed \iso{12}C/\iso{13}C ratios for most of the Galactic C-rich stars.
Whereas most C-rich AGB stars have \iso{12}C/\iso{13}C ratios between about 30 and 80,
there is a small number with \iso{12}C/\iso{13}C ratios below 30
\citep[e.g.][]{lambert86,abia97} that cannot be explained
using the method above. That is, for a minimum \iso{12}C/\iso{13}C ratio of 10 at the tip 
of the RGB the minimum predicted value at 
the tip of the AGB is about 30 (depending on mass and composition). 
A value lower than this will require some form of extra mixing on the AGB.

\subsubsection{Nitrogen isotopic ratios}

Nitrogen isotopic measurements have been made for a small number of evolved stars. Measurements were made for
the cool C star IRC$+$10216, where the \iso{14}N/\iso{15}N ratio was estimated at $> 4400$ \citep{guelin95,kahane00}.  
A tentative value of $\approx 150$ was made for the J-type star Y CVn \citep{olson79}, and lower limits 
(along with two detections) were obtained in eight C stars and two proto-PNe \citep{wannier91}. 
In this last study six of the lower limits ($> 500$) were significantly larger than the ratio found in giant molecular 
clouds (330).  Note that the \citet{wannier91} \iso{14}N/\iso{15}N abundance ratio for Y CVn is 70, and for IRC$+$10216 
it is $5300$.

The most recent measurements by \citet{hedrosa13} were made for a selection of AGB stars of type C, SC, and J, 
where J-type C stars are defined mainly by their low \iso{12}C/\iso{13}C ratio and by the absence of $s$-process 
elements \citep{wallerstein98}. While almost all the data for C-type AGB stars show \iso{14}N/\iso{15}N $\gtrsim 1000$, 
a few C-type AGB stars have N isotopic values close to solar. These are difficult to reconcile with current 
models because known mixing events either increase the \iso{14}N/\iso{15}N ratio by mixing with regions that have 
experienced H burning %decreased the \iso{15}N content
(e.g. FDU, SDU with typical values shown in Table~\ref{fdusduvalues} and extra mixing) 
or leave it largely unchanged because the material mixed to the surface is not primarily from H-burning regions 
(e.g., TDU). Furthermore, some of the SC type AGB stars, which are defined by having C/O $\approx1$, and presumably 
on an evolutionary path that takes them from M-type (O-rich) to C-type (C-rich) should have N isotopic 
ratios similar to C-type AGB stars.
Instead, \citet{hedrosa13} find the SC-type AGB stars to be \iso{15}N-rich (with \iso{14}N/\iso{15}N $\lesssim 1000$), 
regardless of their C isotopic ratios. The reason for this deviation between SC and C-type AGB stars is unclear and 
difficult to understand from a theoretical viewpoint. Furthermore, the J-type stars, whose origin is already a 
mystery, show \iso{14}N/\iso{15}N ratios $\lesssim 1000$.  The origin of the \iso{15}N enrichments is not 
clear but the only way that this isotope can be produced in low-mass AGB stars is via the CNO cycle reaction
\iso{18}O(p,$\alpha$)\iso{15}N, which can take place in the He-burning shell provided there is a supply of protons
(as discussed below in the context of F production).

\subsubsection{The Intershell Oxygen Abundance}\label{sec:intershellOxy}

The short duration of thermal pulses and the low C content of the region
means that the \iso{12}C($\alpha,\gamma$)\iso{16}O reaction produces negligible energy and 
does not produce much \iso{16}O. Canonical stellar evolution calculations of AGB stars find \iso{16}O intershell compositions
of $\lesssim 2$\% \citep[by mass, e.g.,][]{boothroyd88c,karakas10b}. 
Here by ``canonical'' we are referring to model calculations with no convective overshoot of the flash-driven convective 
pocket into the C-O core. \citet{herwig00} does include convective overshoot at the inner border of the flash driven convective zone 
and finds that some C and O from the C-O core is mixed into the intershell. This has the effect of increasing the C and O intershell 
abundances to up to $\approx 40$\% and 20\%, respectively. The inclusion of such overshoot means that the oxygen stellar yields from 
low-mass AGB stars may become significant \citep{pignatari13}. We discuss this further in \S\ref{sec:PNobs}.

\subsubsection{Fluorine}

Fluorine is produced through a complex series of reactions as outlined in detail by \citet{forestini92}, \citet{mowlavi96},
\cite{mowlavi98} and more recently by \citet{lugaro04}. The main reaction pathway involves the production of \iso{15}N which is burnt to 
\iso{19}F via \iso{15}N($\alpha$,$\gamma$)\iso{19}F. The difficulty here is in making \iso{15}N which is destroyed by 
proton captures in the CNO cycles, which means that the composition of the He shell before a thermal pulse will be almost
devoid of this isotope. If protons can be produced by secondary reactions (e.g., \iso{14}N(n,p)\iso{12}C which itself 
requires free neutrons) then the CNO chain reaction \iso{18}O(p,$\alpha$)\iso{15}N can make \iso{15}N. Because F is 
produced in the He intershell the composition in the envelope correlates with the abundance of C and $s$-process 
elements, as shown in Figure~\ref{m3z02-nucleo} (for C).

Fluorine has been observed in AGB stars in our Galaxy and in Local Group Galaxies \citep{jorissen92,lebzelter08,abia09,abia10}, 
PG 1159 post-AGB stars and PNe \citep{werner05,otsuka08}, and in barium stars \citep{alves-brito11}, which 
are hypothesised to have received their C and Ba through mass transfer from a previous AGB companion. 

The observations of \citet{jorissen92} revealed F abundances that were much higher than model predictions 
\citep{forestini92,lugaro04,karakas08}, especially for the SC type AGB stars with C/O $\approx 1$. 
A re-analysis of the F abundance in three Galactic AGB stars (TX Psc, AQ Sgr, and R Scl) by \citet{abia10} 
revealed the cause of the discrepancy to be the model atmospheres used in the original analysis, which did not 
properly take into account
blending with C-bearing molecules. The new abundances are up to 0.8~dex lower, bringing the 
models into agreement with the observations.  Observations of F in the C-rich AGB stars in the LMC cluster 
NGC 1846 however show a discrepancy with models \citep{lebzelter08,kamath12}, with the 
observed F abundance increasing more strongly with the C/O ratio than in the theoretical model.  
While observations of F in extra-galactic C stars show 
most of the stars to be F rich, the models over predict the amount of C relative to the observations; 
see \citet{abia11}, who also suggested possible solutions including the hypothesis that most of the C might be 
trapped in dust grains. 

\subsubsection{Other Species in the Intershell}

There is a wealth of other He-burning products produced as a consequence of thermal pulses including 
\iso{19}F, \iso{22}Ne, \iso{23}Na, \iso{25}Mg, \iso{26}Mg, and \iso{27}Al 
\citep{forestini97,mowlavi99b,herwig00,karakas03a,karakas03b,lugaro04,cristallo09,karakas10a,cristallo11}.
The isotope \iso{22}Ne is produced by the reaction \iso{14}N($\alpha,\gamma$)\iso{18}F, where \iso{18}F $\beta$-decays 
to \iso{18}O allowing for the reaction \iso{18}O($\alpha,\gamma$)\iso{22}Ne. The composition of \iso{22}Ne in the 
intershell is fairly high, at $\approx 2$\%. This is because the abundant \iso{14}N is completely converted into 
\iso{22}Ne during a thermal pulse.  The \iso{22}Ne abundance is predicted to increase by almost an order of magnitude ($\sim 1$~dex) 
in some AGB models \citep{karakas03a}. If the \iso{22}Ne abundance exceeds or is equal to the \iso{20}Ne abundance we 
should expect an observable enhancement in the elemental Ne composition.  The intershell is also enriched in \iso{23}Na 
and \iso{27}Al. Sodium and \iso{27}Al are not He-burning products but are synthesised in the H shell during the 
previous interpulse. Unlike other H-burning products (e.g., \iso{14}N) these are left unburnt by the subsequent TP 
and mixed into the envelope by the next TDU episode. 

\subsubsection{Heavy Magnesium isotopes}

If the peak temperature of the thermal pulse exceeds $300 \times 10^{6}$K, the neutron-rich Mg isotopes, \iso{25}Mg and 
\iso{26}Mg, can be synthesised by the \iso{22}Ne($\alpha,n$)\iso{25}Mg and \iso{22}Ne($\alpha,\gamma$)\iso{26}Mg reactions. 
These two \iso{22}Ne~$+ \, \alpha$ reactions have similar although uncertain rates at He-shell burning temperatures 
\citep{angulo99,karakas06a,longland12,wiescher12}. Owing to the relatively high temperatures required for these two reactions, 
they are predicted to occur efficiently in intermediate-mass AGB stars with masses greater than about $4\Msun$ depending 
on metallicity \citep{karakas03b,karakas06a}. The He intershell of lower mass AGB stars will only reach 300 MK
during the last few thermal pulses (if at all). This means that the \iso{22}Ne($\alpha,n$)\iso{25}Mg reaction is only marginally 
activated near the end of the AGB. 
The \iso{22}Ne($\alpha,n$)\iso{25}Mg is particularly important because it produces free neutrons that can be captured by 
iron-peak elements enabling the $s$-process \citep{iben75,wiescher12}.  It is the dominant neutron-producing 
reaction in the He and C-burning regions of massive stars \citep{the07,heil08,pignatari10} and the dominant neutron 
source in intermediate-mass AGB stars \citep{garcia06,karakas12}. 
We will come back to this reaction when we discuss $s$-process nucleosynthesis in \S\ref{sec:sprocess}.

\subsubsection{Planetary nebulae and post-AGB stars}\label{sec:PNobs}

Comparisons to observations of AGB stars and their progeny can be made for many of the species considered so far.
Comparisons with Ne measured in PG 1159 stars reveal good agreement with \iso{22}Ne intershell abundances found in 
standard models \citep{werner99}. Neon abundances can be reliably measured in PNe so observations of 
these objects can be used as a probe of AGB nucleosynthesis. A correlation is observed to exist between the Ne/H and O/H 
abundance in PNe in the Galaxy, LMC, SMC and M31, within a small but probably real spread 
\citep{kaler78,aller83,henry89,dopita97,stasinska98,leisy06,stanghellini00,stanghellini06,bernard08}. 
While AGB models can produce considerable \iso{22}Ne which results in an overall increase in the elemental Ne 
abundance, this is predicted to occur in only a narrow mass range and the overall agreement with the observations is 
good \citep{marigo01,karakas03a,henry12,shingles13}. 

We can compare AGB predictions to the surface abundance observations of PG 1159-type post-AGB stars, which are thought 
to be in transition from central stars of PNe to white dwarfs \citep{werner09}. These H-deficient objects are quite rare, 
with only about two dozen known, and their atmospheres are mostly composed of He, C, and O
\citep{werner94,werner06,jahn07,werner09}.
Spectroscopic observations of the PG 1159 central stars reveal O mass fractions as high as 
20\% \citep[e.g.,][]{werner06} clearly at odds with standard stellar models.  Spectroscopic observations of 
Ne and F reveal abundances consistent with the models \citep{werner99,werner05}.
The diffusive convective overshooting models of \citet{herwig00} have intershell abundances that are consistent 
with the abundance patterns observed in PG 1159 central stars  \citep[see also, e.g.,][]{miller06,althaus09}, 
as discussed in \S\ref{sec:intershellOxy}. The degenerate thermal pulses found by \citet{frost98b} may 
have a similar effect.  In this case, deep third dredge-up following the degenerate pulse can mix material 
from the C-O core into the envelope, enhancing the envelope in \iso{16}O. 

There is other observational evidence for increased O intershell content. The high [O/Fe] abundances measured 
in post-AGB stars in the Galaxy and Magellanic Clouds are difficult to reconcile with standard O intershell 
abundances of 2\% or less \citep{vanwinkel00,desmedt12}. In particular the low-metallicity SMC post-AGB star 
J004441.04-732136.4 has a [Fe/H] $\approx -1.3$, a low C/O ratio of 1.9, combined with a high [O/Fe] $\approx 1.10$. 
These numbers cannot be explained by canonical stellar evolution models as discussed in \citet{desmedt12}, which produce
very high C/O ratios $\approx 20$ at this low metallicity. One way to reconcile the models and observations 
would be through changing the O intershell abundances. Further evidence for non-standard intershell compositions 
comes from the O isotope ratios measured in evolved red giant stars \citep[see discussion in][]{karakas10b}, 
which show an increase in the \iso{16}O/\iso{17}O and \iso{16}O/\iso{18}O ratios with evolution along the AGB. 
The C/O and \iso{12}C/\iso{13}C ratios measured in the C-rich AGB stars in NGC 1978 are also difficult to 
reconcile with standard models and require higher O intershell compositions of 15\% \citep{kamath12}.  
One source of uncertainty in these conclusions is the large error bars present in the O isotopic data 
measured by Harris, Lambert and collaborators 
\citep{harris83,harris84,harris85a,harris85b,harris87,smith90a}. 

\subsection{Nucleosynthesis from Hot Bottom Burning} \label{sec:hbb}

\begin{figure}
\begin{center}
\includegraphics[height=8cm,angle=270]{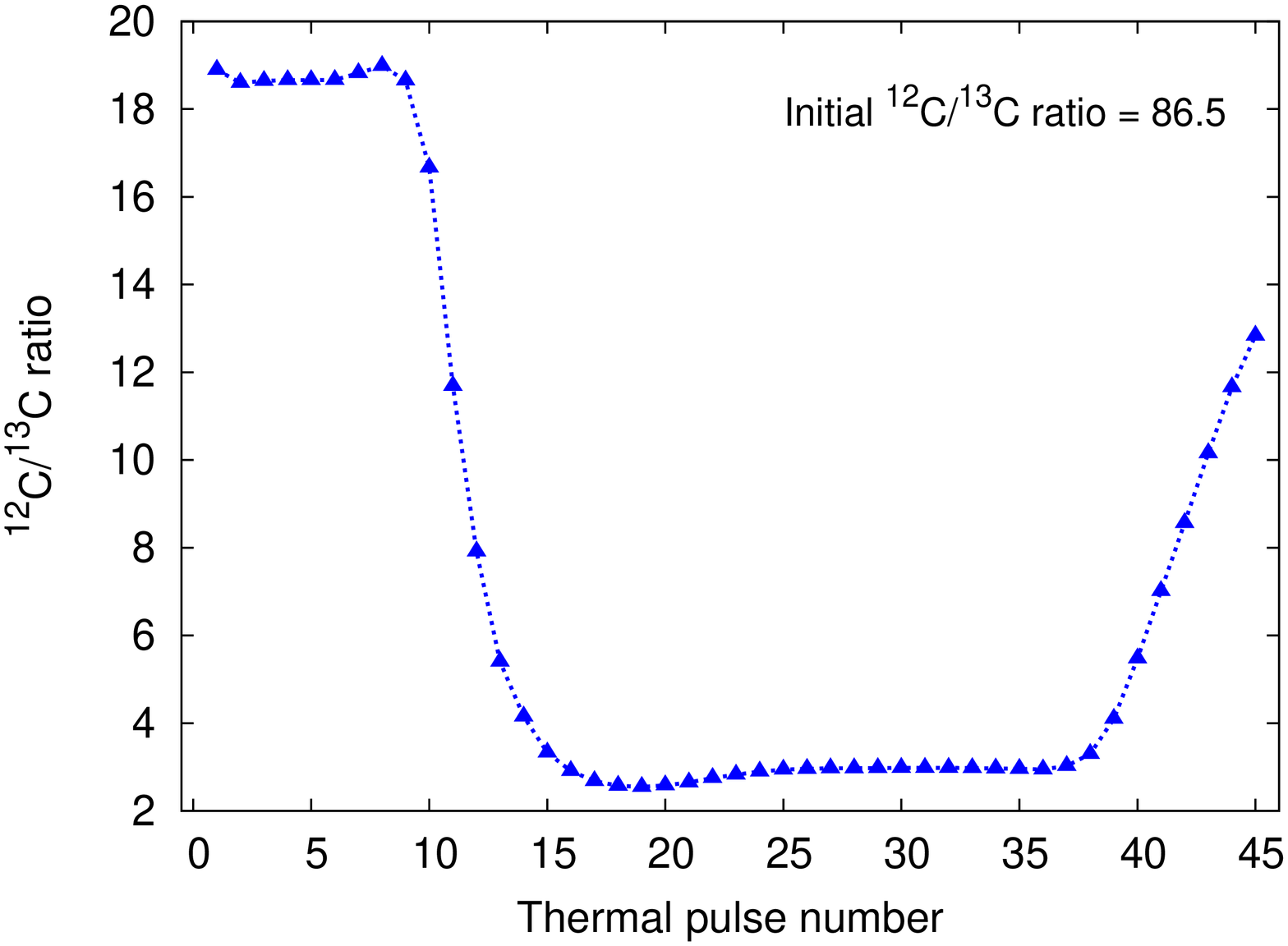}
\includegraphics[height=8cm,angle=270]{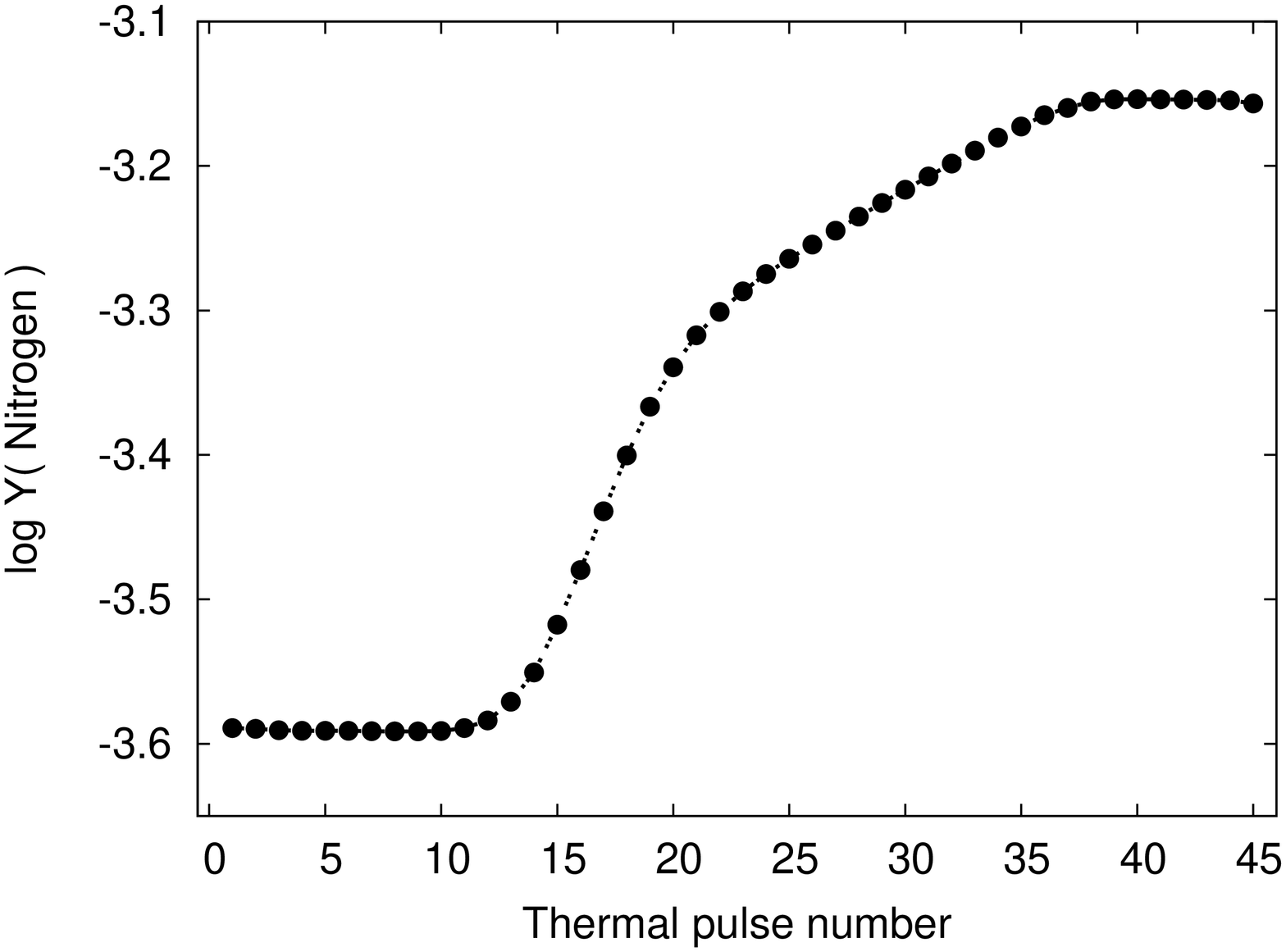}
\caption{The evolution of the \iso{12}C/\iso{13}C ratio and the nitrogen elemental abundance at the 
surface of the 6$\Msun$, $Z = 0.02$ model during the TP-AGB. The ratio is given by number and the abundance of 
nitrogen is in units of $\log_{10}(Y)$, where $Y = X/A$ and $X$ is mass fraction and $A$ is atomic mass. }
\label{m6z02-cno}
\end{center} 
\end{figure}

H-burning occurs primarily via the CNO cycles but also via the Ne-Na and Mg-Al chains if the temperature 
is high enough.  In this section we summarise the main H-burning reactions and their products and 
review the results of HBB that are predicted to be observed at the surface of intermediate-mass AGB stars. 

\subsubsection{C, N, and O}

In \S\ref{sec:FDUabunds} we discussed the CNO cycle in the context of FDU abundance changes. 
During HBB the temperatures at the base of the convective envelope are higher than in the H shell during the first 
ascent of the giant branch, reaching $T_{\rm bce} \gtrsim$ 100 MK  in the lowest metallicity, massive AGB stars 
\citep[with C-O and O-Ne cores, e.g.,][]{karakas07b,ventura09a,siess10,doherty14a}. While these high temperatures 
are normally  associated with He burning, the density at 
the base of the envelope is only a few grams cm$^{-3}$, much lower than in the H shell and other H-burning regions 
(e.g., the central density in the Sun is $\rho_{\odot} \approx 160$ gram cm$^{-3}$; at the base of the H shell the 
typical densities during the interpulse are $\approx 30-40$ gram cm$^{-3}$ in an intermediate-mass AGB star). This 
means that higher temperatures are required for nucleosynthesis and energy production than in a typical H-burning
environment and also partly explains why HBB is more efficient at lower metallicities where the stars are more compact.

During HBB the CN cycle, which results in an increase in the abundance of \iso{13}C and \iso{14}N from the destruction of 
other CNO species, comes into equilibrium quickly.  The isotopes \iso{12}C and \iso{15}N are first destroyed by the CN 
cycle and later the oxygen isotopes \iso{16}O and \iso{18}O are also destroyed to produce \iso{14}N. The abundance of 
\iso{17}O can be enhanced by the CNO bi-cycle, depending on the uncertain rate of the \iso{17}O $+$ p branching 
reactions whereas \iso{19}F is destroyed \citep[e.g.,][]{angulo99,arnould99,iliadis10}.

In Figure~\ref{co-evol} (b) we show the evolution of the C/O ratio at the surface of a 6$\Msun$, $Z = 0.02$ 
model with HBB. The C/O ratio stays below unity for the entire TP-AGB phase and only starts to increase from C/O 
$\lesssim 0.1$ during the final 8 thermal pulses, which is when HBB starts to shut down owing to the erosion of the envelope 
by mass loss.  By the final calculated thermal pulse the C/O ratio is just above the starting (solar) value of 0.5. 
The evolution of the \iso{12}C/\iso{13}C ratio and elemental N abundance in Figure~\ref{m6z02-cno} also 
demonstrates efficient activation of the CNO cycles. The \iso{12}C/\iso{13}C ratio behaves similarly to the C/O 
ratio and stays close to the equilibrium value of $\approx 3$ for much of the TP-AGB. The N abundance is 
seen to increase by almost an order of magnitude, more than would be allowed if the initial C$+$N$+$O was 
consumed to produce \iso{14}N. This is because primary \iso{12}C is mixed from the intershell by the TDU into 
the envelope, where it is converted into N. Note also that the \iso{14}N/\iso{15}N ratio during the 
TP-AGB is $\gtrsim 10,000$, reaching essentially the CN cycle equilibrium value. The O isotopic ratios 
also evolve, where the \iso{16}O/\iso{17}O ratio increases with evolution to a final value of $\approx 465$ 
whereas the \iso{16}O/\iso{18}O ratio increases to above $10^{6}$ as almost all of the available 
\iso{18}O is destroyed.  The elemental O abundance in the 6$\Msun$ model only decreases however by $\approx 0.06$~dex. 

\subsubsection{Ne, Na, Mg, and Al}

\begin{figure*}
\begin{center}
\includegraphics[height=7.5cm,angle=0]{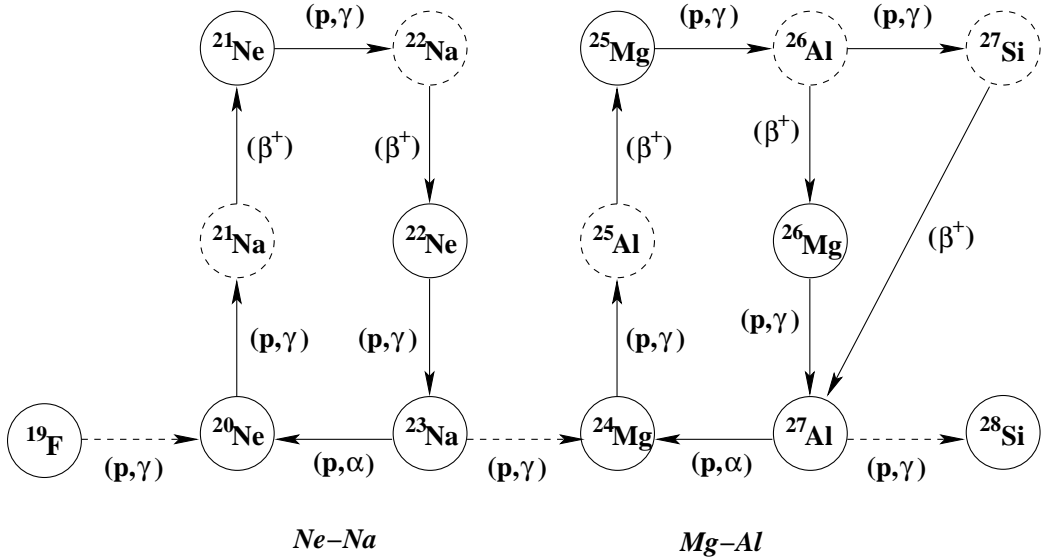}
\caption{Reactions of the Ne-Na and Mg-Al chains. Unstable
isotopes are denoted by dashed circles. From \citet{karakas03a} and based on a similar figure in \citet{arnould99}.}
\label{nenamgal}
\end{center} 
\end{figure*}

In the left part of Figure~\ref{nenamgal} we show the reactions of the Ne-Na chain \citep{rolfs88,arnould99}, where unstable isotopes 
are denoted by dashed circles. The main result of the Ne-Na chain is the production of \iso{23}Na at the expense of the Ne
isotopes, primarily \iso{22}Ne but also \iso{21}Ne, which is the rarest neon isotope.  The production of Na by the Ne-Na 
chain was examined in detail by \citet{mowlavi99b}, who predicted that AGB stars could play an integral 
role in the chemical evolution of Na in the Galaxy. 

The dominant \iso{20}Ne is not significantly altered by H-shell burning, but the destruction of \iso{23}Na at temperatures over 
90 MK can lead to a slight enhancement in the \iso{20}Ne abundance. The rate of \iso{23}Na destruction is important for determining 
Na yields \citep[e.g.,][]{iliadis01,izzard07}.  Whether there is leakage out of the Ne-Na chain into the Mg-Al chain depends 
on the relative rates of the uncertain \iso{23}Na(p,$\alpha$)\iso{20}Ne and \iso{23}Na(p,$\gamma$)\iso{24}Mg reactions. 
\citet{hale04} presented revised rates of both of these reactions and found them to be faster than previous estimates
\citep[e.g.,][]{angulo99,iliadis01} which had a significant impact on Na yields from AGB stars as we discuss in \S\ref{sec:yields}.

Magnesium and aluminium are altered in the H-burning shell via the activation of the Mg-Al chain, whose reactions are shown 
on the right-hand side of Figure~\ref{nenamgal}. This series of reactions involves the radioactive nuclide \iso{26}Al which 
has a ground state $\iso{26}{\rm Al}_{g}$ with a half-life of $\tau_{1/2} = 700,000$ years along with a short-lived 
($\tau_{1/2} = 6.35\,$s) isomeric state $\iso{26}{\rm Al}_{m}$. These have to be considered as separate species since 
they are out of thermal equilibrium at the relevant temperatures \citep{arnould99}.  Hereafter, when we refer to 
\iso{26}Al we are referring to the ground-state, $\iso{26}{\rm Al}_{g}$.  

\begin{figure}
\begin{center}
\includegraphics[height=8cm,angle=90]{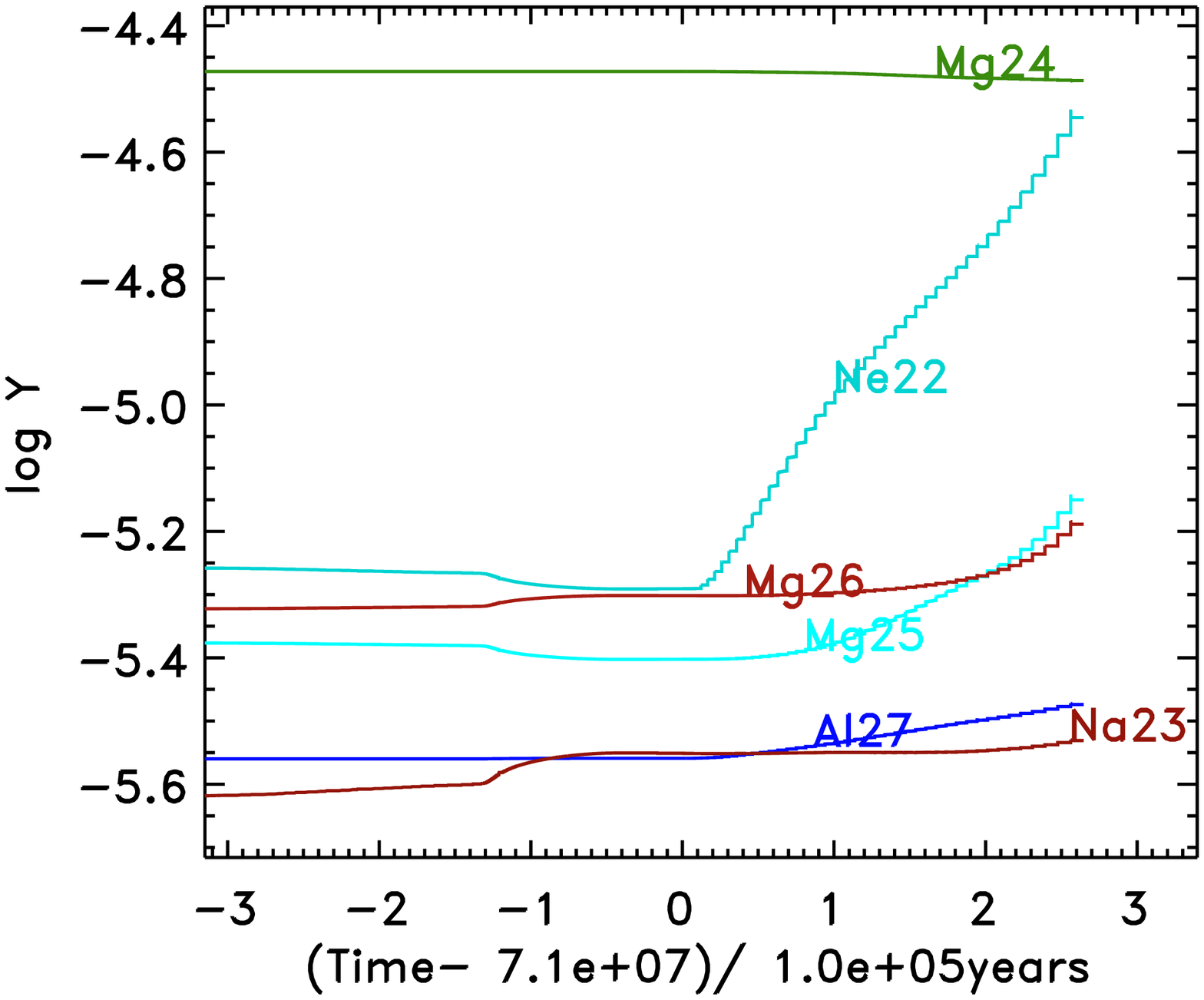}
\includegraphics[height=8cm,angle=90]{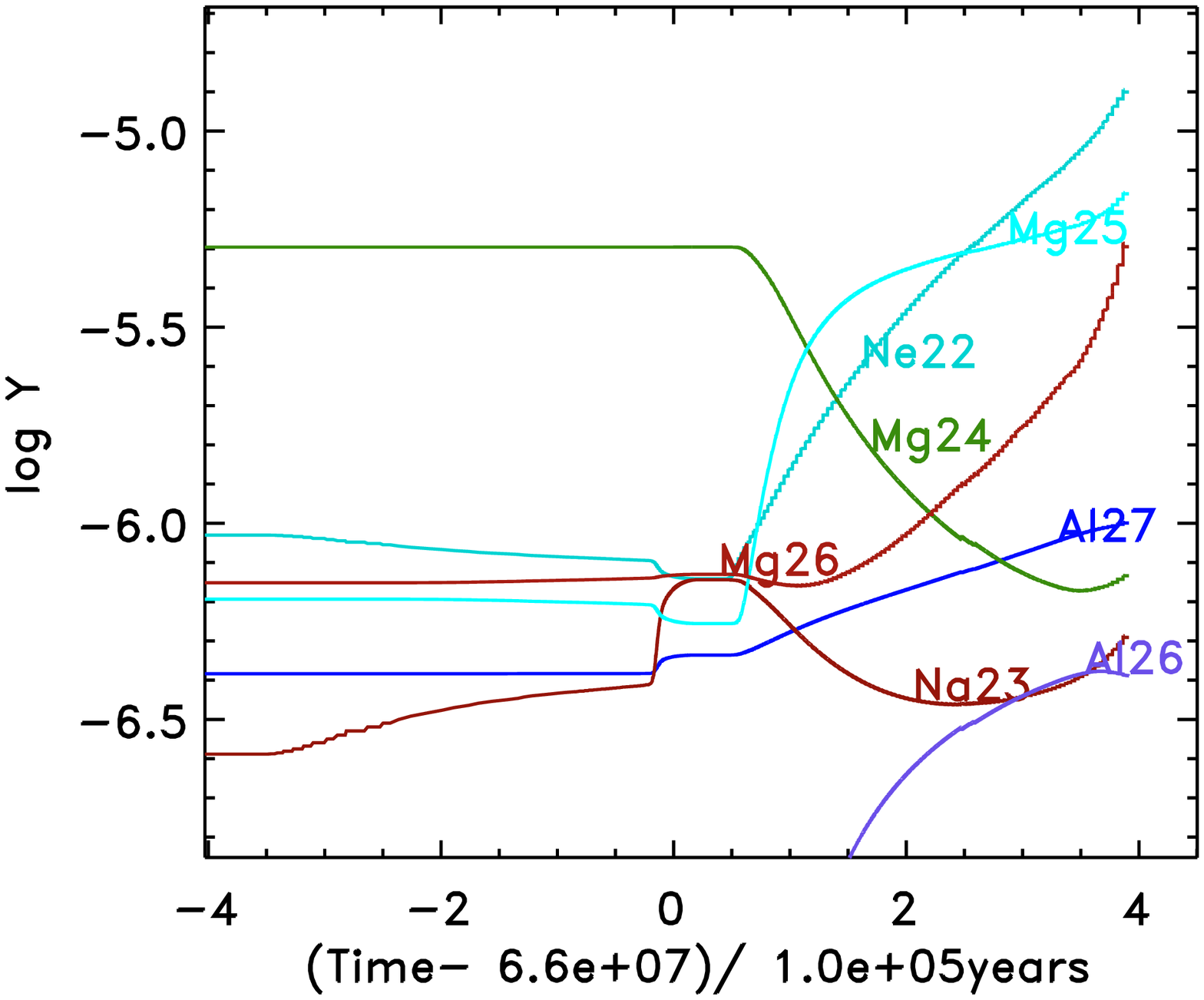}
\caption{The evolution of various species involved in the Ne-Na and Mg-Al chains at the surface of the 6$\Msun$, $Z = 0.02$ 
model (upper panel) and 6$\Msun$, $Z = 0.004$ model (lower panel) during the TP-AGB. Time on the $x$-axis is scaled such that 
$t=0$ is the time at the first thermal pulse.  Abundances on the $y$-axis are in units of $\log_{10} Y$, where $Y = X/A$, 
where $X$ is mass fraction and $A$ is atomic mass. Both calculations used the same set of of reaction rates and scaled 
solar abundances. The 6$\Msun$, $Z = 0.004$ model has been described previously in \citet{karakas10a}.}
\label{m6-nemgal}
\end{center} 
\end{figure}

The first isotope in the Mg-Al chain to be affected is \iso{25}Mg, which is burnt to \iso{26}Al. The $\beta$-decay 
lifetime of \iso{26}Al relative to proton capture generally favours proton capture within the H-burning shell.  
This produces the unstable \iso{27}Si which quickly $\beta$-decays (with a lifetime on the order of a few seconds) to 
\iso{27}Al.  The abundance of \iso{26}Mg is enhanced by the  $\beta$-decay of \iso{26}Al in the H-shell ashes.  
Proton capture on \iso{24}Mg requires higher temperatures than those required for the other reactions in the Mg-Al 
chain but model predictions (e.g., Figure~\ref{m6-nemgal}) suggest that this dominant isotope can be efficiently destroyed by HBB.

In Figure~\ref{m6-nemgal} we show the evolution of various isotopes involved in the Ne-Na and Mg-Al chains at the surface of 
two models of 6$\Msun$. The upper panel shows the predicted nucleosynthesis for the 6$\Msun$, $Z = 0.02$ model we have been 
describing so far, which has a peak $T_{\rm bce}$ of 82 MK. The lower panel shows results from a 6$\Msun$, $Z  =0.004$ model 
which has a peak $T_{\rm bce}$ of 95 MK. While the CNO isotopes for the $6\Msun$, $Z=0.02$ model (Figure~\ref{m6z02-cno}) clearly
show the effects of HBB on the predicted surface abundances, the abundances of heavier isotopes show only marginal activation 
of the Ne-Na and Mg-Al chains. Most of the increase in \iso{22}Ne, \iso{25}Mg, and \iso{26}Mg is from the TDU bringing 
He-shell burning products to the surface. The slight increase in \iso{27}Al ([Al/Fe] $\approx 0.1$ at the tip of the AGB) is 
mostly from Al produced in the H shell and mixed to the surface by the TDU and not from HBB. Sodium barely increases from the 
post-SDU value. In contrast, the lower metallicity 6$\Msun$ model, which is not only hotter but more compact, shows considerable 
destruction of \iso{24}Mg, an increase in \iso{26}Al which can only come from H burning and the Mg-Al chains, and 
variations in \iso{25}Mg and \iso{26}Mg consistent with HBB. Sodium initially increases before being destroyed again by 
proton-capture reactions.

There is a paucity of observational evidence for constraining stellar models of intermediate-mass stars during
their TP-AGB phase. This is partly because there are few stars found at the AGB-luminosity limit near 
$M_{\rm bol} \approx -7$ but also because of the complexity of the model atmospheres required for the interpretation of the spectra.
Evolved intermediate-mass AGB stars are long-period pulsators with low effective temperatures, which means that
the dynamics of the atmosphere must be taken into consideraton \citep{mcsaveney07}.   
The few observations of stars in our Galaxy suggest that most of them, even 
the optically obscured stars, are O-rich and $s$-process rich which points to both efficient HBB and TDU 
\citep{wood83,garcia06,garcia13}. In the Magellanic Clouds, most of the bright AGB stars are also O-rich but 
some dust-obscured objects are also C-rich as we have already noted. 

The study by \citet{mcsaveney07} obtained abundances for a small sample of bright AGB stars for comparison to theoretical models. 
The observations of C, N, and O were a relatively good match to stellar evolution models but no observed enrichments were found 
for Na and Al. We note that the latest nuclear reaction rates suggest much lower Na production than previously calculated 
\citep{karakas10a} so perhaps this is not surprising.  Al production is also predicted to be highly
metallicity dependent with little production in AGB stars with [Fe/H] $\gtrsim -0.7$ \citep{ventura13}.

Predictions such as those presented here \citep[or by others, e.g.,][]{ventura13} that intermediate-mass AGB stars result in 
nucleosynthesis variations in the Ne, Na, Mg, and Al isotopes are particularly useful for comparison to GC
abundance anomalies. All well studied GC show star-to-star abundance variations in C, N, O, F, Na, and 
some show variations in Mg, Al and Si \citep[e.g., ][and references therein]{gratton04,gratton12,carretta09,yong13} and only
a few GCs show variations in iron-peak and heavy elements \citep[e.g., $\omega$ Cen and M22][]{norris95,cjohnson08,dacosta11}.
The lack of star-to-star variations in Fe and Eu have led to the conclusion that core-collapse supernovae did not play a role 
in the self enrichment of these systems and have suggested an important contribution from intermediate-mass AGB stars. 
Globular cluster star abundances show C and N are anti-correlated with each other,
as are O and Na, and (sometimes) Mg and Al. That is the pattern expected if H burning at relatively high temperatures has 
caused the observed abundance patterns \citep{prantzos07}. 

At the highest HBB temperatures, breakout of the Mg-Al cycle can occur via the \iso{27}Al(p,$\gamma$)\iso{28}Si reaction. 
If this occurs then we would expect to see correlations between enhanced Al and Si \citep{ventura11b}. \citet{yong08d} 
find that N abundances in the giant stars of the GC NGC 6752 are positively correlated with Si, Al and Na,
indicating breakout of the Mg-Al chain. \citet{carretta09} also find a spread in the Si abundances 
of the GCs NGC 6752 \citep[see also][]{yong13} and NGC 2808, and a positive correlation between Al and Si.  Models by 
\citet{ventura11b} show that the most Al enriched models are also enriched in Si so it seems that hot H burning 
can produce such a trend, even if the site of the proton-capture nucleosynthesis in GCs is still unknown.

\begin{figure}
\begin{center}
\includegraphics[height=8cm,angle=90]{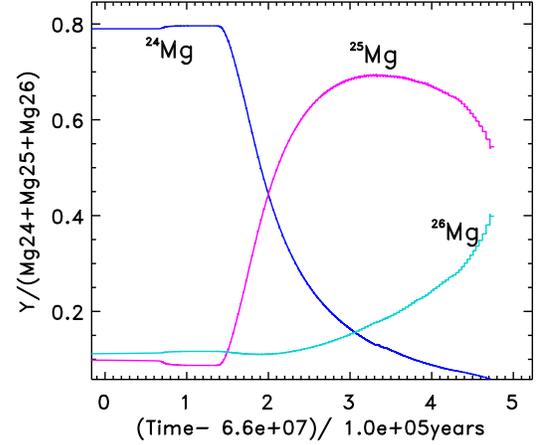}
\caption{The evolution of stable Mg isotopes at the surface of the 6$\Msun$, $Z = 0.004$ model during the TP-AGB. 
Time on the $x$-axis is scaled such that $t=0$ is the time at the first thermal pulse. Abundances on the $y$-axis 
are scaled to the total Mg composition, 
$Y$(\iso{i}Mg)/$\{Y$(\iso{24}Mg)$+Y$(\iso{25}Mg)$+Y$(\iso{26}Mg)$\}$, 
where $Y = X/A$, where $X$ 
is mass fraction and $A$ is atomic mass. The initial Mg isotopic ratios on the main sequence are solar:
\iso{24}Mg/\iso{25}Mg = 7.89 and \iso{24}Mg/\iso{26}Mg = 7.17 \citep[e.g.,][]{asplund09}. By the tip of the TP-AGB, 
the model ratios are \iso{24}Mg/\iso{25}Mg = 0.11 and \iso{24}Mg/\iso{26}Mg = 0.14 indicating that most of the
 \iso{24}Mg has been destroyed by proton captures.}
\label{m6z004-mgisotope}
\end{center} 
\end{figure}

It is possible to obtain isotopic ratios for Mg by using the MgH line 
\citep{guelin95,shetrone96a,shetrone96b,gay00,kahane00,yong03a,yong03b,yong06a,dacosta13}. 
The Mg isotopic ratios therefore become an important probe of the site of the nucleosynthesis that has added to the chemical 
enrichment of GC systems. 
The GCs show an intriguing trend: the stars that are considered normal or ``not polluted'' are (relatively) 
O-rich and  Na-poor, and sometimes Mg-rich and Al-poor. These stars show a near solar Mg isotopic ratio.
The stars that are considered ``polluted'' show O-depletions and sometimes Mg-depletions,
are rich in Na and sometimes Al. For those globular clusters that do show stars with
variations in Mg and Al we find that \iso{24}Mg is depleted at the expense of 
\iso{26}Mg with essentially {\em no} star-to-star variations in \iso{25}Mg \citep[e.g.,][]{dacosta13}.  

In Figure~\ref{m6z004-mgisotope} we show the evolution of the Mg isotopes at the surface of a 6$\Msun$, 
$Z = 0.004$ ([Fe/H] $\approx -0.7$) AGB model. The metallicity of this model matches some of the stars
in $\omega$~Cen studied by \citet{dacosta13} and the sole M~71 star analysed
by \citet{yong06a}, and may help reveal a trend with metallicity.
If we focus just on $\omega$~Cen, then at all metallicities the stars show 
approximately solar ratios for \iso{25}Mg/Mg $\approx 0.1$ \citep[e.g., Fig~9 in][]{dacosta13}.
In contrast, \iso{24}Mg/Mg shows a decrease with [Fe/H], while \iso{26}Mg/Mg shows an increase, with the most
extreme cases showing \iso{24}Mg/Mg $\approx 50$ and \iso{26}Mg/Mg $\approx 40$ at [Fe/H] $\approx -1.5$.
At the metallicity of the 6$\Msun$ AGB model shown in Figure~\ref{m6z004-mgisotope} all observed ratios are 
again approximately solar, in disagreement with model predictions. 

What is particularly unusual about these observed abundance ratios is: (1) at the
low metallicities of the GC stars examined to date, chemical evolution models suggest a 
dominant contribution from core-collapse supernovae that produce mostly
\iso{24}Mg and there should be almost no \iso{25}Mg or \iso{26}Mg \citep[e.g.,][]{kobayashi11a}. 
That is, the normal stars should be completely dominated by \iso{24}Mg (about 97\% of the total Mg) 
and not show a solar Mg isotopic ratio.  (2) H burning in AGB stars or massive stars, regardless of the stellar 
evolution code used, struggles to explain these isotopic ratios and unchanging
\iso{25}Mg abundances without resorting to variations in reaction rates \citep{fenner04,herwig04b,decressin07,demink09,ventura09a}. 
Parametric models can explain the observed abundances but provide few clues as to the physical site responsible \citep{prantzos07}. 
We note that reaction rates involving the Mg and Al species are quite uncertain and the new reaction rates presented by 
\citet{straniero13} may help resolve the issue.

\subsubsection{Lithium}

\begin{figure}
\centering
\includegraphics[height=8cm,angle=90]{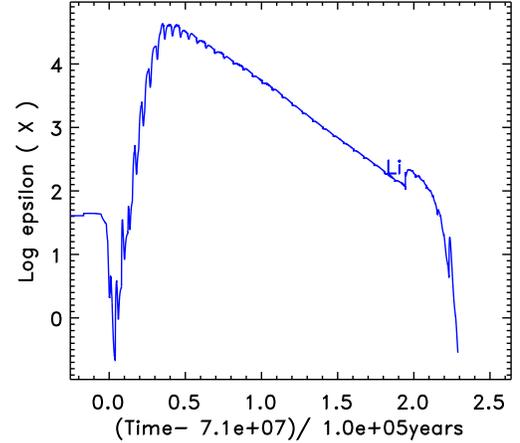}
\caption{
The surface abundance of \iso{7}Li during the TP-AGB phase for a 6$\Msun$, $Z=0.02$ model. The units on the $y$-axis are 
$\log_{10}($n(Li)/n(H)$+12)$ and time on the $x$-axis is scaled such that $t=0$ is the beginning of the TP-AGB. 
The lithium rich phase lasts for about 200,000 years.
\label{m6z02-li7}
}
\end{figure}

The discovery that the brightest AGB stars are rich in Li \citep{smith89,smith90b,plez93,garcia13} 
gave further credibility to  the idea that HBB was actually occurring in intermediate-mass AGB stars.  
The production of \iso{7}Li is thought to occur via the Cameron-Fowler mechanism \citep{cameron71}:  
Some \iso{3}He, created earlier in the evolution (during central H-burning), captures an 
$\alpha$-particle to create \iso{7}Be.  The \iso{7}Be can either 1) capture a proton to complete the 
PPIII chain, or 2) capture an electron to produce \iso{7}Li.  Whether the \iso{7}Be follows path 1) or 
path 2) depends critically on the temperature of the region. Owing to efficient mixing in the convective 
envelope, some of the \iso{7}Be is mixed into a cooler region which prevents proton capture. The \iso{7}Be 
will undergo electron capture instead, producing \iso{7}Li.  The \iso{7}Li is also subject to proton capture 
and is eventually mixed into the hot temperature region and subsequently destroyed.  Once the envelope is 
depleted in \iso{3}He, \iso{7}Li production stops. In Figure~\ref{m6z02-li7} we illustrate the evolution 
of \iso{7}Li at the surface of a 6$\Msun$, $Z = 0.02$ model during the TP-AGB. The Li-rich phase occurs 
when the abundance of Li exceeds $\log \epsilon$(Li)$\gtrsim 2$ and lasts for $\sim$200,000~years for 
the 6$\Msun$, $Z=0.02$ model shown in Figure~\ref{m6z02-li7}.  

Some approximation to time-dependent mixing is required to produce \iso{7}Li in a HBB calculation 
because the nuclear timescale for the reactions involved in the Cameron-Fowler mechanism is similar to 
the convective turnover timescale \citep[see Fig.~2 in][]{boothroyd92}.   
Stellar evolution calculations usually use the diffusion equation to approximate mixing in stellar interiors, 
although we warn that this is only an approximation and that mixing is advective rather than diffusive. 

Stellar evolution models are able to account for the magnitude of the Li-enrichment observed in intermediate-mass 
AGB stars, even though there are considerable modelling uncertainties \citep{vanraai12,garcia13}. \citet{ventura00} 
were able to use the Li-rich phase of bright intermediate-mass AGB stars in the Magellanic Clouds as a 
constraint of mass-loss rates, which are a highly uncertain but important ingredient in stellar evolution modelling. 
\citet{ventura00} concluded that large mass-loss rates of $10^{-4}\Msun/$year are required to fit the 
observations of Li-rich AGB stars in the Magellanic Clouds. Using the Bl\"ocker mass-loss rate \citep{bloecker95}, 
they were able to constrain the $\eta_r$ parameter to $\approx 0.01$, where higher values of $\ge 0.05$ 
lead to too high mass-loss rates when compared to the population of optically visible 
luminous, Li-rich AGB stars in the Magellanic Clouds.

\subsubsection{Type I planetary nebulae}

Type I PNe are defined as a separate class based on both
abundances and morphology. \citet{peimbert87} originally defined Type I PNe to have He/H $> 0.125$, 
N/O $> 0.5$ and to show, in general, bipolar morphologies \citep{peimbert78,peimbert87,peimbert90}. 
\citet{kingsburgh94} propose a modified N threshold (N/O $> 0.8$) based on nuclear processing
constraints and find that Type I PNe constitute about 16\% of their sample.  This fraction 
is close to the fraction of bipolar PNe found by \citet{manchado03}, at 17\%. The bipolars are 
more or less the same as the Type Is, with an average N/O = 1.3. Note that selection effects
are uncertain and can be substantial \citep[e.g., not accounting for selection effects, 
roughly 23\% of the PNe are Type I in the sample of][]{sterling08}.

Type I PNe are associated with a younger, metal-rich population that evolved from 
initial stellar masses of 2--8$\Msun$ \citep{peimbert90,corradi95,pena13}.  
The origin of Type I PNe has been associated with
intermediate-mass stars experiencing HBB \citep{vw96} but the large number of Type I objects 
(roughly 17\%), combined with the very short post-AGB crossing time for $M \ge 4\Msun$ stars 
\citep[e.g.,][]{bloecker95b} suggest that the initial progenitor masses are closer 
to 3$\Msun$.

Standard AGB models of $\approx 3\Msun$ do not produce the high He/H and N/O ratios 
that are typical of Type I PNe (e.g., Figure~\ref{m3z02-nucleo}).  While some Type I PNe
may be associated with binary evolution owing to the high frequency of Type I objects
associated with non-spherical/elliptical morphologies \citep[e.g.,][]{shaw06,stanghellini10} 
some fraction of the Type I PNe will have evolved as essentially single stars. Rotation rates 
peak in main sequence stars of $\gtrsim 3\Msun$ and it has been suggested that 
rotationally induced mixing on the main sequence could be one mechanism to increase the 
post-FDU He and N abundance \citep{karakas09,lagarde12a,miszalski12,stasinska13}.

\subsection{The slow neutron capture process} \label{sec:sprocess}

\begin{figure}
\begin{center}
\includegraphics[width=6cm,angle=270]{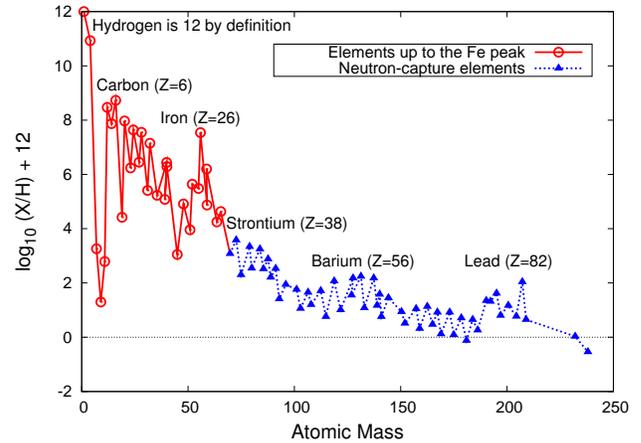}
\caption{Solar abundance distribution using data from \citet{asplund09}. The main features of the abundance distribution
include the hydrogen (proton number, $Z=1$) and helium peak, resulting from Big Bang nucleosynthesis, followed by the
gorge separating helium from carbon where the light elements lithium, beryllium, and boron reside. From carbon there is a continuous
decrease to scandium followed by the iron peak and then a gentle downwards slope to the elements predominantly 
produced by neutron captures. These include elements heavier than zinc and are highlighted in blue. Proton numbers
are also given for a selection of elements.}\label{ss}
\end{center}
\end{figure}

Most heavy nuclei with atomic masses greater than $A > 56$ are formed by neutron addition onto abundant 
Fe-peak elements. The solar abundance distribution shown in Figure~\ref{ss} is characterised by peaks 
that can be explained by:
\begin{enumerate}
\item {\em the slow neutron capture process}, the $s$-process,
\item {\em the rapid neutron capture process}, the $r$-process,
\end{enumerate}

The seminal papers by \citet*{b2fh} and \citet{cameron57} laid down the 
foundations for these processes and \citet{wallerstein97} provides an updated review 
\citep[see also the reviews by][]{meyer94,busso99,herwig05,lattanzio05,kaeppeler11}.

During the $r$-process neutron densities as high as $N_{\rm n} \gtrsim 10^{20}$ neutrons/cm$^{3}$ are produced. 
This means that the timescales for neutron capture are much faster than $\beta$-decay rates. The $r$-process will 
produce isotopes essentially up to the neutron drip line. These isotopes then decay to stable, neutron-rich isotopes once 
the neutron flux is gone. Given the extreme conditions required for the $r$-process, it has been hypothesised to 
occur during supernovae explosions \citep{fryer06,wanajo09,wanajo11,arcones07,arcones11,winteler12} but other 
sites have also been proposed including colliding neutron stars \citep{argast04,korobkin12}, and black 
hole/neutron star mergers \citep{surman08}.  
We refer to \citet{meyer94}, \citet{arnould07}, and \citet{thielemann11} for further details.

\begin{figure*}
\begin{center}
\includegraphics[width=0.5\textheight,angle=0]{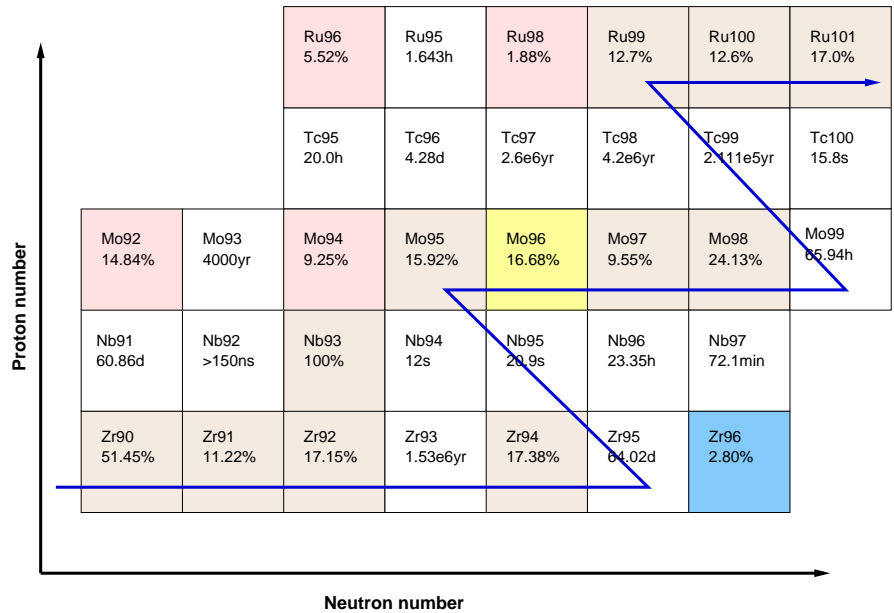}
\caption{Schematic showing the Zr to Ru region of the chart of the nuclides.
Neutron number increases along the $x$-axis and proton number on the $y$-axis. Unstable isotopes are shown as
white squares with the half-life of the ground state. Stable isotopes are shown in colour 
with the solar-system percentage shown (not all isotopes are shown, so the total may not sum to 100). 
The typical $s$-process path that results
from neutron densities typical of the \iso{13}C($\alpha$,n)\iso{16}O neutron source is shown by the thick blue line. 
Under these conditions, the isotope \iso{96}Zr is not reached by the $s$-process and is an $r$-process only 
isotope (in blue).
Similarly, the isotope \iso{96}Mo is an $s$-only isotope (in yellow) because it is shielded from the 
$r$-process by \iso{96}Zr. The isotopes 
that are not reached by neutron capture are shown in pink and are produced either by proton addition or spallation 
under extreme conditions. The unstable isotope \iso{99}Tc is on the main $s$-process path.
}\label{sprocfig}
\end{center}
\end{figure*}

For the rest of this section we will concern ourselves with the $s$-process, which occurs under conditions of 
relatively low neutron densities ($N_{\rm n} \lesssim 10^{8}$ neutrons/cm$^{3}$). In Figure~\ref{sprocfig} we show 
the typical path of the $s$-process through a section of the chart of the nuclides around the Zr to Ru region.  
During the $s$-process the timescale for neutron capture is slower, in general, than the $\beta$-decay rate of 
unstable isotopes. In Figure~\ref{sprocfig} we can see that when neutrons reach a relatively short-lived
isotope, such as \iso{95}Zr with a half-life of approximately 64 days, the isotope will have time to 
decay to \iso{95}Mo instead of capturing another neutron.  The $s$-process will therefore produce isotopes 
along the {\em valley of $\beta$-stability} and is responsible for the production of roughly half of all 
elements heavier than Fe.

While it is common to characterise elements as being produced by one of the neutron-capture
processes (e.g., Eu is an $r$-process element), we stress that most heavy elements are produced 
in part by both the $s$ and the $r$-processes. However, there are some {\em isotopes} that are only
produced by the $s$-process (e.g., \iso{96}Mo in Figure~\ref{sprocfig}) because of shielding from the
$r$-process, while some neutron-rich isotopes cannot be easily reached by the $s$-process (e.g., \iso{96}Zr). 
Some proton-rich heavy isotopes cannot be reached by either the $s$- or $r$-process and are usually rare 
in nature and a small component of the total elemental fraction. Examples include \iso{92,94}Mo, 
shown in Figure~\ref{sprocfig}, which together make up 24\% of elemental Mo in the solar system. 
\citet{arlandini99} used detailed AGB stellar models to provide a breakdown of the solar system isotopic abundance 
distribution according to an origin in either the $r$- or $s$-process 
\citep[their Table 2; see similar breakdowns by][]{goriely99,travaglio04,simmerer04,sneden08}.
Examples of elements where the solar-system abundance fraction is mainly produced by the $s$-process 
include Sr, Y, Zr, Ba, La, and Pb, where 91\% of solar-system Pb was produced in the $s$-process 
\citep[Table 3 in][]{travaglio01a}. Similarly some elements such as Ag, Xe, Eu are predominantly 
produced by the $r$-process \citep[e.g., 97\% of solar-system Eu, see][]{sneden08}.

In Figure~\ref{ss} we show the distribution of elements in our solar system using the latest set of solar abundances 
from \citet{asplund09}. We have highlighted elements heavier than Fe in blue, and an examination of this figure shows 
peaks around Sr, Ba, and Pb (corresponding to atomic masses 88, 137, 207 respectively).
These elements are dominated by nuclei with a magic number of neutrons 
($n = 50, 82, 126$).  Note that for lighter elements there are also peaks at $n = 2, 8, 20$, and 28. 
A nucleus composed of a magic number of protons and a magic number of neutrons is very stable and considered to be ``doubly magic''.  
Examples include \iso{16}O with 8 protons and 8 neutrons, and \iso{208}Pb, with 82 protons and 126 neutrons.
Supernovae produce a considerable amount of \iso{56}Ni, which is doubly magic with 28 protons and 28 neutrons and 
eventually decays to \iso{56}Fe and is the cause of the Fe-peak seen in Figure~\ref{ss}. The stability of nuclei 
with a magic number of nucleons follows from the closed shells in the quantum mechanical model of the nucleus \citep{mayer50}.
In practice, nuclei with a magic number of neutrons are more stable against neutron capture than surrounding nuclei 
because of their low neutron capture cross sections. These nuclei act as  bottlenecks and are consequently 
seen as $s$-process peaks in the abundance distribution.

Theoretically there are two main astrophysical ``sites'' of the $s$-process in nature. The first are AGB stars, 
which are observationally and theoretically confirmed as factories for the production of heavy elements. 
The first evidence that stars and not the Big Bang are responsible for the production of elements heavier than Fe
came from observations by \citet{merrill52} of radioactive Tc in red giant stars. 
The second main $s$-process site is massive stars, where the $s$-process occurs during core He burning and in the
convective C-burning shell, prior to the supernova explosion \citep{the00,the07,pignatari10,frischknecht12}. 
We will concentrate on the $s$-process occurring in AGB stars. 

\subsubsection{Neutron sources in AGB stars}

Neutron capture processes require a source of free neutrons, given that neutrons are unstable and decay in 
about 15 minutes. There are two important neutron sources available during He-shell burning in AGB stars:
\begin{enumerate}
\item ~\iso{14}N($\alpha,\gamma$)\iso{18}F($\beta^{+}\nu$)
\iso{18}O($\alpha,\gamma$)\iso{22}Ne($\alpha,n$)\iso{25}Mg.
\item ~\iso{12}C($p,\gamma$)\iso{13}N($\beta^{+}\nu$)\iso{13}C($\alpha,n$)\iso{16}O.
\end{enumerate}

The \iso{22}Ne($\alpha,n$)\iso{25}Mg reaction was first identified as a neutron source for AGB stars by \citet{cameron60}. 
The intershell region is rich in \iso{14}N from CNO cycling and during a thermal pulse, \iso{14}N can suffer successive 
$\alpha$ captures to produce \iso{22}Ne. If the temperature exceeds $300 \times 10^{6}$~K, \iso{22}Ne starts to capture 
$\alpha$ particles to produce \iso{25}Mg and \iso{26}Mg in almost equal proportions. Neutrons are released by the 
\iso{22}Ne($\alpha,n$)\iso{25}Mg reaction during convective thermal pulses.  Given the high temperatures required for 
the \iso{22}Ne($\alpha,n$)\iso{25}Mg to operate efficiently, it is theoretically 
predicted to be effective in AGB stars with initial masses $\gtrsim 4\Msun$. 

The observational data for the $s$-process mainly comes from ``intrinsic'' low-mass AGB stars and their progeny, 
with initial progenitor masses $\lesssim 4\Msun$. The \iso{22}Ne($\alpha,n$)\iso{25}Mg is not efficient in these
stars and requires operation of the \iso{13}C($\alpha,n$)\iso{16}O neutron source instead.
Observations come from stars with spectral types M, S, SC, C(N), post-AGB stars
and planetary nebulae \citep{smith86b,vanture91,abia01b,abia02,abia08,vanwinkel00,reyniers07a,sterling08,vanaarle13}.
Extrinsic $s$-process rich objects also provide a wealth of observational data and
include barium and CH-type stars, carbon enhanced metal-poor (CEMP) stars, dwarf C stars, and some planetary
nebulae \citep{luck91,allen06a,allen06b,allen07,suda08,suda11,masseron10,pereira12,miszalski13}. 
These observations are consistent with theoretical models covering a broad range in metallicity and mass
\citep[e.g.,][]{hollowell88,gallino98,goriely00,busso01,lugaro03b,karakas07a,karakas09,cristallo09,cristallo11,
karakas10c,bisterzo10,bisterzo11,bisterzo12,lugaro12,pignatari13}.

Efficient activation of the \iso{13}C($\alpha,n$)\iso{16}O reaction requires some \iso{13}C to be present 
in the intershell.  CNO cycling during the previous interpulse phase leaves a small amount of \iso{13}C but 
not enough to account for the $s$-process enrichments of AGB stars \citep[e.g.,][]{gallino98,karakas07a}.   
For the \iso{13}C neutron source to produce enough neutrons to feed the $s$-process there has to be an
additional source of \iso{13}C. This requires the operation of both proton and $\alpha$-capture reactions in
the He intershell, a region normally devoid of protons. 

If some protons are mixed from the convective envelope into the top layers of the He intershell then these protons will 
react with the abundant \iso{12}C to produce \iso{13}C via the CN cycle reactions:
 \iso{12}C(p,$\gamma$)\iso{13}N($\beta^{+}\nu$)\iso{13}C \citep{iben82a}.
This results in a thin layer rich in  \iso{13}C and \iso{14}N 
known as the ``\iso{13}C pocket'' \citep{iben82b}.  \citet{straniero95} discovered that the \iso{13}C nuclei
then burns via ($\alpha$,n) reactions in radiative conditions before the onset of the next thermal pulse.  
The neutrons are released in the \iso{13}C pocket, and the $s$-process occurs between 
thermal pulses in the same layers where the \iso{13}C was produced. When the next convective thermal pulse occurs, 
it ingests this $s$-element rich layer, mixing it over the intershell. 

It does appear that there is a dichotomy in models of the $s$-process in AGB stars. This arises because low-mass
models do not reach temperatures high enough to activate the \iso{22}Ne source and as a consequence the \iso{13}C
neutron source is dominant. At a metallicity of $Z=0.02$ the mass at which the importance of the two neutron sources 
switch is $\simeq 4\Msun$ \citep{goriely04}.  This dichotomy has important implications for the yields of $s$-process 
elements produced by AGB stars of different mass ranges.

The \iso{13}C and the \iso{22}Ne neutron sources produce $s$-process abundance distributions that are very
different from each other.   There are two main reasons for this. The first is that the \iso{13}C 
source operates over long timescales ($\approx 10^{3}$ years), which means that the time integrated neutron fluxes
are high even if the peak neutron densities are lower ($\lesssim 10^{7}$ neutrons/cm$^{3}$) than for the 
\iso{22}Ne source \citep{busso01}. This means the $s$-process can reach isotopes beyond the first $s$-peak at Sr-Y-Zr to Ba and 
Pb \citep{gallino98}. It is for this reason that the \iso{13}C source is responsible for the production of the bulk of the 
$s$-process elements in low-mass AGB stars reaching, as mentioned above, up to Pb at low metallicities \citep{gallino98}. 
In contrast, the \iso{22}Ne source operates on timescales of $\approx 10$ years and even though the peak
neutron densities are high (up to $10^{15}$ neutrons/cm$^{3}$) the time integrated
neutron fluxes are low and the $s$-process will not, in general, reach beyond the first $s$-process peak. 

The second reason for the difference in the predicted distribution is that branching points on the $s$-process
path are activated by the \iso{22}Ne source \citep{abia01b,vanraai12,karakas12}. For example, the amount of Rb
produced during the $s$-process depends on the probability of the two unstable nuclei \iso{85}Kr and \iso{86}Rb
capturing a neutron before decaying. These two isotopes therefore act as ``branching points'' and the probability
of this occurring depends on the local neutron density \citep{beer89}. When the \iso{22}Ne source is active, 
branching points at \iso{85}Kr and \iso{86}Rb are open and \iso{87}Rb is produced. In particular, more Rb is produced
relative to Sr (or Y or Zr). In constrast, during the operation of the \iso{13}C neutron source these branching
points are not efficiently activated and the ratio of Rb to Sr (or Y or Zr) remains less than unity.

\subsubsection{The formation of \iso{13}C pockets} \label{c13pockets}

For the \iso{13}C($\alpha,n$)\iso{16}O reaction to occur efficiently, some partial mixing is required at the border 
between the H-rich envelope and the C-rich intershell. 
This mixing pushes protons into a C-rich region suitable for the production of $^{13}$C.
It is important that there are not too many protons mixed into this region because 
then the CN cycle goes to completion, producing \iso{14}N rather than \iso{13}C.
Now \iso{14}N is a  neutron poison, which means that it is an efficient neutron absorber and will 
change the resulting abundance distribution. 
In the region of the pocket where \iso{14}N is more abundant than 
\iso{13}C, no $s$-process nucleosynthesis occurs because of the dominance of the \iso{14}N(n,p)\iso{14}C reaction over 
neutron captures by Fe-seed nuclei and their progeny.  

In the intermediate-mass AGB stars that experience HBB, the formation of a \iso{13}C pocket may be inhibited by proton 
captures occurring at the hot base of the convective envelope during the TDU, which produces \iso{14}N and not \iso{13}C 
\citep{goriely04}. Extremely deep TDU may also inhibit the activation of the \iso{13}C($\alpha$,n)\iso{16}O reaction by 
penetrating into regions of the stellar core with a low abundance of He \citep{herwig04a}. 
Furthermore, a lack of Tc in the spectra of intermediate-mass AGB stars that are rich in Rb 
is observational evidence that \iso{13}C pocket formation is inhibited \citep{garcia13}.

The details of how the \iso{13}C pocket forms and its shape and extent in mass in the He intershell are still unknown.
These are serious uncertainties and mostly arise from our inability to accurately model convection in stars \citep{busso99}.
Various mechanisms have been proposed including partial mixing from
convective overshoot \citep{herwig00,cristallo09,cristallo11}, rotation
\citep{herwig03,piersanti13}, and gravity waves \citep{denissenkov03a}. 

Progress will come from continued 3-D hydrodynamical simulations of the interface between the envelope 
and the intershell of AGB stars.  The first simulations by \citet{herwig06} have provided some insight into 
the nature of convection during thermal pulses but are still too crude in space and time resolution, 
and cover such a small amount of star time that we are still very limited in any insights into the nature of 
\iso{13}C pocket formation in low-mass AGB stars.

Models that include artificial \iso{13}C pockets produce $s$-process abundance distributions that fit the 
observational data reasonably  well. Free parameters allow us to adjust the features of the mixing zone 
(e.g., the shape of the \iso{13}C profile and its extent in mass) in order to match the observations 
\citep{goriely00,cristallo09,bisterzo10,kamath12,lugaro12}. Some of the best observational
constraints come from post-AGB stars, where the higher photospheric temperatures
allow for more accurate abundance determinations. 
Observations suggest that stochastic variations in the size of the \iso{13}C pocket in AGB stars are present. 
\citet{axel07a} find that Galactic disk objects are reproduced by a spread of a factor of two or three 
in the effectiveness of the \iso{13}C pocket, lower than that found by \citet{busso01}, who needed a 
spread of a factor of about $20$.  Comparisons to lower metallicity post-AGB stars infer spreads
of a factor of 3--6 \citep{axel07b,desmedt12}, while comparison to even lower metallicity CEMP stars 
require spreads of up to a factor of 10 or more \citep{bisterzo12,lugaro12}.

In summary, there is observational evidence that a spread in effectiveness of the \iso{13}C pocket
is needed in theoretical models, but there is no consensus on how large that spread actually is. This
problem indicates a significant lack of understanding of the mechanism(s) responsible for the 
formation of \iso{13}C pockets in AGB stars. In \S\ref{sec:sprocun} we discuss other uncertainties related 
to the modelling of the $s$-process in AGB stars such as stellar rotation.

\subsubsection{The $s$-process in low-mass AGB stars} \label{sec:lowmass}

\begin{figure}
\begin{center}
\includegraphics[width=6cm,angle=270]{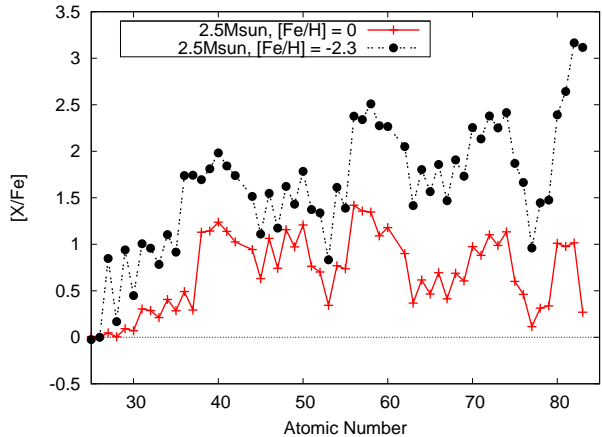}
\caption{Average abundance in the stellar wind (in [X/Fe]) for elements heavier than iron
for AGB models of 2.5$\Msun$ at two different metallicities: $Z=0.0001$ using data published in \citet{lugaro12},
and new predictions for the $Z=0.02$ model. In both models the same size partially mixed zone
is inserted into the post-processing nucleosynthesis 
calculations to produce a \iso{13}C pocket (see text for details). The average abundance
is calculated from the integrated yield of mass expelled into the interstellar medium over
the model star's lifetime.}\label{m2.5}
\end{center}
\end{figure}

%%%%%%%%%%%%%%

%1
We first define the $s$-process indices light ``ls'' and heavy ``hs''.   We choose the three main elements 
belonging to the first $s$-process peak Sr, Y, and Zr to define [ls/Fe]=([Sr/Fe]+[Y/Fe]+[Zr/Fe])/3, and 
three main elements belonging to the second $s$-process peak Ba, La, and Ce to define 
[hs/Fe]=([Ba/Fe]+[La/Fe]+[Ce/Fe])/3.  These are chosen because data for these elements are often available
in observational compilations e.g., CEMP stars
\citep[see the SAGA database and other compilations,][]{suda08,suda11,frebel10,masseron10}.

%2

The ``intrinsic'' ratios [hs/ls] and [Pb/hs] reflect ratios of elements
only produced by AGB stars, differing from e.g., [C/Fe] and [Ba/Fe], because Fe is not produced during AGB nucleosynthesis.
These intrinsic ratios move away from their initial solar values towards their $s$-process values 
after a small number of thermal pulses in low-mass AGB models (in intermediate-mass models this is not necessarily the case
owing to the large dilution of the envelope). The intrinsic ratios are, in a first 
approximation, independent of stellar modelling uncertainties that affect comparison to the observations
including TDU, mass loss, stellar lifetime, and accretion and mixing processes on a binary companion.
Instead, they mostly constrain the nucleosynthesis occurring in the deep layers of the star.

%3
We have seen that some partial mixing is assumed to produce a \iso{13}C pocket 
that provides the neutrons for low-mass stars. Some authors add a partially mixed
region by parameterising an overshoot zone where the mixing velocity falls to 
zero \citep{herwig00,cristallo09,cristallo11}. The models by \citet{bisterzo10}
instead artificially introduce a \iso{13}C pocket into a post-processing code. The details of the 
pocket are free parameters and kept constant from pulse by pulse. Starting from a ``standard case'' 
first adopted by \citet{gallino98} in order to match the $s$-process main component, \citet{bisterzo10}
multiply or divide the \iso{13}C and \iso{14}N abundances in the pocket by different factors.

In our calculations the inclusion of the \iso{13}C pocket was made artificially during the post-processing by forcing 
the code to mix a small amount of protons from the envelope into the intershell \citep[e.g.,][]{karakas07a,lugaro12}.
We assume that the proton abundance in the intershell decreases monotonically from the envelope value 
of $\simeq$ 0.7 to a minimum value of 10$^{-4}$ at a given point in mass located at ``M$_{\rm mix}$'' below 
the base of the envelope. This method is described in detail in \citet{lugaro12} and is very 
similar to that used by \citet{goriely00}. The protons are subsequently captured by \iso{12}C to form a 
\iso{13}C-rich layer during the next interpulse, where the \iso{13}C pocket is much less than $M_{\rm mix}$ 
and is typically about 1/10$^{\rm th}$  the mass of the intershell region.  In the calculations shown in 
Figures~\ref{m2.5} and~\ref{z0001} we set $M_{\rm mix} = 2\times 10^{-3}\Msun$ and this value is held constant 
for each TDU episode. For further details on our method for introducing \iso{13}C pockets 
we refer to \citet{lugaro04}, \citet{karakas10a}, \citet{karakas07a}, and \citet{kamath12}.

%4
In Figure~\ref{m2.5} we show $s$-process predictions from models of 2.5$\Msun$ at two different metallicities:
[Fe/H] = $-2.3$ and [Fe/H] = 0.0.  In both models we manually add a \iso{13}C pocket
into the top layers of the He intershell at the deepest extent of each TDU episode, as discussed above.
The operation of the \iso{13}C($\alpha$,n)\iso{16}O reaction produces [hs/ls] $\ge 0$ for low-mass AGB models,
regardless of metallicity and the size of the \iso{13}C pocket. For example, for the 2.5$\Msun$ models shown in
Figure~\ref{m2.5} the average [hs/ls] in the ejected wind is 0.20 and 0.58, for the $Z=0.02$
and $Z = 0.0001$ models, respectively. Figure~\ref{m2.5} illustrates that the initial metallicity has a 
strong impact on $s$-process abundance predictions, with significantly more Pb produced at lower metallicity.

%5
The metallicity dependence arises
because (most of) the \iso{13}C nuclei needed for the \iso{13}C($\alpha$,n)\iso{16}O are produced from the
H and He initially present in the model star, which is fused into \iso{12}C and then to \iso{13}C. 
That is to say that it is a {\em primary\/} neutron source, and is essentially independent of 
the metallicity of the star, or in other words, largely independent of [Fe/H]. These neutrons are captured 
on the heavy-element seeds, whose number is roughly proportional to [Fe/H], which is again roughly
proportional to the metallicity Z.

This means that the number of time-integrated neutron captures from the \iso{13}C source is proportional
to \iso{13}C/$Z$ \citep{clayton88}. Thus there are more neutron captures, producing heavier elements,
for stars of lower [Fe/H] \citep{busso01}. 
This property allowed \citet{gallino98} to predict the existence of low-metallicity Pb-rich stars, which was 
confirmed by observations of Pb-rich CEMP stars \citep{vaneck01,vaneck03}.  Note that for our 2.5$\Msun$ 
examples, the [Pb/hs] ratio in the ejected stellar wind is $-0.36$ for the $Z=0.02$ model 
and increases to 0.76 for the $Z=0.0001$ model. Here we have focused on predictions for AGB models with 
[Fe/H] $\gtrsim -2.5$; lower metallicity AGB models will be discussed in \S\ref{sec:PIE2}.

%%%%%%%%%%%%%%%%%%%%%%%%%%%%%%%%%%%%%%%%%%%%%%%

\subsubsection{The $s$-process in intermediate-mass AGB stars}

\begin{figure}
\begin{center}
\includegraphics[width=6cm,angle=270]{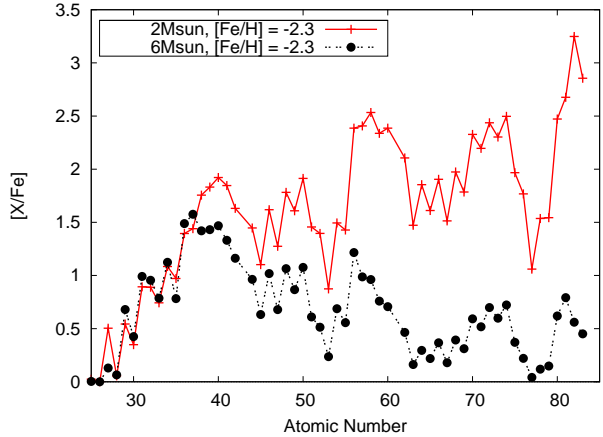}
\caption{Average abundance predicted in the ejected wind (in [X/Fe]) for elements heavier than iron
for AGB models of 2$\Msun$ and 6$\Msun$ at a metallicity of $Z=0.0001$ ([Fe/H] = $-2.3$) using data published in \citet{lugaro12}.
No \iso{13}C pocket is included in the 6$\Msun$ model, while we set $M_{\rm mix} = 2\times 10^{-3}\Msun$ in the
2$\Msun$ case (see \S\ref{sec:lowmass} for details).}\label{z0001}
\end{center}
\end{figure}

\citet{wood83} noted that the brightest, O-rich AGB stars 
in the Magellanic Clouds exhibit strong molecular bands of ZrO, indicating that the atmospheres of these stars are 
enriched in the $s$-process element Zr. \citet{garcia06} and \citet{garcia09} identified several bright, 
O-rich Galactic and Magellanic Cloud AGB stars with significant enrichments of the neutron-capture element Rb. 
These observations support the prediction that efficient TDU and HBB has occurred in these stars  
\citep{garcia07a}.

An enrichment in the element Rb over the elements Sr, Y, and Zr is an important clue that points toward 
the efficient operation of the \iso{22}Ne($\alpha$,n)\iso{25}Mg neutron source in intermediate-mass AGB stars 
\citep{truran77,cosner80,lambert95,abia01b,vanraai12}. Under conditions of 
high neutron densities two branching points open that allow Rb to be synthesised \citep{vanraai12}. 
At $N_{\rm n} = 5 \times 10^{8}$ n/cm$^{3}$, $\approx 86$\% of the neutron flux goes through \iso{85}Kr 
allowing for \iso{85}Kr($n,\gamma$)\iso{86}Kr($n,\gamma$)\iso{87}Kr that decays to \iso{87}Rb. Note that 
\iso{87}Rb has a magic number of neutrons and the probability to capture neutrons is extremely low.
Also, high neutron densities allow neutrons to bypass the branching point at \iso{85}Kr and \iso{86}Rb, allowing for the 
chain \iso{86}Rb($n,\gamma$)\iso{87}Rb. For this reason, the elemental ratios of Rb/Sr and Rb/Zr are indicators 
of the neutron densities, and have been used as evidence that the \iso{13}C($\alpha,n$)\iso{16}O reaction 
is the major neutron source in low-mass AGB stars \citep{lambert95,abia01b}.

In Figure~\ref{z0001} we show the predicted $s$-process abundance pattern from a 6$\Msun$, $Z = 0.0001$ AGB model 
alongside predictions from a low-mass 2$\Msun$, $Z=0.0001$ AGB model. The predictions for the 6$\Msun$ model
are typical of intermediate-mass AGB stars and the operation of the \iso{22}Ne neutron source in that 
elements at the first peak around Rb dominate over elements at the second peak, around Ba. This is in 
contrast to the $s$-process abundance pattern from the 2$\Msun$ model, which is typical of the \iso{13}C 
neutron source operating in a low-mass, low-metallicity AGB model in that it produces significant amounts of 
Sr, Ba, and Pb, where the average ejected [Pb/Fe] $\approx 3.2$ (see also Figure~\ref{m2.5}).

The 6$\Msun$ model produces considerable Rb, where the average [Rb/Fe] = 1.60 in the ejected wind, 
higher than the neighbouring [Sr/Fe] = $1.4$, which gives [Sr/Rb] $< 0$.  The $s$-process indicators, 
[hs/ls] and [Pb/hs] are negative ($-0.38$ and $-0.50$, respectively for the 6$\Msun$, $Z = 0.0001$ model), 
indicating that elements at the first $s$-process peak around Rb-Sr-Y-Zr are predominantly produced.  
Figure~\ref{z0001} highlights why it is important to have $s$-process yields covering a large range in mass, 
not just metallicity as shown in Figure~\ref{m2.5}.

There are few model predictions of $s$-process nucleosynthesis from intermediate-mass AGB stars, in contrast to the
situation for lower mass AGB stars. This is partly because the models experience many thermal pulses and 
computations involving hundreds of isotopes are particularly time-consuming. It is also because there are 
few observational constraints that there is 
much debate as to the occurrence (or not) of the TDU in intermediate-mass AGB stars.

The $s$-process predictions published for intermediate-mass stars include
the following works: \citet{goriely04}, who studied the interplay
between hot TDU episodes and the $s$-process as a function of mass; \citet{karakas09}, who provided 
predictions up to the first $s$-process peak for comparison to Type I PNe;  \citet{lugaro12} who provided predictions
for a mass grid ($M=1$ to $6\Msun$) at $Z=0.0001$ (or [Fe/H] $\approx -2.3$); \citet{karakas12} who extended the
study to $Z = 0.02$ for comparison to the Galactic OH/IR sample; 
\citet{dorazi13b} who presented new predictions for comparison to globular
cluster stars; \citet{pignatari13} who provided NuGrid $s$-process predictions from a 5$\Msun$ model at two 
metallicities ($Z =0.02, 0.01$); and \citet{straniero14} who provide predictions for 4, 5, and 6$\Msun$
models at one metallicity ($Z=0.0003$ and [$\alpha$/Fe] = $+$0.5).
The models by \citet{pignatari13} are calculated with the MESA stellar 
evolution code\footnote{MESA stands for ``Modules for Experiments in Stellar Astrophysics'': {\tt http://mesa.sourceforge.net/}.} 
and include convective overshoot into the C-O core and consequently have higher peak temperatures 
than canonical models without such overshoot. 

\subsection{Proton ingestion episodes: PIEs} \label{sec:PIE2}

%1 Intro
Stellar evolution at low metallicity presents a new phenomenon, which we have briefly discussed in \S\ref{sec:PIE1}.
We saw that during the core flash in very low metallicity models ([Fe/H] $\lesssim -3$) we can find 
contact between the flash-driven convective region, which is rich in He, and the H-rich envelope. 
There are two reasons why this happens preferentially in lower metallicity stars: firstly, the core flash is ignited far off centre,
and relatively close to the H-rich envelope. Secondly, 
there is normally a substantial entropy gradient that acts to keep these two regions separated. But at 
lower metallicity this gradient is greatly reduced, enabling the two convective regions to make contact. This results in H-rich material
being mixed down to temperatures where He is burning almost explosively, at $T \gtrsim 100$MK. There are many
names for these events in the literature, some quite convoluted. We follow \citet{campbell08} and refer 
generically to these events as ``proton ingestion episodes'' or PIEs. We shall define two different kinds below.
We note in passing that 3D simulations show that turbulent entrainment of material could produce similar PIEs at 
higher metallicities \citep{mocak08,mocak09} although this is not found in simple 1D models.

%2 History
It was realised some time ago that PIEs may arise in the evolution of low mass and low metallicity 
stars \citep{dantona82,fujimoto90,hollowell90,cassisi96}. Later work studied how such events depended on the
numerical details \citep{schlattl01} and the possibility of explaining some of the abundances
seen in CEMP stars 
\citep[][]{hollowell90,fujimoto00,chieffi01,schlattl01,schlattl02,picardi04,weiss04,suda04,lau09} 
including possible $s$-process nucleosynthesis \citep{goriely01}.

%3 Defintions, when and where
A PIE is defined as any event that mixes protons into a very hot region, typically at temperatures
where He burns, or is already burning. These naturally occur during a He flash which drives strong convection.
They could be He-core flashes or even He-shell flashes on the AGB. When the H is exposed to very high
temperatures it produces a secondary ``H-flash'' and a PIE results in a ``dual-flash'' event:~a He-flash
rapidly followed by the initiation of a H-flash.
Although both are defined as PIEs, \citet{campbell08} distinguish between them according to
whether the initiating He flash is a result of a core flash or an AGB shell flash: hence we have dual 
core flashes (DCFs) and dual shell flashes (DSFs). 

Note that PIEs are also seen in the ``late hot flasher'' scenario \citep{sweigart97,brown01,cassisi03} and 
the ``born-again'' or ``(very) late thermal pulse'' scenario for explaining objects like Sakurai's Object 
\citep{iben83,asplund97,herwig99}. We do not discuss these phenomena here, but we remind the reader that the 
essential physics is the same, involving dredge-up of CNO-processed material to the 
surface and even neutron captures, as we
discuss below. For a comparison of nucleosynthesis predictions in very late thermal pulses and
data from pre-solar grain analysis see \citet{fujiya13} and \citet{jadhav13}.

%4 1D vs 3D
The largest uncertainty with calculating the evolution during a PIE is how to handle the 
convective mixing. This must be done in a time-dependent way because the convective turnover timescale
is close to the burning timescale for H at the high temperatures found during a PIE.
Usually convective mixing is approximated by solving a diffusion equation, and usually in 1D.
Within this approximation the diffusion coefficient $D$ is naively given by $\frac{1}{3} v l$ where
$v$ is the velocity of the blob and $l$ is its mean-free-path. These values are normally taken from the 
Mixing-Length Theory.  The results are to be taken with all of the caveats that come with a literal 
interpretation of this phenomenological theory.  Note that one of the assumptions of the Mixing-Kength Theory
is that nuclear burning is negligible in the blob of gas, and this is
clearly untrue in the present case. 

Within the 1D diffusive mixing approximation the protons are mixed down to a region 
where they burn fiercely and generate essentially as much energy as the He-burning reactions, which 
initiated the convection in the first place. This leads to a split in the convective zone:~two separate 
convective burning shells develop, the inner one burning He and the outer burning H, with the two separated 
by an initially narrow %narrow (in mass) 
radiative region. Clearly the details of this process will depend on how one calculates the burning and mixing, 
and that is essentially a 3D process as we expect burning plumes to develop, rather than strict spherical 
symmetry. Fortunately, multi-dimensional hydrodynamical simulations have progressed to the stage where 
exploratory calculations are possible \citep[to see simulations of normal shell flash convection see][]{herwig06}. 
We will discuss this further below in \S\ref{sec:3DPIE}.

%6 1D NS
\subsubsection{Overview of 1D PIEs}\label{sec:1DPIE}

Most calculations of PIEs have been done within the paradigm 
of 1D models using the diffusion approximation for mixing 
\citep[for some recent calculations please see][]{campbell08,cristallo09,lau09,campbell10,suda10}.
Following the PIE, dredge-up mixes the products of H and He burning to the
stellar surface, and the surface layers of the models are dramatically enriched in \iso{12}C and \iso{14}N. 
One of the motivations for the study of such events initially was that they share some of the characteristics of 
CEMP stars (see references given above).
Although there has been a reasonable number of calculations of such events, most authors have stopped their calculations
after the PIE rather than go to the end of the star's evolution. It is only if that is done that we can determine
the yields for such stars (see \S\ref{sec:1DPIEy}). Until we have reliable calculations of PIEs, it will be hard to
provide improved yields for stars undergoing PIEs.

\begin{figure}
\centering
\includegraphics[width=1.0\columnwidth]{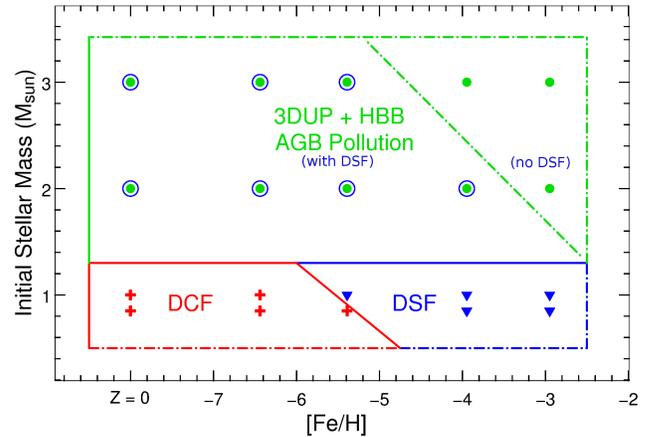}
\caption{Models calculated by \citet{campbell08} in the [Fe/H]-mass plane. Red crosses represent models
where the DCF dominates the nucleosynthesis. Filled blue triangles  show where DSFs dominate. Open blue circles
indicate models that experience DSFs, although they are not the dominant event for those models. Green filled circles
are used where TDU and HBB on the AGB dominate the nucleosynthesis occurring. The models for $Z=0$ are plotted at the 
position of [Fe/H] $=-8$.}
\label{PIEyields}
\end{figure}

\citet{campbell08} calculated the complete evolution for low and zero metallicity stars of low mass, through
to the end of the AGB phase, so that they included the PIEs, third dredge-up, and HBB. They were able to delimit regions 
in mass and metallicity space where specific nucleosynthesis events dominate. For example, the DCF dominate the 
nucleosynthesis  for $M \lesssim 1.5\Msun$ and [Fe/H] $\lesssim -5.$ For higher metallicities, in the same mass range, 
the DSF dominates. Increasing the mass allows for the TDU and HBB to become active, and dominate the
production of elements. For [Fe/H] $\gtrsim -4$ (with a slight mass dependence) the DSF does 
not occur and normal AGB evolution results. The various regions are shown in Figure~\ref{PIEyields} taken from 
\citet{campbell08}.   Note that uncertainties in the location of convective borders, in particular, have a 
substantial effect on the borders of the various regimes shown in the figure. Nevertheless, the agreement 
with a similar plot in \citet{fujimoto00} is very good.

\subsubsection{1D PIEs and the $i$-process}\label{sec:1DPIEs}
With protons and \iso{12}C in abundance one may expect neutrons to be present, provided by the \iso{13}C neutron source. 
This indeed is the case, with calculations showing a ``neutron super-burst'' when convection engulfs the long 
H tail left behind by the H shell \citep{fujimoto00,iwamoto04,campbell10}. 
This tail is longer than normal because of the very low
metallicity in these stars, which means that the pp chains (showing a relatively low temperature dependence) 
play a significant role in the H burning, and 
the shell is more extended than is the case when there are enough catalysts for efficient CNO cycling (which has a 
much higher temperature dependence).
During the peak of the burst we find neutron densities as high as $10^{15}$ neutrons\,cm$^{-3}$
\citep{cristallo09,campbell10,herwig11}.

Similar, but not identical, results were found by \citet{cruz13} and \citet{bertolli13}. 
The neutron densities are higher than expected for the $s$-process, but well short of what is needed for the 
$r$-process. These neutron exposure events have been called the $i$-process, for ``intermediate'' neutron capture. 
These stars are of great interest to those trying to understand the CEMP-r/s stars, which show evidence of 
neutron capture nucleosynthesis that lies between the $s$- and $r$-process extremes 
\citep{lugaro09,lugaro12,bisterzo12,bertolli13}. The patterns predicted by the models
do a reasonably good job of matching the observations, assuming that the observed CEMP-r/s stars result from 
mass transfer in a binary (with appropriate dilution). The fit is not perfect, however, 
with too much Ba \citep{campbell10} and N \citep{cruz13} produced to match HE1327-2326. 
Clearly  the field is ripe for further study, but see caveats concerning 1D models below.
For a study of the (not insignificant) impact of nuclear uncertainties, see \citet{bertolli13}.

\subsubsection{Yields from 1D PIEs}\label{sec:1DPIEy}
%7 Yields
To the best of our knowledge, the only yields available at present for the complete evolution of
low-mass stars, including PIEs and later AGB phases, are those given by \citet{campbell08} and \citet{iwamoto09}.
One reason for the dearth of such calculations is the legitimate question over
the validity of the 1D and Mixing-LengthTheory (MLT) approximations used in most calculations. Further
there is the question of mass-loss rates for low metallicity stars, although 
this is mitigated somewhat by the fact that 
dredge-up events produce substantial contamination of the envelope during 
the star's evolution.  We are only now beginning
to develop hydrodynamical models that may be used for guidance in such cases.

%5 2 zones or 1 zone
%8 2D and 3D models
\subsubsection{Multi-dimensional PIE calculations}\label{sec:3DPIE}
Since PIEs clearly challenge the assumptions used in most 1D stellar models, they have been the subject of
various multi-dimensional hydrodynamical calculations. Extra motivation for this is the work of \citet{herwig01}
who showed that he could match the observed evolutionary timescale of Sakurai's Object if the convective
timescale during the PIE was about 30 to 100 times slower than predicted from the MLT. 

The simulations performed
with the {\sl Djehuty\/} code by \citet{stancliffe11} found significant inhomogeneities in the mixing
process \citep[see also][]{herwig11}, with plumes of H mixed all the way down to the CO core where they burned rapidly. 
These calculations showed no sign of the split into two convective zones that is a universal feature of 1D MLT calculations.
Similarly, \citet{mocak10} found that a calculation that started with a split convective zone soon showed merging into
one zone. The final word is yet to be spoken, but a significant step forward
was made by \citet{herwig13}.
They found an initial period showing one deep convective region, but that
this soon  developed an entropy step that seems to divide two 
mixing zones {\em within\/} a single convective zone. The mixing between the zones seems to be 
inefficient but not negligible.
Indeed, they may later separate, but we cannot be sure because  the 3D simulations covered 
only a relatively short simulation time of $\sim 10$ hours.
It is noteworthy that \citet{herwig11} found that such a history of mixed regions
was able to match the nucleosynthesis of Sakurai's object and that these new results 
are also in agreement with the convection constraints imposed by the evolutionary timescale of Sakurai's object.

It is a common result from these hydrodynamical  models that the radial velocities of the material are significantly 
higher than found in the MLT, by factors of 20--30. Whereas MLT predicts 
velocities of a few km\,s$^{-1}$ the simulations show spherically averaged values of 10-20 km\,s$^{-1}$ with
individual plumes reaching even higher velocities at times \citep{stancliffe11,herwig11}. 
This seems to contradict the 1D models of \citet{herwig01}
who could only match the observed evolutionary timescale of Sakurai's object by decreasing the MLT mixing efficiency by
factors of 30-100. But this need not be the case.  Reducing the convective efficiency
places the H-burning closer to the surface, where the response time to thermal perturbations is shorter. 
This was found to occur in the simulations of \citet{herwig13}, who found the position of the 
split in the convective zone matched the location found by the 1D code using the reduced convective efficiency. 
Further, \citet{herwig11} showed that if there was a single mixed zone
for the first $\sim 900$min of the simulation, before the split into two zones occurs, then the simulation
was able to match the observed abundances also.

In summary, it appears the PIEs are crucial events in the nucleosynthesis of
low mass, low metallicity stars. They may be involved in explaining some of the observations of CEMP stars.
However, we should have some serious concerns about the reliability of the 1D models during this phase.
Although the deviations in temperature from spherical symmetry are small \citep{stancliffe11} there are substantial 
inhomogeneities in the composition in the mixed regions. The details of the PIE and how (and if) the resulting 
convective zone splits into two require much further work. Significant work has been done but we are only at the 
beginning, and are still restricted by numerical resolution. Indeed, to the best of our knowledge there is only 
one calculation that has shown convergence of results with increasing resolution \citep{woodward13}, which is 
crucial for these studies. Most calculations are almost certainly limited by resolution in a way that has not 
been quantified. The fact that these numerical simulations can only cover a few hours of star time is another 
concern:~we need to be confident that the behaviour seen is representative of the behaviour of the star over 
longer timescales. Much work remains to be done, but at least it has started.
%9 Summary

\subsection{Beyond the AGB: Super-AGB stars} \label{sec:superagb}

Stars in a relatively narrow mass range will ignite C off-centre under
degenerate conditions. These stars will experience thermal pulses on the ``super-AGB''.
The upper mass limit for this range is set by those stars that go on to burn Ne and
heavier species, ending with an iron core and dying as a core-collapse supernova (these
are the traditional ``massive'' stars. In the nomenclature introduced in this paper the 
super-AGB stars are the {\em middle intermediate} and {\em massive intermediate} stars, 
as shown in Figure~\ref{masses}.

These stars are particularly difficult to calculate for many reasons, including high
sensitivity to spatial and temporal resolution, a large number of thermal pulses
(hundreds to thousands), and dramatically different final fates (electron-capture
supernova or white dwarf) depending on the uncertain mass loss and its importance
compared to core growth. The first studies of the evolution of super-AGB stars
were by \citet{garcia-berro94}, \citet{ritossa96}, \citet{garcia-berro97}, \citet{iben97},
and \citet{ritossa99}.
These early calculations ignored mass loss and looked only at solar compositions,
as is entirely appropriate for the first explorations. In the last decade researchers have extended
the investigations to cover a wide range of metallicities, now that computers have enabled
us to face the difficult computational task these stars present. 
There has been an increasing number of studies of super-AGB stars in recent years
\citep{gilpons02,gilpons05,gilpons07,siess07,poelarends08,ventura09a,ventura10b,doherty10,ventura11a,ventura11b,karakas12,herwig12,doherty14a,gilpons13,ventura13,jones13,doherty14b}.

\subsubsection{Super-AGB Evolution}

These calculations all agree qualitatively on the evolution of super-AGB stars, although
there are substantial and important quantitative differences. The stars ignite
C off-centre when the temperature reaches 600-650 MK, and develop a
convective shell reaching outward from the ignition point. 
After this C burning dies down and contraction
resumes, secondary convective zones develop and the C burning
eventually  reaches the centre where it ceases before consuming all of the C,
leaving a mass fraction of C of around 1\%. 
Carbon burning continues just outside the inner core
generating more convective flashes when regions of
high C content are traversed. At the completion of C burning we are left with an O-Ne
core surrounded by an inactive C-O shell, as
well as He- and H-burning shells and the large convective envelope.

Quantitative details are harder to agree upon. This is mostly due to uncertainties in
how to treat convection and semiconvection, as much during the core
He burning phase as during C burning. It is the size of the C-O core 
at the end of He burning that is the prime determinant of  
super-AGB evolution during C burning. For a comparison between different codes
and the resulting differences in evolution, and critical mass boundaries 
(see Figure~\ref{masses}) we refer to \citet{poelarends08} and \citet{doherty10}.

During, or soon after, core C burning (which need not go to completion) 
the star will experience second dredge-up. This relatively simple event is
much more complicated in the case of super-AGB stars. Sometimes, usually in the more massive models, 
 we find ``corrosive'' SDU, where the convection extends inward of the He-burning shell with 
the result that it dredges C (and sometimes O) to the surface \citep{doherty14a,gilpons13}. 
Similarly we can find
``dredge-out'' episodes, first encountered by \citet{ritossa96}, where H
diffuses into a He-burning convective zone. This mixes the H
to extreme temperatures, resulting in a H flash. Thus these dredge-out
events are another form of proton-ingestion episode \citep{campbell08}.
As expected, the details of these events, and even their occurrence or not,
depend on the previous evolution, especially the mass of the core, and how the
borders of convection are treated. 

Once the super-AGB stars settle on the TP-AGB their evolution is similar to 
other intermediate-mass AGB stars, showing recurring thermal pulses and HBB, 
with the usual associated nucleosynthesis. The differences here are that the
thermal pulses are not very strong, reaching typically $L_{He} \simeq 10^6\Lsun$
for models without much dredge-up \citep[e.g.,][]{siess07,siess10},
compared with values more like $10^8 L_{\odot}$ where there is deep dredge-up in lower
mass C-O core AGB stars.
These pulses have short duration, and the flash-driven convective
pocket is only in existence for about 6-12 months. The pulses repeat every 100 years or so
with the result that the expected number of pulses can be hundreds to thousands,
depending on the uncertain mass-loss rate. For this reason most of the calculations
to date do not cover the full evolution, but only a few pulses at
the start of the TP-AGB. Detailed models of the full super-AGB evolution
include \citet{pumo08}, \citet{siess10}, \citet{karakas12}, \citet{gilpons13}, 
\citet{ventura13}, and \citet{doherty14a,doherty14b}.

Note that all authors find convergence difficulties near the end of the
evolution of intermediate-mass AGB stars. \citet{lau12} attribute this to
the opacity bump produced by Fe, which has also been noted in models 
of Wolf-Rayet stars by \citet{petrovic06} \citep[but not, apparently, by][]{dray03}.
The convergence problem is believed due to the disappearance of a hydrostatic solution
for large core masses, when a region of super-Eddington luminosity develops
\citep{wood86}. Exactly how 
the star responds to this is uncertain. \citet{lau12} show that there is not enough energy
available to eject the envelope so after some dynamical motion the envelope may
again settle on the star and further evolution may occur. For this reason 
\citet{doherty14a} and \citet{gilpons13} provide yields for two cases: a) assuming the envelope is ejected 
with the composition it had when the instability developed, and b) assuming that 
pulses similar to the last few continue until the envelope is removed by the 
stellar wind. Reality is likely to lie somewhere between these two extremes.

\subsubsection{Third Dredge-Up}

One area that requires further study is the occurrence, or not, of third dredge-up in
super-AGB stars. \citet{siess10} and \citet{ventura11a} find no dredge-up in their detailed models, 
whereas \citet{karakas12}, \citet{doherty14a} and \citet{gilpons13} do find 
reasonably efficient TDU. This is again a sensitive function of mixing details and 
how the numerical calculations are
performed \citep{frost96}. In any event, even if TDU is very efficient, as 
measured by the dredge-up parameter $\lambda$, it is unlikely to be 
very {\em important\/} for most species, because of the masses involved.

In a typical low to intermediate-mass AGB star the intershell region contains
a mass within a factor of two of  $m_{\rm is} \approx 0.01\Msun$ and the increase in core mass 
between thermal pulses is similar with $\Delta M_c \approx 0.01\Msun$. Compare these with
typical values found for super-AGB stars, which are at least two orders of magnitude smaller:
$m_{\rm is}\approx 10^{-4}\Msun$ 
and $\Delta M_{\rm c} \approx 10^{-4}-10^{-5}\Msun$.
Hence dredging up such a small amount, albeit over many pulses, 
has only a small effect on surface abundances.  The total amount dredged 
to the surface is usually less than $\approx 0.1\Msun$ \citep{doherty14a}.
This has a negligible effect, especially when one considers that this material 
is diluted in a much larger envelope than is the case for lower mass AGB stars.

\subsubsection{Super-AGB Nucleosynthesis}

The nucleosynthesis in super-AGB stars is again qualitatively similar to that in 
their lower mass AGB star siblings. Here the intershell is very hot, 
where temperatures can reach well above 400MK, depending on the initial mass and
metallicity.  This results is substantial $s$-processing with $\alpha$-captures on
$^{22}$Ne providing the neutron source. If the TDU proceeds as in the models
of \citet{doherty14a} then we expect this material to be mixed
to the surface. The sizes of the regions involved, however, are small and
from a chemical evolution viewpoint any species that is primarily produced by
dredge-up is unlikely to be substantially produced by super-AGB stars. (Note that this
is not true at very low metallicity, where the small amount of heavy elements
added to the surface can be a significant perturbation on the original content.)

The other main nucleosynthesis pathway for AGB and super-AGB stars is HBB. 
Indeed, for super-AGB stars the temperature at the bottom of the convective 
envelope can exceed 100MK which is high enough for substantial and 
extensive H burning. We expect CNO cycling, in near equilibrium 
conditions, as well as the more exotic Ne-Na and Mg-Al chains to be very active.

Detailed calculations by \citet{siess10} conclude that these stars may produce significant 
amounts of \iso{13}C, \iso{14}N, \iso{17}O, \iso{22}Ne, \iso{23}Na, \iso{25,26}Mg, 
\iso{26}Al, and \iso{60}Fe. These results were confirmed by \citet{doherty14a,doherty14b}, 
and \citet{gilpons13}, who present yields for super-AGB stars \citep[see also][]{ventura13}.

It is because of their strong HBB that these stars have been discussed frequently 
as being implicated in explaining the chemical compositions of globular clusters.
There is an extensive literature on the problem
\citep{cottrell81,pumo08,ventura09a,ventura10b,dercole10,dercole11,ventura11a,ventura11b,dercole12,dantona12,ventura12,ventura13,doherty14b}.
Although super-AGB stars qualitatively have many of the required 
properties for the polluters of GCs, there are still substantial difficulties, and one has to 
tune many inputs to get models that are close to the observations. While super-AGB
stars may be involved in the GC abundance problems, they are not apparently the magic bullet.

\subsection{Final fates of AGB and Super-AGB stars} 

The most massive AGB stars, the lower intermediate mass stars, will end their 
lives as C-O white dwarfs \citep{althaus10}.
The super-AGB stars, however, produce a  richer variety of remnants \citep[e.g.,][]{jones13}.
At the lower mass limit of stars that ignite C, the ignition is in 
the outermost layers, and does not proceed to the centre.
This leads to a class of hybrid white dwarfs that we refer to as CO(Ne)s. 
In these stars a C-O core is surrounded by a shell of $0.1$--$0.4\Msun$
that has seen C burning and is mostly O and Ne.
We note that the 8$\Msun$, $Z =0.02$ model presented here is an example 
of a CO(Ne), and these are also found by other authors 
\citep[Heger, private communication;][]{ventura11a,denissenkov13a}.
Stars with more massive cores will undergo off-centre Ne ignition with convective flashes
\citep{ritossa99,eldridge04}.

Following core He-exhaustion, most of the middle and massive intermediate-mass stars 
have core masses that exceed the Chandrasekhar mass,
which we take to be $M_{\rm Ch} = 1.37\Msun$ as determined by \citet{miyaji80}, \citet{hillebrandt84},
and \citet{nomoto87}. The core  masses we find are typically $2$--$2.5\Msun$. 
Following the SDU (or dredge-out, if it 
occurs) the core mass is reduced to below the critical Chandrasekhar value.
The final fate of these stars is crucially dependent on SDU and dredge-out, for 
if these do not reduce their core mass to below $M_{\rm Ch}$ then 
the star will continue on through the various nuclear burning stages and explode
as an iron core-collapse supernova; i.e. as a true massive star. If the core mass 
is reduced sufficiently by SDU or dredge-out then the fate depends on the competition 
between core growth and mass loss. If the former dominates and the 
core reaches $M_{\rm Ch}$ then an electron-capture supernova will result. If mass loss 
keeps the mass below the critical value then the star ends as an O-Ne WD.

The existence of massive WDs in the Galaxy with oxygen-rich atmospheres 
\citep{gansicke10} lends credence to the idea that at least some of the super-AGB stars
avoid core-collapse supernova explosions. Furthermore, there is evidence that some 
classical novae explode in a binary where the compact companion is an O-Ne
WD \citep[e.g.,][]{jose98,denissenkov13b}.

\begin{figure}
\centering
\includegraphics[height=8cm]{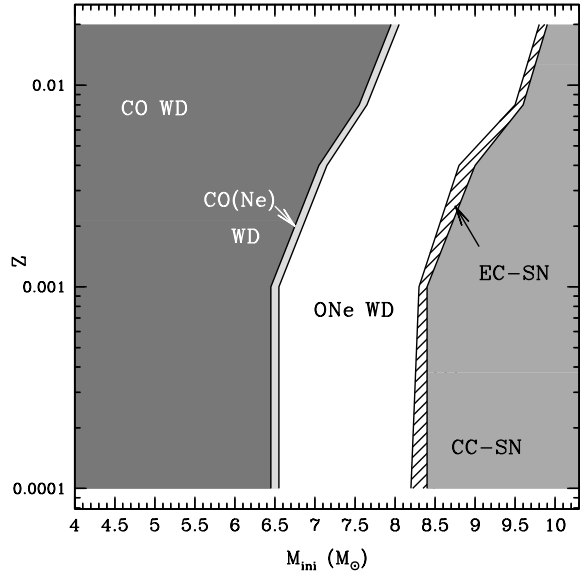}
\caption{Predicted final fates for the SAGB mass range, as a function of metallicity $Z$.
CC-SN refers to core collapse supernovae and EC-SN refers to electron capture supernovae.
The regions of C-O, CO(Ne) and O-Ne white dwarfs are also indicated. See text for details.}
\label{finalfates}
\end{figure}

Figure~\ref{finalfates} is taken from Doherty et al. (2014, in preparation) and shows the predicted fate
for models in the AGB and SAGB mass-range. We note that these results are in reasonably 
good agreement with other authors \citep{siess07,poelarends08} given the sensitivity inherent in the calculations. 
We find the electron-capture supernovae comprise just a few percent of all supernovae in the 
metallicity range $Z=0.02$ to $10^{-4}$.

\section{Major uncertainties} \label{sec:uncerts}

The evolution and nucleosynthesis of low-mass and intermediate-mass stars is significantly
affected by numerical modelling uncertainties as well as uncertainties in the input physics. Here we review
the main uncertainties affecting AGB model calculations including convection, which affects the occurrence
and efficiency of TDU, and mass loss, which determines the AGB lifetime. We also comment on uncertainties
affecting $s$-process element predictions.

\subsection{Convection and the third dredge up} \label{sec:tdu}

Dealing with convection in stellar interiors is one of the major problems of stellar evolution modelling.
There are two main ways that convection affects AGB  evolution and nucleosynthesis. The first is through the way
stellar evolution codes  treat the interface between radiative and convective regions within a stellar model 
\citep{frost96,mowlavi99a}. The second is how the temperature gradient affects the energy transport in
the convective regions.

\subsubsection{Determining the Borders of Convection}

We know from observations that AGB stars experience third dredge up. We unambiguously observe C and Tc-rich AGB 
stars but we still (after 40 years) do not know at which initial 
stellar mass that dredge-up begins. Circumstantial evidence suggests that the minimum mass be about $1.5\Msun$ 
in the Galaxy \citep[e.g.,][]{wallerstein98} but theoretical models mostly struggle to obtain enough TDU at 
this stellar mass without the inclusion of some form of convective overshoot 
\citep{herwig97,mowlavi99a,herwig00,cristallo09,karakas10b,kamath12}.

Observations of external galaxies such as the LMC, SMC, and dwarf spheroidal galaxies show higher numbers
of C-stars than in our Galaxy \citep[e.g.,][]{zijlstra06,sloan08,sloan12}, which is evidence that it 
is easier to obtain  TDU at lower metallicities, for a given mass. On the other hand, \citet{boyer13} found
a lack of C stars in the inner metal-rich region of M31, indicating that there is a metallicity ceiling for 
the operation of TDU. Stellar evolution codes qualitatively agree with these observations.

The main problem for calculating TDU is how we determine the border between a convective and radiative region.
Applying the Schwarzschild criterion for convection ($\nabla_{\rm rad} > \nabla_{\rm ad}$) is too simplistic because 
while the accelerations of blobs at this border is zero, the velocities may be finite. 
This suggests that some overshoot is inevitable and to be expected. The question then is how much? 

While the amount of convective overshoot can be constrained by considering various observations 
\citep{herwig00,cristallo09,weiss09,kamath12}, this provides little insight into the actual 
physics occurring in convective envelopes, given the numerical differences between the various
stellar evolution codes. The fact that \citet{kamath12} found such a diversity in the amount of 
overshoot required to match observations may indicate that this is not the best way to characterise 
the depth of mixing required.  

The problem of determining convective borders in stellar interiors is surely also linked to 
the question of \iso{13}C pocket formation, which can be formed by the inclusion of a partially 
mixed overshoot region, for example.
Uncertainties not only include the formation mechanism but also the size of the pocket in the 
He intershell as well as the shape. When more protons are added the resultant \iso{13}C pocket
is accompanied by a sizeable \iso{14}N pocket, which acts as an efficient neutron absorber and
suppresses the $s$-process. 

The way forward is to consider multi-dimensional simulations which unfortunately have not 
yet advanced sufficiently to answer the problem of overshoot or \iso{13}C pocket formation.

\subsubsection{Structural Changes from Convection}

The second important way that convection affects models is the substantial effect 
on the structure of AGB convective envelopes. 
The treatment of convection in stellar envelopes determines surface properties such as the luminosity and effective
temperature and it has an impact on the efficiency of hot bottom burning \citep{ventura05a}. The most commonly
used treatment of convection is the Mixing-Length Theory which has a free parameter, the 
mixing-length parameter $\alpha$, which is usually set by calibrating a 1$\Msun$, $Z=Z_{\odot}$ stellar model to the Sun's 
present day radius. The parameter $\alpha$ is then assumed to remain constant throughout the star's evolution,
with the same value used for all masses and metallicities.

However,  AGB stars have very different envelope structures to the shallow convection zone found
in our Sun. There is no good reason why the value of $\alpha$ required to fit a standard solar model is appropriate for 
AGB stars. Furthermore \citet{lebzelter07} found evidence for $\alpha$ to increase with evolution along the 
AGB, which suggests that $\alpha$ should not be constant. Increasing $\alpha$ leads to shallower 
temperature gradients which produces higher luminosities and stronger burning in intermediate-mass AGB stars
\citep{ventura05a}.  Convective models other than the Mixing-Length Theory are also used, such as the Full Spectrum
of Turbulence \citep{ventura98,mazzitelli99} applied by e.g., \citet{ventura05a}. 

\subsection{Mass loss}

Dealing with the extent and temporal variation of mass loss in AGB stars is one of the main uncertainties 
in stellar modelling \citep{bloecker95,habing96,ventura05b}.  This is because the mass-loss rates of AGB stars 
are very uncertain and difficult to determine from observations without {\it a priori\/} assumptions about dust mass 
and type, and/or radiative transfer modelling
\citep[e.g.,][]{bedjin88,vw93,habing96,vanloon99b,groen07,groen09,lagadec08,lagadec10,guandalini10,debeck10,riebel12}.
Mass loss on the AGB determines the AGB lifetime and the number of thermal pulses experienced by TP-AGB models. This 
then limits the number of TDU episodes and the duration of HBB; and hence determines the level of chemical enrichment
expected from a population of AGB stellar models at a given metallicity.

In order to calculate stellar yields, mass loss has to be included in the calculation of AGB 
models. The available prescriptions are simple, parameterised formulae that result in mass being removed 
from the envelope smoothly in time, in contrast to observations which suggest that AGB mass loss in 
real stars is clumpy and asymmetric \citep[e.g.,][]{meixner98,dinh08,olofsson10,witt11,paladini12,lomb13}.

The upper part of the AGB in particular is dominated by continuously increasing mass loss \citep{habing96}. 
Observations indicate rates increase from $10^{-7}\Msun$ year$^{-1}$ for short period Mira variables to 
$\approx 10^{-4}\Msun$ year$^{-1}$ for luminous long-period variables including OH/IR stars and Miras 
\citep{groen07,justtanont13}.  These winds are most likely dust and shock driven \citep{winters03}, which 
leads to the star becoming completely enshrouded by dust and visible predominantly in the infra-red 
\citep{habing96,uttenthaler13}. 

One of the biggest uncertainties is the rate of mass loss from low-metallicity 
AGB stars.  Based on theoretical calculations, \citet{mattsson08} conclude that 
low-metallicity C stars have similar mass-loss rates to their metal-rich counterparts. 
Observations showed that mass-loss rates in low metallicity C-rich AGB stars in  nearby galaxies are of a 
similar magnitude to AGB stars in our Galaxy \citep[e.g.,][]{sloan09,lagadec09}.  

The most widely used prescriptions in AGB evolutionary calculations include the \citet{vw93} mass-loss law, 
based on empirical observations of mass-loss rates in C and O-rich AGB stars in the Galaxy and Magellanic Clouds; the 
\citet{bloecker95} formula, based on dynamical calculations of the atmospheres of Mira-like stars; and the Reimer's 
mass-loss prescription \citep{reimers75,kudritzki78}, even though it was originally derived for first ascent giant 
stars and does not predict a superwind \citep{groen12}. Both the \citet{bloecker95} and \citet{reimers75} rates depend
on an uncertain parameter $\eta$ which typically takes values from 0.01 to 10 \citep{ventura00,straniero97,karakas10a}.
\citet{ventura00} found $\eta = 0.01$ by calibrating their intermediate-mass AGB models to Li abundances in 
the LMC. 

Other mass-loss prescriptions for AGB stars are available 
\citep[e.g.,][]{bedjin88,arndt97,wachter02,vanloon05,wachter08,mattsson10}. 
Some of these prescriptions are specifically for C-rich stars e.g., \citet{arndt97} and \citet{wachter02} and
are not appropriate for bright intermediate-mass AGB stars. The theoretical mass-loss rates from \citet{mattsson10} 
for solar-metallicity C stars are available as a FORTRAN routine that can be coupled to a stellar 
evolution code. 

\subsection{Extra mixing in AGB stars} 

A case has been made for some form of slow non-convective mixing to operate in AGB envelopes, in an
analogous situation to the extra mixing operating in first giant branch envelopes (\S\ref{sec:needextramix}
and \S\ref{sec:extramix}). 

The evidence for deep mixing on the AGB comes mainly from O and Al isotope ratios measured
in pre-solar oxide grains, which support the existence of such extra mixing in low-mass ($M\lesssim 1.3\Msun$)
AGB stars \citep{busso10,palmerini09,palmerini11}. The C isotopic ratios measured in AGB stars span a large
range, from very low \iso{12}C/\iso{13}C ratios of $\approx 4$ to a maximum of about
100 \citep{lambert86,abia97}. The sample by \citet{lambert86} has an average value of 58 (without
the J-type C stars whose origin is unknown). The range of \iso{12}C/\iso{13}C ratios 
in AGB stars is similar to that measured in mainstream pre-solar silicon carbide grains, with values
between 40 $\lesssim$ \iso{12}C/\iso{13}C $\lesssim$ 100 with an average \iso{12}C/\iso{13}C $\approx 60$ 
\citep{zinner98}. The lowest values of the \iso{12}C/\iso{13}C ratio in AGB stars suggest that some
small fraction of Galactic disk C-rich AGB stars experience extra mixing. In \S\ref{sec:c12c13ratio}
we summarise the C isotope predictions from AGB models, including the range expected when extra mixing
occurs on the first gaint branch.

Carbon-enhanced metal-poor stars that are $s$-process rich presumably received their C, N
and $s$-process enrichments from a previous AGB companion and their \iso{12}C/\iso{13}C ratios are therefore 
an indicator of extra mixing and nucleosynthesis in the AGB star.  The \iso{12}C/\iso{13}C ratio has been 
measured in CEMP stars covering a range of evolutionary phase, from turn-off stars through to giants. Figure~7 from 
\citet{stancliffe09} illustrates the observed C isotopic ratios of unevolved ($\log g \ge 3$) 
CEMP stars are $\lesssim 10$  \citep[e.g.,][]{cohen04,sivarani06,jonsell06,aoki07,beers07,lucatello11}.  
Such low observed ratios are difficult to reconcile with standard AGB nucleosynthesis models, which produce very high 
\iso{12}C/\iso{13}C ratios ($> 10^{3}$) at low metallicity \citep[e.g.,][]{karakas10a,cristallo11,lugaro12}.  
Such low  \iso{12}C/\iso{13}C ratios could reveal a metallicity 
dependence to the extra-mixing occurring in AGB envelopes.

Nitrogen abundances of CEMP stars also show a spread that is not easily explained  by
canonical models as shown by the population synthesis study by \citet{izzard09}. \citet{stancliffe10} 
studied thermohaline 
mixing in low-metallicity AGB models and found it not strong enough to explain the low C isotopic ratios.
Clearly additional mixing -- whatever the mechanism -- is required in low-metallicity AGB envelopes,
but the need for extra mixing in solar-metallicity C-rich AGB stars is more 
ambiguous as discussed in detail by \citet[][but see \citealt{busso10}]{karakas10b}.

\subsection{Low temperature Opacities}

In recent years there has been considerable effort put into developing accurate low-temperature molecular
opacity tables for stellar evolution calculations. The opacity tables of \citet{alexander94} and later \citet{ferguson05}
included the first detailed treatment of the inclusion of molecules to the total opacity at temperatures
where $T \lesssim 10^{4}$K. These tables were only available for solar or scaled-solar abundance mixtures.
As we have seen, AGB stars experience multiple mixing episodes that alter their envelope compositions, such that
the stars may become C and N-rich and in some cases, the envelope C/O ratio can exceed unity.

\citet{marigo02} showed that at the transition
from C/O $< 1$ to C/O $\ge1$ the dominant source of molecular opacity changes from oxygen-bearing molecules
to C-bearing molecules. In AGB stellar models, this change in opacity leads to a sudden decrease in
the effective temperature and subsequent expansion in radius. These changes to the stellar structure cause
an increase in the rate of mass loss. 
\citet{marigo02} showed that this resulted in shorter AGB lifetimes and therefore smaller stellar yields.
AGB models with HBB can also deplete C and O (while producing N), which causes changes to the stellar structure and nucleosynthesis
\citep{weiss09,ventura09b,ventura10a,fishlock14}. 
Despite claims to the contrary, \citet{constantino14} showed that there was no threshold in [Fe/H] below which
the composition dependent molecular opacities would not produce significant changes.
It is therefore necessary to use low-temperature molecular 
opacity tables that follow the change in C, N, and C/O ratio with time at all masses and compositions.

Besides the scaled-solar tables from \citet{ferguson05}, other tables currently available for stellar evolutionary 
calculations are 1) \citet{lederer09}, who account for an enhancement of C and N compared to the initial 
abundance for various metallicities, and 2) \cite{marigo09}, who provide the {\AE}SOPUS on-line downloadable 
tables\footnote{http://stev.oapd.inaf.it/aesopus}. These tables are available for 
essentially arbitrary variations in C, N, and C/O 
(including enhancements and depletions) for whatever metallicity desired 
and for various choices of the
solar composition.

\subsection{The $s$-process} \label{sec:sprocun}

While the formation mechanism of the \iso{13}C neutron pocket is still the main uncertainty in the
$s$-process models, there are a number of further problems associated with the $s$-process scenario 
discussed previously in \S\ref{sec:sprocess}.

\citet{lugaro12} identified four different regimes of neutron captures that can occur in
theoretical AGB models including: 1) the \iso{22}Ne($\alpha$,n)\iso{25}Mg source operates during  convective 
thermal pulses, 2) the \iso{13}C($\alpha$,n)\iso{16}O reaction burns under radiative conditions, with
the \iso{13}C produced via the inclusion of a \iso{13}C pocket, 3) the  \iso{13}C($\alpha$,n)\iso{16}O reaction
burns under convective conditions during a thermal pulse, with the \iso{13}C produced via the inclusion of a 
\iso{13}C pocket, and 4) the \iso{13}C($\alpha$,n)\iso{16}O reaction operates under convective conditions with 
the \iso{13}C produced via the ingestion of a small number of protons from the tail of the H shell during the 
thermal pulse. 

The mass range at which these regimes occur is model dependent as described in \citet{lugaro12}.
At the metallicity considered in that study ($Z=0.0001$) proton ingestion (Regime 4) dominates the $s$-process
abundance predictions at the lowest masses ($M \le 1\Msun$ at $Z=0.0001$) when no (or small) \iso{13}C pockets
are present.
Proton ingestion is expected to be more important at even lower metallicities, although the mass and 
metallicity range where proton ingestion occurs is still very uncertain and may occur at solar metallicities 
under specific conditions (e.g., Sakurai's Object). Again, the problem comes down to the treatment of convection 
and convective borders in stars. The first multi-dimensional studies are becoming available to guide the 
1D models \citep{stancliffe11,herwig11,herwig13}.

For the first few pulses in low-mass stars, the temperature in the intershell is not large enough for 
efficient radiative burning of the \iso{13}C.
Hence Regime 3, where all or most of the \iso{13}C is burnt under convective conditions during the next thermal pulse 
is possibly a common occurrence during  the first few thermal pulses for  all low-mass AGB 
stars of $\lesssim 2\Msun$, regardless of $Z$ \citep[see also][]{cristallo09,lugaro12}.  
The main result of Regime 3 is that the overall neutron exposure (i.e., total number of free neutrons) decreases 
owing to ingestion of the neutron poison \iso{14}N alongside the \iso{13}C from the H-shell ashes. 
Also during Regime 3 the neutron density increases owing to the short timescale of thermal pulses 
(order $10^{2}$ years) relative to radiative burning during the interpulse ($\approx 10^{4}$ years). 

The occurrence of \iso{13}C ingestion during thermal pulses is strongly connected to the uncertainties related to the onset 
of TDU in the lowest-mass AGB stars \citep[e.g., the initial stellar mass for the onset of the TDU at a given $Z$, the
efficiency of TDU as a function of stellar mass, core mass, metallicity;]
[]{frost96,straniero97,mowlavi99a,karakas02,stancliffe07a,karakas10b}.  

If stars as low as $\approx 1.2\Msun$ experience TDU at solar metallicities then we expect Regime 3 to be dominant in these 
stars for the first few thermal pulses, before the He shell region has heated up to sufficient temperature to ignite 
\iso{13}C($\alpha$,n)\iso{16}O under radiative conditions. If these stars only experience a few TDU episodes, 
then this regime will dominate the stellar yields of $s$-process elements.

All stars rotate but the effect of rotation on stellar structure in general, and the $s$-process 
in particular, is still poorly known. The angular velocity
profile inside AGB stars may produce a strong shear instability between the 
contracting core and the expanding envelope. 
Unlike the (partial) mixing postulated to come from overshoot during TDU, this shear layer does not
disappear at the end of TDU.
This is expected to result in continuous mixing of protons into the top layers of the He intershell, 
resulting in the 
complete operation of the CN cycle and the production of a higher abundance of the neutron-poison \iso{14}N 
instead of \iso{13}C. This has been shown to lower the neutron exposure and suppress the formation of $s$-process
elements \citep{herwig03,piersanti13}. The effect of magnetic fields \citep{suijs08} and gravity waves may
modify the angular momentum in the star and reduce the mixing between core and envelope but has not been
considered in detailed stellar evolution models so far. 

We have mentioned convective overshoot at the bottom of the thermal pulse in the context of O intershell abundances.
Such overshoot can lead to increased temperatures and the activation of the \iso{22}Ne 
neutron source at lower initial stellar mass than canonical models with no overshoot. While it has been shown
for a 3$\Msun$ star that such overshoot produces Zr isotopic ratios inconsistent with those measured in stardust
grains \citep{lugaro03b}, other observations such as oxygen in post-AGB stars suggest that such overshoot occurs.
While the first $s$-process yields from models with overshoot into the C-O core are becoming available \citep{pignatari13}
a comprehensive study on the effect of such overshoot on the $s$-process is still lacking.

\subsection{Binary evolution}

Most stars (roughly 60\%) exist in binary or multiple systems \citep{duq91}. Not all stars in binary systems (or higher
order multiples) will be close enough to interact and hence they will evolve essentially as single stars. 
The fate of these stars and
their contribution to the enrichment of the Galaxy is determined by their initial mass and metallicity. For binary
stars that are close enough to interact, there are many more variables that determine the type of interaction 
including the orbital parameters of the system and the mass ratio between the two stars. 

Possible interactions include mass transfer via Roche Lobe Overflow (RLOF), which can lead to a common envelope
and possible stellar merger. For example, the warm R-type C stars are all single stars which has led various
authors to propose that they must be the result of stellar mergers \citep{mcclure97a,izzard07}.  If the stars do not merge 
during the common envelope phase the orbital period will be 
dramatically shortened, allowing for later mass transfer. The details of %mass transfer via RLOF and 
common envelope evolution are complex and not well understood \citep[e.g.,][]{taam10}.

Regardless of the final outcome of the common envelope, the evolution of the two stars will be significantly 
altered from a single stellar evolution channel.   Common envelopes may truncate the evolution of the more 
massive star on the first giant branch, which means it will never become an AGB star. Clearly the nucleosynthesis yields 
will be significantly altered from the expectations from  single stellar evolution \citep{izzardthesis,izzard06}.

Interactions can come in other forms. For example if one of the stars is on the AGB and 
has a strong wind, then some of that wind may be transferred to the companion. That wind may then contain the 
products of AGB nucleosynthesis, which will later be observed on the surface of the lower mass companion. Stellar
wind accretion is thought to be the dominant mechanism to produce barium and CH-type stars 
\citep{mcclure83,mcclure90,mcclure97b,boffin88,han95,karakas00,izzard10,miszalski13}, as well as carbon
and nitrogen-enhanced metal-poor stars \citep{beers05,lucatello05,izzard09,pols12,abate13}. 

\section{Chemical enrichment from AGB stars} \label{sec:yields}

Stellar yields are a key ingredient in models of the chemical evolution of galaxies and stellar systems
\citep{tinsley80,gibson97a,romano10,nomoto13}. Core collapse supernova explosions release vast 
quantities of $\alpha$-elements (e.g., O, Mg, Si, Ti) and Fe-peak elements into the Galaxy on relatively 
short timescales ($\lesssim 10^{7}$ years). Classical nova explosions \citep{romano03a,romano10} and rapidly 
rotating massive stars may also be an important source of C, N and heavy elements 
at the earliest times \citep{chiappini03,chiappini06,hirschi07,frischknecht12}. 

Binary systems that explode as Type Ia supernovae are also responsible for producing substantial metals, 
mostly in the form of Fe \citep[e.g.,][]{seit10}, although Type Ia explosions typically 
take place on much longer timescales from a few hundred million years to a few Gyr \citep{matt86}.  
There is evidence that galaxies had a small number of prompt Type Ia explosions, which take place on much 
shorter timescales of $\approx 100$~Myr since the beginning of star formation \citep{bonaparte13}. 
Together, massive stars, that explode as Type II supernovae, and Type Ia supernovae dominate the chemical 
evolution of many elements and  for this reason AGB stars were until recently largely ignored in models 
of chemical enrichment \citep{matt89,timmes95,gibson97a,kobayashi06}. 

However, low and intermediate-mass stars are a common inhabitant of galaxies and stellar systems and produce 
a copious amount of the gas and dust seen in the interstellar medium. In the last decade there has been considerable 
progress in our understanding of the nucleosynthesis of AGB stars. We now know that they produce considerable amounts 
of C, N, F, Na, the neutron-rich isotopes of Ne, Mg, as well as  Na, Al and heavy elements produced by the $s$-process. 
For a complete picture of the chemical evolution of galaxies, stellar yields from low and intermediate-mass 
stars must be included.

\subsection{Stellar yields from AGB stars}

\begin{table*}[t]
\begin{center}
\caption{List of AGB yields available. We only include detailed AGB evolutionary studies that include
yields; not just surface abundance predictions, and we include studies with more than one mass.
We list the range of masses and metallicities for each study and we note if they include $s$-process 
element predictions.}
\begin{tabular}{lcccc}
\hline 
Reference &  Mass Range & Metallicity Range & $s$-process? & Downloadable\\
          &  (in $\Msun$) &  (in mass fraction, $Z$) &  &  tables? \\
\hline 
\hline
\citet{fenner04}    & 2.5--6.5 & [Fe/H] = $-1.4$ &  No &  No \\
\citet{herwig04b}   & 2.0--6.0 & $1 \times 10^{-4}$ & No & Yes \\
\citet{karakas07b}  & 1.0--6.0  & $1\times 10^{-4}, 4, 8\times 10^{-3}, 0.02$ &  No & Yes \\
\citet{campbell08}  & 1.0--3.0  & $Z=0$, [Fe/H] $-6.5,-5.45,-4,-3$ & No & Yes \\
\citet{iwamoto09}   & 1.0--8.0  & $Z = 2\times 10^{-5}$ & No & No \\
\citet{karakas10a}  & 1.0--6.0  & $1\times 10^{-4},4, 8\times 10^{-3}, 0.02$ &  No & Yes \\
\citet{siess10}$^{(\rm a)}$  & 7.5--10.5 & $1\times 10^{-4}$ to 0.02 &   No & Yes \\
\citet{cristallo11}$^{(\rm b)}$ & 1.3--3.0  & $1\times 10^{-4}$ to 0.02 & Yes  & Yes \\
\citet{ventura13}   & 1.5--8.0  & $3\times 10^{-4}, 10^{-3}, 0.008$ & No & No   \\
\citet{gilpons13}$^{(\rm c)}$   & 4.0--9.0    & $1\times 10^{-5}$ &  No & Yes \\
\citet{pignatari13} & 1.5, 3.0, 5.0 & $0.01, 0.02$ & Yes & Yes \\
\citet{karakas14} & 1.7, 2.36 & $3, 6\times 10^{-4}$ & Yes & Yes \\
\citet{ventura14} & 1--8.0 & $4\times 10^{-3}$ & No & No \\
\citet{doherty14a}  & 6.5--9.0 & $0.004, 0.008, 0.02$ & No & Yes \\
\citet{doherty14b} & 6.5--7.5  & $0.001, 1\times 10^{-4}$ & No & Yes \\
\citet{straniero14} & 4--6 & 0.0003, [$\alpha$/Fe]=$+$0.5 & Yes & Yes \\
\hline
\hline
\end{tabular}
\label{tab:yields}
\medskip\\
$({\rm a})$ Yields for six metallicities are provided with the range noted in the table.\\
$({\rm b})$ Yields for nine metallicities are provided with the range noted in the table.\\
$({\rm c})$ Downloadable tables are surface abundance predictions; yields are given in their Table 4.\\
\end{center}
\end{table*}

\citet{renzini81} produced the first set of stellar yields from low to intermediate-mass stars. These CNO 
yields were calculated with a fully synthetic evolutionary algorithm, which included HBB, TDU and
mass loss via the \citet{reimers75} formula. Further contributions using synthetic AGB models have been made by 
\citet{marigo96}, \citet{vandenhoek97}, \citet{marigo01}, \citet{izzard04b}, and \citet{gavilan05}. 
The biggest difference between 
the recent calculations listed above and those of \citet{renzini81} is in the improved parameterisations 
of the AGB phase of evolution, based on detailed models with improved input physics. The latest
synthetic models also use parameterisations that depend on the initial metallicity, which is something that 
Renzini \& Voli's calculation did not do.

The increasing speed of modern computers means that the problem of running large grids of stellar models 
becomes time consuming, rather than impossible. There are now various compilations in the
literature for yields from detailed AGB stellar models.
Table~\ref{tab:yields} compiles these yields along with the mass and metallicity range, tells if they include 
predictions for $s$-process elements, and if downloadable yield tables are provided. 

There are other studies of low and intermediate-mass stars that include stellar yields but are not
included in Table~\ref{tab:yields} for the following reasons. \citet{stancliffe07a} 
perform a detailed study of the uncertaities affecting yields of low-mass AGB stars but only at 
one mass (1.5$\Msun$, $Z=0.008$). \citet{church09} present full $s$-process yields but also only
for one mass (3$\Msun$, $Z=0.02$). We include only the most recent stellar yield predictions 
calculated by Ventura and collaborators from 2013 and 2014. Earlier calculations
\citep[e.g.,][]{ventura01,ventura02,ventura08a,ventura09a,ventura09b,ventura11a} either use older
input physics or cover a smaller range in mass and metallicity.  There are also many papers that
include surface abundances predictions for AGB models 
\citep[a highly incomplete list includes the following 
examples:][]{kahane00,chieffi01,abia02,lebzelter08,weiss09,campbell10,bisterzo10,kamath12,dorazi13b} 
but not stellar yields; these are not included in Table~\ref{tab:yields}.

The yields from \citet{lagarde11} are not included in Table~\ref{tab:yields}
because they only include one isotope, \iso{3}He, although for a grid of low and intermediate-mass models 
covering a large range in metallicity. \citet{charbonnel10} examine the effects of 
extra mixing and rotation on the light-element yields of model stars up to 4$\Msun$ but only
provide yields for \iso{7}Li; we therefore do not include these in Table~\ref{tab:yields}.

There are fewer yields for low-metallicity AGB stars below [Fe/H] $\le -3$ because of the difficulty
of calculating the stellar evolution owing to the added complexity of proton ingestion episodes and 
mixing during the core He flash.  \citet{iwamoto09} and \citet{campbell08} present yields for 
low-metallicity AGB stars of [Fe/H] $\le -3$, and \citet{gilpons13} provide 
surface abundance predictions for intermediate-mass AGB and super-AGB stars at $Z=10^{-5}$ between 
$M=4\Msun$ to 9$\Msun$. \citet{chieffi01} present calculations of $Z=0$ intermediate-mass (4--8$\Msun$)
AGB stars but no tabulated yields.  No yields of $s$-process elements are available from very 
low-metallicity AGB models at the present \citep[although see][]{campbell10,cruz13}. 

Ideally, chemical evolution modellers would like a self-consistent set of AGB stellar yields covering 
a range in mass from $\approx 0.8\Msun$ to the limit for core collapse supernovae,
$\approx 10\Msun$, for a broad range of metallicities, and for all elements from H through to Bi.
In this context, none of the AGB yield sets mentioned is complete.

Some of these studies provide yields for a limited number of masses but for all elements up to bismuth 
\citep[e.g.,][]{cristallo11,pignatari13}.  The yields by \citet{siess10}, \citet{doherty14a,doherty14b}, and \citet{ventura13}
focus on light elements (up to Fe) for the mass range of stars that become super-AGB stars. The 
largest grid of detailed stellar yields for low and intermediate-mass AGB stars in terms of the 
range of masses (1--6$\Msun$), metallicities ($Z=0.0001,0.004,0.008,0.02$), and number of species 
(light elements up to Fe) is still \citet{karakas10a} but the yields of \citet{ventura13} 
cover almost the same range of masses, elements and metallicities (but not including solar).

\subsection{Summary of elements produced by low and intermediate-mass stars}

The stellar yields shown here are calculated according to
\begin{equation}
 M_{\rm i} = \int_{0}^{\tau} \left[ X(i) - X_{0} (i)\right] 
\frac{d M}{dt} dt,
\label{yield-eqn}
\end{equation}
where $M_{i}$ is the yield of species $i$ (in solar masses), $dM/dt$ is the current mass-loss rate, $X(i)$ and $X_{0} (i)$ 
refer to the current and initial mass fraction of species $i$, and $\tau$ is the total lifetime of the stellar model. The 
yield can be negative, in the case where the element is destroyed, and positive if it is produced.

\begin{figure*}
\begin{center}
\begin{tabular}{cc}
\includegraphics[width=5cm, angle=270]{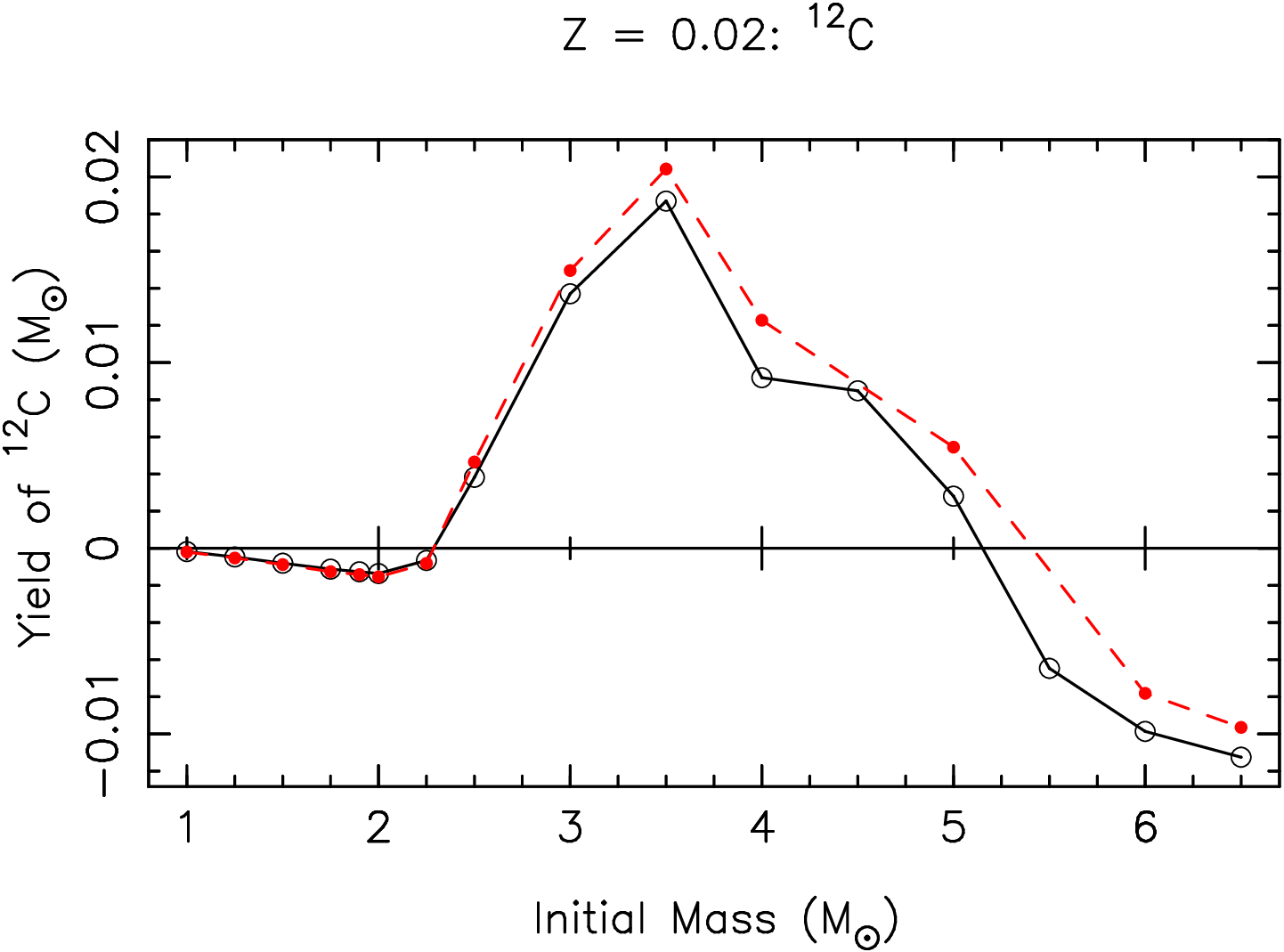} &
\includegraphics[width=5cm, angle=270]{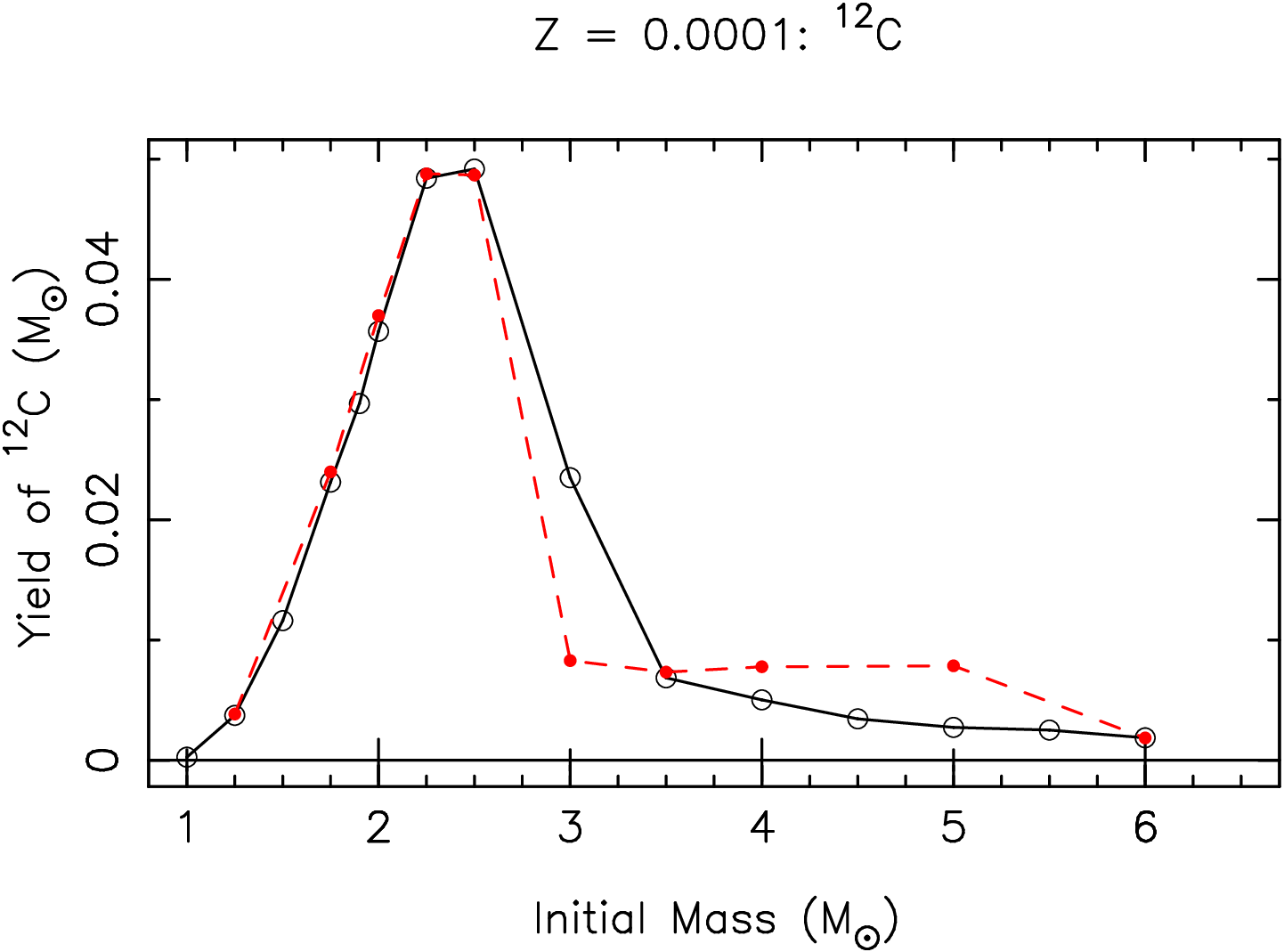} \\
\includegraphics[width=5cm, angle=270]{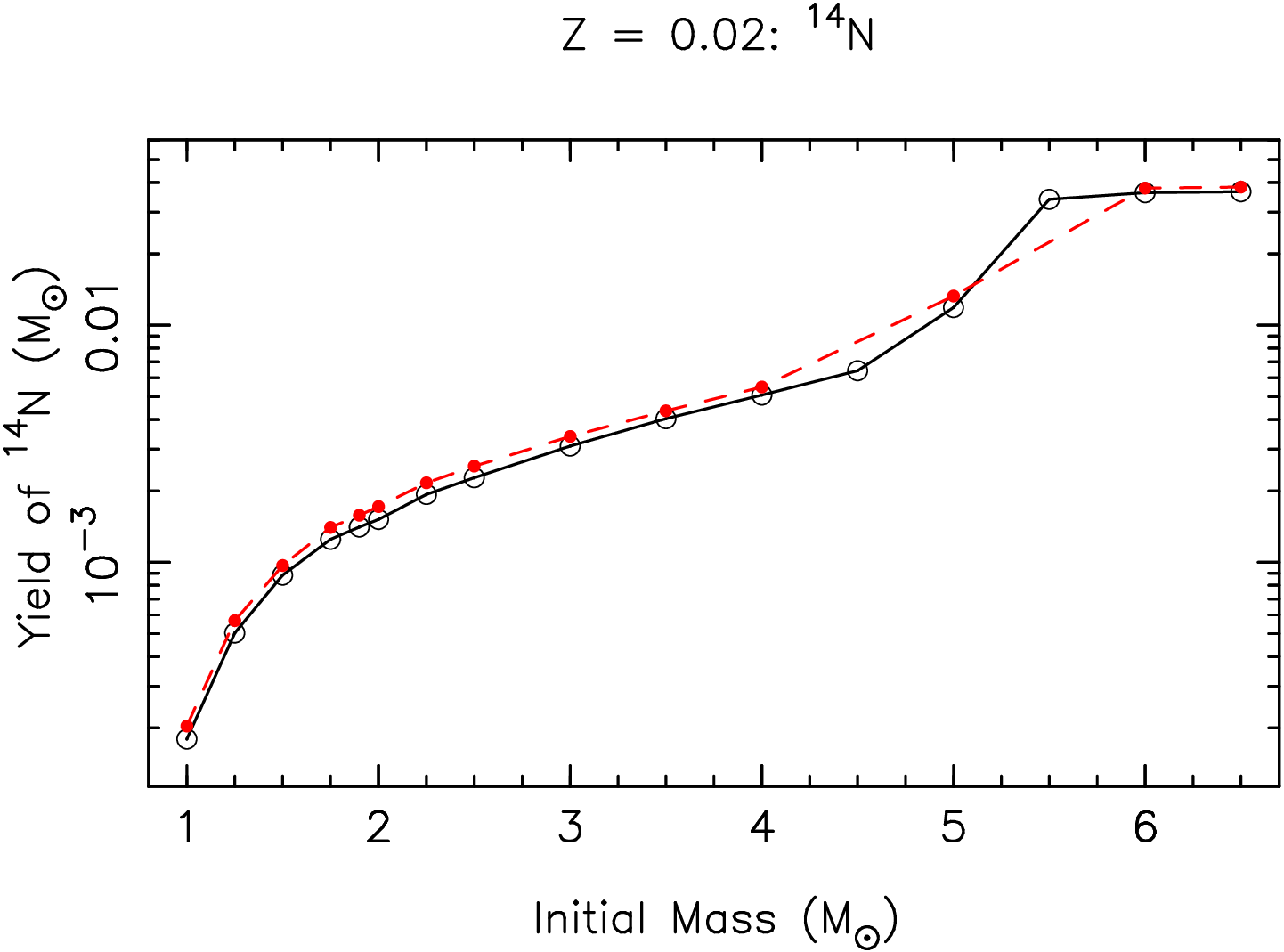} &
\includegraphics[width=5cm, angle=270]{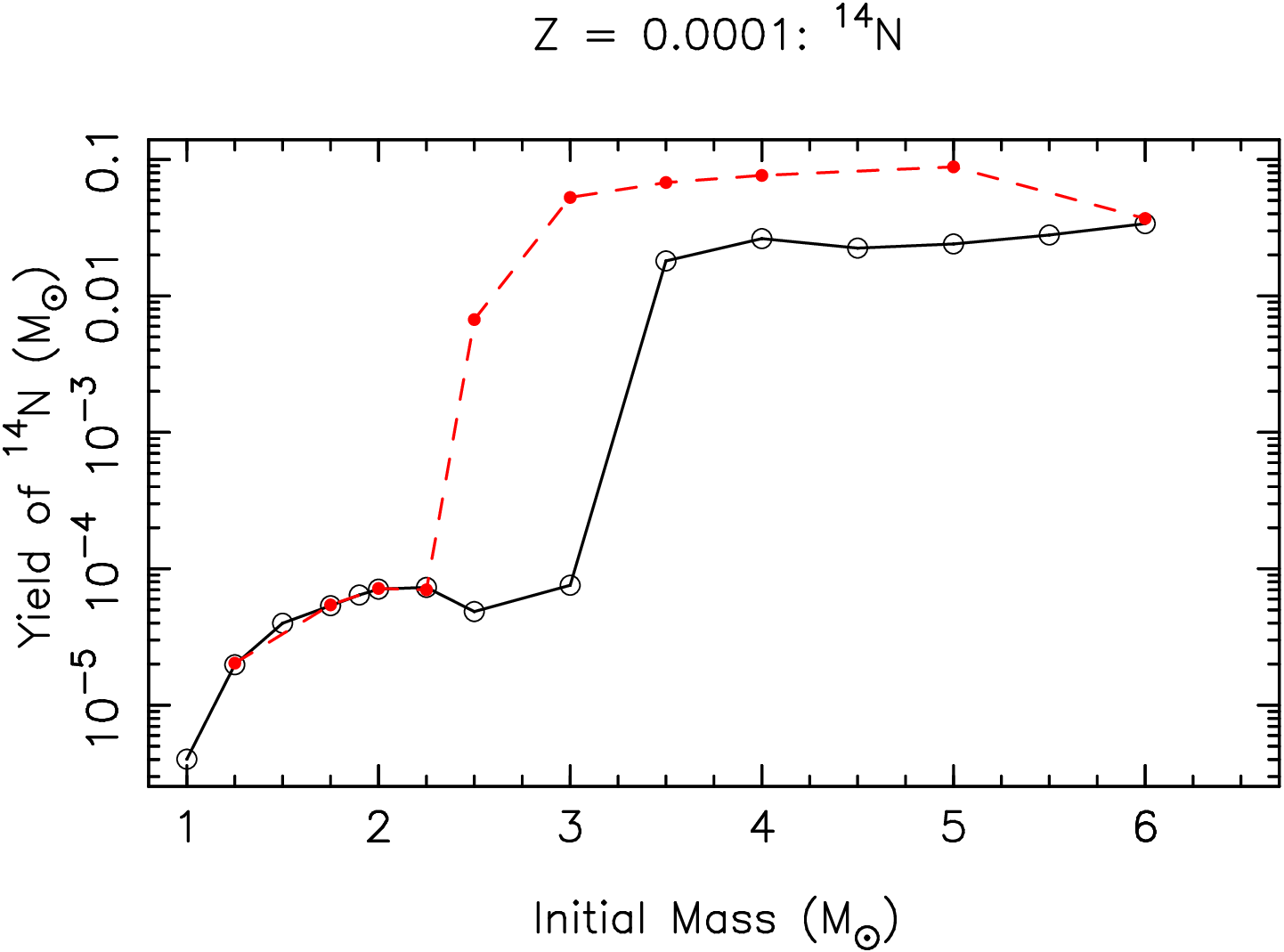} \\
\includegraphics[width=5cm, angle=270]{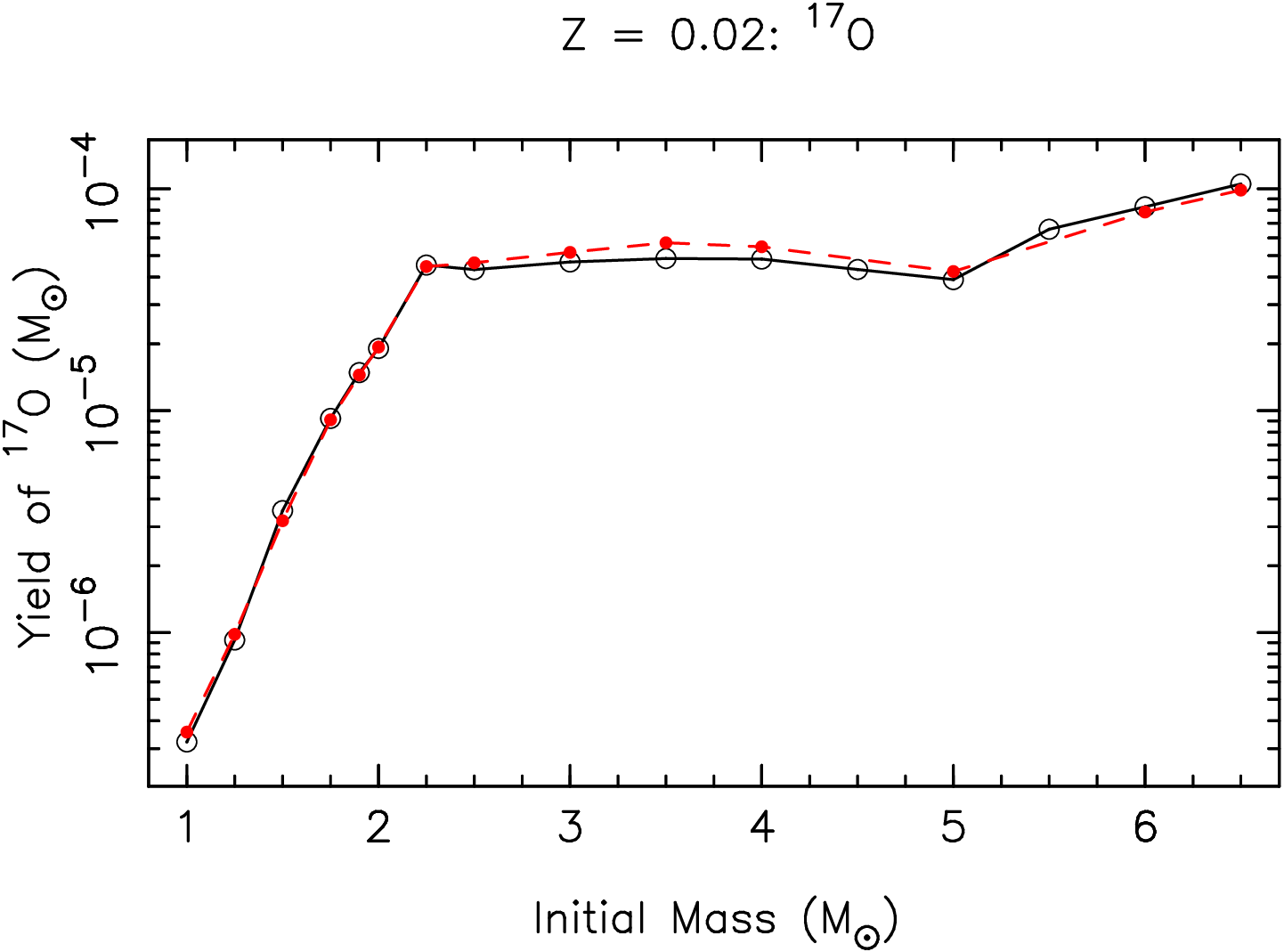} &
\includegraphics[width=5cm, angle=270]{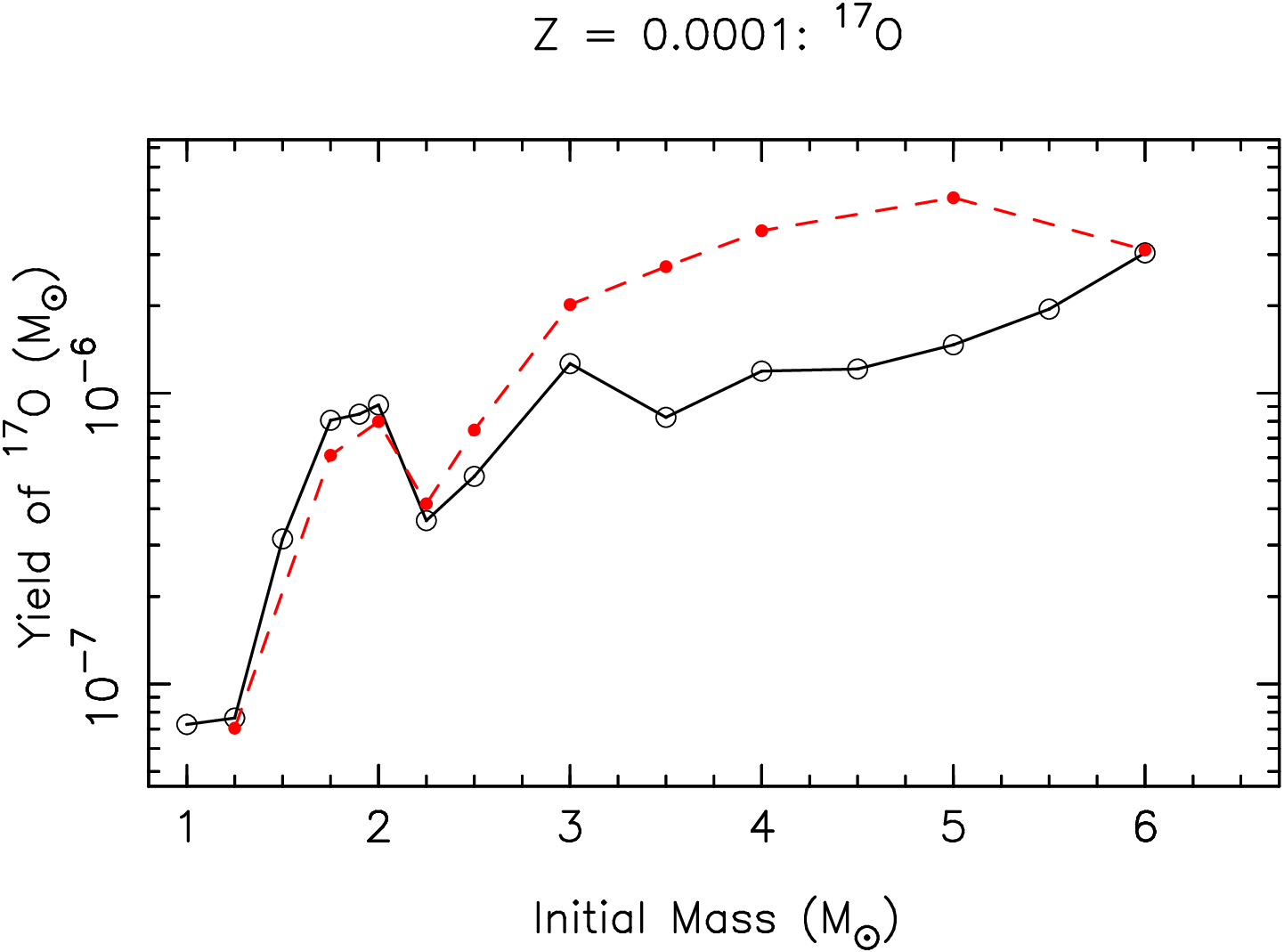} \\
\includegraphics[width=5cm, angle=270]{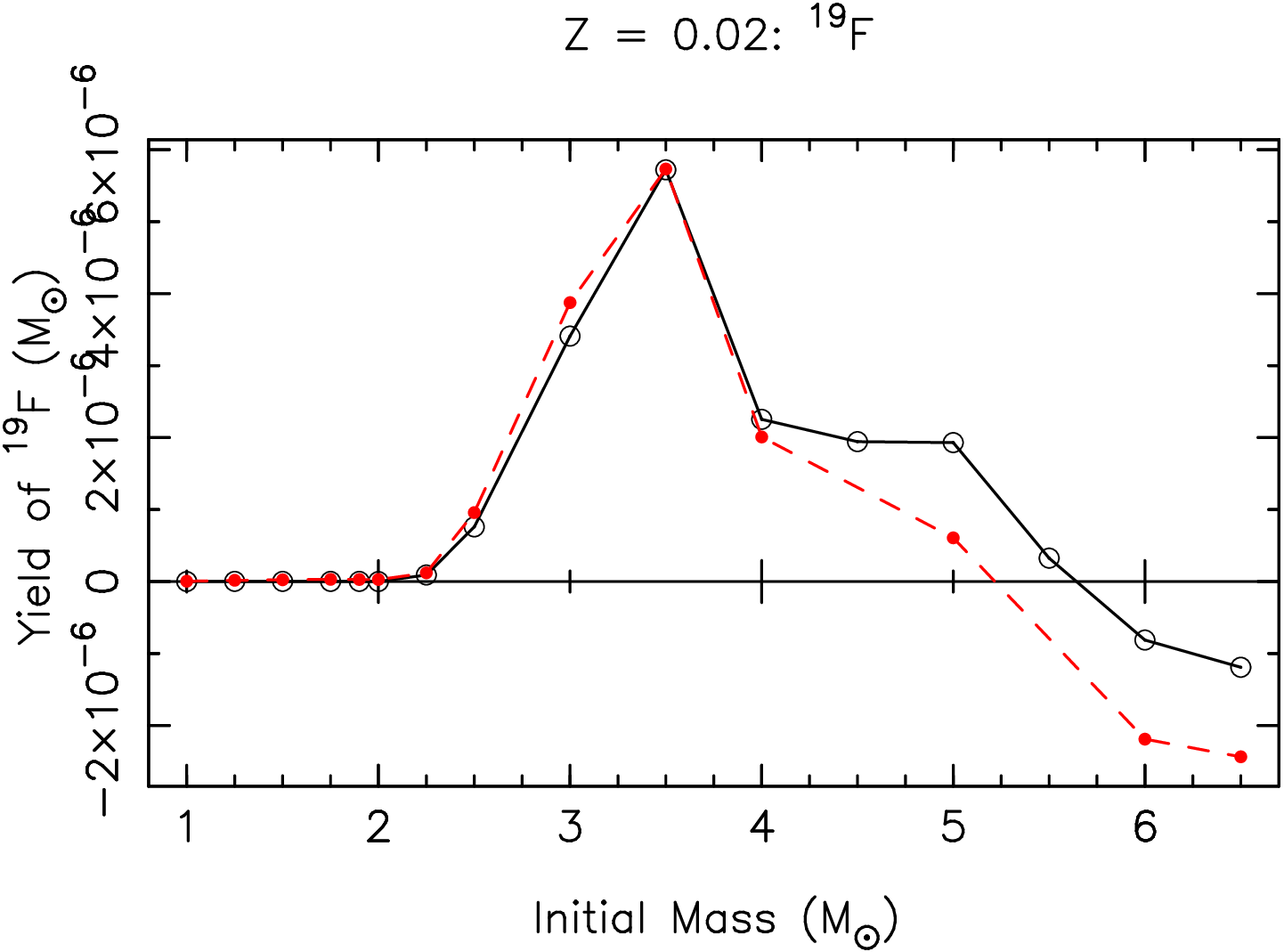} &
\includegraphics[width=5cm, angle=270]{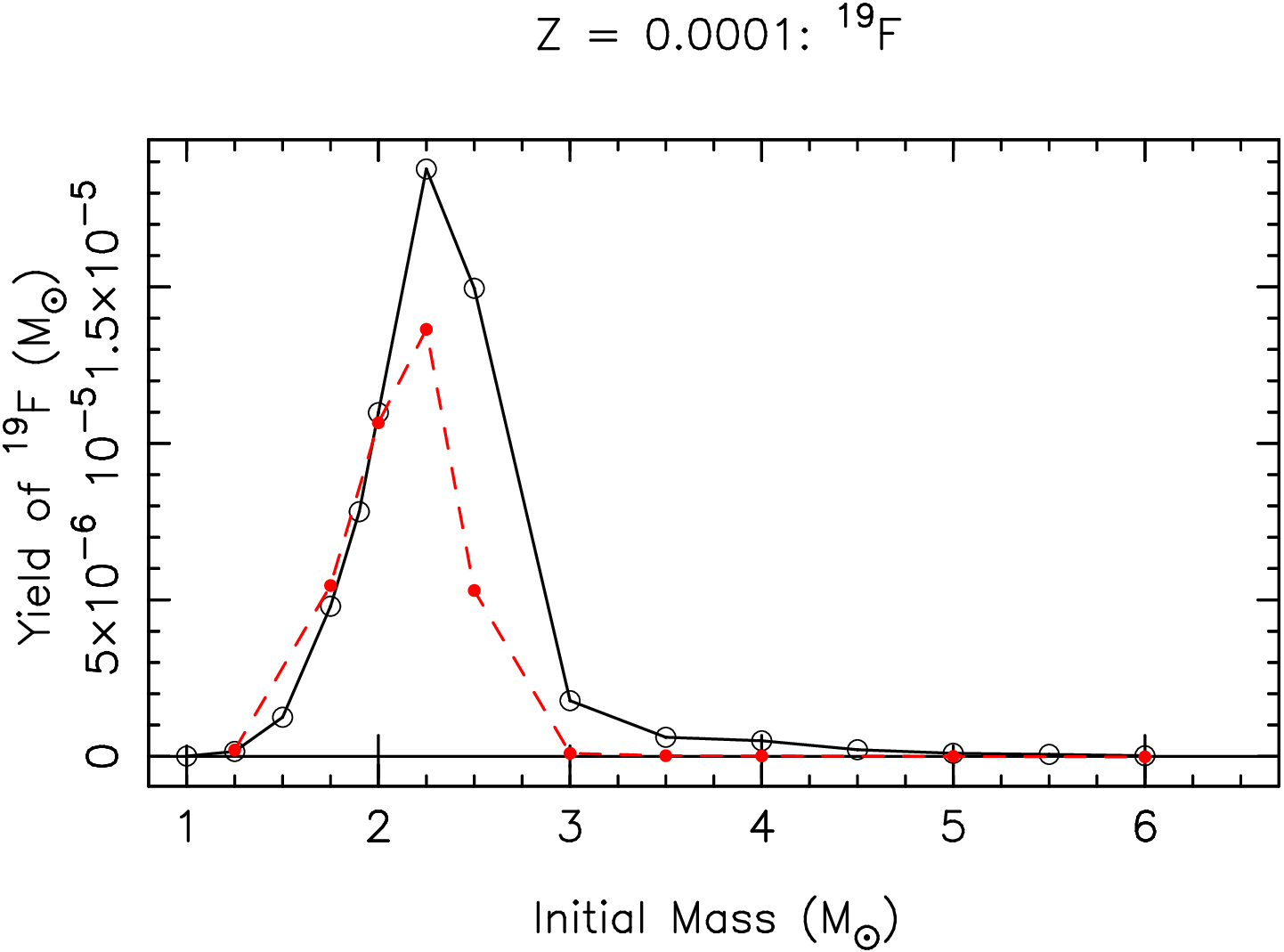} \\
\end{tabular}
\caption{Stellar yields of \iso{12}C, \iso{14}N, \iso{17}O, and \iso{19}F as a function of
the initial mass for models of $Z=0.02$ (left-hand panels) and $Z=0.0001$ (right-hand panels)
from \citet{karakas10a}. The solid line and open circles show results for the updated yields; 
the dashed line and closed circles show results from \citet{karakas07b}. The updated yields from
\citet{karakas10a} use scaled-solar abundances, whereas the yields from \citet{karakas07b}
used non-solar C, N, and O to reflect the composition of the LMC and SMC. Reaction rates were
also updated, which mostly affected \iso{19}F and \iso{23}Na. Also, we used Reimer's mass loss
on the AGB in the $M \ge 3\Msun$, $Z=0.0001$ models from \citet{karakas10a}, 
whereas in \citet{karakas07b} we used \citet{vw93} on the AGB.}
\label{fig-yields}
\end{center}
\end{figure*}

\subsubsection{Lithium}

It is still an open question whether low and intermediate-mass stars contribute to the production of \iso{7}Li 
in the Galaxy \citep{romano01,travaglio01b,prantzos12}.  \citet{prantzos12} concluded that primordial 
nucleosynthesis can produce at most only about 30\% of the solar Li and that stellar sources 
(red giants, AGB stars, novae) must be responsible for at least half.
Current stellar yields of Li from AGB stars do not support this production. The stellar yields by e.g., \citet{karakas10a} 
show that only a narrow mass range of intermediate-mass AGB stars produce more Li than they destroy. This occurs when the
Li produced from HBB takes place at the period of highest mass loss. At $Z=0.02$ this occurs at $\approx 5\Msun$.
Super-AGB stars are also a possible source of Li \citep{ventura10b,ventura13,siess10,doherty14a}.

There are many uncertainties involved in the production of \iso{7}Li in AGB models, including the mass-loss rates and 
the treatment of convective mixing \citep{ventura05a,ventura05b,iwamoto09}.  The stellar Li content
initially rises dramatically from production through the Cameron-Fowler mechanism, but it then 
decreases slowly as the Li is cycled through the hot bottom of the envelope, resulting in its gradual destruction.
Mass-loss rates for AGB stars, such as the formulae
given by \citet{vw93} and \citet{bloecker95}, have a superwind phase which occurs during the final few thermal pulses. 
The superwind phase results in a period of rapid mass loss, and most of the convective envelope is lost during this 
time. Thus the composition of the envelope at the start of the superwind phase critically determines the contribution 
that AGB stars make to the enrichment of the interstellar medium.  By adjusting the mass-loss formula, one can 
manipulate the Li yield.  In Figure~\ref{m6z02-li7} most of the \iso{7}Li has been 
destroyed by the time the superwind phase starts. Other factors that may influence the yields of Li include 
the presence of a binary companion, rotation, and the efficiency of extra mixing on the first and asymptotic
giant branches \citep{charbonnel10,lagarde12a}.

\subsubsection{Carbon, Nitrogen, Oxygen}

AGB stars are one of the most important sources of \iso{12}C in the Galaxy. An estimate of the contribution of \iso{12}C 
from AGB stars suggests that they produce roughly one third of the Galaxy's inventory of \iso{12}C, providing roughly 
the same amount as core-collapse supernovae and Wolf-Rayet stars \citep{dray03}. These quantitative estimates
are hindered by uncertainties in the depth and onset of the third dredge-up. Figure~\ref{fig-yields} shows yields from
\citet{karakas10a} for \iso{12}C, \iso{14}N, \iso{17}O, and \iso{19}F for two metallicities. 
At $Z=0.02$ production is dominated by models of about $3\Msun$, with no C production for models
below about $2.5\Msun$. While it is difficult to determine masses for Galactic C stars, estimates point to
stars with initial masses as low as about $1.5\Msun$ \citep{wallerstein98} for the Galaxy. This suggests
that C production is underestimated in the yields by \citet{karakas10a}. 

The isotopes \iso{13}C and \iso{14}N are produced by the CNO cycles and mixed to the surface by 
first and second dredge-up prior to the AGB, and by HBB during the AGB. 
Figure~\ref{fig-yields} shows that the yields of N are dominated by intermediate-mass stars 
that experience HBB \citep[see also][]{frost98a,chieffi01,pols12}. Chemical evolution models with 
AGB yields show that low-metallicity intermediate-mass stars play an essential role in the production of 
N along with massive rotating stars \citep[e.g.,][]{fenner04,romano10,kobayashi11a}. 

Canonical AGB models do not produce substantial quantities of elemental O and  stellar yields 
from such models are generally negligible, except at the lowest metallicities 
\citep[e.g., at $Z \le 10^{-4}$][]{karakas07b,campbell08,karakas10a,cristallo11}.  Intermediate-mass stars of low
metallicity can destroy a significant amount of \iso{16}O by HBB such that the surface oxgyen abundance
decreases by 0.5--1.0 dex, depending on the stellar model \citep{ventura13}.
The stellar yields by \citet{pignatari13}, which include diffusive 
convective overshoot into the C-O core, suggest that low-mass AGB stars may be an important source of \iso{16}O
in the Universe. Chemical evolution models that use these yields are needed to test the idea. 
 
The only O isotope produced by canonical models is \iso{17}O which is produced by the CNO cycle during HBB.
\citet{kobayashi11a} examined the evolution of the isotopic \iso{16}O/\iso{17}O and \iso{16}O/\iso{18}O ratios 
taking into account the contributions from Type II SNe and AGB stars. It was found that while the solar 
\iso{16}O/\iso{18}O ratio is well matched by current yields, the present-day ratio for \iso{16}O/\iso{17}O 
was too low, indicating an over-production of \iso{17}O by AGB models. This may put constraints on the 
rates of the \iso{17}O $+$ p reactions, which are uncertain at stellar energies \citep[e.g.,][]{chafa07,sergi10}.

\subsubsection{Fluorine}

The cosmic origin of fluorine is not yet completely understood. Core collapse supernovae \citep{woosley95} 
and stellar winds from Wolf Rayet stars \citep{meynet00} are both predicted to release F-enriched material
into the ISM, alongside AGB stars \citep{renda04}. Observationally, AGB stars and their progeny 
(e.g., post-AGB stars, planetary nebulae) are the only confirmed site of F production 
\citep{jorissen92,werner05,zhang05,pandey06,schuler07,abia10,lucatello11}, with no clear indication for enhanced 
F abundances resulting from the $\nu$-process in a region shaped by past SNe \citep{federman05}.  
\citet{recio-blanco12} noted that AGB stars are likely the dominant source of F in the cool main-sequence 
dwarfs they observed in the solar neighbourhood.

Figure~\ref{fig-yields} shows that F production is coupled with C production.
Observations also show a clear correlation between [F/O] content and C/O in AGB stars 
\citep[][and Figure~\ref{m3z02-nucleo}]{jorissen92}.
The stellar yields follow a similar function in mass and metallicity space. 
This means that the uncertainties that are the most significant
for C similarly affect F, although with the added complication that the reaction rates involved in
F production in the He shell are rather uncertain as discussed in \S\ref{sec:tps}.

\citet{kobayashi11b} provide the most recent estimates of the chemical evolution of F using updated AGB 
yields as well as the latest $\nu$-process yields from core-collapse SNe \citep[see also][]{sanchez12}.
The model by \citet{kobayashi11b} was able to reproduce the F
abundances observed in field stars covering a range of metallicities, with SNe dominating production at
the lowest metallicities (here using O as the tracer, [O/H] $\lesssim -1.2$), followed by a rapid increase 
from AGB stars at around [O/H] $\approx -0.5$. 

\subsubsection{From Neon to Iron}

The yields of elemental Ne from AGB stars are generally small, except in the case when substantial 
\iso{22}Ne is produced during thermal pulses \citep{karakas03a}. 
\citet{kobayashi11a} found that the contribution of AGB stars was essential for matching the 
solar Ne isotopic ratios \citep[see also][]{gibson05}. Without AGB stars the contribution from 
SNe dominate and produce too much \iso{20}Ne relative to the neutron-rich \iso{21}Ne 
and \iso{22}Ne. 

\begin{figure}
\begin{center}
\includegraphics[width=0.5\columnwidth, angle=270]{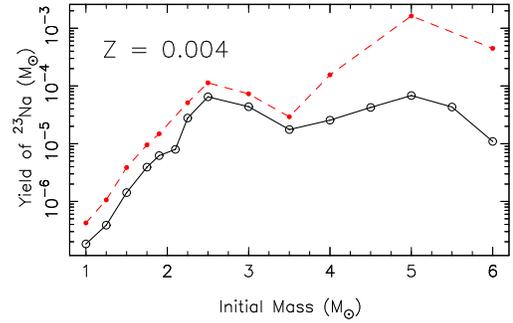}
\caption{Stellar yields of \iso{23}Na as a function of 
the initial mass for models of $Z=0.004$. The solid line and open
circles show results from \citet{karakas10a}, while the dashed line and closed circles show results 
from \citet{karakas07b}.}\label{fig-nayields}
\end{center}
\end{figure}

Intermediate-mass AGB stars produce some Na via the Ne-Na chain 
\citep{forestini97,mowlavi99b,ventura13}, although production is highly dependent on the uncertain rates of the 
\iso{23}Na $+$ p reactions \citep{hale04,izzard07}. The AGB models used by \citet{fenner04} produced copious
sodium and lead to much larger [Na/Fe] abundances compared to observations of globular cluster stars 
\citep[see also][]{gibson07}.  The stellar models in \citet{karakas10a} produced between 
$\sim 6$ to 30 times less Na compared to the stellar models in \citet{karakas07b} as a result of using updated 
reaction rates.  Figure~\ref{fig-nayields} shows the difference for the $Z = 0.004$ models.  Using the updated 
yields from \citet{karakas10a}, \citet{kobayashi11a} found that AGB stars do not noticeably affect the chemical 
evolution of Na in the Milky Way Galaxy.

The yields of Na and Al from AGB stars are also critically dependent on model assumptions and in particular on
the convective model and temperature structure of the envelope \citep{ventura05a}. While the Na and Al yields of 
\citet{karakas10a} are reasonably small, the yields from \citet{ventura13} suggest that intermediate-mass AGB 
and super-AGB stars may be substantial producers of Na and Al at low metallicities \citep{ventura11b}.

The neutron-rich isotopes of Mg are produced by intermediate-mass AGB stars alongside core collapse SNe. 
The amounts of \iso{25}Mg and \iso{26}Mg produced by low-metallicity intermediate-mass AGB stars can be enough 
to affect the galactic chemical evolution of these isotopes.  \citet{fenner03} found that the contribution of AGB stars was 
essential to explain the Mg isotopic ratios observed in cool evolved field stars \citep{gay00,yong03b}.
\citet{kobayashi11a} noted that their chemical evolution model predicted higher than present-day solar ratios
for \iso{24}Mg/\iso{25,26}Mg using yields from AGB stars and SNe and concluded that AGB stars (or some
other source, such as Wolf Rayet stars) need to produce more \iso{25}Mg and \iso{26}Mg.

Phosphorus and Sc can also be produced in small quantities by AGB stars as a consequence of neutron captures
in the He intershell \citep{smith87,karakas12}. Most of the other intermediate-mass elements including Si, Cl, 
Ar, K, Mn, are not significantly produced by AGB nucleosynthesis except for small isotopic shifts caused by neutron
captures \citep{karakas09}. The predicted isotopic shifts, which include increases in the neutron-rich
\iso{29,30}Si, can be compared to measurements of the Si isotopes in pre-solar silicon carbide 
grains. We refer to \citet{lugaro99} and references therein for details \citep[see also][]{zinner06,zinner08,lewis13}.

\subsubsection{Heavy elements produced by the $s$-process}

The contribution from AGB stars is crucial to understand the origin and evolution of elements heavier than 
iron. About half of all heavy elements are produced by the $s$-process, and most of those elements are produced 
by AGB stars. One current uncertainty is the Galactic epoch at which AGB stars begin contributing toward the
bulk Galactic chemical evolution of elements. \citet{simmerer04} suggested this epoch occurred
around [Fe/H] $\gtrsim -1$, but that AGB stars can contribute inhomogeneously (locally) from [Fe/H] $\gtrsim -3$. 
This is also the metallicity at which CEMP stars with $s$-process elements become more common, compared to 
the CEMP stars without neutron-capture element overabundances, which dominate at lower metallicities 
\citep{beers05,sneden08,frebel13}. 

Unfortunately, stellar yields from AGB stars that include predictions for  heavy elements are even 
more incomplete than for light elements. The yields by \citet{cristallo11} include a complete network of
elements to Bi for masses to 3$\Msun$, these were extended to 4, 5, and 6$\Msun$ AGB models for one
metallicity ($Z=0.0003$) by \citet{straniero14};  \citet{pignatari13} include yields for three 
AGB masses at two metallicities ($Z=0.01$ and 0.02); and Fishlock et al. (2014, in preparation) present 
yields for $M=1\Msun$ to $7\Msun$ at one metallicity, $Z=0.001$.  \citet{lugaro12} present 
tables of stellar abundance predictions as a function of thermal pulse number from  models from 0.9$\Msun$ to 6$\Msun$ 
for only one metallicity ($Z = 0.0001$ or [Fe/H] = $-2.3$). The predictions from \citet{lugaro12} 
are not included in Table~\ref{tab:yields} because integrated yields are not provided. 

For super-AGB stars the situation is even worse. 
The only $s$-process calculations currently published are for a single 9$\Msun$, $Z = 0.02$ model in 
\citet{karakas12} and only for a limited nuclear network up to Mo. No yield tables were included with
that study.  \citet{wanajo11} calculate $r$-process yields from electron-capture SNe, which have 
evolved from super-AGB stars with massive O-Ne cores. 

There have been various chemical evolution models that focus on the evolution of the 
neutron-capture elements and the contribution of AGB stars 
\citep[e.g.,][]{raiteri99,travaglio99,travaglio01a,travaglio04,fenner06,serminato09,hansen13}.
In these models, the yields of $s$-process elements are included by extrapolating from the existing 
models, especially for intermediate-mass AGB stars where there are no or few existing theoretical 
predictions.

We comment on the production of $s$-process elements from 
intermediate-mass AGB stars. While the contribution from low-mass AGB stars
to the chemical evolution of Ba and Pb is well supported by models and observations \citep{travaglio01a},
the contribution from intermediate-mass AGB stars has for some time been seen as minimal. 
For example, \citet{travaglio04} estimate that intermediate-mass stars contribute $\approx 8$, 
6, 6, 1, and 5\% toward the solar-system composition of Sr, Y, Zr, Nb, and Mo, respectively.
However, observational evidence suggests that intermediate-mass AGB stars produce substantial
amounts of Rb \citep{garcia06}. Chemical evolution models are required, with a complete set of 
intermediate-mass and super-AGB yields, to quantitatively assess the impact of intermediate-mass stars
on the chemical evolution of Rb. 

We finish with a discussion of another uncertainty on stellar yield predictions: the effect of helium
enrichment. \citet{karakas14} study the effect of helium enrichment on AGB evolution and 
nucleosynthesis for two masses ($M=1.7, 2.36\Msun$) at two metallicities appropriate for 
the GC $\omega$ Centauri ($Z=0.0003, 0.0006$, which is roughly [Fe/H] $\approx -1.8$ 
and $-1.4$, respectively). 
An increase of $\Delta Y = 0.10$ at a given mass decreases the yields of C by up to 
$\approx 60$\%, of F by up to 80\%, and the yields of the $s$-process elements 
Ba and La by $\approx 45$\%. The main reason is that an increase of 
$\Delta Y = 0.10$ leads to roughly a factor of 3 decrease in the amount of dredged
up material during the AGB.   The lifetimes of He-enriched models are significantly 
shorter than their counterparts with primordial He content, 
which means that they will contribute to the 
chemical evolution of a system sooner.

It may not be enough to simply evolve grids of 
stellar evolutionary sequences covering a range in mass and metallicity. Variations in the 
helium mass fraction have a significant impact on the stellar yields and may be an important 
third parameter. This reminds us of the days
before the primordial He abundance was determined, and
stellar models were typically published with a spread of $Y$ values.

\section{Summary and outlook}

% summary
Stellar yields are a key ingredient in chemical evolution
models.  Low and intermediate mass stars are an integral part of 
galaxies and help shape their evolution, gas and dust content, as well as their integrated light.
Even stars as low as 0.9$\Msun$ can,  at low metallicity, contribute to the 
chemical evolution of elements.  The days of 
only considering supernovae are over. However, for low and intermediate-mass stars to be
included, theoretical predictions from stars covering a large range in mass 
and metallicity need to be calculated.

In this review we have discussed the various mixing processes that affect the surface
composition and yields of stars less massive than about 10$\Msun$. These recurrent mixing events
can significantly change the surface composition of the envelope, with the richest nucleosynthesis
occurring during the AGB phase of evolution. AGB stars are observed to show
enrichments in C, N, F, and heavy elements synthesized by the $s$-process. AGB stars release their 
nucleosynthesis products through stellar outflows or winds, in contrast to 
massive stars that explode as core-collapse SNe.

% gaps in knowledge
Supercomputers have allowed the calculation of
stellar yields from detailed (but still single!) AGB models
covering large ranges in mass and composition. While significant progress 
has been made over the past decade, there are still crucial gaps, especially for elements 
produced by the $s$-process for all mass and metallicity ranges. This is mostly because
many nuclear species (on the order of hundreds) are required to accurately model the 
$s$-process and the computational time required is still significant (e.g., months of 
supercomputer time on a single CPU is required for an intermediate-mass AGB model of low metallicity).

Gaps in our knowledge are also apparent for AGB stars of very low metallicity 
(e.g., [Fe/H] $\le -3$). More theoretical effort is needed to address these gaps, especially
because current and new surveys (e.g., SEGUE, GALAH, APOGEE, and GAIA-ESO) will provide 
spectra of hundreds of thousands of stars in all regions of our Milky Way Galaxy including
in the metal-poor halo.   These huge surveys are
going to drive dramatic improvements in the reliability of stellar models, by providing
data that shows inconsistencies and errors in our current understanding. 
Detailed nucleosynthesis models of AGB stars and SNe at low metallicity will be 
required in order to disentangle their history or to provide insight into the nature 
of the Galaxy at the earliest times.

% uncertainties and present progress
Many significant uncertainties affect the stellar yield calculations, such as 
convection and mass loss, and these in turn affect the accuracy and reliability of 
chemical evolution model predictions. Convection has proven to be a persistent
problem in 1D stellar evolution calculations. While we have better observations with
which to constrain convection and convective borders in AGB models to calibrate any 
given stellar evolution code, we are only slowly improving our understanding of the physics 
of convection in stellar interiors \citep{meakin07,arnett09,viallet13}.

The Spitzer Space Telescope has provided important insight into the nature of mass
loss in evolved stars. We have learnt that mass-loss rates are not necessarily smaller
at low metallicity owing to the copious dredge-up of primary C.
We also presumably have a better understanding of the theory of mass loss, at least for C-rich
AGB stars and progress is being made for O-rich AGB stars as well.

Non-standard physics such as rotation and thermohaline mixing are now starting to be 
included in stellar evolutionary calculations and the first yields are appearing, albeit
only for a small number of isotopes. Chemical evolution calculations using these yields
show the importance of these physical phenomena on the evolution of light species such
as \iso{3}He, \iso{7}Li, and the C isotopes. Ideally these calculations should be
extended to include all species affected by extra mixing.

% the future 
Where will we be in the next 5 to 20 years? Future effort must go into understanding how 
convection operates in stellar interiors.  This is singly the most important and crucial uncertainty 
and one that requires multi-dimensional calculations on supercomputers. 
Advances driven by supercomputers will reveal insights into the nature of \iso{13}C pocket
formation in low-mass AGB stars as well as help solve the puzzle of the O abundances observed
in post-AGB stars (e.g., is there really overshoot into the C-O core?). We still have some way 
to go to unravel these puzzles!

Supercomputers will also help drive advances in our understanding of rotation and 
magnetic fields in stellar interiors, as well as non-convective extra mixing processes. 
While progress has been made in understanding how thermohaline mixing operates in red 
giant envelopes, we still do not know if thermohaline mixing is efficient in AGB stars.
Some form of non-convective mixing is needed to drive changes that we
know occur in the envelopes of low-metallicity AGB stars  (e.g., low observed \iso{12}C/\iso{13}C 
ratios compared to AGB yields).

The greatest understanding of mass loss from evolved stars will be driven by observations
from e.g., ALMA and JWST. ALMA  is already starting to probe the clumpy nature 
of mass loss from evolved stars and supergiants.  While thermonuclear reaction rates
are probably the least of our worries for low and intermediate-mass stars, we know that
some key rates (e.g., those that destroy \iso{23}Na and the neutron-producing reactions)
are still highly uncertain and can effect stellar yields. New experimental facilities 
such as the Facility for Rare Isotope Beams being built at the University of Michigan 
will provide new experimental data.

Stellar yields from populations of binaries covering a range of metallicities are
desperately needed. Most stars are in binaries and many will interact. The interactions
can lead to dramatic outcomes such as Type Ia SNe, which play an essential role 
in chemical evolution (and cosmology), but also less energetic outcomes such as novae,
symbiotic stars, barium and CH stars, and CEMP stars. Binary 
evolution will also change the yields from a single stellar population but exactly how
still needs to be determined. 

% new surveys and end
In the next 10 years there will be an explosion of new stellar abundance data
driven by new surveys and instruments (e.g., the GALAH survey using HERMES on the AAT, the
GAIA-ESO survey, LAMOST, APOGEE etc).  These data will help answer big questions facing 
astronomy  including how stars evolve and produce elements and how 
the elements are ejected to enrich the Universe,  as well as questions related to the formation 
and evolution of galaxies. These tremendous investments in astronomical infrastructure 
will pay the largest dividends when augmented by complementary theoretical and modelling 
research.

\begin{acknowledgements}

The authors would like to thank the Editors for their patience in waiting for this review
and the referee for providing constructive comments on the manuscript.
We would also like to thank George Angelou, Harriet Dinerstein, Carolyn Doherty, Cherie Fishlock, 
Brad Gibson, Falk Herwig, Robert Izzard, Maria Lugaro, Brent Miszalski, David Nataf, 
and Richard Stancliffe for help in writing this review. 
A.I.K. is grateful to the ARC for support through a Future Fellowship (FT110100475).
This work was partially supported by ARC grants DP120101815, DP1095368 and DP0877317.
\end{acknowledgements}

% UNCOMMENT THE LINES BELOW IF YOU WISH TO USE BIBTEX
\bibliographystyle{apj}
\bibliography{mnemonic,library}

\begin{thebibliography}{645}
\expandafter\ifx\csname natexlab\endcsname\relax\def\natexlab#1{#1}\fi

\bibitem[{{Abate} {et~al.}(2013){Abate}, {Pols}, {Izzard}, {Mohamed}, \& {de
  Mink}}]{abate13}
{Abate}, C., {Pols}, O.~R., {Izzard}, R.~G., {Mohamed}, S.~S., \& {de Mink},
  S.~E. 2013, A\&A, 552, A26

\bibitem[{{Abia} {et~al.}(2001){Abia}, {Busso}, {Gallino}, {Dom{\'{\i}}nguez},
  {Straniero}, \& {Isern}}]{abia01b}
{Abia}, C., {Busso}, M., {Gallino}, R., {Dom{\'{\i}}nguez}, I., {Straniero},
  O., \& {Isern}, J. 2001, ApJ, 559, 1117

\bibitem[{{Abia} {et~al.}(2010){Abia}, {Cunha}, {Cristallo}, {de Laverny},
  {Dom{\'{\i}}nguez}, {Eriksson}, {Gialanella}, {Hinkle}, {Imbriani},
  {Recio-Blanco}, {Smith}, {Straniero}, \& {Wahlin}}]{abia10}
{Abia}, C., {Cunha}, K., {Cristallo}, S., {de Laverny}, P., {Dom{\'{\i}}nguez},
  I., {Eriksson}, K., {Gialanella}, L., {Hinkle}, K., {Imbriani}, G.,
  {Recio-Blanco}, A., {Smith}, V.~V., {Straniero}, O., \& {Wahlin}, R. 2010,
  ApJL, 715, L94

\bibitem[{{Abia} {et~al.}(2011){Abia}, {Cunha}, {Cristallo}, {de Laverny},
  {Dom{\'{\i}}nguez}, {Recio-Blanco}, {Smith}, \& {Straniero}}]{abia11}
{Abia}, C., {Cunha}, K., {Cristallo}, S., {de Laverny}, P., {Dom{\'{\i}}nguez},
  I., {Recio-Blanco}, A., {Smith}, V.~V., \& {Straniero}, O. 2011, ApJL, 737,
  L8

\bibitem[{{Abia} {et~al.}(2008){Abia}, {de Laverny}, \& {Wahlin}}]{abia08}
{Abia}, C., {de Laverny}, P., \& {Wahlin}, R. 2008, A\&A, 481, 161

\bibitem[{{Abia} {et~al.}(2002){Abia}, {Dom{\'{\i}}nguez}, {Gallino}, {Busso},
  {Masera}, {Straniero}, {de Laverny}, {Plez}, \& {Isern}}]{abia02}
{Abia}, C., {Dom{\'{\i}}nguez}, I., {Gallino}, R., {Busso}, M., {Masera}, S.,
  {Straniero}, O., {de Laverny}, P., {Plez}, B., \& {Isern}, J. 2002, ApJ, 579,
  817

\bibitem[{{Abia} \& {Isern}(1997)}]{abia97}
{Abia}, C. \& {Isern}, J. 1997, MNRAS, 289, L11

\bibitem[{{Abia} {et~al.}(2009){Abia}, {Recio-Blanco}, {de Laverny},
  {Cristallo}, {Dom{\'{\i}}nguez}, \& {Straniero}}]{abia09}
{Abia}, C., {Recio-Blanco}, A., {de Laverny}, P., {Cristallo}, S.,
  {Dom{\'{\i}}nguez}, I., \& {Straniero}, O. 2009, ApJ, 694, 971

\bibitem[{{Alcal{\'a}} {et~al.}(2011){Alcal{\'a}}, {Biazzo}, {Covino},
  {Frasca}, \& {Bedin}}]{alcala11}
{Alcal{\'a}}, J.~M., {Biazzo}, K., {Covino}, E., {Frasca}, A., \& {Bedin},
  L.~R. 2011, A\&A, 531, L12

\bibitem[{{Alexander} \& {Ferguson}(1994)}]{alexander94}
{Alexander}, D.~R. \& {Ferguson}, J.~W. 1994, ApJ, 437, 879

\bibitem[{{Allen} \& {Barbuy}(2006{\natexlab{a}})}]{allen06a}
{Allen}, D.~M. \& {Barbuy}, B. 2006{\natexlab{a}}, A\&A, 454, 895

\bibitem[{{Allen} \& {Barbuy}(2006{\natexlab{b}})}]{allen06b}
---. 2006{\natexlab{b}}, A\&A, 454, 917

\bibitem[{{Allen} \& {Porto de Mello}(2007)}]{allen07}
{Allen}, D.~M. \& {Porto de Mello}, G.~F. 2007, A\&A, 474, 221

\bibitem[{{Aller} \& {Czyzak}(1983)}]{aller83}
{Aller}, L.~H. \& {Czyzak}, S.~J. 1983, ApJS, 51, 211

\bibitem[{{Althaus} {et~al.}(2010){Althaus}, {C{\'o}rsico}, {Isern}, \&
  {Garc{\'{\i}}a-Berro}}]{althaus10}
{Althaus}, L.~G., {C{\'o}rsico}, A.~H., {Isern}, J., \& {Garc{\'{\i}}a-Berro},
  E. 2010, A\&AR, 18, 471

\bibitem[{{Althaus} {et~al.}(2009){Althaus}, {Panei}, {Miller Bertolami},
  {Garc{\'{\i}}a-Berro}, {C{\'o}rsico}, {Romero}, {Kepler}, \&
  {Rohrmann}}]{althaus09}
{Althaus}, L.~G., {Panei}, J.~A., {Miller Bertolami}, M.~M.,
  {Garc{\'{\i}}a-Berro}, E., {C{\'o}rsico}, A.~H., {Romero}, A.~D., {Kepler},
  S.~O., \& {Rohrmann}, R.~D. 2009, ApJ, 704, 1605

\bibitem[{{Alves-Brito} {et~al.}(2011){Alves-Brito}, {Karakas}, {Yong},
  {Mel{\'e}ndez}, \& {V{\'a}squez}}]{alves-brito11}
{Alves-Brito}, A., {Karakas}, A.~I., {Yong}, D., {Mel{\'e}ndez}, J., \&
  {V{\'a}squez}, S. 2011, A\&A, 536, A40

\bibitem[{{Angelou} {et~al.}(2011){Angelou}, {Church}, {Stancliffe},
  {Lattanzio}, \& {Smith}}]{angelou11}
{Angelou}, G.~C., {Church}, R.~P., {Stancliffe}, R.~J., {Lattanzio}, J.~C., \&
  {Smith}, G.~H. 2011, ApJ, 728, 79

\bibitem[{{Angelou} {et~al.}(2012){Angelou}, {Stancliffe}, {Church},
  {Lattanzio}, \& {Smith}}]{angelou12}
{Angelou}, G.~C., {Stancliffe}, R.~J., {Church}, R.~P., {Lattanzio}, J.~C., \&
  {Smith}, G.~H. 2012, ApJ, 749, 128

\bibitem[{{Angulo} {et~al.}(1999){Angulo}, {Arnould}, {Rayet}, {Descouvemont},
  {Baye}, {Leclercq-Willain}, {Coc}, {Barhoumi}, {Aguer}, {Rolfs}, {Kunz},
  {Hammer}, {Mayer}, {Paradellis}, {Kossionides}, \& {Chronidou}}]{angulo99}
{Angulo}, C., {Arnould}, M., {Rayet}, M., {Descouvemont}, P., {Baye}, D.,
  {Leclercq-Willain}, C., {Coc}, A., {Barhoumi}, S., {Aguer}, P., {Rolfs}, C.,
  {Kunz}, R., {Hammer}, J.~W., {Mayer}, A., {Paradellis}, T., {Kossionides},
  S., \& {Chronidou}, C. 1999, Nucl. Phys. A, 656, 3

\bibitem[{{Aoki} {et~al.}(2007){Aoki}, {Beers}, {Christlieb}, {Norris}, {Ryan},
  \& {Tsangarides}}]{aoki07}
{Aoki}, W., {Beers}, T.~C., {Christlieb}, N., {Norris}, J.~E., {Ryan}, S.~G.,
  \& {Tsangarides}, S. 2007, ApJ, 655, 492

\bibitem[{{Arcones} {et~al.}(2007){Arcones}, {Janka}, \& {Scheck}}]{arcones07}
{Arcones}, A., {Janka}, H.-T., \& {Scheck}, L. 2007, A\&A, 467, 1227

\bibitem[{{Arcones} \& {Montes}(2011)}]{arcones11}
{Arcones}, A. \& {Montes}, F. 2011, ApJ, 731, 5

\bibitem[{{Argast} {et~al.}(2004){Argast}, {Samland}, {Thielemann}, \&
  {Qian}}]{argast04}
{Argast}, D., {Samland}, M., {Thielemann}, F.-K., \& {Qian}, Y.-Z. 2004, A\&A,
  416, 997

\bibitem[{{Arlandini} {et~al.}(1999){Arlandini}, {K{\"a}ppeler}, {Wisshak},
  {Gallino}, {Lugaro}, {Busso}, \& {Straniero}}]{arlandini99}
{Arlandini}, C., {K{\"a}ppeler}, F., {Wisshak}, K., {Gallino}, R., {Lugaro},
  M., {Busso}, M., \& {Straniero}, O. 1999, ApJ, 525, 886

\bibitem[{{Arndt} {et~al.}(1997){Arndt}, {Fleischer}, \& {Sedlmayr}}]{arndt97}
{Arndt}, T.~U., {Fleischer}, A.~J., \& {Sedlmayr}, E. 1997, A\&A, 327, 614

\bibitem[{{Arnett} {et~al.}(2009){Arnett}, {Meakin}, \& {Young}}]{arnett09}
{Arnett}, D., {Meakin}, C., \& {Young}, P.~A. 2009, ApJ, 690, 1715

\bibitem[{{Arnould} {et~al.}(1999){Arnould}, {Goriely}, \&
  {Jorissen}}]{arnould99}
{Arnould}, M., {Goriely}, S., \& {Jorissen}, A. 1999, A\&A, 347, 572

\bibitem[{{Arnould} {et~al.}(2007){Arnould}, {Goriely}, \&
  {Takahashi}}]{arnould07}
{Arnould}, M., {Goriely}, S., \& {Takahashi}, K. 2007, Phys. Rep., 450, 97

\bibitem[{{Asplund} {et~al.}(2009){Asplund}, {Grevesse}, {Sauval}, \&
  {Scott}}]{asplund09}
{Asplund}, M., {Grevesse}, N., {Sauval}, A.~J., \& {Scott}, P. 2009, ARA\&A,
  47, 481

\bibitem[{{Asplund} {et~al.}(1997){Asplund}, {Gustafsson}, {Lambert}, \&
  {Kameswara Rao}}]{asplund97}
{Asplund}, M., {Gustafsson}, B., {Lambert}, D.~L., \& {Kameswara Rao}, N. 1997,
  A\&A, 321, L17

\bibitem[{{Balser} {et~al.}(2007){Balser}, {Rood}, \& {Bania}}]{balser07}
{Balser}, D.~S., {Rood}, R.~T., \& {Bania}, T.~M. 2007, Science, 317, 1171

\bibitem[{{Bania} {et~al.}(2002){Bania}, {Rood}, \& {Balser}}]{bania02}
{Bania}, T.~M., {Rood}, R.~T., \& {Balser}, D.~S. 2002, Nature, 415, 54

\bibitem[{{Bedijn}(1988)}]{bedjin88}
{Bedijn}, P.~J. 1988, A\&A, 205, 105

\bibitem[{{Beer} \& {Macklin}(1989)}]{beer89}
{Beer}, H. \& {Macklin}, R.~L. 1989, ApJ, 339, 962

\bibitem[{{Beers} \& {Christlieb}(2005)}]{beers05}
{Beers}, T.~C. \& {Christlieb}, N. 2005, ARA\&A, 43, 531

\bibitem[{{Beers} {et~al.}(2007){Beers}, {Sivarani}, {Marsteller}, {Lee},
  {Rossi}, \& {Plez}}]{beers07}
{Beers}, T.~C., {Sivarani}, T., {Marsteller}, B., {Lee}, Y., {Rossi}, S., \&
  {Plez}, B. 2007, AJ, 133, 1193

\bibitem[{{Bellman} {et~al.}(2001){Bellman}, {Briley}, {Smith}, \&
  {Claver}}]{bellman01}
{Bellman}, S., {Briley}, M.~M., {Smith}, G.~H., \& {Claver}, C.~F. 2001, PASP,
  113, 326

\bibitem[{{Bernard-Salas} {et~al.}(2008){Bernard-Salas}, {Pottasch},
  {Gutenkunst}, {Morris}, \& {Houck}}]{bernard08}
{Bernard-Salas}, J., {Pottasch}, S.~R., {Gutenkunst}, S., {Morris}, P.~W., \&
  {Houck}, J.~R. 2008, ApJ, 672, 274

\bibitem[{{Bertelli} {et~al.}(1986{\natexlab{a}}){Bertelli}, {Bressan},
  {Chiosi}, \& {Angerer}}]{bertelli86}
{Bertelli}, G., {Bressan}, A., {Chiosi}, C., \& {Angerer}, K.
  1986{\natexlab{a}}, A\&AS, 66, 191

\bibitem[{{Bertelli} {et~al.}(1986{\natexlab{b}}){Bertelli}, {Bressan},
  {Chiosi}, \& {Angerer}}]{bertelli86b}
---. 1986{\natexlab{b}}, Mem. Soc. Astron. Ital., 57, 427

\bibitem[{{Bertolli} {et~al.}(2013){Bertolli}, {Pignatari}, \&
  {Lawano}}]{bertolli13}
{Bertolli}, M. G.and~{Herwig}, F., {Pignatari}, M., \& {Lawano}, T. 2013, ArXiv
  e-prints

\bibitem[{{Bisterzo} {et~al.}(2010){Bisterzo}, {Gallino}, {Straniero},
  {Cristallo}, \& {K{\"a}ppeler}}]{bisterzo10}
{Bisterzo}, S., {Gallino}, R., {Straniero}, O., {Cristallo}, S., \&
  {K{\"a}ppeler}, F. 2010, MNRAS, 404, 1529

\bibitem[{{Bisterzo} {et~al.}(2011){Bisterzo}, {Gallino}, {Straniero},
  {Cristallo}, \& {K{\"a}ppeler}}]{bisterzo11}
---. 2011, MNRAS, 418, 284

\bibitem[{{Bisterzo} {et~al.}(2012){Bisterzo}, {Gallino}, {Straniero},
  {Cristallo}, \& {K{\"a}ppeler}}]{bisterzo12}
---. 2012, MNRAS, 422, 849

\bibitem[{{Bjork} \& {Chaboyer}(2006)}]{bjork06}
{Bjork}, S.~R. \& {Chaboyer}, B. 2006, ApJ, 641, 1102

\bibitem[{{Bl{\"o}cker}(1995)}]{bloecker95}
{Bl{\"o}cker}, T. 1995, A\&A, 297, 727

\bibitem[{{Bl{\"o}cker} \& {Schoenberner}(1991)}]{bloecker91}
{Bl{\"o}cker}, T. \& {Schoenberner}, D. 1991, A\&A, 244, L43

\bibitem[{{Bloecker}(1995)}]{bloecker95b}
{Bloecker}, T. 1995, A\&A, 299, 755

\bibitem[{{Boffin} \& {Jorissen}(1988)}]{boffin88}
{Boffin}, H.~M.~J. \& {Jorissen}, A. 1988, A\&A, 205, 155

\bibitem[{{Bonaparte} {et~al.}(2013){Bonaparte}, {Matteucci}, {Recchi},
  {Spitoni}, {Pipino}, \& {Grieco}}]{bonaparte13}
{Bonaparte}, I., {Matteucci}, F., {Recchi}, S., {Spitoni}, E., {Pipino}, A., \&
  {Grieco}, V. 2013, MNRAS, 435, 2460

\bibitem[{{Bona{\v c}i{\'c} Marinovi{\'c}} {et~al.}(2007{\natexlab{a}}){Bona{\v
  c}i{\'c} Marinovi{\'c}}, {Izzard}, {Lugaro}, \& {Pols}}]{axel07a}
{Bona{\v c}i{\'c} Marinovi{\'c}}, A., {Izzard}, R.~G., {Lugaro}, M., \& {Pols},
  O.~R. 2007{\natexlab{a}}, A\&A, 469, 1013

\bibitem[{{Bona{\v c}i{\'c} Marinovi{\'c}} {et~al.}(2007{\natexlab{b}}){Bona{\v
  c}i{\'c} Marinovi{\'c}}, {Lugaro}, {Reyniers}, \& {van Winckel}}]{axel07b}
{Bona{\v c}i{\'c} Marinovi{\'c}}, A., {Lugaro}, M., {Reyniers}, M., \& {van
  Winckel}, H. 2007{\natexlab{b}}, A\&A, 472, L1

\bibitem[{{Boothroyd} \& {Sackmann}(1988)}]{boothroyd88c}
{Boothroyd}, A.~I. \& {Sackmann}, I.-J. 1988, ApJ, 328, 653

\bibitem[{{Boothroyd} \& {Sackmann}(1992)}]{boothroyd92}
---. 1992, ApJL, 393, L21

\bibitem[{{Boothroyd} \& {Sackmann}(1999)}]{boothroyd99}
---. 1999, ApJ, 510, 232

\bibitem[{{Boothroyd} {et~al.}(1993){Boothroyd}, {Sackmann}, \&
  {Ahern}}]{boothroyd93}
{Boothroyd}, A.~I., {Sackmann}, I.-J., \& {Ahern}, S.~C. 1993, ApJ, 416, 762

\bibitem[{{Boothroyd} {et~al.}(1994){Boothroyd}, {Sackmann}, \&
  {Wasserburg}}]{boothroyd94}
{Boothroyd}, A.~I., {Sackmann}, I.-J., \& {Wasserburg}, G.~J. 1994, ApJL, 430,
  L77

\bibitem[{{Boothroyd} {et~al.}(1995){Boothroyd}, {Sackmann}, \&
  {Wasserburg}}]{boothroyd95}
---. 1995, ApJL, 442, L21

\bibitem[{{Boyer} {et~al.}(2013){Boyer}, {Girardi}, {Marigo}, {Williams},
  {Aringer}, {Nowotny}, {Rosenfield}, {Dorman}, {Guhathakurta}, {Dalcanton},
  {Melbourne}, {Olsen}, \& {Weisz}}]{boyer13}
{Boyer}, M.~L., {Girardi}, L., {Marigo}, P., {Williams}, B.~F., {Aringer}, B.,
  {Nowotny}, W., {Rosenfield}, P., {Dorman}, C.~E., {Guhathakurta}, P.,
  {Dalcanton}, J.~J., {Melbourne}, J.~L., {Olsen}, K.~A.~G., \& {Weisz}, D.~R.
  2013, ApJ, 774, 83

\bibitem[{{Bragaglia} {et~al.}(2001){Bragaglia}, {Carretta}, {Gratton}, {Tosi},
  {Bonanno}, {Bruno}, {Cal{\`i}}, {Claudi}, {Cosentino}, {Desidera},
  {Farisato}, {Rebeschini}, \& {Scuderi}}]{bragaglia01}
{Bragaglia}, A., {Carretta}, E., {Gratton}, R.~G., {Tosi}, M., {Bonanno}, G.,
  {Bruno}, P., {Cal{\`i}}, A., {Claudi}, R., {Cosentino}, R., {Desidera}, S.,
  {Farisato}, G., {Rebeschini}, M., \& {Scuderi}, S. 2001, AJ, 121, 327

\bibitem[{{Bressan} {et~al.}(1993){Bressan}, {Fagotto}, {Bertelli}, \&
  {Chiosi}}]{bressan93}
{Bressan}, A., {Fagotto}, F., {Bertelli}, G., \& {Chiosi}, C. 1993, A\&AS, 100,
  647

\bibitem[{{Brown} {et~al.}(2013){Brown}, {Garaud}, \& {Stellmach}}]{brown13}
{Brown}, J.~M., {Garaud}, P., \& {Stellmach}, S. 2013, ApJ, 768, 34

\bibitem[{{Brown} {et~al.}(2001){Brown}, {Sweigart}, {Lanz}, {Landsman}, \&
  {Hubeny}}]{brown01}
{Brown}, T.~M., {Sweigart}, A.~V., {Lanz}, T., {Landsman}, W.~B., \& {Hubeny},
  I. 2001, ApJ, 562, 368

\bibitem[{{Buonanno} {et~al.}(1985){Buonanno}, {Corsi}, \& {Fusi
  Pecci}}]{buonanno85}
{Buonanno}, R., {Corsi}, C.~E., \& {Fusi Pecci}, F. 1985, A\&A, 145, 97

\bibitem[{{Burbidge} {et~al.}(1957){Burbidge}, {Burbidge}, {Fowler}, \&
  {Hoyle}}]{b2fh}
{Burbidge}, E.~M., {Burbidge}, G.~R., {Fowler}, W.~A., \& {Hoyle}, F. 1957,
  Rev. Mod. Phys., 29, 547

\bibitem[{{Busso} {et~al.}(2001){Busso}, {Gallino}, {Lambert}, {Travaglio}, \&
  {Smith}}]{busso01}
{Busso}, M., {Gallino}, R., {Lambert}, D.~L., {Travaglio}, C., \& {Smith},
  V.~V. 2001, ApJ, 557, 802

\bibitem[{{Busso} {et~al.}(1999){Busso}, {Gallino}, \& {Wasserburg}}]{busso99}
{Busso}, M., {Gallino}, R., \& {Wasserburg}, G.~J. 1999, ARA\&A, 37, 239

\bibitem[{{Busso} {et~al.}(2010){Busso}, {Palmerini}, {Maiorca}, {Cristallo},
  {Straniero}, {Abia}, {Gallino}, \& {La Cognata}}]{busso10}
{Busso}, M., {Palmerini}, S., {Maiorca}, E., {Cristallo}, S., {Straniero}, O.,
  {Abia}, C., {Gallino}, R., \& {La Cognata}, M. 2010, ApJL, 717, L47

\bibitem[{{Busso} {et~al.}(2007){Busso}, {Wasserburg}, {Nollett}, \&
  {Calandra}}]{busso07b}
{Busso}, M., {Wasserburg}, G.~J., {Nollett}, K.~M., \& {Calandra}, A. 2007,
  ApJ, 671, 802

\bibitem[{{Buzzoni} {et~al.}(1983){Buzzoni}, {Pecci}, {Buonanno}, \&
  {Corsi}}]{buzzoni83}
{Buzzoni}, A., {Pecci}, F.~F., {Buonanno}, R., \& {Corsi}, C.~E. 1983, A\&A,
  128, 94

\bibitem[{{Cameron}(1957)}]{cameron57}
{Cameron}, A.~G.~W. 1957, PASP, 69, 201

\bibitem[{{Cameron}(1960)}]{cameron60}
---. 1960, AJ, 65, 485

\bibitem[{{Cameron} \& {Fowler}(1971)}]{cameron71}
{Cameron}, A.~G.~W. \& {Fowler}, W.~A. 1971, ApJ, 164, 111

\bibitem[{{Campbell} {et~al.}(2013){Campbell}, {D'Orazi}, {Yong},
  {Constantino}, {Lattanzio}, {Stancliffe}, {Angelou}, {Wylie-de Boer}, \&
  {Grundahl}}]{campbell13}
{Campbell}, S.~W., {D'Orazi}, V., {Yong}, D., {Constantino}, T.~N.,
  {Lattanzio}, J.~C., {Stancliffe}, R.~J., {Angelou}, G.~C., {Wylie-de Boer},
  E.~C., \& {Grundahl}, F. 2013, Nature, 498, 198

\bibitem[{{Campbell} \& {Lattanzio}(2008)}]{campbell08}
{Campbell}, S.~W. \& {Lattanzio}, J.~C. 2008, A\&A, 490, 769

\bibitem[{{Campbell} {et~al.}(2010){Campbell}, {Lugaro}, \&
  {Karakas}}]{campbell10}
{Campbell}, S.~W., {Lugaro}, M., \& {Karakas}, A.~I. 2010, A\&A, 522, L6

\bibitem[{{Cannon}(1993)}]{cannon93}
{Cannon}, R.~C. 1993, MNRAS, 263, 817

\bibitem[{{Cantiello} \& {Langer}(2010)}]{cantiello10}
{Cantiello}, M. \& {Langer}, N. 2010, A\&A, 521, A9

\bibitem[{{Canuto}(2011{\natexlab{a}})}]{canuto11a}
{Canuto}, V.~M. 2011{\natexlab{a}}, A\&A, 528, A76

\bibitem[{{Canuto}(2011{\natexlab{b}})}]{canuto11b}
---. 2011{\natexlab{b}}, A\&A, 528, A77

\bibitem[{{Canuto}(2011{\natexlab{c}})}]{canuto11c}
---. 2011{\natexlab{c}}, A\&A, 528, A78

\bibitem[{{Canuto}(2011{\natexlab{d}})}]{canuto11d}
---. 2011{\natexlab{d}}, A\&A, 528, A79

\bibitem[{{Canuto}(2011{\natexlab{e}})}]{canuto11e}
---. 2011{\natexlab{e}}, A\&A, 528, A80

\bibitem[{{Caputo} {et~al.}(1989){Caputo}, {Chieffi}, {Tornambe}, {Castellani},
  \& {Pulone}}]{caputo89}
{Caputo}, F., {Chieffi}, A., {Tornambe}, A., {Castellani}, V., \& {Pulone}, L.
  1989, ApJ, 340, 241

\bibitem[{{Carlberg} {et~al.}(2009){Carlberg}, {Majewski}, \&
  {Arras}}]{carlberg09}
{Carlberg}, J.~K., {Majewski}, S.~R., \& {Arras}, P. 2009, ApJ, 700, 832

\bibitem[{{Carlberg} {et~al.}(2010){Carlberg}, {Smith}, {Cunha}, {Majewski}, \&
  {Rood}}]{carlberg10}
{Carlberg}, J.~K., {Smith}, V.~V., {Cunha}, K., {Majewski}, S.~R., \& {Rood},
  R.~T. 2010, ApJL, 723, L103

\bibitem[{{Carretta} {et~al.}(2009){Carretta}, {Bragaglia}, {Gratton}, \&
  {Lucatello}}]{carretta09}
{Carretta}, E., {Bragaglia}, A., {Gratton}, R., \& {Lucatello}, S. 2009, A\&A,
  505, 139

\bibitem[{{Cassisi} {et~al.}(1996){Cassisi}, {Castellani}, \&
  {Tornambe}}]{cassisi96}
{Cassisi}, S., {Castellani}, V., \& {Tornambe}, A. 1996, ApJ, 459, 298

\bibitem[{{Cassisi} {et~al.}(2011){Cassisi}, {Mar{\'{\i}}n-Franch}, {Salaris},
  {Aparicio}, {Monelli}, \& {Pietrinferni}}]{cassisi11}
{Cassisi}, S., {Mar{\'{\i}}n-Franch}, A., {Salaris}, M., {Aparicio}, A.,
  {Monelli}, M., \& {Pietrinferni}, A. 2011, A\&A, 527, A59

\bibitem[{{Cassisi} \& {Salaris}(1997)}]{cassisi97}
{Cassisi}, S. \& {Salaris}, M. 1997, MNRAS, 285, 593

\bibitem[{{Cassisi} {et~al.}(2002){Cassisi}, {Salaris}, \& {Bono}}]{cassisi02}
{Cassisi}, S., {Salaris}, M., \& {Bono}, G. 2002, ApJ, 565, 1231

\bibitem[{{Cassisi} {et~al.}(2003){Cassisi}, {Schlattl}, {Salaris}, \&
  {Weiss}}]{cassisi03}
{Cassisi}, S., {Schlattl}, H., {Salaris}, M., \& {Weiss}, A. 2003, ApJL, 582,
  L43

\bibitem[{{Castellani} {et~al.}(1985){Castellani}, {Chieffi}, {Tornambe}, \&
  {Pulone}}]{castellani85}
{Castellani}, V., {Chieffi}, A., {Tornambe}, A., \& {Pulone}, L. 1985, ApJ,
  296, 204

\bibitem[{{Castellani} {et~al.}(1971{\natexlab{a}}){Castellani}, {Giannone}, \&
  {Renzini}}]{castellani71a}
{Castellani}, V., {Giannone}, P., \& {Renzini}, A. 1971{\natexlab{a}}, ApSpSci,
  10, 340

\bibitem[{{Castellani} {et~al.}(1971{\natexlab{b}}){Castellani}, {Giannone}, \&
  {Renzini}}]{castellani71b}
---. 1971{\natexlab{b}}, ApSpSci, 10, 355

\bibitem[{{Catelan}(2000)}]{catelan00}
{Catelan}, M. 2000, ApJ, 531, 826

\bibitem[{{Chafa} {et~al.}(2007){Chafa}, {Tatischeff}, {Aguer}, {Barhoumi},
  {Coc}, {Garrido}, {Hernanz}, {Jos{\'e}}, {Kiener}, {Lefebvre-Schuhl},
  {Ouichaoui}, {S{\'e}r{\'e}ville}, \& {Thibaud}}]{chafa07}
{Chafa}, A., {Tatischeff}, V., {Aguer}, P., {Barhoumi}, S., {Coc}, A.,
  {Garrido}, F., {Hernanz}, M., {Jos{\'e}}, J., {Kiener}, J.,
  {Lefebvre-Schuhl}, A., {Ouichaoui}, S., {S{\'e}r{\'e}ville}, N.~D., \&
  {Thibaud}, J.-P. 2007, Phys. Rev. C, 75, 035810

\bibitem[{{Chanam{\'e}} {et~al.}(2005){Chanam{\'e}}, {Pinsonneault}, \&
  {Terndrup}}]{chaname05}
{Chanam{\'e}}, J., {Pinsonneault}, M., \& {Terndrup}, D.~M. 2005, ApJ, 631, 540

\bibitem[{{Charbonnel}(1994)}]{charbonnel94}
{Charbonnel}, C. 1994, A\&A, 282, 811

\bibitem[{{Charbonnel}(1995)}]{charbonnel95}
---. 1995, ApJL, 453, L41

\bibitem[{{Charbonnel} \& {Balachandran}(2000)}]{charbal00}
{Charbonnel}, C. \& {Balachandran}, S.~C. 2000, A\&A, 359, 563

\bibitem[{{Charbonnel} \& {Lagarde}(2010)}]{charbonnel10}
{Charbonnel}, C. \& {Lagarde}, N. 2010, A\&A, 522, A10

\bibitem[{{Charbonnel} \& {Zahn}(2007{\natexlab{a}})}]{charbonnel07b}
{Charbonnel}, C. \& {Zahn}, J.-P. 2007{\natexlab{a}}, A\&A, 476, L29

\bibitem[{{Charbonnel} \& {Zahn}(2007{\natexlab{b}})}]{charbonnel07a}
---. 2007{\natexlab{b}}, A\&A, 467, L15

\bibitem[{{Chiappini} {et~al.}(2006){Chiappini}, {Hirschi}, {Meynet},
  {Ekstr{\"o}m}, {Maeder}, \& {Matteucci}}]{chiappini06}
{Chiappini}, C., {Hirschi}, R., {Meynet}, G., {Ekstr{\"o}m}, S., {Maeder}, A.,
  \& {Matteucci}, F. 2006, A\&A, 449, L27

\bibitem[{{Chiappini} {et~al.}(2003){Chiappini}, {Matteucci}, \&
  {Meynet}}]{chiappini03}
{Chiappini}, C., {Matteucci}, F., \& {Meynet}, G. 2003, A\&A, 410, 257

\bibitem[{{Chieffi} {et~al.}(2001){Chieffi}, {Dom{\'{\i}}nguez}, {Limongi}, \&
  {Straniero}}]{chieffi01}
{Chieffi}, A., {Dom{\'{\i}}nguez}, I., {Limongi}, M., \& {Straniero}, O. 2001,
  ApJ, 554, 1159

\bibitem[{{Church} {et~al.}(2009){Church}, {Cristallo}, {Lattanzio},
  {Stancliffe}, {Straniero}, \& {Cannon}}]{church09}
{Church}, R.~P., {Cristallo}, S., {Lattanzio}, J.~C., {Stancliffe}, R.~J.,
  {Straniero}, O., \& {Cannon}, R.~C. 2009, PASA, 26, 217

\bibitem[{{Clayton}(1988)}]{clayton88}
{Clayton}, D.~D. 1988, MNRAS, 234, 1

\bibitem[{{Coc} {et~al.}(2004){Coc}, {Vangioni-Flam}, {Descouvemont},
  {Adahchour}, \& {Angulo}}]{coc04}
{Coc}, A., {Vangioni-Flam}, E., {Descouvemont}, P., {Adahchour}, A., \&
  {Angulo}, C. 2004, ApJ, 600, 544

\bibitem[{{Cohen} {et~al.}(2005){Cohen}, {Briley}, \& {Stetson}}]{cohen05b}
{Cohen}, J.~G., {Briley}, M.~M., \& {Stetson}, P.~B. 2005, AJ, 130, 1177

\bibitem[{{Cohen} {et~al.}(2004){Cohen}, {Christlieb}, {McWilliam}, {Shectman},
  {Thompson}, {Wasserburg}, {Ivans}, {Dehn}, {Karlsson}, \&
  {Melendez}}]{cohen04}
{Cohen}, J.~G., {Christlieb}, N., {McWilliam}, A., {Shectman}, S., {Thompson},
  I., {Wasserburg}, G.~J., {Ivans}, I., {Dehn}, M., {Karlsson}, T., \&
  {Melendez}, J. 2004, ApJ, 612, 1107

\bibitem[{{Cohen} \& {Drake}(2014)}]{cohen14}
{Cohen}, O. \& {Drake}, J.~J. 2014, ApJ, 783, 55

\bibitem[{{Constantino} {et~al.}(2014){Constantino}, {Campbell}, {Gil-Pons}, \&
  {Lattanzio}}]{constantino14}
{Constantino}, T., {Campbell}, S., {Gil-Pons}, P., \& {Lattanzio}, J.~C. 2014,
  ApJ, in press

\bibitem[{{Corradi} \& {Schwarz}(1995)}]{corradi95}
{Corradi}, R.~L.~M. \& {Schwarz}, H.~E. 1995, A\&A, 293, 871

\bibitem[{{Cosner} {et~al.}(1980){Cosner}, {Iben}, \& {Truran}}]{cosner80}
{Cosner}, K., {Iben}, Jr., I., \& {Truran}, J.~W. 1980, ApJL, 238, L91

\bibitem[{{Cottrell} \& {Da Costa}(1981)}]{cottrell81}
{Cottrell}, P.~L. \& {Da Costa}, G.~S. 1981, ApJL, 245, L79

\bibitem[{{Cristallo} {et~al.}(2011){Cristallo}, {Piersanti}, {Straniero},
  {Gallino}, {Dom{\'{\i}}nguez}, {Abia}, {Di Rico}, {Quintini}, \&
  {Bisterzo}}]{cristallo11}
{Cristallo}, S., {Piersanti}, L., {Straniero}, O., {Gallino}, R.,
  {Dom{\'{\i}}nguez}, I., {Abia}, C., {Di Rico}, G., {Quintini}, M., \&
  {Bisterzo}, S. 2011, ApJS, 197, 17

\bibitem[{{Cristallo} {et~al.}(2009){Cristallo}, {Straniero}, {Gallino},
  {Piersanti}, {Dom{\'{\i}}nguez}, \& {Lederer}}]{cristallo09}
{Cristallo}, S., {Straniero}, O., {Gallino}, R., {Piersanti}, L.,
  {Dom{\'{\i}}nguez}, I., \& {Lederer}, M.~T. 2009, ApJ, 696, 797

\bibitem[{{Cruz} {et~al.}(2013){Cruz}, {Serenelli}, \& {Weiss}}]{cruz13}
{Cruz}, M.~A., {Serenelli}, A., \& {Weiss}, A. 2013, A\&A, 559, A4

\bibitem[{{Cyburt} {et~al.}(2008){Cyburt}, {Fields}, \& {Olive}}]{cyburt08}
{Cyburt}, R.~H., {Fields}, B.~D., \& {Olive}, K.~A. 2008, JCAP, 11, 12

\bibitem[{{Da Costa} \& {Marino}(2011)}]{dacosta11}
{Da Costa}, G.~S. \& {Marino}, A.~F. 2011, PASA, 28, 28

\bibitem[{{Da Costa} {et~al.}(2013){Da Costa}, {Norris}, \& {Yong}}]{dacosta13}
{Da Costa}, G.~S., {Norris}, J.~E., \& {Yong}, D. 2013, ApJ, 769, 8

\bibitem[{{D'Antona} {et~al.}(2002){D'Antona}, {Caloi}, {Montalb{\'a}n},
  {Ventura}, \& {Gratton}}]{dantona02}
{D'Antona}, F., {Caloi}, V., {Montalb{\'a}n}, J., {Ventura}, P., \& {Gratton},
  R. 2002, A\&A, 395, 69

\bibitem[{{D'Antona} {et~al.}(2012){D'Antona}, {D'Ercole}, {Carini},
  {Vesperini}, \& {Ventura}}]{dantona12}
{D'Antona}, F., {D'Ercole}, A., {Carini}, R., {Vesperini}, E., \& {Ventura}, P.
  2012, MNRAS, 426, 1710

\bibitem[{{D'Antona} \& {Mazzitelli}(1982)}]{dantona82}
{D'Antona}, F. \& {Mazzitelli}, I. 1982, A\&A, 115, L1

\bibitem[{{De Beck} {et~al.}(2010){De Beck}, {Decin}, {de Koter}, {Justtanont},
  {Verhoelst}, {Kemper}, \& {Menten}}]{debeck10}
{De Beck}, E., {Decin}, L., {de Koter}, A., {Justtanont}, K., {Verhoelst}, T.,
  {Kemper}, F., \& {Menten}, K.~M. 2010, A\&A, 523, A18

\bibitem[{{de Mink} {et~al.}(2009){de Mink}, {Pols}, {Langer}, \&
  {Izzard}}]{demink09}
{de Mink}, S.~E., {Pols}, O.~R., {Langer}, N., \& {Izzard}, R.~G. 2009, A\&A,
  507, L1

\bibitem[{{De Smedt} {et~al.}(2012){De Smedt}, {Van Winckel}, {Karakas},
  {Siess}, {Goriely}, \& {Wood}}]{desmedt12}
{De Smedt}, K., {Van Winckel}, H., {Karakas}, A.~I., {Siess}, L., {Goriely},
  S., \& {Wood}, P.~R. 2012, A\&A, 541, A67

\bibitem[{{Dearborn}(1992)}]{dearborn92}
{Dearborn}, D.~S.~P. 1992, Phys. Rep., 210, 367

\bibitem[{{Dearborn} {et~al.}(2006){Dearborn}, {Lattanzio}, \&
  {Eggleton}}]{dearborn06}
{Dearborn}, D.~S.~P., {Lattanzio}, J.~C., \& {Eggleton}, P.~P. 2006, ApJ, 639,
  405

\bibitem[{{Decressin} {et~al.}(2007){Decressin}, {Charbonnel}, \&
  {Meynet}}]{decressin07}
{Decressin}, T., {Charbonnel}, C., \& {Meynet}, G. 2007, A\&A, 475, 859

\bibitem[{{Demarque} \& {Mengel}(1971)}]{demarque71}
{Demarque}, P. \& {Mengel}, J.~G. 1971, ApJ, 164, 317

\bibitem[{{Denissenkov}(2010)}]{denissenkov10}
{Denissenkov}, P.~A. 2010, ApJ, 723, 563

\bibitem[{{Denissenkov} {et~al.}(2006){Denissenkov}, {Chaboyer}, \&
  {Li}}]{denissenkov06}
{Denissenkov}, P.~A., {Chaboyer}, B., \& {Li}, K. 2006, ApJ, 641, 1087

\bibitem[{{Denissenkov} \& {Herwig}(2004)}]{denher04}
{Denissenkov}, P.~A. \& {Herwig}, F. 2004, ApJ, 612, 1081

\bibitem[{{Denissenkov} {et~al.}(2013{\natexlab{a}}){Denissenkov}, {Herwig},
  {Truran}, \& {Paxton}}]{denissenkov13a}
{Denissenkov}, P.~A., {Herwig}, F., {Truran}, J.~W., \& {Paxton}, B.
  2013{\natexlab{a}}, ApJ, 772, 37

\bibitem[{{Denissenkov} \& {Merryfield}(2011)}]{denissenkov11}
{Denissenkov}, P.~A. \& {Merryfield}, W.~J. 2011, ApJL, 727, L8

\bibitem[{{Denissenkov} {et~al.}(2009){Denissenkov}, {Pinsonneault}, \&
  {MacGregor}}]{denissenkov09}
{Denissenkov}, P.~A., {Pinsonneault}, M., \& {MacGregor}, K.~B. 2009, ApJ, 696,
  1823

\bibitem[{{Denissenkov} \& {Tout}(2000)}]{denissenkov00}
{Denissenkov}, P.~A. \& {Tout}, C.~A. 2000, MNRAS, 316, 395

\bibitem[{{Denissenkov} \& {Tout}(2003)}]{denissenkov03a}
---. 2003, MNRAS, 340, 722

\bibitem[{{Denissenkov} {et~al.}(2013{\natexlab{b}}){Denissenkov}, {Truran},
  {Pignatari}, {Trappitsch}, {Ritter}, {Herwig}, {Battino}, \&
  {Setoodehnia}}]{denissenkov13b}
{Denissenkov}, P.~A., {Truran}, J.~W., {Pignatari}, M., {Trappitsch}, R.,
  {Ritter}, C., {Herwig}, F., {Battino}, U., \& {Setoodehnia}, K.
  2013{\natexlab{b}}, {MNRAS}, submitted

\bibitem[{{D'Ercole} {et~al.}(2012){D'Ercole}, {D'Antona}, {Carini},
  {Vesperini}, \& {Ventura}}]{dercole12}
{D'Ercole}, A., {D'Antona}, F., {Carini}, R., {Vesperini}, E., \& {Ventura}, P.
  2012, MNRAS, 423, 1521

\bibitem[{{D'Ercole} {et~al.}(2010){D'Ercole}, {D'Antona}, {Ventura},
  {Vesperini}, \& {McMillan}}]{dercole10}
{D'Ercole}, A., {D'Antona}, F., {Ventura}, P., {Vesperini}, E., \& {McMillan},
  S.~L.~W. 2010, MNRAS, 407, 854

\bibitem[{{D'Ercole} {et~al.}(2011){D'Ercole}, {D'Antona}, \&
  {Vesperini}}]{dercole11}
{D'Ercole}, A., {D'Antona}, F., \& {Vesperini}, E. 2011, MNRAS, 415, 1304

\bibitem[{{Despain}(1981)}]{despain81}
{Despain}, K.~H. 1981, ApJ, 251, 639

\bibitem[{{Deupree}(1984)}]{deupree84}
{Deupree}, R.~G. 1984, ApJ, 287, 268

\bibitem[{{Deupree}(1986)}]{deupree86}
---. 1986, ApJ, 303, 649

\bibitem[{{Deupree}(1996)}]{deupree96}
---. 1996, ApJ, 471, 377

\bibitem[{{Deupree} \& {Wallace}(1987)}]{deupree87}
{Deupree}, R.~G. \& {Wallace}, R.~K. 1987, ApJ, 317, 724

\bibitem[{{Di Cecco} {et~al.}(2010){Di Cecco}, {Bono}, {Stetson},
  {Pietrinferni}, {Becucci}, {Cassisi}, {Degl'Innocenti}, {Iannicola}, {Prada
  Moroni}, {Buonanno}, {Calamida}, {Caputo}, {Castellani}, {Corsi}, {Ferraro},
  {Dall'Ora}, {Monelli}, {Nonino}, {Piersimoni}, {Pulone}, {Romaniello},
  {Salaris}, {Walker}, \& {Zoccali}}]{dicecco10}
{Di Cecco}, A., {Bono}, G., {Stetson}, P.~B., {Pietrinferni}, A., {Becucci},
  R., {Cassisi}, S., {Degl'Innocenti}, S., {Iannicola}, G., {Prada Moroni},
  P.~G., {Buonanno}, R., {Calamida}, A., {Caputo}, F., {Castellani}, M.,
  {Corsi}, C.~E., {Ferraro}, I., {Dall'Ora}, M., {Monelli}, M., {Nonino}, M.,
  {Piersimoni}, A.~M., {Pulone}, L., {Romaniello}, M., {Salaris}, M., {Walker},
  A.~R., \& {Zoccali}, M. 2010, ApJ, 712, 527

\bibitem[{{Dinh-V-Trung} \& {Lim}(2008)}]{dinh08}
{Dinh-V-Trung} \& {Lim}, J. 2008, ApJ, 678, 303

\bibitem[{{Doherty} {et~al.}(2014{\natexlab{a}}){Doherty}, {Gil-Pons}, {Lau},
  {Lattanzio}, \& {Siess}}]{doherty14a}
{Doherty}, C.~L., {Gil-Pons}, P., {Lau}, H.~H.~B., {Lattanzio}, J.~C., \&
  {Siess}, L. 2014{\natexlab{a}}, MNRAS, 437, 195

\bibitem[{{Doherty} {et~al.}(2014{\natexlab{b}}){Doherty}, {Gil-Pons}, {Lau},
  {Lattanzio}, {Siess}, \& {Campbell}}]{doherty14b}
{Doherty}, C.~L., {Gil-Pons}, P., {Lau}, H.~H.~B., {Lattanzio}, J.~C., {Siess},
  L., \& {Campbell}, S.~W. 2014{\natexlab{b}}, MNRAS, in press

\bibitem[{{Doherty} {et~al.}(2010){Doherty}, {Siess}, {Lattanzio}, \&
  {Gil-Pons}}]{doherty10}
{Doherty}, C.~L., {Siess}, L., {Lattanzio}, J.~C., \& {Gil-Pons}, P. 2010,
  MNRAS, 401, 1453

\bibitem[{{Dominguez} {et~al.}(1999){Dominguez}, {Chieffi}, {Limongi}, \&
  {Straniero}}]{dominguez99}
{Dominguez}, I., {Chieffi}, A., {Limongi}, M., \& {Straniero}, O. 1999, ApJ,
  524, 226

\bibitem[{{Dopita} {et~al.}(1997){Dopita}, {Vassiliadis}, {Wood},
  {Meatheringham}, {Harrington}, {Bohlin}, {Ford}, {Stecher}, \&
  {Maran}}]{dopita97}
{Dopita}, M.~A., {Vassiliadis}, E., {Wood}, P.~R., {Meatheringham}, S.~J.,
  {Harrington}, J.~P., {Bohlin}, R.~C., {Ford}, H.~C., {Stecher}, T.~P., \&
  {Maran}, S.~P. 1997, ApJ, 474, 188

\bibitem[{{D'Orazi} {et~al.}(2013){D'Orazi}, {Campbell}, {Lugaro}, {Lattanzio},
  {Pignatari}, \& {Carretta}}]{dorazi13b}
{D'Orazi}, V., {Campbell}, S.~W., {Lugaro}, M., {Lattanzio}, J.~C.,
  {Pignatari}, M., \& {Carretta}, E. 2013, MNRAS, 433, 366

\bibitem[{{Dorman} \& {Rood}(1993)}]{dorman93}
{Dorman}, B. \& {Rood}, R.~T. 1993, ApJ, 409, 387

\bibitem[{{Drake} {et~al.}(2011){Drake}, {Ball}, {Eldridge}, {Ness}, \&
  {Stancliffe}}]{drake11}
{Drake}, J.~J., {Ball}, B., {Eldridge}, J.~J., {Ness}, J.-U., \& {Stancliffe},
  R.~J. 2011, AJ, 142, 144

\bibitem[{{Dray} {et~al.}(2003){Dray}, {Tout}, {Karakas}, \&
  {Lattanzio}}]{dray03}
{Dray}, L.~M., {Tout}, C.~A., {Karakas}, A.~I., \& {Lattanzio}, J.~C. 2003,
  MNRAS, 338, 973

\bibitem[{{Duquennoy} \& {Mayor}(1991)}]{duq91}
{Duquennoy}, A. \& {Mayor}, M. 1991, A\&A, 248, 485

\bibitem[{{Eggenberger} {et~al.}(2012){Eggenberger}, {Haemmerl{\'e}}, {Meynet},
  \& {Maeder}}]{eggenberger12}
{Eggenberger}, P., {Haemmerl{\'e}}, L., {Meynet}, G., \& {Maeder}, A. 2012,
  A\&A, 539, A70

\bibitem[{{Eggleton} {et~al.}(2006){Eggleton}, {Dearborn}, \&
  {Lattanzio}}]{eggleton06}
{Eggleton}, P.~P., {Dearborn}, D.~S.~P., \& {Lattanzio}, J.~C. 2006, Science,
  314, 1580

\bibitem[{{Eggleton} {et~al.}(2008){Eggleton}, {Dearborn}, \&
  {Lattanzio}}]{eggleton08}
---. 2008, ApJ, 677, 581

\bibitem[{{Ekstr{\"o}m} {et~al.}(2012){Ekstr{\"o}m}, {Georgy}, {Eggenberger},
  {Meynet}, {Mowlavi}, {Wyttenbach}, {Granada}, {Decressin}, {Hirschi},
  {Frischknecht}, {Charbonnel}, \& {Maeder}}]{ekstrom12}
{Ekstr{\"o}m}, S., {Georgy}, C., {Eggenberger}, P., {Meynet}, G., {Mowlavi},
  N., {Wyttenbach}, A., {Granada}, A., {Decressin}, T., {Hirschi}, R.,
  {Frischknecht}, U., {Charbonnel}, C., \& {Maeder}, A. 2012, A\&A, 537, A146

\bibitem[{{El Eid}(1994)}]{eleid94}
{El Eid}, M.~F. 1994, A\&A, 285, 915

\bibitem[{{El Eid} \& {Champagne}(1995)}]{eleid95}
{El Eid}, M.~F. \& {Champagne}, A.~E. 1995, ApJ, 451, 298

\bibitem[{{Eldridge} \& {Tout}(2004)}]{eldridge04}
{Eldridge}, J.~J. \& {Tout}, C.~A. 2004, MNRAS, 353, 87

\bibitem[{{Fagotto} {et~al.}(1994){Fagotto}, {Bressan}, {Bertelli}, \&
  {Chiosi}}]{fagotto94}
{Fagotto}, F., {Bressan}, A., {Bertelli}, G., \& {Chiosi}, C. 1994, A\&AS, 104,
  365

\bibitem[{{Federman} {et~al.}(2005){Federman}, {Sheffer}, {Lambert}, \&
  {Smith}}]{federman05}
{Federman}, S.~R., {Sheffer}, Y., {Lambert}, D.~L., \& {Smith}, V.~V. 2005,
  ApJ, 619, 884

\bibitem[{{Fenner} {et~al.}(2004){Fenner}, {Campbell}, {Karakas}, {Lattanzio},
  \& {Gibson}}]{fenner04}
{Fenner}, Y., {Campbell}, S., {Karakas}, A.~I., {Lattanzio}, J.~C., \&
  {Gibson}, B.~K. 2004, MNRAS, 353, 789

\bibitem[{{Fenner} {et~al.}(2006){Fenner}, {Gibson}, {Gallino}, \&
  {Lugaro}}]{fenner06}
{Fenner}, Y., {Gibson}, B.~K., {Gallino}, R., \& {Lugaro}, M. 2006, ApJ, 646,
  184

\bibitem[{{Fenner} {et~al.}(2003){Fenner}, {Gibson}, {Lee}, {Karakas},
  {Lattanzio}, {Chieffi}, {Limongi}, \& {Yong}}]{fenner03}
{Fenner}, Y., {Gibson}, B.~K., {Lee}, H.-c., {Karakas}, A.~I., {Lattanzio},
  J.~C., {Chieffi}, A., {Limongi}, M., \& {Yong}, D. 2003, PASA, 20, 340

\bibitem[{{Ferguson} {et~al.}(2005){Ferguson}, {Alexander}, {Allard}, {Barman},
  {Bodnarik}, {Hauschildt}, {Heffner-Wong}, \& {Tamanai}}]{ferguson05}
{Ferguson}, J.~W., {Alexander}, D.~R., {Allard}, F., {Barman}, T., {Bodnarik},
  J.~G., {Hauschildt}, P.~H., {Heffner-Wong}, A., \& {Tamanai}, A. 2005, ApJ,
  623, 585

\bibitem[{{Ferraro} {et~al.}(2006){Ferraro}, {Mucciarelli}, {Carretta}, \&
  {Origlia}}]{ferraro06}
{Ferraro}, F.~R., {Mucciarelli}, A., {Carretta}, E., \& {Origlia}, L. 2006,
  ApJL, 645, L33

\bibitem[{{Fishlock} {et~al.}(2014){Fishlock}, {Karakas}, \&
  {Stancliffe}}]{fishlock14}
{Fishlock}, C.~K., {Karakas}, A.~I., \& {Stancliffe}, R.~J. 2014, MNRAS

\bibitem[{{Forestini} \& {Charbonnel}(1997)}]{forestini97}
{Forestini}, M. \& {Charbonnel}, C. 1997, A\&AS, 123, 241

\bibitem[{{Forestini} {et~al.}(1992){Forestini}, {Goriely}, {Jorissen}, \&
  {Arnould}}]{forestini92}
{Forestini}, M., {Goriely}, S., {Jorissen}, A., \& {Arnould}, M. 1992, A\&A,
  261, 157

\bibitem[{{Frebel}(2010)}]{frebel10}
{Frebel}, A. 2010, Astronomische Nachrichten, 331, 474

\bibitem[{{Frebel} \& {Norris}(2013)}]{frebel13}
{Frebel}, A. \& {Norris}, J.~E. 2013, {Metal-Poor Stars and the Chemical
  Enrichment of the Universe} (Springer Science$+$Business Media Dordrecht), 55

\bibitem[{{Freeman} \& {Bland-Hawthorn}(2002)}]{freeman02}
{Freeman}, K. \& {Bland-Hawthorn}, J. 2002, ARA\&A, 40, 487

\bibitem[{{Frischknecht} {et~al.}(2012){Frischknecht}, {Hirschi}, \&
  {Thielemann}}]{frischknecht12}
{Frischknecht}, U., {Hirschi}, R., \& {Thielemann}, F.-K. 2012, A\&A, 538, L2

\bibitem[{{Frogel} {et~al.}(1990){Frogel}, {Mould}, \& {Blanco}}]{frogel90}
{Frogel}, J.~A., {Mould}, J., \& {Blanco}, V.~M. 1990, ApJ, 352, 96

\bibitem[{{Frost} \& {Lattanzio}(1996{\natexlab{a}})}]{frost96proc}
{Frost}, C. \& {Lattanzio}, J. 1996{\natexlab{a}}, ArXiv Astrophysics e-prints,
  astro-ph/9601017

\bibitem[{{Frost} {et~al.}(1998{\natexlab{a}}){Frost}, {Cannon}, {Lattanzio},
  {Wood}, \& {Forestini}}]{frost98a}
{Frost}, C.~A., {Cannon}, R.~C., {Lattanzio}, J.~C., {Wood}, P.~R., \&
  {Forestini}, M. 1998{\natexlab{a}}, A\&A, 332, L17

\bibitem[{{Frost} \& {Lattanzio}(1996{\natexlab{b}})}]{frost96}
{Frost}, C.~A. \& {Lattanzio}, J.~C. 1996{\natexlab{b}}, ApJ, 473, 383

\bibitem[{{Frost} {et~al.}(1998{\natexlab{b}}){Frost}, {Lattanzio}, \&
  {Wood}}]{frost98b}
{Frost}, C.~A., {Lattanzio}, J.~C., \& {Wood}, P.~R. 1998{\natexlab{b}}, ApJ,
  500, 355

\bibitem[{{Fryer} {et~al.}(2006){Fryer}, {Herwig}, {Hungerford}, \&
  {Timmes}}]{fryer06}
{Fryer}, C.~L., {Herwig}, F., {Hungerford}, A., \& {Timmes}, F.~X. 2006, ApJL,
  646, L131

\bibitem[{{Fujimoto} {et~al.}(1990){Fujimoto}, {Iben}, \&
  {Hollowell}}]{fujimoto90}
{Fujimoto}, M.~Y., {Iben}, Jr., I., \& {Hollowell}, D. 1990, ApJ, 349, 580

\bibitem[{{Fujimoto} {et~al.}(2000){Fujimoto}, {Ikeda}, \& {Iben}}]{fujimoto00}
{Fujimoto}, M.~Y., {Ikeda}, Y., \& {Iben}, Jr., I. 2000, ApJL, 529, L25

\bibitem[{{Fujiya} {et~al.}(2013){Fujiya}, {Hoppe}, {Zinner}, {Pignatari}, \&
  {Herwig}}]{fujiya13}
{Fujiya}, W., {Hoppe}, P., {Zinner}, E., {Pignatari}, M., \& {Herwig}, F. 2013,
  ApJL, 776, L29

\bibitem[{{Fusi Pecci} {et~al.}(1990){Fusi Pecci}, {Ferraro}, {Crocker},
  {Rood}, \& {Buonanno}}]{fusipecci90}
{Fusi Pecci}, F., {Ferraro}, F.~R., {Crocker}, D.~A., {Rood}, R.~T., \&
  {Buonanno}, R. 1990, A\&A, 238, 95

\bibitem[{{Gallet} \& {Bouvier}(2013)}]{gallet13}
{Gallet}, F. \& {Bouvier}, J. 2013, A\&A, 556, A36

\bibitem[{{Gallino} {et~al.}(1998){Gallino}, {Arlandini}, {Busso}, {Lugaro},
  {Travaglio}, {Straniero}, {Chieffi}, \& {Limongi}}]{gallino98}
{Gallino}, R., {Arlandini}, C., {Busso}, M., {Lugaro}, M., {Travaglio}, C.,
  {Straniero}, O., {Chieffi}, A., \& {Limongi}, M. 1998, ApJ, 497, 388

\bibitem[{{G{\"a}nsicke} {et~al.}(2010){G{\"a}nsicke}, {Koester}, {Girven},
  {Marsh}, \& {Steeghs}}]{gansicke10}
{G{\"a}nsicke}, B.~T., {Koester}, D., {Girven}, J., {Marsh}, T.~R., \&
  {Steeghs}, D. 2010, Science, 327, 188

\bibitem[{{Garcia-Berro} \& {Iben}(1994)}]{garcia-berro94}
{Garcia-Berro}, E. \& {Iben}, I. 1994, ApJ, 434, 306

\bibitem[{{Garcia-Berro} {et~al.}(1997){Garcia-Berro}, {Ritossa}, \&
  {Iben}}]{garcia-berro97}
{Garcia-Berro}, E., {Ritossa}, C., \& {Iben}, Jr., I. 1997, ApJ, 485, 765

\bibitem[{{Garc{\'{\i}}a-Hern{\'a}ndez}
  {et~al.}(2006){Garc{\'{\i}}a-Hern{\'a}ndez}, {Garc{\'{\i}}a-Lario}, {Plez},
  {D'Antona}, {Manchado}, \& {Trigo-Rodr{\'{\i}}guez}}]{garcia06}
{Garc{\'{\i}}a-Hern{\'a}ndez}, D.~A., {Garc{\'{\i}}a-Lario}, P., {Plez}, B.,
  {D'Antona}, F., {Manchado}, A., \& {Trigo-Rodr{\'{\i}}guez}, J.~M. 2006,
  Science, 314, 1751

\bibitem[{{Garc{\'{\i}}a-Hern{\'a}ndez}
  {et~al.}(2007){Garc{\'{\i}}a-Hern{\'a}ndez}, {Garc{\'{\i}}a-Lario}, {Plez},
  {Manchado}, {D'Antona}, {Lub}, \& {Habing}}]{garcia07a}
{Garc{\'{\i}}a-Hern{\'a}ndez}, D.~A., {Garc{\'{\i}}a-Lario}, P., {Plez}, B.,
  {Manchado}, A., {D'Antona}, F., {Lub}, J., \& {Habing}, H. 2007, A\&A, 462,
  711

\bibitem[{{Garc{\'{\i}}a-Hern{\'a}ndez}
  {et~al.}(2009){Garc{\'{\i}}a-Hern{\'a}ndez}, {Manchado}, {Lambert}, {Plez},
  {Garc{\'{\i}}a-Lario}, {D'Antona}, {Lugaro}, {Karakas}, \& {van
  Raai}}]{garcia09}
{Garc{\'{\i}}a-Hern{\'a}ndez}, D.~A., {Manchado}, A., {Lambert}, D.~L., {Plez},
  B., {Garc{\'{\i}}a-Lario}, P., {D'Antona}, F., {Lugaro}, M., {Karakas},
  A.~I., \& {van Raai}, M.~A. 2009, ApJL, 705, L31

\bibitem[{{Garc{\'{\i}}a-Hern{\'a}ndez}
  {et~al.}(2013){Garc{\'{\i}}a-Hern{\'a}ndez}, {Zamora}, {Yag{\"u}e},
  {Uttenthaler}, {Karakas}, {Lugaro}, {Ventura}, \& {Lambert}}]{garcia13}
{Garc{\'{\i}}a-Hern{\'a}ndez}, D.~A., {Zamora}, O., {Yag{\"u}e}, A.,
  {Uttenthaler}, S., {Karakas}, A.~I., {Lugaro}, M., {Ventura}, P., \&
  {Lambert}, D.~L. 2013, A\&A, 555, L3

\bibitem[{{Gavil{\'a}n} {et~al.}(2005){Gavil{\'a}n}, {Buell}, \&
  {Moll{\'a}}}]{gavilan05}
{Gavil{\'a}n}, M., {Buell}, J.~F., \& {Moll{\'a}}, M. 2005, A\&A, 432, 861

\bibitem[{{Gay} \& {Lambert}(2000)}]{gay00}
{Gay}, P.~L. \& {Lambert}, D.~L. 2000, ApJ, 533, 260

\bibitem[{{Gibson}(1997)}]{gibson97a}
{Gibson}, B.~K. 1997, MNRAS, 290, 471

\bibitem[{{Gibson}(2007)}]{gibson07}
{Gibson}, B.~K. 2007, in IAU Symposium, Vol. 241, IAU Symposium, ed.
  A.~{Vazdekis} \& R.~{Peletier}, 161--164

\bibitem[{{Gibson} {et~al.}(2005){Gibson}, {Fenner}, \& {Kiessling}}]{gibson05}
{Gibson}, B.~K., {Fenner}, Y., \& {Kiessling}, A. 2005, Nuclear Physics A, 758,
  259

\bibitem[{{Gil-Pons} {et~al.}(2013){Gil-Pons}, {Doherty}, {Lau}, {Campbell},
  {Suda}, {Guilani}, {Guti{\'e}rrez}, \& {Lattanzio}}]{gilpons13}
{Gil-Pons}, P., {Doherty}, C.~L., {Lau}, H., {Campbell}, S.~W., {Suda}, T.,
  {Guilani}, S., {Guti{\'e}rrez}, J., \& {Lattanzio}, J.~C. 2013, A\&A, 557,
  A106

\bibitem[{{Gil-Pons} \& {Garc{\'{\i}}a-Berro}(2002)}]{gilpons02}
{Gil-Pons}, P. \& {Garc{\'{\i}}a-Berro}, E. 2002, A\&A, 396, 589

\bibitem[{{Gil-Pons} {et~al.}(2007){Gil-Pons}, {Guti{\'e}rrez}, \&
  {Garc{\'{\i}}a-Berro}}]{gilpons07}
{Gil-Pons}, P., {Guti{\'e}rrez}, J., \& {Garc{\'{\i}}a-Berro}, E. 2007, A\&A,
  464, 667

\bibitem[{{Gil-Pons} {et~al.}(2005){Gil-Pons}, {Suda}, {Fujimoto}, \&
  {Garc{\'{\i}}a-Berro}}]{gilpons05}
{Gil-Pons}, P., {Suda}, T., {Fujimoto}, M.~Y., \& {Garc{\'{\i}}a-Berro}, E.
  2005, A\&A, 433, 1037

\bibitem[{{Gilroy}(1989)}]{gilroy89}
{Gilroy}, K.~K. 1989, ApJ, 347, 835

\bibitem[{{Gilroy} \& {Brown}(1991)}]{gilroy91}
{Gilroy}, K.~K. \& {Brown}, J.~A. 1991, ApJ, 371, 578

\bibitem[{{Gingold}(1976)}]{gingold76}
{Gingold}, R.~A. 1976, ApJ, 204, 116

\bibitem[{{Girardi} {et~al.}(1995){Girardi}, {Chiosi}, {Bertelli}, \&
  {Bressan}}]{girardi95}
{Girardi}, L., {Chiosi}, C., {Bertelli}, G., \& {Bressan}, A. 1995, A\&A, 298,
  87

\bibitem[{{Girardi} {et~al.}(2009){Girardi}, {Rubele}, \& {Kerber}}]{girardi09}
{Girardi}, L., {Rubele}, S., \& {Kerber}, L. 2009, MNRAS, 394, L74

\bibitem[{{Gloeckler} \& {Geiss}(1996)}]{gloeckler96}
{Gloeckler}, G. \& {Geiss}, J. 1996, Nature, 381, 210

\bibitem[{{Goriely}(1999)}]{goriely99}
{Goriely}, S. 1999, A\&A, 342, 881

\bibitem[{{Goriely} \& {Mowlavi}(2000)}]{goriely00}
{Goriely}, S. \& {Mowlavi}, N. 2000, A\&A, 362, 599

\bibitem[{{Goriely} \& {Siess}(2001)}]{goriely01}
{Goriely}, S. \& {Siess}, L. 2001, A\&A, 378, L25

\bibitem[{{Goriely} \& {Siess}(2004)}]{goriely04}
---. 2004, A\&A, 421, L25

\bibitem[{{Gratton} {et~al.}(2004){Gratton}, {Sneden}, \&
  {Carretta}}]{gratton04}
{Gratton}, R., {Sneden}, C., \& {Carretta}, E. 2004, ARA\&A, 42, 385

\bibitem[{{Gratton} {et~al.}(2012){Gratton}, {Carretta}, \&
  {Bragaglia}}]{gratton12}
{Gratton}, R.~G., {Carretta}, E., \& {Bragaglia}, A. 2012, ARA\&A, 20, 50

\bibitem[{{Gratton} {et~al.}(2000){Gratton}, {Sneden}, {Carretta}, \&
  {Bragaglia}}]{gratton00}
{Gratton}, R.~G., {Sneden}, C., {Carretta}, E., \& {Bragaglia}, A. 2000, A\&A,
  354, 169

\bibitem[{{Groenewegen}(2004)}]{groen04}
{Groenewegen}, M.~A.~T. 2004, A\&A, 425, 595

\bibitem[{{Groenewegen}(2012)}]{groen12}
---. 2012, A\&A, 540, A32

\bibitem[{{Groenewegen} \& {de Jong}(1993)}]{groen93}
{Groenewegen}, M.~A.~T. \& {de Jong}, T. 1993, A\&A, 267, 410

\bibitem[{{Groenewegen} {et~al.}(2009{\natexlab{a}}){Groenewegen}, {Sloan},
  {Soszy{\'n}ski}, \& {Petersen}}]{groen07}
{Groenewegen}, M.~A.~T., {Sloan}, G.~C., {Soszy{\'n}ski}, I., \& {Petersen},
  E.~A. 2009{\natexlab{a}}, A\&A, 506, 1277

\bibitem[{{Groenewegen} {et~al.}(2009{\natexlab{b}}){Groenewegen}, {Sloan},
  {Soszy{\'n}ski}, \& {Petersen}}]{groen09}
---. 2009{\natexlab{b}}, A\&A, 506, 1277

\bibitem[{{Grossman} {et~al.}(1993){Grossman}, {Narayan}, \&
  {Arnett}}]{grossman93}
{Grossman}, S.~A., {Narayan}, R., \& {Arnett}, D. 1993, ApJ, 407, 284

\bibitem[{{Grossman} \& {Taam}(1996)}]{grossman96}
{Grossman}, S.~A. \& {Taam}, R.~E. 1996, MNRAS, 283, 1165

\bibitem[{{Guandalini}(2010)}]{guandalini10}
{Guandalini}, R. 2010, A\&A, 513, A4

\bibitem[{{Guelin} {et~al.}(1995){Guelin}, {Forestini}, {Valiron}, {Ziurys},
  {Anderson}, {Cernicharo}, \& {Kahane}}]{guelin95}
{Guelin}, M., {Forestini}, M., {Valiron}, P., {Ziurys}, L.~M., {Anderson},
  M.~A., {Cernicharo}, J., \& {Kahane}, C. 1995, A\&A, 297, 183

\bibitem[{{Guzman-Ramirez} {et~al.}(2013){Guzman-Ramirez}, {Pineda},
  {Zijlstra}, {Stancliffe}, \& {Karakas}}]{guzman13}
{Guzman-Ramirez}, L., {Pineda}, J.~E., {Zijlstra}, A.~A., {Stancliffe}, R., \&
  {Karakas}, A. 2013, MNRAS, 432, 793

\bibitem[{{Habing}(1996)}]{habing96}
{Habing}, H.~J. 1996, A\&AR, 7, 97

\bibitem[{{Halabi} {et~al.}(2012){Halabi}, {El Eid}, \& {Champagne}}]{halabi12}
{Halabi}, G.~M., {El Eid}, M.~F., \& {Champagne}, A. 2012, ApJ, 761, 10

\bibitem[{{Hale} {et~al.}(2004){Hale}, {Champagne}, {Iliadis}, {Hansper},
  {Powell}, \& {Blackmon}}]{hale04}
{Hale}, S.~E., {Champagne}, A.~E., {Iliadis}, C., {Hansper}, V.~Y., {Powell},
  D.~C., \& {Blackmon}, J.~C. 2004, Phys. Rev. C, 70, 045802

\bibitem[{{Hamdani} {et~al.}(2000){Hamdani}, {North}, {Mowlavi}, {Raboud}, \&
  {Mermilliod}}]{hamdani00}
{Hamdani}, S., {North}, P., {Mowlavi}, N., {Raboud}, D., \& {Mermilliod}, J.-C.
  2000, A\&A, 360

\bibitem[{{Han} {et~al.}(1995){Han}, {Eggleton}, {Podsiadlowski}, \&
  {Tout}}]{han95}
{Han}, Z., {Eggleton}, P.~P., {Podsiadlowski}, P., \& {Tout}, C.~A. 1995,
  MNRAS, 277

\bibitem[{{Hansen} {et~al.}(2013){Hansen}, {Bergemann}, {Cescutti}, {Fran{\c
  c}ois}, {Arcones}, {Karakas}, {Lind}, \& {Chiappini}}]{hansen13}
{Hansen}, C.~J., {Bergemann}, M., {Cescutti}, G., {Fran{\c c}ois}, P.,
  {Arcones}, A., {Karakas}, A.~I., {Lind}, K., \& {Chiappini}, C. 2013, A\&A,
  551, A57

\bibitem[{{Harris} {et~al.}(1983){Harris}, {Fowler}, {Caughlan}, \&
  {Zimmerman}}]{harris83}
{Harris}, M.~J., {Fowler}, W.~A., {Caughlan}, G.~R., \& {Zimmerman}, B.~A.
  1983, ARA\&A, 21, 165

\bibitem[{{Harris} \& {Lambert}(1984)}]{harris84}
{Harris}, M.~J. \& {Lambert}, D.~L. 1984, ApJ, 285, 674

\bibitem[{{Harris} {et~al.}(1987){Harris}, {Lambert}, {Hinkle}, {Gustafsson},
  \& {Eriksson}}]{harris87}
{Harris}, M.~J., {Lambert}, D.~L., {Hinkle}, K.~H., {Gustafsson}, B., \&
  {Eriksson}, K. 1987, ApJ, 316, 294

\bibitem[{{Harris} {et~al.}(1985{\natexlab{a}}){Harris}, {Lambert}, \&
  {Smith}}]{harris85a}
{Harris}, M.~J., {Lambert}, D.~L., \& {Smith}, V.~V. 1985{\natexlab{a}}, ApJ,
  292, 620

\bibitem[{{Harris} {et~al.}(1985{\natexlab{b}}){Harris}, {Lambert}, \&
  {Smith}}]{harris85b}
---. 1985{\natexlab{b}}, ApJ, 299, 375

\bibitem[{{Hedrosa} {et~al.}(2013){Hedrosa}, {Abia}, {Busso}, {Cristallo},
  {Dom{\'{\i}}nguez}, {Palmerini}, {Plez}, \& {Straniero}}]{hedrosa13}
{Hedrosa}, R.~P., {Abia}, C., {Busso}, M., {Cristallo}, S., {Dom{\'{\i}}nguez},
  I., {Palmerini}, S., {Plez}, B., \& {Straniero}, O. 2013, ApJL, 768, L11

\bibitem[{{Heger} {et~al.}(2000){Heger}, {Langer}, \& {Woosley}}]{heger00}
{Heger}, A., {Langer}, N., \& {Woosley}, S.~E. 2000, ApJ, 528, 368

\bibitem[{{Heil} {et~al.}(2008){Heil}, {K{\"a}ppeler}, {Uberseder}, {Gallino},
  \& {Pignatari}}]{heil08}
{Heil}, M., {K{\"a}ppeler}, F., {Uberseder}, E., {Gallino}, R., \& {Pignatari},
  M. 2008, Phys. Rev. C, 77, 015808

\bibitem[{{Henry}(1989)}]{henry89}
{Henry}, R.~B.~C. 1989, MNRAS, 241, 453

\bibitem[{{Henry} {et~al.}(2012){Henry}, {Speck}, {Karakas}, {Ferland}, \&
  {Maguire}}]{henry12}
{Henry}, R.~B.~C., {Speck}, A., {Karakas}, A.~I., {Ferland}, G.~J., \&
  {Maguire}, M. 2012, ApJ, 749, 61

\bibitem[{{Herwig}(2000)}]{herwig00}
{Herwig}, F. 2000, A\&A, 360, 952

\bibitem[{{Herwig}(2001)}]{herwig01}
---. 2001, Ap\&SS, 275, 15

\bibitem[{{Herwig}(2004{\natexlab{a}})}]{herwig04a}
---. 2004{\natexlab{a}}, ApJ, 605, 425

\bibitem[{{Herwig}(2004{\natexlab{b}})}]{herwig04b}
---. 2004{\natexlab{b}}, ApJS, 155, 651

\bibitem[{{Herwig}(2005)}]{herwig05}
---. 2005, ARA\&A, 43, 435

\bibitem[{{Herwig} {et~al.}(1999){Herwig}, {Bl{\"o}cker}, {Langer}, \&
  {Driebe}}]{herwig99}
{Herwig}, F., {Bl{\"o}cker}, T., {Langer}, N., \& {Driebe}, T. 1999, A\&A, 349,
  L5

\bibitem[{{Herwig} {et~al.}(1997){Herwig}, {Bloecker}, {Schoenberner}, \& {El
  Eid}}]{herwig97}
{Herwig}, F., {Bloecker}, T., {Schoenberner}, D., \& {El Eid}, M. 1997, A\&A,
  324, L81

\bibitem[{{Herwig} {et~al.}(2006){Herwig}, {Freytag}, {Hueckstaedt}, \&
  {Timmes}}]{herwig06}
{Herwig}, F., {Freytag}, B., {Hueckstaedt}, R.~M., \& {Timmes}, F.~X. 2006,
  ApJ, 642, 1057

\bibitem[{{Herwig} {et~al.}(2003){Herwig}, {Langer}, \& {Lugaro}}]{herwig03}
{Herwig}, F., {Langer}, N., \& {Lugaro}, M. 2003, ApJ, 593, 1056

\bibitem[{{Herwig} {et~al.}(2011){Herwig}, {Pignatari}, {Woodward}, {Porter},
  {Rockefeller}, {Fryer}, {Bennett}, \& {Hirschi}}]{herwig11}
{Herwig}, F., {Pignatari}, M., {Woodward}, P.~R., {Porter}, D.~H.,
  {Rockefeller}, G., {Fryer}, C.~L., {Bennett}, M., \& {Hirschi}, R. 2011, ApJ,
  727, 89

\bibitem[{{Herwig} {et~al.}(2012){Herwig}, {VandenBerg}, {Navarro}, {Ferguson},
  \& {Paxton}}]{herwig12}
{Herwig}, F., {VandenBerg}, D.~A., {Navarro}, J.~F., {Ferguson}, J., \&
  {Paxton}, B. 2012, ApJ, 757, 132

\bibitem[{{Herwig} {et~al.}(2013){Herwig}, {Woodward}, {Lin}, {Knox}, \&
  {Fryer}}]{herwig13}
{Herwig}, F., {Woodward}, P.~R., {Lin}, P.-H., {Knox}, M., \& {Fryer}, C. 2013,
  {ApJL}, submitted

\bibitem[{{Hillebrandt} {et~al.}(1984){Hillebrandt}, {Nomoto}, \&
  {Wolff}}]{hillebrandt84}
{Hillebrandt}, W., {Nomoto}, K., \& {Wolff}, R.~G. 1984, A\&A, 133, 175

\bibitem[{{Hirschi}(2007)}]{hirschi07}
{Hirschi}, R. 2007, A\&A, 461, 571

\bibitem[{{Hollowell} \& {Iben}(1988)}]{hollowell88}
{Hollowell}, D. \& {Iben}, Jr., I. 1988, ApJL, 333, L25

\bibitem[{{Hollowell} {et~al.}(1990){Hollowell}, {Iben}, \&
  {Fujimoto}}]{hollowell90}
{Hollowell}, D., {Iben}, Jr., I., \& {Fujimoto}, M.~Y. 1990, ApJ, 351, 245

\bibitem[{{Iben}(1975)}]{iben75}
{Iben}, Jr., I. 1975, ApJ, 196, 549

\bibitem[{{Iben}(1981)}]{iben81}
{Iben}, Jr., I. 1981, in Astrophysics and Space Science Library, Vol.~88,
  Physical Processes in Red Giants, ed. I.~{Iben}, Jr. \& A.~{Renzini}, 3--24

\bibitem[{{Iben}(1991)}]{iben91}
{Iben}, Jr., I. 1991, in IAU Symp. 145: Evolution of Stars: the Photospheric
  Abundance Connection, ed. G.~{Michaud} \& A.~V. {Tutukov}, 257

\bibitem[{{Iben} {et~al.}(1983){Iben}, {Kaler}, {Truran}, \&
  {Renzini}}]{iben83}
{Iben}, Jr., I., {Kaler}, J.~B., {Truran}, J.~W., \& {Renzini}, A. 1983, ApJ,
  264, 605

\bibitem[{{Iben} \& {Renzini}(1982{\natexlab{a}})}]{iben82b}
{Iben}, Jr., I. \& {Renzini}, A. 1982{\natexlab{a}}, ApJL, 263, L23

\bibitem[{{Iben} \& {Renzini}(1982{\natexlab{b}})}]{iben82a}
---. 1982{\natexlab{b}}, ApJL, 259, L79

\bibitem[{{Iben} {et~al.}(1997){Iben}, {Ritossa}, \& {Garcia-Berro}}]{iben97}
{Iben}, Jr., I., {Ritossa}, C., \& {Garcia-Berro}, E. 1997, ApJ, 489, 772

\bibitem[{{Iliadis} {et~al.}(2001){Iliadis}, {D'Auria}, {Starrfield},
  {Thompson}, \& {Wiescher}}]{iliadis01}
{Iliadis}, C., {D'Auria}, J.~M., {Starrfield}, S., {Thompson}, W.~J., \&
  {Wiescher}, M. 2001, ApJS, 134, 151

\bibitem[{{Iliadis} {et~al.}(2010){Iliadis}, {Longland}, {Champagne}, {Coc}, \&
  {Fitzgerald}}]{iliadis10}
{Iliadis}, C., {Longland}, R., {Champagne}, A.~E., {Coc}, A., \& {Fitzgerald},
  R. 2010, Nuclear Physics A, 841, 31

\bibitem[{{Imbriani} {et~al.}(2001){Imbriani}, {Limongi}, {Gialanella},
  {Terrasi}, {Straniero}, \& {Chieffi}}]{imbriani01}
{Imbriani}, G., {Limongi}, M., {Gialanella}, L., {Terrasi}, F., {Straniero},
  O., \& {Chieffi}, A. 2001, ApJ, 558, 903

\bibitem[{{Iwamoto}(2009)}]{iwamoto09}
{Iwamoto}, N. 2009, PASA, 26, 145

\bibitem[{{Iwamoto} {et~al.}(2004){Iwamoto}, {Kajino}, {Mathews}, {Fujimoto},
  \& {Aoki}}]{iwamoto04}
{Iwamoto}, N., {Kajino}, T., {Mathews}, G.~J., {Fujimoto}, M.~Y., \& {Aoki}, W.
  2004, ApJ, 602, 377

\bibitem[{{Izzard} {et~al.}(2006){Izzard}, {Karakas}, {Lugaro}, \&
  {Tout}}]{izzard06}
{Izzard}, R.~G.and~{Dray}, L.~M., {Karakas}, A.~I., {Lugaro}, M., \& {Tout},
  C.~A. 2006, A\&A, 460, 565

\bibitem[{{Izzard}(2004)}]{izzardthesis}
{Izzard}, R.~G. 2004, PhD thesis, University of Cambridge

\bibitem[{{Izzard} {et~al.}(2010){Izzard}, {Dermine}, \& {Church}}]{izzard10}
{Izzard}, R.~G., {Dermine}, T., \& {Church}, R.~P. 2010, A\&A, 523, A10

\bibitem[{{Izzard} {et~al.}(2009){Izzard}, {Glebbeek}, {Stancliffe}, \&
  {Pols}}]{izzard09}
{Izzard}, R.~G., {Glebbeek}, E., {Stancliffe}, R.~J., \& {Pols}, O.~R. 2009,
  A\&A, 508, 1359

\bibitem[{{Izzard} {et~al.}(2007){Izzard}, {Lugaro}, {Karakas}, {Iliadis}, \&
  {van Raai}}]{izzard07}
{Izzard}, R.~G., {Lugaro}, M., {Karakas}, A.~I., {Iliadis}, C., \& {van Raai},
  M. 2007, A\&A, 466, 641

\bibitem[{{Izzard} {et~al.}(2004){Izzard}, {Tout}, {Karakas}, \&
  {Pols}}]{izzard04b}
{Izzard}, R.~G., {Tout}, C.~A., {Karakas}, A.~I., \& {Pols}, O.~R. 2004, MNRAS,
  350, 407

\bibitem[{{Jacobson} {et~al.}(2007){Jacobson}, {Friel}, \&
  {Pilachowski}}]{jacobson07}
{Jacobson}, H.~R., {Friel}, E.~D., \& {Pilachowski}, C.~A. 2007, AJ, 134, 1216

\bibitem[{{Jadhav} {et~al.}(2013){Jadhav}, {Pignatari}, {Herwig}, {Zinner},
  {Gallino}, \& {Huss}}]{jadhav13}
{Jadhav}, M., {Pignatari}, M., {Herwig}, F., {Zinner}, E., {Gallino}, R., \&
  {Huss}, G.~R. 2013, ApJL, 777, L27

\bibitem[{{Jahn} {et~al.}(2007){Jahn}, {Rauch}, {Reiff}, {Werner}, {Kruk}, \&
  {Herwig}}]{jahn07}
{Jahn}, D., {Rauch}, T., {Reiff}, E., {Werner}, K., {Kruk}, J.~W., \& {Herwig},
  F. 2007, A\&A, 462, 281

\bibitem[{{Jeffries} {et~al.}(2013){Jeffries}, {Naylor}, {Mayne}, {Bell}, \&
  {Littlefair}}]{jeffries13}
{Jeffries}, R.~D., {Naylor}, T., {Mayne}, N.~J., {Bell}, C.~P.~M., \&
  {Littlefair}, S.~P. 2013, MNRAS, 434, 2438

\bibitem[{{Johnson} {et~al.}(2008){Johnson}, {Pilachowski}, {Simmerer}, \&
  {Schwenk}}]{cjohnson08}
{Johnson}, C.~I., {Pilachowski}, C.~A., {Simmerer}, J., \& {Schwenk}, D. 2008,
  ApJ, 681, 1505

\bibitem[{{Jones} {et~al.}(2013){Jones}, {Hirschi}, {Nomoto}, {Fischer},
  {Timmes}, {Herwig}, {Paxton}, {Toki}, {Suzuki}, {Mart{\'{\i}}nez-Pinedo},
  {Lam}, \& {Bertolli}}]{jones13}
{Jones}, S., {Hirschi}, R., {Nomoto}, K., {Fischer}, T., {Timmes}, F.~X.,
  {Herwig}, F., {Paxton}, B., {Toki}, H., {Suzuki}, T.,
  {Mart{\'{\i}}nez-Pinedo}, G., {Lam}, Y.~H., \& {Bertolli}, M.~G. 2013, ApJ,
  772, 150

\bibitem[{{Jonsell} {et~al.}(2006){Jonsell}, {Barklem}, {Gustafsson},
  {Christlieb}, {Hill}, {Beers}, \& {Holmberg}}]{jonsell06}
{Jonsell}, K., {Barklem}, P.~S., {Gustafsson}, B., {Christlieb}, N., {Hill},
  V., {Beers}, T.~C., \& {Holmberg}, J. 2006, A\&A, 451, 651

\bibitem[{{Jorissen} {et~al.}(1992){Jorissen}, {Smith}, \&
  {Lambert}}]{jorissen92}
{Jorissen}, A., {Smith}, V.~V., \& {Lambert}, D.~L. 1992, A\&A, 261, 164

\bibitem[{{Jose} \& {Hernanz}(1998)}]{jose98}
{Jose}, J. \& {Hernanz}, M. 1998, ApJ, 494, 680

\bibitem[{{Justtanont} {et~al.}(2013){Justtanont}, {Teyssier}, {Barlow},
  {Matsuura}, {Swinyard}, {Waters}, \& {Yates}}]{justtanont13}
{Justtanont}, K., {Teyssier}, D., {Barlow}, M.~J., {Matsuura}, M., {Swinyard},
  B., {Waters}, L.~B.~F.~M., \& {Yates}, J. 2013, A\&A, 556, A101

\bibitem[{{Kahane} {et~al.}(2000){Kahane}, {Dufour}, {Busso}, {Gallino},
  {Lugaro}, {Forestini}, \& {Straniero}}]{kahane00}
{Kahane}, C., {Dufour}, E., {Busso}, M., {Gallino}, R., {Lugaro}, M.,
  {Forestini}, M., \& {Straniero}, O. 2000, A\&A, 357, 669

\bibitem[{{Kaler}(1978)}]{kaler78}
{Kaler}, J.~B. 1978, ApJ, 225, 527

\bibitem[{{Kalirai} {et~al.}(2014){Kalirai}, {Marigo}, \&
  {Tremblay}}]{kalirai14}
{Kalirai}, J.~S., {Marigo}, P., \& {Tremblay}, P.-E. 2014, ApJ, 782, 17

\bibitem[{{Kamath} {et~al.}(2012){Kamath}, {Karakas}, \& {Wood}}]{kamath12}
{Kamath}, D., {Karakas}, A.~I., \& {Wood}, P.~R. 2012, ApJ, 746, 20

\bibitem[{{Kamath} {et~al.}(2010){Kamath}, {Wood}, {Soszy{\'n}ski}, \&
  {Lebzelter}}]{kamath10}
{Kamath}, D., {Wood}, P.~R., {Soszy{\'n}ski}, I., \& {Lebzelter}, T. 2010,
  MNRAS, 408, 522

\bibitem[{{K{\"a}ppeler} {et~al.}(2011){K{\"a}ppeler}, {Gallino}, {Bisterzo},
  \& {Aoki}}]{kaeppeler11}
{K{\"a}ppeler}, F., {Gallino}, R., {Bisterzo}, S., \& {Aoki}, W. 2011, Reviews
  of Modern Physics, 83, 157

\bibitem[{{Karakas}(2003)}]{karakasThesis}
{Karakas}, A.~I. 2003, PhD thesis, Monash University

\bibitem[{{Karakas}(2010)}]{karakas10a}
---. 2010, MNRAS, 403, 1413

\bibitem[{{Karakas} {et~al.}(2010){Karakas}, {Campbell}, \&
  {Stancliffe}}]{karakas10b}
{Karakas}, A.~I., {Campbell}, S.~W., \& {Stancliffe}, R.~J. 2010, ApJ, 713, 374

\bibitem[{{Karakas} {et~al.}(2006{\natexlab{a}}){Karakas}, {Fenner}, {Sills},
  {Campbell}, \& {Lattanzio}}]{karakas06b}
{Karakas}, A.~I., {Fenner}, Y., {Sills}, A., {Campbell}, S.~W., \& {Lattanzio},
  J.~C. 2006{\natexlab{a}}, ApJ, 652, 1240

\bibitem[{{Karakas} {et~al.}(2012){Karakas}, {Garc{\'{\i}}a-Hern{\'a}ndez}, \&
  {Lugaro}}]{karakas12}
{Karakas}, A.~I., {Garc{\'{\i}}a-Hern{\'a}ndez}, D.~A., \& {Lugaro}, M. 2012,
  ApJ, 751, 8

\bibitem[{{Karakas} \& {Lattanzio}(2003{\natexlab{a}})}]{karakas03a}
{Karakas}, A.~I. \& {Lattanzio}, J.~C. 2003{\natexlab{a}}, PASA, 20, 393

\bibitem[{{Karakas} \& {Lattanzio}(2003{\natexlab{b}})}]{karakas03b}
---. 2003{\natexlab{b}}, PASA, 20, 279

\bibitem[{{Karakas} \& {Lattanzio}(2007)}]{karakas07b}
---. 2007, PASA, 24, 103

\bibitem[{{Karakas} {et~al.}(2002){Karakas}, {Lattanzio}, \&
  {Pols}}]{karakas02}
{Karakas}, A.~I., {Lattanzio}, J.~C., \& {Pols}, O.~R. 2002, PASA, 19, 515

\bibitem[{{Karakas} {et~al.}(2008){Karakas}, {Lee}, {Lugaro}, {G{\"o}rres}, \&
  {Wiescher}}]{karakas08}
{Karakas}, A.~I., {Lee}, H.~Y., {Lugaro}, M., {G{\"o}rres}, J., \& {Wiescher},
  M. 2008, ApJ, 676, 1254

\bibitem[{{Karakas} \& {Lugaro}(2010)}]{karakas10c}
{Karakas}, A.~I. \& {Lugaro}, M. 2010, PASA, 27, 227

\bibitem[{{Karakas} {et~al.}(2007){Karakas}, {Lugaro}, \&
  {Gallino}}]{karakas07a}
{Karakas}, A.~I., {Lugaro}, M., \& {Gallino}, R. 2007, ApJL, 656, L73

\bibitem[{{Karakas} {et~al.}(2006{\natexlab{b}}){Karakas}, {Lugaro},
  {Wiescher}, {Goerres}, \& {Ugalde}}]{karakas06a}
{Karakas}, A.~I., {Lugaro}, M., {Wiescher}, M., {Goerres}, J., \& {Ugalde}, C.
  2006{\natexlab{b}}, ApJ, 643, 471

\bibitem[{{Karakas} {et~al.}(2014){Karakas}, {Marino}, \& {Nataf}}]{karakas14}
{Karakas}, A.~I., {Marino}, A.~F., \& {Nataf}, D.~M. 2014, ApJ, 784, 32

\bibitem[{{Karakas} {et~al.}(2000){Karakas}, {Tout}, \&
  {Lattanzio}}]{karakas00}
{Karakas}, A.~I., {Tout}, C.~A., \& {Lattanzio}, J.~C. 2000, MNRAS, 316, 689

\bibitem[{{Karakas} {et~al.}(2009){Karakas}, {van Raai}, {Lugaro}, {Sterling},
  \& {Dinerstein}}]{karakas09}
{Karakas}, A.~I., {van Raai}, M.~A., {Lugaro}, M., {Sterling}, N.~C., \&
  {Dinerstein}, H.~L. 2009, ApJ, 690, 1130

\bibitem[{{Kawaler}(1988)}]{kawaler88}
{Kawaler}, S.~D. 1988, ApJ, 333, 236

\bibitem[{{Keller} {et~al.}(2007){Keller}, {Schmidt}, {Bessell}, {Conroy},
  {Francis}, {Granlund}, {Kowald}, {Oates}, {Martin-Jones}, {Preston},
  {Tisserand}, {Vaccarella}, \& {Waterson}}]{keller07}
{Keller}, S.~C., {Schmidt}, B.~P., {Bessell}, M.~S., {Conroy}, P.~G.,
  {Francis}, P., {Granlund}, A., {Kowald}, E., {Oates}, A.~P., {Martin-Jones},
  T., {Preston}, T., {Tisserand}, P., {Vaccarella}, A., \& {Waterson}, M.~F.
  2007, PASA, 24, 1

\bibitem[{{Kingsburgh} \& {Barlow}(1994)}]{kingsburgh94}
{Kingsburgh}, R.~L. \& {Barlow}, M.~J. 1994, MNRAS, 271, 257

\bibitem[{{Kippenhahn} {et~al.}(1980){Kippenhahn}, {Ruschenplatt}, \&
  {Thomas}}]{kippenhahn80}
{Kippenhahn}, R., {Ruschenplatt}, G., \& {Thomas}, H. 1980, A\&A, 91, 175

\bibitem[{{Kippenhahn} \& {Weigert}(1990)}]{kippwei90}
{Kippenhahn}, R. \& {Weigert}, A. 1990, {Stellar Structure and Evolution}
  (Springer-Verlag)

\bibitem[{{Kobayashi} {et~al.}(2011{\natexlab{a}}){Kobayashi}, {Izutani},
  {Karakas}, {Yoshida}, {Yong}, \& {Umeda}}]{kobayashi11b}
{Kobayashi}, C., {Izutani}, N., {Karakas}, A.~I., {Yoshida}, T., {Yong}, D., \&
  {Umeda}, H. 2011{\natexlab{a}}, ApJL, 739, L57

\bibitem[{{Kobayashi} {et~al.}(2011{\natexlab{b}}){Kobayashi}, {Karakas}, \&
  {Umeda}}]{kobayashi11a}
{Kobayashi}, C., {Karakas}, A.~I., \& {Umeda}, H. 2011{\natexlab{b}}, MNRAS,
  414, 3231

\bibitem[{{Kobayashi} {et~al.}(2006){Kobayashi}, {Umeda}, {Nomoto}, {Tominaga},
  \& {Ohkubo}}]{kobayashi06}
{Kobayashi}, C., {Umeda}, H., {Nomoto}, K., {Tominaga}, N., \& {Ohkubo}, T.
  2006, ApJ, 653, 1145

\bibitem[{{Korobkin} {et~al.}(2012){Korobkin}, {Rosswog}, {Arcones}, \&
  {Winteler}}]{korobkin12}
{Korobkin}, O., {Rosswog}, S., {Arcones}, A., \& {Winteler}, C. 2012, MNRAS,
  426, 1940

\bibitem[{{Kudritzki} \& {Reimers}(1978)}]{kudritzki78}
{Kudritzki}, R.~P. \& {Reimers}, D. 1978, A\&A, 70, 227

\bibitem[{{Kumar} {et~al.}(2011){Kumar}, {Reddy}, \& {Lambert}}]{kumar11}
{Kumar}, Y.~B., {Reddy}, B.~E., \& {Lambert}, D.~L. 2011, ApJL, 730, L12

\bibitem[{{Lagadec} \& {Zijlstra}(2008)}]{lagadec08}
{Lagadec}, E. \& {Zijlstra}, A.~A. 2008, MNRAS, 390, L59

\bibitem[{{Lagadec} {et~al.}(2010){Lagadec}, {Zijlstra}, {Mauron}, {Fuller},
  {Josselin}, {Sloan}, \& {Riggs}}]{lagadec10}
{Lagadec}, E., {Zijlstra}, A.~A., {Mauron}, N., {Fuller}, G., {Josselin}, E.,
  {Sloan}, G.~C., \& {Riggs}, A.~J.~E. 2010, MNRAS, 403, 1331

\bibitem[{{Lagadec} {et~al.}(2009){Lagadec}, {Zijlstra}, {Sloan}, {Wood},
  {Matsuura}, {Bernard-Salas}, {Blommaert}, {Cioni}, {Feast}, {Groenewegen},
  {Hony}, {Menzies}, {van Loon}, \& {Whitelock}}]{lagadec09}
{Lagadec}, E., {Zijlstra}, A.~A., {Sloan}, G.~C., {Wood}, P.~R., {Matsuura},
  M., {Bernard-Salas}, J., {Blommaert}, J.~A.~D.~L., {Cioni}, M.-R.~L.,
  {Feast}, M.~W., {Groenewegen}, M.~A.~T., {Hony}, S., {Menzies}, J.~W., {van
  Loon}, J.~T., \& {Whitelock}, P.~A. 2009, MNRAS, 396, 598

\bibitem[{{Lagarde} {et~al.}(2011){Lagarde}, {Charbonnel}, {Decressin}, \&
  {Hagelberg}}]{lagarde11}
{Lagarde}, N., {Charbonnel}, C., {Decressin}, T., \& {Hagelberg}, J. 2011,
  A\&A, 536, A28

\bibitem[{{Lagarde} {et~al.}(2012){Lagarde}, {Romano}, {Charbonnel}, {Tosi},
  {Chiappini}, \& {Matteucci}}]{lagarde12a}
{Lagarde}, N., {Romano}, D., {Charbonnel}, C., {Tosi}, M., {Chiappini}, C., \&
  {Matteucci}, F. 2012, A\&A, 542, A62

\bibitem[{{Lambert} {et~al.}(1986){Lambert}, {Gustafsson}, {Eriksson}, \&
  {Hinkle}}]{lambert86}
{Lambert}, D.~L., {Gustafsson}, B., {Eriksson}, K., \& {Hinkle}, K.~H. 1986,
  ApJS, 62, 373

\bibitem[{{Lambert} {et~al.}(1995){Lambert}, {Smith}, {Busso}, {Gallino}, \&
  {Straniero}}]{lambert95}
{Lambert}, D.~L., {Smith}, V.~V., {Busso}, M., {Gallino}, R., \& {Straniero},
  O. 1995, ApJ, 450, 302

\bibitem[{{Langer} \& {Hoffman}(1995)}]{langer95}
{Langer}, G.~E. \& {Hoffman}, R.~D. 1995, PASP, 107, 1177

\bibitem[{{Langer}(2012)}]{langer12}
{Langer}, N. 2012, ARA\&A, 50, 107

\bibitem[{{Lattanzio} {et~al.}(1996){Lattanzio}, {Frost}, {Cannon}, \&
  {Wood}}]{lattanzio96}
{Lattanzio}, J., {Frost}, C., {Cannon}, R., \& {Wood}, P.~R. 1996, Memorie
  della Societa Astronomica Italiana, 67, 729

\bibitem[{{Lattanzio}(1986)}]{lattanzio86}
{Lattanzio}, J.~C. 1986, ApJ, 311, 708

\bibitem[{{Lattanzio}(1989)}]{lattanzio89}
---. 1989, ApJL, 344, L25

\bibitem[{{Lattanzio}(1991)}]{lattanzio91a}
---. 1991, ApJS, 76, 215

\bibitem[{{Lattanzio}(1992)}]{lattanzio92}
---. 1992, PASA, 10, 120

\bibitem[{{Lattanzio} \& {Lugaro}(2005)}]{lattanzio05}
{Lattanzio}, J.~C. \& {Lugaro}, M.~A. 2005, Nuclear Physics A, 758, 477

\bibitem[{{Lattanzio} {et~al.}(1991){Lattanzio}, {Vallenari}, {Bertelli}, \&
  {Chiosi}}]{lattanzio91b}
{Lattanzio}, J.~C., {Vallenari}, A., {Bertelli}, G., \& {Chiosi}, C. 1991,
  A\&A, 250, 340

\bibitem[{{Lau} {et~al.}(2012){Lau}, {Gil-Pons}, {Doherty}, \&
  {Lattanzio}}]{lau12}
{Lau}, H.~H.~B., {Gil-Pons}, P., {Doherty}, C., \& {Lattanzio}, J. 2012, A\&A,
  542, A1

\bibitem[{{Lau} {et~al.}(2009){Lau}, {Stancliffe}, \& {Tout}}]{lau09}
{Lau}, H.~H.~B., {Stancliffe}, R.~J., \& {Tout}, C.~A. 2009, MNRAS, 396, 1046

\bibitem[{{Lebzelter} \& {Hron}(2003)}]{lebzelter03}
{Lebzelter}, T. \& {Hron}, J. 2003, A\&A, 411, 533

\bibitem[{{Lebzelter} {et~al.}(2008){Lebzelter}, {Lederer}, {Cristallo},
  {Hinkle}, {Straniero}, \& {Aringer}}]{lebzelter08}
{Lebzelter}, T., {Lederer}, M.~T., {Cristallo}, S., {Hinkle}, K.~H.,
  {Straniero}, O., \& {Aringer}, B. 2008, A\&A, 486, 511

\bibitem[{{Lebzelter} \& {Wood}(2007)}]{lebzelter07}
{Lebzelter}, T. \& {Wood}, P.~R. 2007, A\&A, 475, 643

\bibitem[{{Lederer} \& {Aringer}(2009)}]{lederer09}
{Lederer}, M.~T. \& {Aringer}, B. 2009, A\&A, 494, 403

\bibitem[{{Lederer} {et~al.}(2009){Lederer}, {Lebzelter}, {Cristallo},
  {Straniero}, {Hinkle}, \& {Aringer}}]{lederer09b}
{Lederer}, M.~T., {Lebzelter}, T., {Cristallo}, S., {Straniero}, O., {Hinkle},
  K.~H., \& {Aringer}, B. 2009, A\&A, 502, 913

\bibitem[{{Leisy} \& {Dennefeld}(2006)}]{leisy06}
{Leisy}, P. \& {Dennefeld}, M. 2006, A\&A, 456, 451

\bibitem[{{Lewis} {et~al.}(2013){Lewis}, {Lugaro}, {Gibson}, \&
  {Pilkington}}]{lewis13}
{Lewis}, K.~M., {Lugaro}, M., {Gibson}, B.~K., \& {Pilkington}, K. 2013, ApJL,
  768, L19

\bibitem[{{Lind} {et~al.}(2009){Lind}, {Primas}, {Charbonnel}, {Grundahl}, \&
  {Asplund}}]{lind09}
{Lind}, K., {Primas}, F., {Charbonnel}, C., {Grundahl}, F., \& {Asplund}, M.
  2009, A\&A, 503, 545

\bibitem[{{Little-Marenin} \& {Little}(1979)}]{little-marenin79}
{Little-Marenin}, I.~R. \& {Little}, S.~J. 1979, AJ, 84, 1374

\bibitem[{{Lodders} \& {Amari}(2005)}]{lodders05}
{Lodders}, K. \& {Amari}, S. 2005, Chemie der Erde: Geochemistry, 65, 93

\bibitem[{{Lombaert} {et~al.}(2013){Lombaert}, {Decin}, {de Koter},
  {Blommaert}, {Royer}, {De Beck}, {de Vries}, {Khouri}, \& {Min}}]{lomb13}
{Lombaert}, R., {Decin}, L., {de Koter}, A., {Blommaert}, J.~A.~D.~L., {Royer},
  P., {De Beck}, E., {de Vries}, B.~L., {Khouri}, T., \& {Min}, M. 2013, A\&A,
  554, A142

\bibitem[{{Longland} {et~al.}(2012){Longland}, {Iliadis}, \&
  {Karakas}}]{longland12}
{Longland}, R., {Iliadis}, C., \& {Karakas}, A.~I. 2012, Phys. Rev. C, 85,
  065809

\bibitem[{{Lucatello} {et~al.}(2005){Lucatello}, {Gratton}, {Beers}, \&
  {Carretta}}]{lucatello05}
{Lucatello}, S., {Gratton}, R.~G., {Beers}, T.~C., \& {Carretta}, E. 2005, ApJ,
  625, 833

\bibitem[{{Lucatello} {et~al.}(2011){Lucatello}, {Masseron}, {Johnson},
  {Pignatari}, \& {Herwig}}]{lucatello11}
{Lucatello}, S., {Masseron}, T., {Johnson}, J.~A., {Pignatari}, M., \&
  {Herwig}, F. 2011, ApJ, 729, 40

\bibitem[{{Luck} \& {Bond}(1991)}]{luck91}
{Luck}, R.~E. \& {Bond}, H.~E. 1991, ApJS, 77, 515

\bibitem[{{Lugaro} {et~al.}(2009){Lugaro}, {Campbell}, \& {de Mink}}]{lugaro09}
{Lugaro}, M., {Campbell}, S.~W., \& {de Mink}, S.~E. 2009, PASA, 26, 322

\bibitem[{{Lugaro} {et~al.}(2003){Lugaro}, {Davis}, {Gallino}, {Pellin},
  {Straniero}, \& {K{\"a}ppeler}}]{lugaro03b}
{Lugaro}, M., {Davis}, A.~M., {Gallino}, R., {Pellin}, M.~J., {Straniero}, O.,
  \& {K{\"a}ppeler}, F. 2003, ApJ, 593, 486

\bibitem[{{Lugaro} {et~al.}(2012){Lugaro}, {Karakas}, {Stancliffe}, \&
  {Rijs}}]{lugaro12}
{Lugaro}, M., {Karakas}, A.~I., {Stancliffe}, R.~J., \& {Rijs}, C. 2012, ApJ,
  747, 2

\bibitem[{{Lugaro} {et~al.}(2004){Lugaro}, {Ugalde}, {Karakas}, {G{\"o}rres},
  {Wiescher}, {Lattanzio}, \& {Cannon}}]{lugaro04}
{Lugaro}, M., {Ugalde}, C., {Karakas}, A.~I., {G{\"o}rres}, J., {Wiescher}, M.,
  {Lattanzio}, J.~C., \& {Cannon}, R.~C. 2004, ApJ, 615, 934

\bibitem[{{Lugaro} {et~al.}(1999){Lugaro}, {Zinner}, {Gallino}, \&
  {Amari}}]{lugaro99}
{Lugaro}, M., {Zinner}, E., {Gallino}, R., \& {Amari}, S. 1999, ApJ, 527, 369

\bibitem[{{Maceroni} {et~al.}(2002){Maceroni}, {Testa}, {Plez}, {Garc{\'{\i}}a
  Lario}, \& {D'Antona}}]{maceroni02}
{Maceroni}, C., {Testa}, V., {Plez}, B., {Garc{\'{\i}}a Lario}, P., \&
  {D'Antona}, F. 2002, A\&A, 395, 179

\bibitem[{{Mackey} {et~al.}(2008){Mackey}, {Broby Nielsen}, {Ferguson}, \&
  {Richardson}}]{mackey08}
{Mackey}, A.~D., {Broby Nielsen}, P., {Ferguson}, A.~M.~N., \& {Richardson},
  J.~C. 2008, ApJL, 681, L17

\bibitem[{{Maeder} \& {Meynet}(2010)}]{maeder10}
{Maeder}, A. \& {Meynet}, G. 2010, NewAstR, 54, 32

\bibitem[{{Maeder} {et~al.}(2013){Maeder}, {Meynet}, {Lagarde}, \&
  {Charbonnel}}]{maeder13}
{Maeder}, A., {Meynet}, G., {Lagarde}, N., \& {Charbonnel}, C. 2013, A\&A, 553,
  A1

\bibitem[{{Maeder} \& {Zahn}(1998)}]{maeder98}
{Maeder}, A. \& {Zahn}, J.-P. 1998, A\&A, 334, 1000

\bibitem[{{Manchado}(2003)}]{manchado03}
{Manchado}, A. 2003, in IAU Symposium, Vol. 209, Planetary Nebulae: Their
  Evolution and Role in the Universe, ed. S.~{Kwok}, M.~{Dopita}, \&
  R.~{Sutherland}, 431

\bibitem[{{Maraston}(2005)}]{maraston05}
{Maraston}, C. 2005, MNRAS, 362, 799

\bibitem[{{Maraston} {et~al.}(2006){Maraston}, {Daddi}, {Renzini}, {Cimatti},
  {Dickinson}, {Papovich}, {Pasquali}, \& {Pirzkal}}]{maraston06}
{Maraston}, C., {Daddi}, E., {Renzini}, A., {Cimatti}, A., {Dickinson}, M.,
  {Papovich}, C., {Pasquali}, A., \& {Pirzkal}, N. 2006, ApJ, 652, 85

\bibitem[{{Marigo}(2001)}]{marigo01}
{Marigo}, P. 2001, A\&A, 370, 194

\bibitem[{{Marigo}(2002)}]{marigo02}
---. 2002, A\&A, 387, 507

\bibitem[{{Marigo} \& {Aringer}(2009)}]{marigo09}
{Marigo}, P. \& {Aringer}, B. 2009, A\&A, 508, 1539

\bibitem[{{Marigo} {et~al.}(1996){Marigo}, {Bressan}, \& {Chiosi}}]{marigo96}
{Marigo}, P., {Bressan}, A., \& {Chiosi}, C. 1996, A\&A, 313, 545

\bibitem[{{Marigo} {et~al.}(2013){Marigo}, {Bressan}, {Nanni}, {Girardi}, \&
  {Pumo}}]{colibri}
{Marigo}, P., {Bressan}, A., {Nanni}, A., {Girardi}, L., \& {Pumo}, M.~L. 2013,
  MNRAS

\bibitem[{{Marigo} \& {Girardi}(2007)}]{marigo07}
{Marigo}, P. \& {Girardi}, L. 2007, A\&A, 469, 239

\bibitem[{{Marigo} {et~al.}(1999){Marigo}, {Girardi}, \& {Bressan}}]{marigo99}
{Marigo}, P., {Girardi}, L., \& {Bressan}, A. 1999, A\&A, 344, 123

\bibitem[{{Marigo} {et~al.}(2001){Marigo}, {Girardi}, {Chiosi}, \&
  {Wood}}]{marigo01b}
{Marigo}, P., {Girardi}, L., {Chiosi}, C., \& {Wood}, P.~R. 2001, A\&A, 371,
  152

\bibitem[{{Martell} {et~al.}(2008){Martell}, {Smith}, \& {Briley}}]{martell08}
{Martell}, S.~L., {Smith}, G.~H., \& {Briley}, M.~M. 2008, AJ, 136, 2522

\bibitem[{{Masseron} {et~al.}(2010){Masseron}, {Johnson}, {Plez}, {van Eck},
  {Primas}, {Goriely}, \& {Jorissen}}]{masseron10}
{Masseron}, T., {Johnson}, J.~A., {Plez}, B., {van Eck}, S., {Primas}, F.,
  {Goriely}, S., \& {Jorissen}, A. 2010, A\&A, 509, A93

\bibitem[{{Mathis}(2013)}]{mathis13}
{Mathis}, S. 2013, in Lecture Notes in Physics, Berlin Springer Verlag, Vol.
  865, Lecture Notes in Physics, Berlin Springer Verlag, ed. M.~{Goupil},
  K.~{Belkacem}, C.~{Neiner}, F.~{Ligni{\`e}res}, \& J.~J. {Green}, 23

\bibitem[{{Matteucci} \& {Francois}(1989)}]{matt89}
{Matteucci}, F. \& {Francois}, P. 1989, MNRAS, 239, 885

\bibitem[{{Matteucci} \& {Greggio}(1986)}]{matt86}
{Matteucci}, F. \& {Greggio}, L. 1986, A\&A, 154, 279

\bibitem[{{Mattsson} {et~al.}(2010){Mattsson}, {Wahlin}, \&
  {H{\"o}fner}}]{mattsson10}
{Mattsson}, L., {Wahlin}, R., \& {H{\"o}fner}, S. 2010, A\&A, 509, A14

\bibitem[{{Mattsson} {et~al.}(2008){Mattsson}, {Wahlin}, {H{\"o}fner}, \&
  {Eriksson}}]{mattsson08}
{Mattsson}, L., {Wahlin}, R., {H{\"o}fner}, S., \& {Eriksson}, K. 2008, A\&A,
  484, L5

\bibitem[{{Mayer}(1950)}]{mayer50}
{Mayer}, M.~G. 1950, Physical Review, 78, 16

\bibitem[{{Mazzitelli} {et~al.}(1999){Mazzitelli}, {D'Antona}, \&
  {Ventura}}]{mazzitelli99}
{Mazzitelli}, I., {D'Antona}, F., \& {Ventura}, P. 1999, A\&A, 348, 846

\bibitem[{{McClure}(1983)}]{mcclure83}
{McClure}, R.~D. 1983, ApJ, 268, 264

\bibitem[{{McClure}(1997{\natexlab{a}})}]{mcclure97b}
---. 1997{\natexlab{a}}, PASP, 109, 536

\bibitem[{{McClure}(1997{\natexlab{b}})}]{mcclure97a}
---. 1997{\natexlab{b}}, PASP, 109, 256

\bibitem[{{McClure} \& {Woodsworth}(1990)}]{mcclure90}
{McClure}, R.~D. \& {Woodsworth}, A.~W. 1990, ApJ, 352, 709

\bibitem[{{McSaveney} {et~al.}(2007){McSaveney}, {Wood}, {Scholz}, {Lattanzio},
  \& {Hinkle}}]{mcsaveney07}
{McSaveney}, J.~A., {Wood}, P.~R., {Scholz}, M., {Lattanzio}, J.~C., \&
  {Hinkle}, K.~H. 2007, {MNRAS}, in press

\bibitem[{{Meakin} \& {Arnett}(2007)}]{meakin07}
{Meakin}, C.~A. \& {Arnett}, D. 2007, ApJ, 667, 448

\bibitem[{{Meixner} {et~al.}(1998){Meixner}, {Campbell}, {Welch}, \&
  {Likkel}}]{meixner98}
{Meixner}, M., {Campbell}, M.~T., {Welch}, W.~J., \& {Likkel}, L. 1998, ApJ,
  509, 392

\bibitem[{{Melbourne} {et~al.}(2012){Melbourne}, {Williams}, {Dalcanton},
  {Rosenfield}, {Girardi}, {Marigo}, {Weisz}, {Dolphin}, {Boyer}, {Olsen},
  {Skillman}, \& {Seth}}]{melbourne12}
{Melbourne}, J., {Williams}, B.~F., {Dalcanton}, J.~J., {Rosenfield}, P.,
  {Girardi}, L., {Marigo}, P., {Weisz}, D., {Dolphin}, A., {Boyer}, M.~L.,
  {Olsen}, K., {Skillman}, E., \& {Seth}, A.~C. 2012, ApJ, 748, 47

\bibitem[{{Merrill}(1952)}]{merrill52}
{Merrill}, S.~P.~W. 1952, ApJ, 116, 21

\bibitem[{{Mestel}(1953)}]{mestel53}
{Mestel}, L. 1953, MNRAS, 113, 716

\bibitem[{{Meyer}(1994)}]{meyer94}
{Meyer}, B.~S. 1994, ARA\&A, 32, 153

\bibitem[{{Meynet} \& {Arnould}(2000)}]{meynet00}
{Meynet}, G. \& {Arnould}, M. 2000, A\&A, 355, 176

\bibitem[{{Miglio} {et~al.}(2012){Miglio}, {Brogaard}, {Stello}, {Chaplin},
  {D'Antona}, {Montalb{\'a}n}, {Basu}, {Bressan}, {Grundahl}, {Pinsonneault},
  {Serenelli}, {Elsworth}, {Hekker}, {Kallinger}, {Mosser}, {Ventura},
  {Bonanno}, {Noels}, {Silva Aguirre}, {Szabo}, {Li}, {McCauliff}, {Middour},
  \& {Kjeldsen}}]{miglio12}
{Miglio}, A., {Brogaard}, K., {Stello}, D., {Chaplin}, W.~J., {D'Antona}, F.,
  {Montalb{\'a}n}, J., {Basu}, S., {Bressan}, A., {Grundahl}, F.,
  {Pinsonneault}, M., {Serenelli}, A.~M., {Elsworth}, Y., {Hekker}, S.,
  {Kallinger}, T., {Mosser}, B., {Ventura}, P., {Bonanno}, A., {Noels}, A.,
  {Silva Aguirre}, V., {Szabo}, R., {Li}, J., {McCauliff}, S., {Middour},
  C.~K., \& {Kjeldsen}, H. 2012, MNRAS, 419, 2077

\bibitem[{{Mikolaitis} {et~al.}(2010){Mikolaitis}, {Tautvai{\v s}ien{\.e}},
  {Gratton}, {Bragaglia}, \& {Carretta}}]{miko10}
{Mikolaitis}, {\v S}., {Tautvai{\v s}ien{\.e}}, G., {Gratton}, R., {Bragaglia},
  A., \& {Carretta}, E. 2010, MNRAS, 407, 1866

\bibitem[{{Milam} {et~al.}(2009){Milam}, {Woolf}, \& {Ziurys}}]{milam09}
{Milam}, S.~N., {Woolf}, N.~J., \& {Ziurys}, L.~M. 2009, ApJ, 690, 837

\bibitem[{{Miller Bertolami} \& {Althaus}(2006)}]{miller06}
{Miller Bertolami}, M.~M. \& {Althaus}, L.~G. 2006, A\&A, 454, 845

\bibitem[{{Milone} {et~al.}(2009){Milone}, {Bedin}, {Piotto}, \&
  {Anderson}}]{milone09}
{Milone}, A.~P., {Bedin}, L.~R., {Piotto}, G., \& {Anderson}, J. 2009, A\&A,
  497, 755

\bibitem[{{Milone} {et~al.}(2012{\natexlab{a}}){Milone}, {Marino}, {Piotto},
  {Bedin}, {Anderson}, {Aparicio}, {Cassisi}, \& {Rich}}]{milone12a}
{Milone}, A.~P., {Marino}, A.~F., {Piotto}, G., {Bedin}, L.~R., {Anderson}, J.,
  {Aparicio}, A., {Cassisi}, S., \& {Rich}, R.~M. 2012{\natexlab{a}}, ApJ, 745,
  27

\bibitem[{{Milone} {et~al.}(2012{\natexlab{b}}){Milone}, {Piotto}, {Bedin},
  {Cassisi}, {Anderson}, {Marino}, {Pietrinferni}, \& {Aparicio}}]{milone12c}
{Milone}, A.~P., {Piotto}, G., {Bedin}, L.~R., {Cassisi}, S., {Anderson}, J.,
  {Marino}, A.~F., {Pietrinferni}, A., \& {Aparicio}, A. 2012{\natexlab{b}},
  A\&A, 537, A77

\bibitem[{{Milone} {et~al.}(2012{\natexlab{c}}){Milone}, {Piotto}, {Bedin},
  {King}, {Anderson}, {Marino}, {Bellini}, {Gratton}, {Renzini}, {Stetson},
  {Cassisi}, {Aparicio}, {Bragaglia}, {Carretta}, {D'Antona}, {Di Criscienzo},
  {Lucatello}, {Monelli}, \& {Pietrinferni}}]{milone12b}
{Milone}, A.~P., {Piotto}, G., {Bedin}, L.~R., {King}, I.~R., {Anderson}, J.,
  {Marino}, A.~F., {Bellini}, A., {Gratton}, R., {Renzini}, A., {Stetson},
  P.~B., {Cassisi}, S., {Aparicio}, A., {Bragaglia}, A., {Carretta}, E.,
  {D'Antona}, F., {Di Criscienzo}, M., {Lucatello}, S., {Monelli}, M., \&
  {Pietrinferni}, A. 2012{\natexlab{c}}, ApJ, 744, 58

\bibitem[{{Mirouh} {et~al.}(2012){Mirouh}, {Garaud}, {Stellmach}, {Traxler}, \&
  {Wood}}]{mirouh12}
{Mirouh}, G.~M., {Garaud}, P., {Stellmach}, S., {Traxler}, A.~L., \& {Wood},
  T.~S. 2012, ApJ, 750, 61

\bibitem[{{Mishenina} {et~al.}(2006){Mishenina}, {Bienaym{\'e}}, {Gorbaneva},
  {Charbonnel}, {Soubiran}, {Korotin}, \& {Kovtyukh}}]{mishenina06}
{Mishenina}, T.~V., {Bienaym{\'e}}, O., {Gorbaneva}, T.~I., {Charbonnel}, C.,
  {Soubiran}, C., {Korotin}, S.~A., \& {Kovtyukh}, V.~V. 2006, A\&A, 456, 1109

\bibitem[{{Miszalski} {et~al.}(2012){Miszalski}, {Boffin}, {Frew}, {Acker},
  {K{\"o}ppen}, {Moffat}, \& {Parker}}]{miszalski12}
{Miszalski}, B., {Boffin}, H.~M.~J., {Frew}, D.~J., {Acker}, A., {K{\"o}ppen},
  J., {Moffat}, A.~F.~J., \& {Parker}, Q.~A. 2012, MNRAS, 419, 39

\bibitem[{{Miszalski} {et~al.}(2013){Miszalski}, {Boffin}, {Jones}, {Karakas},
  {K{\"o}ppen}, {Tyndall}, {Mohamed}, {Rodr{\'{\i}}guez-Gil}, \&
  {Santander-Garc{\'{\i}}a}}]{miszalski13}
{Miszalski}, B., {Boffin}, H.~M.~J., {Jones}, D., {Karakas}, A.~I.,
  {K{\"o}ppen}, J., {Tyndall}, A.~A., {Mohamed}, S.~S., {Rodr{\'{\i}}guez-Gil},
  P., \& {Santander-Garc{\'{\i}}a}, M. 2013, MNRAS

\bibitem[{{Miyaji} {et~al.}(1980){Miyaji}, {Nomoto}, {Yokoi}, \&
  {Sugimoto}}]{miyaji80}
{Miyaji}, S., {Nomoto}, K., {Yokoi}, K., \& {Sugimoto}, D. 1980, PASJ, 32, 303

\bibitem[{{Moc{\'a}k} {et~al.}(2010){Moc{\'a}k}, {Campbell}, {M{\"u}ller}, \&
  {Kifonidis}}]{mocak10}
{Moc{\'a}k}, M., {Campbell}, S.~W., {M{\"u}ller}, E., \& {Kifonidis}, K. 2010,
  A\&A, 520, A114

\bibitem[{{Moc{\'a}k} {et~al.}(2008){Moc{\'a}k}, {M{\"u}ller}, {Weiss}, \&
  {Kifonidis}}]{mocak08}
{Moc{\'a}k}, M., {M{\"u}ller}, E., {Weiss}, A., \& {Kifonidis}, K. 2008, A\&A,
  490, 265

\bibitem[{{Moc{\'a}k} {et~al.}(2009){Moc{\'a}k}, {M{\"u}ller}, {Weiss}, \&
  {Kifonidis}}]{mocak09}
---. 2009, A\&A, 501, 659

\bibitem[{{Moc{\'a}k} {et~al.}(2011){Moc{\'a}k}, {Siess}, \&
  {M{\"u}ller}}]{mocak11}
{Moc{\'a}k}, M., {Siess}, L., \& {M{\"u}ller}, E. 2011, A\&A, 533, A53

\bibitem[{{Monaco} {et~al.}(2011){Monaco}, {Villanova}, {Moni Bidin},
  {Carraro}, {Geisler}, {Bonifacio}, {Gonzalez}, {Zoccali}, \&
  {Jilkova}}]{monaco11}
{Monaco}, L., {Villanova}, S., {Moni Bidin}, C., {Carraro}, G., {Geisler}, D.,
  {Bonifacio}, P., {Gonzalez}, O.~A., {Zoccali}, M., \& {Jilkova}, L. 2011,
  A\&A, 529, A90

\bibitem[{{Mouhcine} \& {Lan{\c c}on}(2002)}]{mouchine02}
{Mouhcine}, M. \& {Lan{\c c}on}, A. 2002, A\&A, 393, 149

\bibitem[{{Mowlavi}(1999{\natexlab{a}})}]{mowlavi99a}
{Mowlavi}, N. 1999{\natexlab{a}}, A\&A, 344, 617

\bibitem[{{Mowlavi}(1999{\natexlab{b}})}]{mowlavi99b}
---. 1999{\natexlab{b}}, A\&A, 350, 73

\bibitem[{{Mowlavi} {et~al.}(1996){Mowlavi}, {Jorissen}, \&
  {Arnould}}]{mowlavi96}
{Mowlavi}, N., {Jorissen}, A., \& {Arnould}, M. 1996, A\&A, 311, 803

\bibitem[{{Mowlavi} {et~al.}(1998){Mowlavi}, {Jorissen}, \&
  {Arnould}}]{mowlavi98}
---. 1998, A\&A, 334, 153

\bibitem[{{Mucciarelli} {et~al.}(2008){Mucciarelli}, {Carretta}, {Origlia}, \&
  {Ferraro}}]{muccia08}
{Mucciarelli}, A., {Carretta}, E., {Origlia}, L., \& {Ferraro}, F.~R. 2008, AJ,
  136, 375

\bibitem[{{Mucciarelli} {et~al.}(2006){Mucciarelli}, {Origlia}, {Ferraro},
  {Maraston}, \& {Testa}}]{mucciarelli06}
{Mucciarelli}, A., {Origlia}, L., {Ferraro}, F.~R., {Maraston}, C., \& {Testa},
  V. 2006, ApJ, 646, 939

\bibitem[{{Nataf} {et~al.}(2013){Nataf}, {Gould}, {Pinsonneault}, \&
  {Udalski}}]{nataf13}
{Nataf}, D.~M., {Gould}, A.~P., {Pinsonneault}, M.~H., \& {Udalski}, A. 2013,
  ApJ, 766, 77

\bibitem[{{Nollett} {et~al.}(2003){Nollett}, {Busso}, \&
  {Wasserburg}}]{nollett03}
{Nollett}, K.~M., {Busso}, M., \& {Wasserburg}, G.~J. 2003, ApJ, 582, 1036

\bibitem[{{Nomoto}(1984)}]{nomoto87}
{Nomoto}, K. 1984, ApJ, 277, 791

\bibitem[{{Nomoto} {et~al.}(2013){Nomoto}, {Kobayashi}, \&
  {Tominaga}}]{nomoto13}
{Nomoto}, K., {Kobayashi}, C., \& {Tominaga}, N. 2013, ARA\&A, 51, 457

\bibitem[{{Nordhaus} {et~al.}(2008){Nordhaus}, {Busso}, {Wasserburg},
  {Blackman}, \& {Palmerini}}]{nordhaus08}
{Nordhaus}, J., {Busso}, M., {Wasserburg}, G.~J., {Blackman}, E.~G., \&
  {Palmerini}, S. 2008, ApJL, 684, L29

\bibitem[{{Norris}(2004)}]{norris04}
{Norris}, J.~E. 2004, ApJL, 612, L25

\bibitem[{{Norris} \& {Da Costa}(1995)}]{norris95}
{Norris}, J.~E. \& {Da Costa}, G.~S. 1995, ApJ, 447, 680

\bibitem[{{Olofsson} {et~al.}(2010){Olofsson}, {Maercker}, {Eriksson},
  {Gustafsson}, \& {Sch{\"o}ier}}]{olofsson10}
{Olofsson}, H., {Maercker}, M., {Eriksson}, K., {Gustafsson}, B., \&
  {Sch{\"o}ier}, F. 2010, A\&A, 515, A27

\bibitem[{{Olson} \& {Richter}(1979)}]{olson79}
{Olson}, B.~I. \& {Richter}, H.~B. 1979, ApJ, 227, 534

\bibitem[{{Origlia} {et~al.}(2008){Origlia}, {Valenti}, \& {Rich}}]{origlia08}
{Origlia}, L., {Valenti}, E., \& {Rich}, R.~M. 2008, MNRAS, 388, 1419

\bibitem[{{Otsuka} {et~al.}(2008){Otsuka}, {Izumiura}, {Tajitsu}, \&
  {Hyung}}]{otsuka08}
{Otsuka}, M., {Izumiura}, H., {Tajitsu}, A., \& {Hyung}, S. 2008, ApJL, 682,
  L105

\bibitem[{{Paczy{\'n}ski}(1970)}]{pacz70}
{Paczy{\'n}ski}, B. 1970, Acta Astronomica, 20, 47

\bibitem[{{Palacios} {et~al.}(2001){Palacios}, {Charbonnel}, \&
  {Forestini}}]{palacios01}
{Palacios}, A., {Charbonnel}, C., \& {Forestini}, M. 2001, A\&A, 375, L9

\bibitem[{{Palacios} {et~al.}(2006){Palacios}, {Charbonnel}, {Talon}, \&
  {Siess}}]{palacios06}
{Palacios}, A., {Charbonnel}, C., {Talon}, S., \& {Siess}, L. 2006, A\&A, 453,
  261

\bibitem[{{Palacios} {et~al.}(2003){Palacios}, {Talon}, {Charbonnel}, \&
  {Forestini}}]{palacios03}
{Palacios}, A., {Talon}, S., {Charbonnel}, C., \& {Forestini}, M. 2003, A\&A,
  399, 603

\bibitem[{{Paladini} {et~al.}(2012){Paladini}, {Sacuto}, {Klotz}, {Ohnaka},
  {Wittkowski}, {Nowotny}, {Jorissen}, \& {Hron}}]{paladini12}
{Paladini}, C., {Sacuto}, S., {Klotz}, D., {Ohnaka}, K., {Wittkowski}, M.,
  {Nowotny}, W., {Jorissen}, A., \& {Hron}, J. 2012, A\&A, 544, L5

\bibitem[{{Palla} {et~al.}(2000){Palla}, {Bachiller}, {Stanghellini}, {Tosi},
  \& {Galli}}]{palla00}
{Palla}, F., {Bachiller}, R., {Stanghellini}, L., {Tosi}, M., \& {Galli}, D.
  2000, A\&A, 355, 69

\bibitem[{{Palla} {et~al.}(2002){Palla}, {Galli}, {Marconi}, {Stanghellini}, \&
  {Tosi}}]{palla02}
{Palla}, F., {Galli}, D., {Marconi}, A., {Stanghellini}, L., \& {Tosi}, M.
  2002, ApJL, 568, L57

\bibitem[{{Palmerini} {et~al.}(2009){Palmerini}, {Busso}, {Maiorca}, \&
  {Guandalini}}]{palmerini09}
{Palmerini}, S., {Busso}, M., {Maiorca}, E., \& {Guandalini}, R. 2009, PASA,
  26, 161

\bibitem[{{Palmerini} {et~al.}(2011){Palmerini}, {La Cognata}, {Cristallo}, \&
  {Busso}}]{palmerini11}
{Palmerini}, S., {La Cognata}, M., {Cristallo}, S., \& {Busso}, M. 2011, ApJ,
  729, 3

\bibitem[{{Pandey}(2006)}]{pandey06}
{Pandey}, G. 2006, ApJL, 648, L143

\bibitem[{{Pe{\~n}a} {et~al.}(2013){Pe{\~n}a}, {Rechy-Garc{\'{\i}}a}, \&
  {Garc{\'{\i}}a-Rojas}}]{pena13}
{Pe{\~n}a}, M., {Rechy-Garc{\'{\i}}a}, J.~S., \& {Garc{\'{\i}}a-Rojas}, J.
  2013, Revista Mexicana de Astronomia y Astrofisica, 49, 87

\bibitem[{{Peimbert}(1978)}]{peimbert78}
{Peimbert}, M. 1978, in IAU Symposium, Vol.~76, Planetary Nebulae, ed.
  Y.~{Terzian}, 215--223

\bibitem[{{Peimbert}(1990)}]{peimbert90}
{Peimbert}, M. 1990, Reports of Progress in Physics, 53, 1559

\bibitem[{{Peimbert} \& {Torres-Peimbert}(1987)}]{peimbert87}
{Peimbert}, M. \& {Torres-Peimbert}, S. 1987, Revista Mexicana de Astronomia y
  Astrofisica, vol.~14, 14, 540

\bibitem[{{Pereira} {et~al.}(2012){Pereira}, {Gallino}, \&
  {Bisterzo}}]{pereira12}
{Pereira}, C.~B., {Gallino}, R., \& {Bisterzo}, S. 2012, A\&A, 538, A48

\bibitem[{{Petrovic} {et~al.}(2006){Petrovic}, {Pols}, \&
  {Langer}}]{petrovic06}
{Petrovic}, J., {Pols}, O., \& {Langer}, N. 2006, A\&A, 450, 219

\bibitem[{{Picardi} {et~al.}(2004){Picardi}, {Chieffi}, {Limongi}, {Pisanti},
  {Miele}, {Mangano}, \& {Imbriani}}]{picardi04}
{Picardi}, I., {Chieffi}, A., {Limongi}, M., {Pisanti}, O., {Miele}, G.,
  {Mangano}, G., \& {Imbriani}, G. 2004, ApJ, 609, 1035

\bibitem[{{Piersanti} {et~al.}(2013){Piersanti}, {Cristallo}, \&
  {Straniero}}]{piersanti13}
{Piersanti}, L., {Cristallo}, S., \& {Straniero}, O. 2013, ApJ, 774, 98

\bibitem[{{Pignatari} {et~al.}(2010){Pignatari}, {Gallino}, {Heil}, {Wiescher},
  {K{\"a}ppeler}, {Herwig}, \& {Bisterzo}}]{pignatari10}
{Pignatari}, M., {Gallino}, R., {Heil}, M., {Wiescher}, M., {K{\"a}ppeler}, F.,
  {Herwig}, F., \& {Bisterzo}, S. 2010, ApJ, 710, 1557

\bibitem[{{Pignatari} {et~al.}(2013){Pignatari}, {Herwig}, {Hirschi},
  {Bennett}, {Rockefeller}, {Fryer}, {Timmes}, {Heger}, {Jones}, {Battino},
  {Ritter}, {Dotter}, {Trappitsch}, {Diehl}, {Frischknecht}, {Hungerford},
  {Magkotsios}, {Travaglio}, \& {Young}}]{pignatari13}
{Pignatari}, M., {Herwig}, F., {Hirschi}, R., {Bennett}, M., {Rockefeller}, G.,
  {Fryer}, C., {Timmes}, F.~X., {Heger}, A., {Jones}, S., {Battino}, U.,
  {Ritter}, C., {Dotter}, A., {Trappitsch}, R., {Diehl}, S., {Frischknecht},
  U., {Hungerford}, A., {Magkotsios}, G., {Travaglio}, C., \& {Young}, P. 2013,
  ApJS, submitted

\bibitem[{{Pilachowski} {et~al.}(1993){Pilachowski}, {Sneden}, \&
  {Booth}}]{pilachowski93}
{Pilachowski}, C.~A., {Sneden}, C., \& {Booth}, J. 1993, ApJ, 407, 699

\bibitem[{{Pilachowski} {et~al.}(1996){Pilachowski}, {Sneden}, {Kraft}, \&
  {Langer}}]{pilachowski96}
{Pilachowski}, C.~A., {Sneden}, C., {Kraft}, R.~P., \& {Langer}, G.~E. 1996,
  AJ, 112, 545

\bibitem[{{Piotto} {et~al.}(2012){Piotto}, {Milone}, {Anderson}, {Bedin},
  {Bellini}, {Cassisi}, {Marino}, {Aparicio}, \& {Nascimbeni}}]{piotto12}
{Piotto}, G., {Milone}, A.~P., {Anderson}, J., {Bedin}, L.~R., {Bellini}, A.,
  {Cassisi}, S., {Marino}, A.~F., {Aparicio}, A., \& {Nascimbeni}, V. 2012,
  ApJ, 760, 39

\bibitem[{{Piotto} {et~al.}(2005){Piotto}, {Villanova}, {Bedin}, {Gratton},
  {Cassisi}, {Momany}, {Recio-Blanco}, {Lucatello}, {Anderson}, {King},
  {Pietrinferni}, \& {Carraro}}]{piotto05}
{Piotto}, G., {Villanova}, S., {Bedin}, L.~R., {Gratton}, R., {Cassisi}, S.,
  {Momany}, Y., {Recio-Blanco}, A., {Lucatello}, S., {Anderson}, J., {King},
  I.~R., {Pietrinferni}, A., \& {Carraro}, G. 2005, ApJ, 621, 777

\bibitem[{{Plez} {et~al.}(1993){Plez}, {Smith}, \& {Lambert}}]{plez93}
{Plez}, B., {Smith}, V.~V., \& {Lambert}, D.~L. 1993, ApJ, 418, 812

\bibitem[{{Poelarends} {et~al.}(2008){Poelarends}, {Herwig}, {Langer}, \&
  {Heger}}]{poelarends08}
{Poelarends}, A.~J.~T., {Herwig}, F., {Langer}, N., \& {Heger}, A. 2008, ApJ,
  675, 614

\bibitem[{{Pols} {et~al.}(2012){Pols}, {Izzard}, {Stancliffe}, \&
  {Glebbeek}}]{pols12}
{Pols}, O.~R., {Izzard}, R.~G., {Stancliffe}, R.~J., \& {Glebbeek}, E. 2012,
  A\&A, 547, A76

\bibitem[{{Prantzos}(2012)}]{prantzos12}
{Prantzos}, N. 2012, A\&A, 542, A67

\bibitem[{{Prantzos} {et~al.}(2007){Prantzos}, {Charbonnel}, \&
  {Iliadis}}]{prantzos07}
{Prantzos}, N., {Charbonnel}, C., \& {Iliadis}, C. 2007, A\&A, 470, 179

\bibitem[{{Pumo} {et~al.}(2008){Pumo}, {D'Antona}, \& {Ventura}}]{pumo08}
{Pumo}, M.~L., {D'Antona}, F., \& {Ventura}, P. 2008, ApJL, 672, L25

\bibitem[{{Raiteri} {et~al.}(1999){Raiteri}, {Villata}, {Gallino}, {Busso}, \&
  {Cravanzola}}]{raiteri99}
{Raiteri}, C.~M., {Villata}, M., {Gallino}, R., {Busso}, M., \& {Cravanzola},
  A. 1999, ApJL, 518, L91

\bibitem[{{Recio-Blanco} {et~al.}(2012){Recio-Blanco}, {de Laverny}, {Worley},
  {Santos}, {Melo}, \& {Israelian}}]{recio-blanco12}
{Recio-Blanco}, A., {de Laverny}, P., {Worley}, C., {Santos}, N.~C., {Melo},
  C., \& {Israelian}, G. 2012, A\&A, 538, A117

\bibitem[{{Reimers}(1975)}]{reimers75}
{Reimers}, D. 1975, {Circumstellar envelopes and mass loss of red giant stars}
  (Problems in stellar atmospheres and envelopes.), 229--256

\bibitem[{{Renda} {et~al.}(2004){Renda}, {Fenner}, {Gibson}, {Karakas},
  {Lattanzio}, {Campbell}, {Chieffi}, {Cunha}, \& {Smith}}]{renda04}
{Renda}, A., {Fenner}, Y., {Gibson}, B.~K., {Karakas}, A.~I., {Lattanzio},
  J.~C., {Campbell}, S., {Chieffi}, A., {Cunha}, K., \& {Smith}, V.~V. 2004,
  MNRAS, 354, 575

\bibitem[{{Renzini} \& {Fusi Pecci}(1988)}]{renzini88}
{Renzini}, A. \& {Fusi Pecci}, F. 1988, ARA\&A, 26, 199

\bibitem[{{Renzini} \& {Voli}(1981)}]{renzini81}
{Renzini}, A. \& {Voli}, M. 1981, A\&A, 94, 175

\bibitem[{{Reyniers} {et~al.}(2007){Reyniers}, {Abia}, {van Winckel}, {Lloyd
  Evans}, {Decin}, {Eriksson}, \& {Pollard}}]{reyniers07a}
{Reyniers}, M., {Abia}, C., {van Winckel}, H., {Lloyd Evans}, T., {Decin}, L.,
  {Eriksson}, K., \& {Pollard}, K.~R. 2007, A\&A, 461, 641

\bibitem[{{Riebel} {et~al.}(2012){Riebel}, {Srinivasan}, {Sargent}, \&
  {Meixner}}]{riebel12}
{Riebel}, D., {Srinivasan}, S., {Sargent}, B., \& {Meixner}, M. 2012, ApJ, 753,
  71

\bibitem[{{Riello} {et~al.}(2003){Riello}, {Cassisi}, {Piotto}, {Recio-Blanco},
  {De Angeli}, {Salaris}, {Pietrinferni}, {Bono}, \& {Zoccali}}]{riello03}
{Riello}, M., {Cassisi}, S., {Piotto}, G., {Recio-Blanco}, A., {De Angeli}, F.,
  {Salaris}, M., {Pietrinferni}, A., {Bono}, G., \& {Zoccali}, M. 2003, A\&A,
  410, 553

\bibitem[{{Ritossa} {et~al.}(1996){Ritossa}, {Garcia-Berro}, \&
  {Iben}}]{ritossa96}
{Ritossa}, C., {Garcia-Berro}, E., \& {Iben}, Jr., I. 1996, ApJ, 460, 489

\bibitem[{{Ritossa} {et~al.}(1999){Ritossa}, {Garc{\'{\i}}a-Berro}, \&
  {Iben}}]{ritossa99}
{Ritossa}, C., {Garc{\'{\i}}a-Berro}, E., \& {Iben}, Jr., I. 1999, ApJ, 515,
  381

\bibitem[{{Rolfs} \& {Rodney}(1988)}]{rolfs88}
{Rolfs}, C.~E. \& {Rodney}, W.~S. 1988, {Cauldrons in the cosmos: Nuclear
  astrophysics} (University of Chicago Press)

\bibitem[{{Romano} {et~al.}(2010){Romano}, {Karakas}, {Tosi}, \&
  {Matteucci}}]{romano10}
{Romano}, D., {Karakas}, A.~I., {Tosi}, M., \& {Matteucci}, F. 2010, A\&A, 522,
  A32

\bibitem[{{Romano} \& {Matteucci}(2003)}]{romano03a}
{Romano}, D. \& {Matteucci}, F. 2003, MNRAS, 342, 185

\bibitem[{{Romano} {et~al.}(2001){Romano}, {Matteucci}, {Ventura}, \&
  {D'Antona}}]{romano01}
{Romano}, D., {Matteucci}, F., {Ventura}, P., \& {D'Antona}, F. 2001, A\&A,
  374, 646

\bibitem[{{Rosenblum} {et~al.}(2011){Rosenblum}, {Garaud}, {Traxler}, \&
  {Stellmach}}]{rosenblum11}
{Rosenblum}, E., {Garaud}, P., {Traxler}, A., \& {Stellmach}, S. 2011, ApJ,
  731, 66

\bibitem[{{Rubin} {et~al.}(2004){Rubin}, {Ferland}, {Chollet}, \&
  {Horstmeyer}}]{rubin04}
{Rubin}, R.~H., {Ferland}, G.~J., {Chollet}, E.~E., \& {Horstmeyer}, R. 2004,
  ApJ, 605, 784

\bibitem[{{Sackmann} \& {Boothroyd}(1999)}]{sackboo99}
{Sackmann}, I.-J. \& {Boothroyd}, A.~I. 1999, ApJ, 510, 217

\bibitem[{{S{\'a}nchez-Bl{\'a}zquez} {et~al.}(2012){S{\'a}nchez-Bl{\'a}zquez},
  {Marcolini}, {Gibson}, {Karakas}, {Pilkington}, \& {Calura}}]{sanchez12}
{S{\'a}nchez-Bl{\'a}zquez}, P., {Marcolini}, A., {Gibson}, B.~K., {Karakas},
  A.~I., {Pilkington}, K., \& {Calura}, F. 2012, MNRAS, 419, 1376

\bibitem[{{Santrich} {et~al.}(2013){Santrich}, {Pereira}, \&
  {Drake}}]{santrich13}
{Santrich}, O.~J.~K., {Pereira}, C.~B., \& {Drake}, N.~A. 2013, A\&A, 554, A2

\bibitem[{{Scalo} {et~al.}(1975){Scalo}, {Despain}, \& {Ulrich}}]{scalo75}
{Scalo}, J.~M., {Despain}, K.~H., \& {Ulrich}, R.~K. 1975, ApJ, 196, 805

\bibitem[{{Schlattl} {et~al.}(2001){Schlattl}, {Cassisi}, {Salaris}, \&
  {Weiss}}]{schlattl01}
{Schlattl}, H., {Cassisi}, S., {Salaris}, M., \& {Weiss}, A. 2001, ApJ, 559,
  1082

\bibitem[{{Schlattl} {et~al.}(2002){Schlattl}, {Salaris}, {Cassisi}, \&
  {Weiss}}]{schlattl02}
{Schlattl}, H., {Salaris}, M., {Cassisi}, S., \& {Weiss}, A. 2002, A\&A, 395,
  77

\bibitem[{{Schr{\"o}der} \& {Cuntz}(2005)}]{schroder05}
{Schr{\"o}der}, K.-P. \& {Cuntz}, M. 2005, ApJL, 630, L73

\bibitem[{{Schr{\"o}der} \& {Cuntz}(2007)}]{schroder07}
---. 2007, A\&A, 465, 593

\bibitem[{{Schuler} {et~al.}(2007){Schuler}, {Cunha}, {Smith}, {Sivarani},
  {Beers}, \& {Lee}}]{schuler07}
{Schuler}, S.~C., {Cunha}, K., {Smith}, V.~V., {Sivarani}, T., {Beers}, T.~C.,
  \& {Lee}, Y.~S. 2007, ApJL, 667, L81

\bibitem[{{Schuler} {et~al.}(2009){Schuler}, {King}, \& {The}}]{schuler09}
{Schuler}, S.~C., {King}, J.~R., \& {The}, L.-S. 2009, ApJ, 701, 837

\bibitem[{{Seitenzahl} {et~al.}(2010){Seitenzahl}, {R{\"o}pke}, {Fink}, \&
  {Pakmor}}]{seit10}
{Seitenzahl}, I.~R., {R{\"o}pke}, F.~K., {Fink}, M., \& {Pakmor}, R. 2010,
  MNRAS, 407, 2297

\bibitem[{{Sergi} {et~al.}(2010){Sergi}, {Spitaleri}, {La Cognata}, {Coc},
  {Mukhamedzhanov}, {Burjan}, {Cherubini}, {Crucill{\'a}}, {Gulino},
  {Hammache}, {Hons}, {Irgaziev}, {Kiss}, {Kroha}, {Lamia}, {Pizzone},
  {Puglia}, {Rapisarda}, {Romano}, {de S{\'e}r{\'e}ville}, {Somorjai},
  {Tudisco}, \& {Tumino}}]{sergi10}
{Sergi}, M.~L., {Spitaleri}, C., {La Cognata}, M., {Coc}, A., {Mukhamedzhanov},
  A., {Burjan}, S.~V., {Cherubini}, S., {Crucill{\'a}}, V., {Gulino}, M.,
  {Hammache}, F., {Hons}, Z., {Irgaziev}, B., {Kiss}, G.~G., {Kroha}, V.,
  {Lamia}, L., {Pizzone}, R.~G., {Puglia}, S.~M.~R., {Rapisarda}, G.~G.,
  {Romano}, S., {de S{\'e}r{\'e}ville}, N., {Somorjai}, E., {Tudisco}, S., \&
  {Tumino}, A. 2010, Phys. Rev. C, 82, 032801

\bibitem[{{Serminato} {et~al.}(2009){Serminato}, {Gallino}, {Travaglio},
  {Bisterzo}, \& {Straniero}}]{serminato09}
{Serminato}, A., {Gallino}, R., {Travaglio}, C., {Bisterzo}, S., \&
  {Straniero}, O. 2009, PASA, 26, 153

\bibitem[{{Shaw} {et~al.}(2006){Shaw}, {Stanghellini}, {Villaver}, \&
  {Mutchler}}]{shaw06}
{Shaw}, R.~A., {Stanghellini}, L., {Villaver}, E., \& {Mutchler}, M. 2006,
  ApJS, 167, 201

\bibitem[{{Shetrone}(1996{\natexlab{a}})}]{shetrone96a}
{Shetrone}, M.~D. 1996{\natexlab{a}}, AJ, 112, 1517

\bibitem[{{Shetrone}(1996{\natexlab{b}})}]{shetrone96b}
---. 1996{\natexlab{b}}, AJ, 112, 2639

\bibitem[{{Shingles} \& {Karakas}(2013)}]{shingles13}
{Shingles}, L.~J. \& {Karakas}, A.~I. 2013, MNRAS, 431, 2861

\bibitem[{{Siess}(2007)}]{siess07}
{Siess}, L. 2007, A\&A, 476, 893

\bibitem[{{Siess}(2010)}]{siess10}
---. 2010, A\&A, 512, A10

\bibitem[{{Siess} \& {Livio}(1999{\natexlab{a}})}]{siess99a}
{Siess}, L. \& {Livio}, M. 1999{\natexlab{a}}, MNRAS, 304, 925

\bibitem[{{Siess} \& {Livio}(1999{\natexlab{b}})}]{siess99b}
---. 1999{\natexlab{b}}, MNRAS, 308, 1133

\bibitem[{{Siess} {et~al.}(2002){Siess}, {Livio}, \& {Lattanzio}}]{siess02}
{Siess}, L., {Livio}, M., \& {Lattanzio}, J. 2002, ApJ, 570, 329

\bibitem[{{Simmerer} {et~al.}(2004){Simmerer}, {Sneden}, {Cowan}, {Collier},
  {Woolf}, \& {Lawler}}]{simmerer04}
{Simmerer}, J., {Sneden}, C., {Cowan}, J.~J., {Collier}, J., {Woolf}, V.~M., \&
  {Lawler}, J.~E. 2004, ApJ, 617, 1091

\bibitem[{{Sivarani} {et~al.}(2006){Sivarani}, {Beers}, {Bonifacio}, {Molaro},
  {Cayrel}, {Herwig}, {Spite}, {Spite}, {Plez}, {Andersen}, {Barbuy},
  {Depagne}, {Hill}, {Fran{\c c}ois}, {Nordstr{\"o}m}, \&
  {Primas}}]{sivarani06}
{Sivarani}, T., {Beers}, T.~C., {Bonifacio}, P., {Molaro}, P., {Cayrel}, R.,
  {Herwig}, F., {Spite}, M., {Spite}, F., {Plez}, B., {Andersen}, J., {Barbuy},
  B., {Depagne}, E., {Hill}, V., {Fran{\c c}ois}, P., {Nordstr{\"o}m}, B., \&
  {Primas}, F. 2006, A\&A, 459, 125

\bibitem[{{Sloan} {et~al.}(2008){Sloan}, {Kraemer}, {Wood}, {Zijlstra},
  {Bernard-Salas}, {Devost}, \& {Houck}}]{sloan08}
{Sloan}, G.~C., {Kraemer}, K.~E., {Wood}, P.~R., {Zijlstra}, A.~A.,
  {Bernard-Salas}, J., {Devost}, D., \& {Houck}, J.~R. 2008, ApJ, 686, 1056

\bibitem[{{Sloan} {et~al.}(2012){Sloan}, {Matsuura}, {Lagadec}, {van Loon},
  {Kraemer}, {McDonald}, {Groenewegen}, {Wood}, {Bernard-Salas}, \&
  {Zijlstra}}]{sloan12}
{Sloan}, G.~C., {Matsuura}, M., {Lagadec}, E., {van Loon}, J.~T., {Kraemer},
  K.~E., {McDonald}, I., {Groenewegen}, M.~A.~T., {Wood}, P.~R.,
  {Bernard-Salas}, J., \& {Zijlstra}, A.~A. 2012, ApJ, 752, 140

\bibitem[{{Sloan} {et~al.}(2009){Sloan}, {Matsuura}, {Zijlstra}, {Lagadec},
  {Groenewegen}, {Wood}, {Szyszka}, {Bernard-Salas}, \& {van Loon}}]{sloan09}
{Sloan}, G.~C., {Matsuura}, M., {Zijlstra}, A.~A., {Lagadec}, E.,
  {Groenewegen}, M.~A.~T., {Wood}, P.~R., {Szyszka}, C., {Bernard-Salas}, J.,
  \& {van Loon}, J.~T. 2009, Science, 323, 353

\bibitem[{{Smiljanic}(2012)}]{smiljanic12}
{Smiljanic}, R. 2012, MNRAS, 422, 1562

\bibitem[{{Smiljanic} {et~al.}(2009){Smiljanic}, {Gauderon}, {North}, {Barbuy},
  {Charbonnel}, \& {Mowlavi}}]{smiljanic09}
{Smiljanic}, R., {Gauderon}, R., {North}, P., {Barbuy}, B., {Charbonnel}, C.,
  \& {Mowlavi}, N. 2009, A\&A, 502, 267

\bibitem[{{Smith} \& {Martell}(2003)}]{gsmith03}
{Smith}, G.~H. \& {Martell}, S.~L. 2003, PASP, 115, 1211

\bibitem[{{Smith} \& {Tout}(1992)}]{gsmith92}
{Smith}, G.~H. \& {Tout}, C.~A. 1992, MNRAS, 256, 449

\bibitem[{{Smith} \& {Lambert}(1986)}]{smith86b}
{Smith}, V.~V. \& {Lambert}, D.~L. 1986, ApJ, 311, 843

\bibitem[{{Smith} \& {Lambert}(1988)}]{smith88}
---. 1988, ApJ, 333, 219

\bibitem[{{Smith} \& {Lambert}(1989)}]{smith89}
---. 1989, ApJL, 345, L75

\bibitem[{{Smith} \& {Lambert}(1990{\natexlab{a}})}]{smith90b}
---. 1990{\natexlab{a}}, ApJL, 361, L69

\bibitem[{{Smith} \& {Lambert}(1990{\natexlab{b}})}]{smith90a}
---. 1990{\natexlab{b}}, ApJS, 72, 387

\bibitem[{{Smith} {et~al.}(1987){Smith}, {Lambert}, \& {McWilliam}}]{smith87}
{Smith}, V.~V., {Lambert}, D.~L., \& {McWilliam}, A. 1987, ApJ, 320, 862

\bibitem[{{Sneden} {et~al.}(2008){Sneden}, {Cowan}, \& {Gallino}}]{sneden08}
{Sneden}, C., {Cowan}, J.~J., \& {Gallino}, R. 2008, ARA\&A, 46, 241

\bibitem[{{Spiegel}(1972)}]{spiegel72}
{Spiegel}, E.~A. 1972, ARA\&A, 10, 261

\bibitem[{{Stancliffe}(2010)}]{stancliffe10}
{Stancliffe}, R.~J. 2010, MNRAS, 403, 505

\bibitem[{{Stancliffe} {et~al.}(2009){Stancliffe}, {Church}, {Angelou}, \&
  {Lattanzio}}]{stancliffe09}
{Stancliffe}, R.~J., {Church}, R.~P., {Angelou}, G.~C., \& {Lattanzio}, J.~C.
  2009, MNRAS, 396, 2313

\bibitem[{{Stancliffe} {et~al.}(2011){Stancliffe}, {Dearborn}, {Lattanzio},
  {Heap}, \& {Campbell}}]{stancliffe11}
{Stancliffe}, R.~J., {Dearborn}, D.~S.~P., {Lattanzio}, J.~C., {Heap}, S.~A.,
  \& {Campbell}, S.~W. 2011, ApJ, 742, 121

\bibitem[{{Stancliffe} \& {Glebbeek}(2008)}]{stancliffe08}
{Stancliffe}, R.~J. \& {Glebbeek}, E. 2008, MNRAS, 389, 1828

\bibitem[{{Stancliffe} {et~al.}(2005){Stancliffe}, {Izzard}, \&
  {Tout}}]{stancliffe05a}
{Stancliffe}, R.~J., {Izzard}, R.~G., \& {Tout}, C.~A. 2005, MNRAS, 356, L1

\bibitem[{{Stancliffe} \& {Jeffery}(2007)}]{stancliffe07a}
{Stancliffe}, R.~J. \& {Jeffery}, C.~S. 2007, MNRAS, 375, 1280

\bibitem[{{Stanghellini} {et~al.}(2006){Stanghellini}, {Guerrero}, {Cunha},
  {Manchado}, \& {Villaver}}]{stanghellini06}
{Stanghellini}, L., {Guerrero}, M.~A., {Cunha}, K., {Manchado}, A., \&
  {Villaver}, E. 2006, ApJ, 651, 898

\bibitem[{{Stanghellini} \& {Haywood}(2010)}]{stanghellini10}
{Stanghellini}, L. \& {Haywood}, M. 2010, ApJ, 714, 1096

\bibitem[{{Stanghellini} {et~al.}(2000){Stanghellini}, {Shaw}, {Balick}, \&
  {Blades}}]{stanghellini00}
{Stanghellini}, L., {Shaw}, R.~A., {Balick}, B., \& {Blades}, J.~C. 2000, ApJL,
  534, L167

\bibitem[{{Stasi{\'n}ska} {et~al.}(2013){Stasi{\'n}ska}, {Pe{\~n}a},
  {Bresolin}, \& {Tsamis}}]{stasinska13}
{Stasi{\'n}ska}, G., {Pe{\~n}a}, M., {Bresolin}, F., \& {Tsamis}, Y.~G. 2013,
  A\&A, 552, A12

\bibitem[{{Stasi{\'n}ska} {et~al.}(1998){Stasi{\'n}ska}, {Richer}, \&
  {McCall}}]{stasinska98}
{Stasi{\'n}ska}, G., {Richer}, M.~G., \& {McCall}, M.~L. 1998, A\&A, 336, 667

\bibitem[{{Sterling} \& {Dinerstein}(2008)}]{sterling08}
{Sterling}, N.~C. \& {Dinerstein}, H.~L. 2008, ApJS, 174, 158

\bibitem[{{Straniero} {et~al.}(1997){Straniero}, {Chieffi}, {Limongi}, {Busso},
  {Gallino}, \& {Arlandini}}]{straniero97}
{Straniero}, O., {Chieffi}, A., {Limongi}, M., {Busso}, M., {Gallino}, R., \&
  {Arlandini}, C. 1997, ApJ, 478, 332

\bibitem[{{Straniero} {et~al.}(2014){Straniero}, {Cristallo}, \&
  {Piersanti}}]{straniero14}
{Straniero}, O., {Cristallo}, S., \& {Piersanti}, L. 2014, ApJ, 785, 77

\bibitem[{{Straniero} {et~al.}(2003{\natexlab{a}}){Straniero},
  {Dom{\'{\i}}nguez}, {Cristallo}, \& {Gallino}}]{straniero03}
{Straniero}, O., {Dom{\'{\i}}nguez}, I., {Cristallo}, R., \& {Gallino}, R.
  2003{\natexlab{a}}, PASA, 20, 389

\bibitem[{{Straniero} {et~al.}(2003{\natexlab{b}}){Straniero},
  {Dom{\'{\i}}nguez}, {Imbriani}, \& {Piersanti}}]{straniero03b}
{Straniero}, O., {Dom{\'{\i}}nguez}, I., {Imbriani}, G., \& {Piersanti}, L.
  2003{\natexlab{b}}, ApJ, 583, 878

\bibitem[{{Straniero} {et~al.}(1995){Straniero}, {Gallino}, {Busso}, {Chiefei},
  {Raiteri}, {Limongi}, \& {Salaris}}]{straniero95}
{Straniero}, O., {Gallino}, R., {Busso}, M., {Chiefei}, A., {Raiteri}, C.~M.,
  {Limongi}, M., \& {Salaris}, M. 1995, ApJ, 440, L85

\bibitem[{{Straniero} {et~al.}(2013){Straniero}, {Imbriani}, {Strieder},
  {Bemmerer}, {Broggini}, {Caciolli}, {Corvisiero}, {Costantini}, {Cristallo},
  \& {et. al.}}]{straniero13}
{Straniero}, O., {Imbriani}, G., {Strieder}, F., {Bemmerer}, D., {Broggini},
  C., {Caciolli}, A., {Corvisiero}, P., {Costantini}, H., {Cristallo}, S., \&
  {et. al.} 2013, ApJ, 763, 100

\bibitem[{{Suda} {et~al.}(2004){Suda}, {Aikawa}, {Machida}, {Fujimoto}, \&
  {Iben}}]{suda04}
{Suda}, T., {Aikawa}, M., {Machida}, M.~N., {Fujimoto}, M.~Y., \& {Iben}, Jr.,
  I. 2004, ApJ, 611, 476

\bibitem[{{Suda} \& {Fujimoto}(2010)}]{suda10}
{Suda}, T. \& {Fujimoto}, M.~Y. 2010, MNRAS, 405, 177

\bibitem[{{Suda} {et~al.}(2007){Suda}, {Fujimoto}, \& {Itoh}}]{suda07}
{Suda}, T., {Fujimoto}, M.~Y., \& {Itoh}, N. 2007, ApJ, 667, 1206

\bibitem[{{Suda} {et~al.}(2008){Suda}, {Katsuta}, {Yamada}, {Suwa}, {Ishizuka},
  {Komiya}, {Sorai}, {Aikawa}, \& {Fujimoto}}]{suda08}
{Suda}, T., {Katsuta}, Y., {Yamada}, S., {Suwa}, T., {Ishizuka}, C., {Komiya},
  Y., {Sorai}, K., {Aikawa}, M., \& {Fujimoto}, M.~Y. 2008, PASJ, 60, 1159

\bibitem[{{Suda} {et~al.}(2011){Suda}, {Yamada}, {Katsuta}, {Komiya},
  {Ishizuka}, {Aoki}, \& {Fujimoto}}]{suda11}
{Suda}, T., {Yamada}, S., {Katsuta}, Y., {Komiya}, Y., {Ishizuka}, C., {Aoki},
  W., \& {Fujimoto}, M.~Y. 2011, MNRAS, 412, 843

\bibitem[{{Suijs} {et~al.}(2008){Suijs}, {Langer}, {Poelarends}, {Yoon},
  {Heger}, \& {Herwig}}]{suijs08}
{Suijs}, M.~P.~L., {Langer}, N., {Poelarends}, A., {Yoon}, S., {Heger}, A., \&
  {Herwig}, F. 2008, A\&A, 481, L87

\bibitem[{{Surman} {et~al.}(2008){Surman}, {McLaughlin}, {Ruffert}, {Janka}, \&
  {Hix}}]{surman08}
{Surman}, R., {McLaughlin}, G.~C., {Ruffert}, M., {Janka}, H.-T., \& {Hix},
  W.~R. 2008, ApJL, 679, L117

\bibitem[{{Sweigart}(1978)}]{sweigart78}
{Sweigart}, A.~V. 1978, in IAU Symposium, Vol.~80, The HR Diagram - The 100th
  Anniversary of Henry Norris Russell, ed. A.~G.~D. {Philip} \& D.~S. {Hayes},
  333--343

\bibitem[{{Sweigart}(1997)}]{sweigart97}
{Sweigart}, A.~V. 1997, ApJL, 474, L23

\bibitem[{{Sweigart} \& {Demarque}(1973)}]{sweidem73}
{Sweigart}, A.~V. \& {Demarque}, P. 1973, in Astrophysics and Space Science
  Library, Vol.~36, IAU Colloq. 21: Variable Stars in Globular Clusters and in
  Related Systems, ed. J.~D. {Fernie}, 221

\bibitem[{{Sweigart} \& {Mengel}(1979)}]{sweigart79}
{Sweigart}, A.~V. \& {Mengel}, J.~G. 1979, ApJ, 229, 624

\bibitem[{{Taam} \& {Ricker}(2010)}]{taam10}
{Taam}, R.~E. \& {Ricker}, P.~M. 2010, NewAstR, 54, 65

\bibitem[{{Tassoul}(2007)}]{tassoul07}
{Tassoul}, J.-L. 2007, {Stellar Rotation} (Cambridge University Press)

\bibitem[{{Tassoul} \& {Tassoul}(1995)}]{tassoul95}
{Tassoul}, M. \& {Tassoul}, J.-L. 1995, ApJ, 440, 789

\bibitem[{{Tautvai{\v s}ien{\.e}} {et~al.}(2013){Tautvai{\v s}ien{\.e}},
  {Barisevi{\v c}ius}, {Chorniy}, {Ilyin}, \& {Puzeras}}]{taut13}
{Tautvai{\v s}ien{\.e}}, G., {Barisevi{\v c}ius}, G., {Chorniy}, Y., {Ilyin},
  I., \& {Puzeras}, E. 2013, MNRAS, 430, 621

\bibitem[{{The} {et~al.}(2000){The}, {El Eid}, \& {Meyer}}]{the00}
{The}, L.-S., {El Eid}, M.~F., \& {Meyer}, B.~S. 2000, ApJ, 533, 998

\bibitem[{{The} {et~al.}(2007){The}, {El Eid}, \& {Meyer}}]{the07}
---. 2007, ApJ, 655, 1058

\bibitem[{{Thielemann} {et~al.}(2011){Thielemann}, {Arcones}, {K{\"a}ppeli},
  {Liebend{\"o}rfer}, {Rauscher}, {Winteler}, {Fr{\"o}hlich}, {Dillmann},
  {Fischer}, {Martinez-Pinedo}, {Langanke}, {Farouqi}, {Kratz}, {Panov}, \&
  {Korneev}}]{thielemann11}
{Thielemann}, F.-K., {Arcones}, A., {K{\"a}ppeli}, R., {Liebend{\"o}rfer}, M.,
  {Rauscher}, T., {Winteler}, C., {Fr{\"o}hlich}, C., {Dillmann}, I.,
  {Fischer}, T., {Martinez-Pinedo}, G., {Langanke}, K., {Farouqi}, K., {Kratz},
  K.-L., {Panov}, I., \& {Korneev}, I.~K. 2011, Progress in Particle and
  Nuclear Physics, 66, 346

\bibitem[{{Timmes} {et~al.}(1995){Timmes}, {Woosley}, \& {Weaver}}]{timmes95}
{Timmes}, F.~X., {Woosley}, S.~E., \& {Weaver}, T.~A. 1995, ApJS, 98, 617

\bibitem[{{Tinsley}(1980)}]{tinsley80}
{Tinsley}, B.~M. 1980, Fundam. Cosmic Phys., 5, 287

\bibitem[{{Tonini} {et~al.}(2009){Tonini}, {Maraston}, {Devriendt}, {Thomas},
  \& {Silk}}]{tonini09}
{Tonini}, C., {Maraston}, C., {Devriendt}, J., {Thomas}, D., \& {Silk}, J.
  2009, MNRAS, 396, L36

\bibitem[{{Travaglio} {et~al.}(1999){Travaglio}, {Galli}, {Gallino}, {Busso},
  {Ferrini}, \& {Straniero}}]{travaglio99}
{Travaglio}, C., {Galli}, D., {Gallino}, R., {Busso}, M., {Ferrini}, F., \&
  {Straniero}, O. 1999, ApJ, 521, 691

\bibitem[{{Travaglio} {et~al.}(2004){Travaglio}, {Gallino}, {Arnone}, {Cowan},
  {Jordan}, \& {Sneden}}]{travaglio04}
{Travaglio}, C., {Gallino}, R., {Arnone}, E., {Cowan}, J., {Jordan}, F., \&
  {Sneden}, C. 2004, ApJ, 601, 864

\bibitem[{{Travaglio} {et~al.}(2001{\natexlab{a}}){Travaglio}, {Gallino},
  {Busso}, \& {Gratton}}]{travaglio01a}
{Travaglio}, C., {Gallino}, R., {Busso}, M., \& {Gratton}, R.
  2001{\natexlab{a}}, ApJ, 549, 346

\bibitem[{{Travaglio} {et~al.}(2001{\natexlab{b}}){Travaglio}, {Randich},
  {Galli}, {Lattanzio}, {Elliott}, {Forestini}, \& {Ferrini}}]{travaglio01b}
{Travaglio}, C., {Randich}, S., {Galli}, D., {Lattanzio}, J., {Elliott}, L.~M.,
  {Forestini}, M., \& {Ferrini}, F. 2001{\natexlab{b}}, ApJ, 559, 909

\bibitem[{{Traxler} {et~al.}(2011){Traxler}, {Garaud}, \&
  {Stellmach}}]{traxler11}
{Traxler}, A., {Garaud}, P., \& {Stellmach}, S. 2011, ApJL, 728, L29

\bibitem[{{Truran} \& {Iben}(1977)}]{truran77}
{Truran}, J.~W. \& {Iben}, Jr., I. 1977, ApJ, 216, 797

\bibitem[{{Ulrich}(1972)}]{ulrich72}
{Ulrich}, R.~K. 1972, ApJ, 172, 165

\bibitem[{{Uttenthaler}(2013)}]{uttenthaler13}
{Uttenthaler}, S. 2013, A\&A, 556, A38

\bibitem[{{Valenti} {et~al.}(2011){Valenti}, {Origlia}, \& {Rich}}]{valenti11}
{Valenti}, E., {Origlia}, L., \& {Rich}, R.~M. 2011, MNRAS, 414, 2690

\bibitem[{{Valle} {et~al.}(2013){Valle}, {Dell'Omodarme}, {Prada Moroni}, \&
  {Degl'Innocenti}}]{valle13}
{Valle}, G., {Dell'Omodarme}, M., {Prada Moroni}, P.~G., \& {Degl'Innocenti},
  S. 2013, A\&A, 549, A50

\bibitem[{{van Aarle} {et~al.}(2013){van Aarle}, {Van Winckel}, {De Smedt},
  {Kamath}, \& {Wood}}]{vanaarle13}
{van Aarle}, E., {Van Winckel}, H., {De Smedt}, K., {Kamath}, D., \& {Wood},
  P.~R. 2013, A\&A, 554, A106

\bibitem[{{van den Hoek} \& {Groenewegen}(1997)}]{vandenhoek97}
{van den Hoek}, L.~B. \& {Groenewegen}, M.~A.~T. 1997, A\&AS, 123, 305

\bibitem[{{Van Eck} {et~al.}(2001){Van Eck}, {Goriely}, {Jorissen}, \&
  {Plez}}]{vaneck01}
{Van Eck}, S., {Goriely}, S., {Jorissen}, A., \& {Plez}, B. 2001, Nature, 412,
  793

\bibitem[{{Van Eck} {et~al.}(2003){Van Eck}, {Goriely}, {Jorissen}, \&
  {Plez}}]{vaneck03}
---. 2003, A\&A, 404, 291

\bibitem[{{Van Eck} \& {Jorissen}(1999)}]{vaneck99}
{Van Eck}, S. \& {Jorissen}, A. 1999, A\&A, 345, 127

\bibitem[{{van Loon} {et~al.}(2005){van Loon}, {Cioni}, {Zijlstra}, \&
  {Loup}}]{vanloon05}
{van Loon}, J.~T., {Cioni}, M.-R.~L., {Zijlstra}, A.~A., \& {Loup}, C. 2005,
  A\&A, 438, 273

\bibitem[{{van Loon} {et~al.}(1999{\natexlab{a}}){van Loon}, {Groenewegen}, {de
  Koter}, {Trams}, {Waters}, {Zijlstra}, {Whitelock}, \& {Loup}}]{vanloon99b}
{van Loon}, J.~T., {Groenewegen}, M.~A.~T., {de Koter}, A., {Trams}, N.~R.,
  {Waters}, L.~B.~F.~M., {Zijlstra}, A.~A., {Whitelock}, P.~A., \& {Loup}, C.
  1999{\natexlab{a}}, A\&A, 351, 559

\bibitem[{{van Loon} {et~al.}(1999{\natexlab{b}}){van Loon}, {Zijlstra}, \&
  {Groenewegen}}]{vanloon99a}
{van Loon}, J.~T., {Zijlstra}, A.~A., \& {Groenewegen}, M.~A.~T.
  1999{\natexlab{b}}, A\&A, 346, 805

\bibitem[{{van Raai} {et~al.}(2012){van Raai}, {Lugaro}, {Karakas},
  {Garcia-Hernandez}, \& {Yong}}]{vanraai12}
{van Raai}, M.~A., {Lugaro}, M., {Karakas}, A.~I., {Garcia-Hernandez}, D.~A.,
  \& {Yong}, D. 2012, A\&A, accepted

\bibitem[{{Van Winckel} \& {Reyniers}(2000)}]{vanwinkel00}
{Van Winckel}, H. \& {Reyniers}, M. 2000, A\&A, 354, 135

\bibitem[{{Vanture} {et~al.}(2007){Vanture}, {Smith}, {Lutz}, {Wallerstein},
  {Lambert}, \& {Gonzalez}}]{vanture07}
{Vanture}, A.~D., {Smith}, V.~V., {Lutz}, J., {Wallerstein}, G., {Lambert}, D.,
  \& {Gonzalez}, G. 2007, PASP, 119, 147

\bibitem[{{Vanture} {et~al.}(1991){Vanture}, {Wallerstein}, {Brown}, \&
  {Bazan}}]{vanture91}
{Vanture}, A.~D., {Wallerstein}, G., {Brown}, J.~A., \& {Bazan}, G. 1991, ApJ,
  381, 278

\bibitem[{{Vassiliadis} {et~al.}(1996){Vassiliadis}, {Dopita}, {Bohlin},
  {Harrington}, {Ford}, {Meatheringham}, {Wood}, {Stecher}, \& {Maran}}]{vw96}
{Vassiliadis}, E., {Dopita}, M.~A., {Bohlin}, R.~C., {Harrington}, J.~P.,
  {Ford}, H.~C., {Meatheringham}, S.~J., {Wood}, P.~R., {Stecher}, T.~P., \&
  {Maran}, S.~P. 1996, ApJS, 105, 375

\bibitem[{{Vassiliadis} \& {Wood}(1993)}]{vw93}
{Vassiliadis}, E. \& {Wood}, P.~R. 1993, ApJ, 413, 641

\bibitem[{{Ventura} {et~al.}(2011){Ventura}, {Carini}, \&
  {D'Antona}}]{ventura11b}
{Ventura}, P., {Carini}, R., \& {D'Antona}, F. 2011, MNRAS, 415, 3865

\bibitem[{{Ventura} {et~al.}(2014){Ventura}, {Criscienzo}, {D'Antona},
  {Vesperini}, {Tailo}, {Dell'Agli}, \& {D'Ercole}}]{ventura14}
{Ventura}, P., {Criscienzo}, M.~D., {D'Antona}, F., {Vesperini}, E., {Tailo},
  M., {Dell'Agli}, F., \& {D'Ercole}, A. 2014, MNRAS, 437, 3274

\bibitem[{{Ventura} \& {D'Antona}(2005{\natexlab{a}})}]{ventura05a}
{Ventura}, P. \& {D'Antona}, F. 2005{\natexlab{a}}, A\&A, 431, 279

\bibitem[{{Ventura} \& {D'Antona}(2005{\natexlab{b}})}]{ventura05b}
---. 2005{\natexlab{b}}, A\&A, 439, 1075

\bibitem[{{Ventura} \& {D'Antona}(2008)}]{ventura08a}
---. 2008, A\&A, 479, 805

\bibitem[{{Ventura} \& {D'Antona}(2009)}]{ventura09a}
---. 2009, A\&A, 499, 835

\bibitem[{{Ventura} \& {D'Antona}(2010)}]{ventura10b}
---. 2010, MNRAS, 402, L72

\bibitem[{{Ventura} \& {D'Antona}(2011)}]{ventura11a}
---. 2011, MNRAS, 410, 2760

\bibitem[{{Ventura} {et~al.}(2012){Ventura}, {D'Antona}, {Di Criscienzo},
  {Carini}, {D'Ercole}, \& {vesperini}}]{ventura12}
{Ventura}, P., {D'Antona}, F., {Di Criscienzo}, M., {Carini}, R., {D'Ercole},
  A., \& {vesperini}, E. 2012, ApJL, 761, L30

\bibitem[{{Ventura} {et~al.}(2000){Ventura}, {D'Antona}, \&
  {Mazzitelli}}]{ventura00}
{Ventura}, P., {D'Antona}, F., \& {Mazzitelli}, I. 2000, A\&A, 363, 605

\bibitem[{{Ventura} {et~al.}(2002){Ventura}, {D'Antona}, \&
  {Mazzitelli}}]{ventura02}
---. 2002, A\&A, 393, 215

\bibitem[{{Ventura} {et~al.}(2001){Ventura}, {D'Antona}, {Mazzitelli}, \&
  {Gratton}}]{ventura01}
{Ventura}, P., {D'Antona}, F., {Mazzitelli}, I., \& {Gratton}, R. 2001, ApJL,
  550, L65

\bibitem[{{Ventura} {et~al.}(2013){Ventura}, {Di Criscienzo}, {Carini}, \&
  {D'Antona}}]{ventura13}
{Ventura}, P., {Di Criscienzo}, M., {Carini}, R., \& {D'Antona}, F. 2013,
  MNRAS, 431, 3642

\bibitem[{{Ventura} \& {Marigo}(2009)}]{ventura09b}
{Ventura}, P. \& {Marigo}, P. 2009, MNRAS, 399, L54

\bibitem[{{Ventura} \& {Marigo}(2010)}]{ventura10a}
---. 2010, MNRAS, 408, 2476

\bibitem[{{Ventura} {et~al.}(1998){Ventura}, {Zeppieri}, {Mazzitelli}, \&
  {D'Antona}}]{ventura98}
{Ventura}, P., {Zeppieri}, A., {Mazzitelli}, I., \& {D'Antona}, F. 1998, A\&A,
  334, 953

\bibitem[{{Viallet} {et~al.}(2013){Viallet}, {Meakin}, {Arnett}, \&
  {Mocak}}]{viallet13}
{Viallet}, M., {Meakin}, C., {Arnett}, D., \& {Mocak}, M. 2013, ApJ, 769, 1

\bibitem[{{Vlemmings} {et~al.}(2013){Vlemmings}, {Maercker}, {Lindqvist},
  {Mohamed}, {Olofsson}, {Ramstedt}, {Brunner}, {Groenewegen}, {Kerschbaum}, \&
  {Wittkowski}}]{vlemmings13}
{Vlemmings}, W.~H.~T., {Maercker}, M., {Lindqvist}, M., {Mohamed}, S.,
  {Olofsson}, H., {Ramstedt}, S., {Brunner}, M., {Groenewegen}, M.~A.~T.,
  {Kerschbaum}, F., \& {Wittkowski}, M. 2013, A\&A, 556, L1

\bibitem[{{Wachlin} {et~al.}(2011){Wachlin}, {Miller Bertolami}, \&
  {Althaus}}]{wachlin11}
{Wachlin}, F.~C., {Miller Bertolami}, M.~M., \& {Althaus}, L.~G. 2011, A\&A,
  533, A139

\bibitem[{{Wachter} {et~al.}(2002){Wachter}, {Schr{\"o}der}, {Winters},
  {Arndt}, \& {Sedlmayr}}]{wachter02}
{Wachter}, A., {Schr{\"o}der}, K.-P., {Winters}, J.~M., {Arndt}, T.~U., \&
  {Sedlmayr}, E. 2002, A\&A, 384, 452

\bibitem[{{Wachter} {et~al.}(2008){Wachter}, {Winters}, {Schr{\"o}der}, \&
  {Sedlmayr}}]{wachter08}
{Wachter}, A., {Winters}, J.~M., {Schr{\"o}der}, K.-P., \& {Sedlmayr}, E. 2008,
  A\&A, 486, 497

\bibitem[{{Wagenhuber} \& {Groenewegen}(1998)}]{wagenhuber98}
{Wagenhuber}, J. \& {Groenewegen}, M.~A.~T. 1998, A\&A, 340, 183

\bibitem[{{Wallerstein} {et~al.}(1997){Wallerstein}, {Iben}, {Parker},
  {Boesgaard}, {Hale}, {Champagne}, {Barnes}, {K{\"a}ppeler}, {Smith},
  {Hoffman}, {Timmes}, {Sneden}, {Boyd}, {Meyer}, \& {Lambert}}]{wallerstein97}
{Wallerstein}, G., {Iben}, I.~J., {Parker}, P., {Boesgaard}, A.~M., {Hale},
  G.~M., {Champagne}, A.~E., {Barnes}, C.~A., {K{\"a}ppeler}, F., {Smith},
  V.~V., {Hoffman}, R.~D., {Timmes}, F.~X., {Sneden}, C., {Boyd}, R.~N.,
  {Meyer}, B.~S., \& {Lambert}, D.~L. 1997, Reviews of Modern Physics, 69, 995

\bibitem[{{Wallerstein} \& {Knapp}(1998)}]{wallerstein98}
{Wallerstein}, G. \& {Knapp}, G.~R. 1998, ARA\&A, 36, 369

\bibitem[{{Wanajo} {et~al.}(2011){Wanajo}, {Janka}, \& {M{\"u}ller}}]{wanajo11}
{Wanajo}, S., {Janka}, H.-T., \& {M{\"u}ller}, B. 2011, ApJL, 726, L15

\bibitem[{{Wanajo} {et~al.}(2009){Wanajo}, {Nomoto}, {Janka}, {Kitaura}, \&
  {M{\"u}ller}}]{wanajo09}
{Wanajo}, S., {Nomoto}, K., {Janka}, H.-T., {Kitaura}, F.~S., \& {M{\"u}ller},
  B. 2009, ApJ, 695, 208

\bibitem[{{Wannier} {et~al.}(1991){Wannier}, {Andersson}, {Olofsson}, {Ukita},
  \& {Young}}]{wannier91}
{Wannier}, P.~G., {Andersson}, B.-G., {Olofsson}, H., {Ukita}, N., \& {Young},
  K. 1991, ApJ, 380, 593

\bibitem[{{Wasserburg} {et~al.}(1995){Wasserburg}, {Boothroyd}, \&
  {Sackmann}}]{wasserburg95}
{Wasserburg}, G.~J., {Boothroyd}, A.~I., \& {Sackmann}, I.-J. 1995, ApJL, 447,
  L37

\bibitem[{{Weiss} \& {Ferguson}(2009)}]{weiss09}
{Weiss}, A. \& {Ferguson}, J.~W. 2009, A\&A, 508, 1343

\bibitem[{{Weiss} {et~al.}(2004){Weiss}, {Schlattl}, {Salaris}, \&
  {Cassisi}}]{weiss04}
{Weiss}, A., {Schlattl}, H., {Salaris}, M., \& {Cassisi}, S. 2004, A\&A, 422,
  217

\bibitem[{{Werner} \& {Herwig}(2006)}]{werner06}
{Werner}, K. \& {Herwig}, F. 2006, Publ. Astron. Soc. Pac., 118, 183

\bibitem[{{Werner} \& {Rauch}(1994)}]{werner94}
{Werner}, K. \& {Rauch}, T. 1994, A\&A, 284, L5

\bibitem[{{Werner} {et~al.}(2005){Werner}, {Rauch}, \& {Kruk}}]{werner05}
{Werner}, K., {Rauch}, T., \& {Kruk}, J.~W. 2005, A\&A, 433, 641

\bibitem[{{Werner} {et~al.}(2009){Werner}, {Rauch}, {Reiff}, \&
  {Kruk}}]{werner09}
{Werner}, K., {Rauch}, T., {Reiff}, E., \& {Kruk}, J.~W. 2009, ApSS, 320, 159

\bibitem[{{Werner} \& {Wolff}(1999)}]{werner99}
{Werner}, K. \& {Wolff}, B. 1999, A\&A, 347, L9

\bibitem[{{Wiescher} {et~al.}(2012){Wiescher}, {K{\"a}ppeler}, \&
  {Langanke}}]{wiescher12}
{Wiescher}, M., {K{\"a}ppeler}, F., \& {Langanke}, K. 2012, ARA\&A, 50, 165

\bibitem[{{Winteler} {et~al.}(2012){Winteler}, {K{\"a}ppeli}, {Perego},
  {Arcones}, {Vasset}, {Nishimura}, {Liebend{\"o}rfer}, \&
  {Thielemann}}]{winteler12}
{Winteler}, C., {K{\"a}ppeli}, R., {Perego}, A., {Arcones}, A., {Vasset}, N.,
  {Nishimura}, N., {Liebend{\"o}rfer}, M., \& {Thielemann}, F.-K. 2012, ApJL,
  750, L22

\bibitem[{{Winters} {et~al.}(2003){Winters}, {Le Bertre}, {Jeong}, {Nyman}, \&
  {Epchtein}}]{winters03}
{Winters}, J.~M., {Le Bertre}, T., {Jeong}, K.~S., {Nyman}, L.-{\AA}., \&
  {Epchtein}, N. 2003, A\&A, 409, 715

\bibitem[{{Wittkowski} {et~al.}(2011){Wittkowski}, {Boboltz}, {Ireland},
  {Karovicova}, {Ohnaka}, {Scholz}, {van Wyk}, {Whitelock}, {Wood}, \&
  {Zijlstra}}]{witt11}
{Wittkowski}, M., {Boboltz}, D.~A., {Ireland}, M., {Karovicova}, I., {Ohnaka},
  K., {Scholz}, M., {van Wyk}, F., {Whitelock}, P., {Wood}, P.~R., \&
  {Zijlstra}, A.~A. 2011, A\&A, 532, L7

\bibitem[{{Wood}(1997)}]{wood97}
{Wood}, P.~R. 1997, in IAU Symposium, Vol. 180, Planetary Nebulae, ed. H.~J.
  {Habing} \& H.~J.~G.~L.~M. {Lamers}, 297

\bibitem[{{Wood} {et~al.}(1983){Wood}, {Bessell}, \& {Fox}}]{wood83}
{Wood}, P.~R., {Bessell}, M.~S., \& {Fox}, M.~W. 1983, ApJ, 272, 99

\bibitem[{{Wood} \& {Faulkner}(1986)}]{wood86}
{Wood}, P.~R. \& {Faulkner}, D.~J. 1986, ApJ, 307, 659

\bibitem[{{Woodward} {et~al.}(2013){Woodward}, {Herwig}, \& {Lin}}]{woodward13}
{Woodward}, P.~R., {Herwig}, F., \& {Lin}, P.-H. 2013, ArXiv e-prints

\bibitem[{{Woosley} \& {Weaver}(1995)}]{woosley95}
{Woosley}, S.~E. \& {Weaver}, T.~A. 1995, ApJS, 101, 181

\bibitem[{{Yanny} {et~al.}(2009){Yanny}, {Rockosi}, {Newberg}, {Knapp},
  {Adelman-McCarthy}, {Alcorn}, {Allam}, {Allende Prieto}, \& {An}}]{yanny09}
{Yanny}, B., {Rockosi}, C., {Newberg}, H.~J., {Knapp}, G.~R.,
  {Adelman-McCarthy}, J.~K., {Alcorn}, B., {Allam}, S., {Allende Prieto}, C.,
  \& {An}, D. 2009, AJ, 137, 4377

\bibitem[{{Yee} \& {Jensen}(2010)}]{yee10}
{Yee}, J.~C. \& {Jensen}, E.~L.~N. 2010, ApJ, 711, 303

\bibitem[{{Yong} {et~al.}(2006){Yong}, {Aoki}, \& {Lambert}}]{yong06a}
{Yong}, D., {Aoki}, W., \& {Lambert}, D.~L. 2006, ApJ, 638, 1018

\bibitem[{{Yong} {et~al.}(2008){Yong}, {Grundahl}, {Johnson}, \&
  {Asplund}}]{yong08d}
{Yong}, D., {Grundahl}, F., {Johnson}, J.~A., \& {Asplund}, M. 2008, ApJ, 684,
  1159

\bibitem[{{Yong} {et~al.}(2003{\natexlab{a}}){Yong}, {Grundahl}, {Lambert},
  {Nissen}, \& {Shetrone}}]{yong03a}
{Yong}, D., {Grundahl}, F., {Lambert}, D.~L., {Nissen}, P.~E., \& {Shetrone},
  M.~D. 2003{\natexlab{a}}, A\&A, 402, 985

\bibitem[{{Yong} {et~al.}(2003{\natexlab{b}}){Yong}, {Lambert}, \&
  {Ivans}}]{yong03b}
{Yong}, D., {Lambert}, D.~L., \& {Ivans}, I.~I. 2003{\natexlab{b}}, ApJ, 599,
  1357

\bibitem[{{Yong} {et~al.}(2013){Yong}, {Mel{\'e}ndez}, {Grundahl}, {Roederer},
  {Norris}, {Milone}, {Marino}, {Coelho}, {McArthur}, {Lind}, {Collet}, \&
  {Asplund}}]{yong13}
{Yong}, D., {Mel{\'e}ndez}, J., {Grundahl}, F., {Roederer}, I.~U., {Norris},
  J.~E., {Milone}, A.~P., {Marino}, A.~F., {Coelho}, P., {McArthur}, B.~E.,
  {Lind}, K., {Collet}, R., \& {Asplund}, M. 2013, MNRAS, 434, 3542

\bibitem[{{Zahn}(1992)}]{zahn92}
{Zahn}, J.-P. 1992, A\&A, 265, 115

\bibitem[{{Zhang} \& {Liu}(2005)}]{zhang05}
{Zhang}, Y. \& {Liu}, X.-W. 2005, ApJL, 631, L61

\bibitem[{{Zijlstra} {et~al.}(2006){Zijlstra}, {Gesicki}, {Walsh},
  {P{\'e}quignot}, {van Hoof}, \& {Minniti}}]{zijlstra06}
{Zijlstra}, A.~A., {Gesicki}, K., {Walsh}, J.~R., {P{\'e}quignot}, D., {van
  Hoof}, P.~A.~M., \& {Minniti}, D. 2006, MNRAS, 369, 875

\bibitem[{{Zinner}(1998)}]{zinner98}
{Zinner}, E. 1998, Annual Review of Earth and Planetary Sciences, 26, 147

\bibitem[{{Zinner}(2008)}]{zinner08}
---. 2008, PASA, 25, 7

\bibitem[{{Zinner} {et~al.}(2006){Zinner}, {Nittler}, {Gallino}, {Karakas},
  {Lugaro}, {Straniero}, \& {Lattanzio}}]{zinner06}
{Zinner}, E., {Nittler}, L.~R., {Gallino}, R., {Karakas}, A.~I., {Lugaro}, M.,
  {Straniero}, O., \& {Lattanzio}, J.~C. 2006, ApJ, 650, 350

\end{thebibliography}

\end{document}